\documentclass[letterpaper, 11pt]{elsarticle}

\usepackage{subcaption}

\usepackage{mathtools}

\usepackage{array}
\usepackage{indentfirst}
\usepackage{enumitem}
\usepackage{tabularx} 
\usepackage{tikz}
\usepackage{amsmath,amsfonts,amssymb,bm}
\usepackage{pgfgantt,rotating}
\usepackage{float}
\usepackage{subcaption} 
\usepackage[margin=1in]{geometry}  
\usepackage{graphicx}
\usepackage{booktabs}
\usepackage{threeparttable}
\usepackage[dvipsnames]{xcolor}
\usepackage{caption}
\usepackage{siunitx}
\usepackage{sidecap}
\usepackage{wrapfig}
\usepackage{bold-extra}
\usepackage{appendix}
\usepackage{amsmath}
\usepackage{booktabs,tabularx,multirow}
\newcolumntype{Y}{>{\centering\arraybackslash}X}
\newcommand{\noin}{\noindent}
\newcommand{\beq}{\begin{equation}}
\newcommand{\eeq}{\end{equation}}
\newcommand{\bgqar}{\begin{eqnarray}}
\newcommand{\enqar}{\end{eqnarray}}
\newcommand{\bgqarn}{\begin{eqnarray*}}
\newcommand{\enqarn}{\end{eqnarray*}}
\newcommand{\bgary}{\begin{array}}
\newcommand{\enary}{\end{array}}

\graphicspath{{figures/}}

\begin{document}

\begin{frontmatter}


\title{A Unified Statistical Framework for Multicopter Propeller Damage Diagnosis Based on Functionally Pooled Models and Bayesian Quantification: Experimental Flight Test Assessment}

\author{Shinan Huang}
\author{Jingxi Zhu}
\author{Fotis Kopsaftopoulos\corref{cor1}}
\cortext[cor1]{Corresponding author. Email: kopsaf@rpi.edu}

\affiliation{organization={Intelligent Structural Systems Laboratory (ISSL), Department of Mechanical, Aerospace and Nuclear Engineering, Rensselaer Polytechnic Institute},
             city={Troy},
             state={NY},
             country={USA}}

\begin{abstract}
In this work, a stochastic time series-based framework is introduced for multicopter propeller damage diagnosis via functionally pooled autoregressive (FP-AR) models. The framework addresses the complete damage diagnosis problem---detection, identification (isolation), and quantification---through multi-signal analysis of standard inertial measurement unit (IMU) data, eliminating the need for additional sensing modalities. Building upon the stochastic global identification paradigm, the proposed approach uses functional data pooling to compactly represent system dynamics under varying operating conditions, enabling efficient parameter estimation from short data records. A unified statistical damage diagnosis framework is established that integrates (i) damage detection via hypothesis testing on prediction residuals, (ii) damage mode identification through model selection criteria, and (iii) damage magnitude estimation via a novel Bayesian quantification scheme that provides explicit posterior uncertainty quantification. The framework is experimentally validated through an extensive series of flight tests on a custom-built hexacopter under realistic outdoor conditions with complex figure-eight trajectories and ambient wind disturbances. Six IMU channels (three-axis acceleration and angular velocity) are analyzed across multiple motors and damage levels, demonstrating consistent cross-flight performance without ad-hoc retuning. The Bayesian damage quantification approach is compared against standard batch identification methods, showing superior uncertainty characterization for risk-aware decision making. This work constitutes the first comprehensive experimental assessment of FP model-based fault diagnosis in multicopters, establishing a data-efficient, interpretable, and statistically rigorous framework for UAV structural health monitoring that bridges the gap between model-based transparency and data-driven adaptability.
\end{abstract}


\end{frontmatter}



\section*{Important conventions and symbols}

\noin Scalars are denoted by italic letters, vectors by bold lower-case letters, and matrices by bold upper-case letters. Matrix transposition is denoted by $^{\top}$.

\noin A hat $(\hat{\cdot})$ denotes an estimate, while an overbar $(\bar{\cdot})$ denotes a reference or pooled quantity.

\noin Parentheses $(\cdot)$ denote functions of real variables, e.g., $a_i(k)$. Brackets $[\cdot]$ denote discrete-time signals, e.g., $y[t]$, $t=1,2,\ldots$.

\noin The discrete-time index is $t=1,2,\ldots$, with physical time given by $(t-1)T_s$, where $T_s$ is the sampling period.

\noin The measured response is denoted by $y[t]$, the AR order by $n_a$, the regressor by $\varphi[t]$, and the design matrix by $\boldsymbol{\Phi}$.

\noin The damage condition is characterized by the continuous damage level $k$ and the discrete damage mode $m$, where $m$ designates the index of the damaged motor. The normalized damage level is denoted by $\tilde{k}$.

\noin The FP-AR model parameters are denoted by $a_i(k)$ and are expanded in basis functions $G_j(\tilde{k})$, with the projection coefficients $a_{ij}$ collected in $\boldsymbol{\theta}$.

\noin The innovation (prediction error) is denoted by $e[t]$, and $\hat e[t]$ denotes the estimated residual. The theoretical innovations variance is denoted by $\sigma^2(\cdot)$, while the residual variance, following model estimation, is denoted by $\sigma_e^2(\cdot)$, with arguments indicating conditioning variables.

\noin The residual sum of squares, Bayesian information criterion, and Ljung--Box statistic are denoted by $\mathrm{RSS}$, $\mathrm{BIC}$, and $Q$, respectively.

\noin Segment-wise quantities are indexed by $(s)$, and pooled estimates are denoted by $\hat{\theta}_{\mathrm{pool}}$ when applicable.


\section*{Acronyms}

\noin\begin{tabular}{lcl} 
AccX & : & Accelerometer—X axis (body-frame longitudinal acceleration)\\
AccY & : & Accelerometer—Y axis (body-frame lateral acceleration)\\
AccZ & : & Accelerometer—Z axis (body-frame vertical acceleration)\\
AR   & : & Autoregressive  \\
ARMA & : & Autoregressive moving average \\
ARMAX  & : & Autoregressive moving average with exogenous excitation  \\
ARX  & : & Autoregressive with exogenous excitation  \\
BIC  & : & Bayesian information criterion  \\
CFD  & : & Computational fluid dynamics \\
CNN & : & Convolutional Neural Network\\
DoF  & : & Degree of freedom\\
EKF & : & Extended Kalman Filter\\
ESCs & : & Electronic Speed Controllers\\
FEM  & : & Finite element model \\
FP   & : & Functionally pooled  \\
FRF  & : & Frequency response function  \\
GA   & : & Genetic algorithm  \\
GyrX & : & Gyroscope—X axis (body-frame roll rate)\\
GyrY & : & Gyroscope—Y axis (body-frame pitch rate)\\
GyrZ & : & Gyroscope—Z axis (body-frame yaw rate)\\
HALE & : & High altitude long endurance \\
iid  & : & identically independently distributed  \\
IMM & : & Interacting Multiple Model\\
IMU & : & Inertial Measurement Unit\\
LSTM & : & Long Short-Term Memory\\
MA   & : & Moving average \\
MEMS & : & Micro-electro-mechanical systems \\
NLS  & : & Nonlinear least squares  \\
OLS  & : & Ordinary least squares  \\
PSD  & : & Power Spectral Density \\
PWM & : & Pulse-Width Modulation\\
RLS & : & Recursive Least Squares\\
RSS  & : & Residual sum of squares  \\
SHM  & : & Structural health monitoring  \\
SPP  & : & Samples per parameter \\
SSS  & : & Signal sum of squares  \\
UAV  & : & Unmanned aerial vehicle \\
UKF & : & Unscented Kalman Filter\\
VAE & : & Variational Autoencoder\\
VFP  & : & Vector-dependent functionally pooled  \\
WLS  & : & Weighted least squares  \\
X    & : & Exogenous\\

\end{tabular} 


\tableofcontents 

\section{Introduction} \label{sec:intro}

Small unmanned aerial vehicles (UAVs), particularly multirotors, are rapidly expanding in delivery, inspection, agriculture, public safety, and mapping tasks~\cite{Watts2012UASRemoteSensing, Nex2014UAV3DMapping}. As missions become longer, denser, and more diverse, UAV fleets face increasing reliability and in-flight safety risks arising from component wear, actuator faults, and operating disturbances~\cite{Shraim,Wild}. At the scale of autonomous operations, conventional scheduled maintenance alone is insufficient, since faults can emerge between service intervals and propagate quickly during flight. This requires continuous onboard health monitoring and diagnostics capable of detecting early faults, damage, and degradation, and supporting safe autonomy~\cite{Puchalskidrones6110330}.

UAV fault detection and isolation (FDI) methods are broadly organized into model-based and data-driven families. Model-based approaches derive analytical representations from physics and control models to construct residuals or track parameter changes, including parity and structured-residual schemes~\cite{gertler2017fault,chen2012robust,isermann2005fault}, observers and filters~\cite{Luenberger1971, BlankeEtAl2016,wang2019disturbance}, multiple-model and interacting multiple model (IMM) estimation~\cite{blom2002interacting,BarShalomLi2001}, online parameter estimation via recursive least squares (RLS), extended Kalman filter (EKF), or unscented Kalman filter (UKF)~\cite{JulierUhlmann1997, Simon2006}. These methods offer interpretability, formal isolability analyses, and good data efficiency. However, their performance depends on accurate multicopter and aerodynamic models and careful tuning, and may degrade under unmodeled dynamics, payload variations, and changing flight regimes. 

In contrast, data-driven approaches infer faults directly from the measurements. Representative approaches include: (i) feature-based statistical learning, where descriptors extracted from vibration data (time--frequency features, autoregressive model parameters, and nonlinear complexity metrics such as permutation entropy) are paired with classifiers or probabilistic regressors~\cite{rao2025real,dutta2022multicopter,dutta2020statistical,Yan12,Hios2014,Sakel2016,baldini2023real}; (ii) probabilistic modeling and change detection using Bayesian regression, Gaussian processes, and change-point tests on residuals or health indicators~\cite{Venkatasubramanian2003FDIReview,Basseville1993Abrupt}; (iii) deep representation learning with convolutional neural networks (CNNs), long short-term memory (LSTM) networks, autoencoders, and transformer architectures, often combined with domain adaptation or self-supervision to capture nonlinear spatiotemporal structure and multi-sensor context~\cite{machines9090197,Chalapathy2019DeepAD,zhu2023review, dutta2022time}; and (iv) hybrid physics-aware models that embed gray-box constraints to improve interpretability and robustness to distribution shift~\cite{Puchalskidrones6110330}. These advances broaden the toolbox and improve accuracy across diverse sensing configurations. Nevertheless, such gains raise two persistent questions that cut across techniques: how to maintain data efficiency and interpretability as models grow more complex, and how to validate performance reliably beyond simplified simulation and experimental settings.

Within the family of data-driven FDI methods for UAVs, recent work has improved data efficiency and model interpretability through several representative approaches. Semi-supervised detectors with dynamic thresholding leverage large unlabeled datasets to learn decision rules and reduce false alarms, thereby decreasing labeled-data requirements~\cite{Memarzadeh,bell2022anomaly}. Signal front-end enhancement techniques, such as adaptive dual-tree complex wavelet denoising, reduce noise and preserve short fault signatures, increasing the usable information per sample and the feature separability~\cite{he2025computationally}. Physical metastructure racks that reshape vibration transmission paths enable single sensors to extract more discriminative features, thereby increasing the information density at the source~\cite{zhang2025metastructure}. Multi-task knowledge sharing amortizes supervision across missions and channels by reusing learned representations, improving accuracy under scarce labeling conditions on fixed-wing platforms~\cite{zhang2024fw}. Together, these strategies enable higher robustness across flights while using fewer labels and sensors.

Beyond data efficiency and model interpretability, the thorough assessment of the methods via realistic experimental testing and their validation breadth remain crucial concerns. These limitations are not specific to data-driven methods; model-based and hybrid schemes are often assessed under simulation or simplified flight regimes (e.g., hovering), leaving the generalization to long missions and complex maneuvers insufficiently characterized. A substantial portion of recent UAV FDI studies still rely on simulation or semi-physical testbeds, which facilitate repeatability but leave open questions about generalization to realistic flight scenarios and extended, cross-regime missions.

In practice, the vast majority of UAV FDI method evaluations report real-flight tests primarily under controlled settings, such as hovering or short missions with conventional controllers, as in~\cite{baldini2023real,10912747,Memarzadeh,wang2019disturbance}. These approaches demonstrate real-time feasibility, but focus on simplified operating conditions. For broader benchmarking, RFlyMAD and ALFA are open-source benchmark UAV fault datasets that provide labeled multi-sensor flight logs for evaluating fault detection and anomaly detection methods on multicopter and fixed-wing platforms, respectively. RFlyMAD combines simulation-in-the-loop (SIL) and hardware-in-the-loop (HIL) with real flights, although the majority of data originate from SIL/HIL runs~\cite{le2025rflymad}, while ALFA embeds faults and anomalies directly into flight logs to allow systematic and repeatable comparisons~\cite{keipour2021alfa}.

Motivated by the above gaps in data efficiency and the need for parsimonious but interpretable models, previous work by the authors and co-workers has focused on statistical time-series models for multicopter rotor fault detection and identification~\cite{dutta2020statistical,dutta2022multicopter,dutta2022rotor}. These models offer partial representations of the flight dynamics that are identified from a small number of vibration signals rather than relying on a full analytical model of the multicopter. In an initial, preliminary study~\cite{dutta2020statistical}, scalar autoregressive (AR) models and multivariate vector autoregressive (VAR) models were identified from simulated flight signals (roll--pitch--yaw response in simulated 5 m/s forward flight) under healthy and faulty conditions, and were shown to achieve high accuracy in fault detection and identification via residual-based statistical decision-making schemes. That was the first study in which statistical time series methods were introduced and evaluated for multicopter FDI, although using simulated data sets. A subsequent simulation-based study extended this framework by adopting vector autoregressive models with exogenous excitation (VARX), where the control signals are explicitly used as inputs. The hexacopter was still commanded to fly forward at 5 m/s, but Dryden-type atmospheric turbulence was introduced to emulate realistic wind disturbances. Under these conditions, the resulting VARX residuals improved the separability of the rotor faults and enhanced the robustness across all levels of turbulence~\cite{dutta2022multicopter}. A preliminary experimental assessment of the developed AR- and VAR-model-based methods and corresponding statistical FDI schemes was presented in a recent study by the team~\cite{dutta2022rotor}. This work addressed statistical FDI on quadcopter and hexacopter platforms via a series of hovering flight tests. In this real-flight, but rather limited, setting, the residual-based statistical decision schemes based on these models maintained very good accuracy and robustness in tackling rotor-fault detection and identification, highlighting the potential and practical applicability of the developed framework~\cite{dutta2022rotor}. Overall, these studies establish residual-based AR/VAR/VARX modeling and the associated statistical decision-making schemes as a potentially effective approach for rotor fault detection and identification using only onboard inertial measurement unit (IMU) measurements.

Although AR/VAR/VARX-based frameworks provide effective FDI under constant flight conditions, the resulting models represent a single operating point (e.g., hovering or constant velocity) and a single health state (specific motor fault with fixed magnitude), lacking a compact representation across multiple states. To address this limitation, functionally pooled (FP) stochastic models enable global system representation under varying conditions through parameter functions that explicitly depend on the operating state vector~\cite{Sakellariou-Fassois07a, Fassois2013, Kopsaftopoulos-Fassois06, Kopsaftopoulos-Fassois13, Sakel2016,Kopsaftopoulos-etal-MSSP18}. Model identification is based on finite-length time series segments sampled across the admissible state space. Originally developed for structural health monitoring (SHM) applications~\cite{Kopsaftopoulos-Fassois13,Sakel2016,sakaris2016time,zhou2023bayesian,ahmed2023local}, FP models and extensions to vector-dependent operating states have also demonstrated effectiveness for aerospace structures operating under multiple flight states~\cite{Kopsaftopoulos-etal-MSSP18,james2019data}. For multicopter FDI specifically, prior simulation-based proof-of-concept studies constructed vector-dependent FP-AR (VFP-AR) models with parameters dependent on forward velocity, gross weight, turbulence intensity, and rotor fault magnitude, thus enabling detection, identification, and quantification tasks using IMU measurements~\cite{dutta2020rotor_vfs,duttaunified_vfs}. However, a comprehensive experimental assessment of FP model-based multicopter damage diagnosis under realistic flight conditions remains absent, motivating the present critical, flight-test-based assessment.

Extending this line of research, recent preliminary experimental studies by the authors have focused on propeller damage detection, moving beyond generic rotor anomalies, and have expanded validation from simple hovering to realistic flight trajectories, such as circular and figure-eight patterns~\cite{Huang2024DataDrivenPH,huangfunctional,huang2025multicopter}. Using short IMU data windows (4 s), the FP-AR representation resulted in interpretable damage-dependent parameters and demonstrated cross-flight consistency (i.e., generalization across flights and varying wind conditions) suitable for onboard decision making. Building on these foundations, this paper presents a unified and interpretable FP-AR time series framework for UAV propeller damage diagnosis. The framework extends prior FP-AR schemes by integrating cross-segment model parameter pooling and uncertainty estimation, while proposing a novel Bayesian damage quantification approach to improve robustness and provide explicit posterior uncertainty characterization. Furthermore, this study provides a comprehensive experimental assessment via a series of multicopter flight tests that took place over several different days, thus accounting for environmental (temperature, wind, gusts) and operational (different damaged propellers) uncertainties. The principal contributions and novel aspects of this work are as follows:

\begin{enumerate}
    \item \textit{First comprehensive experimental assessment of FP model-based diagnosis in multicopters.} This work constitutes the first thorough experimental evaluation of FP models for multicopter damage diagnosis, assessed through an extensive series of real-world flight tests under realistic outdoor conditions with ambient wind disturbances and complex trajectories. Models are trained and validated using repeated flight tests conducted across different days and varying ambient conditions. The identified pooled parameters demonstrate stability without ad-hoc retuning, indicating robustness to environmental drift, wind disturbances, and operating-point variations. The framework is assessed using long-duration ($\sim$250 s) missions following a 101-waypoint figure-eight trajectory, moving substantially beyond simple hovering to test performance under realistic flight conditions.

    \item \textit{Standard IMU-based approach without additional sensors.} The framework relies exclusively on standard IMU signals available in all modern flight controllers, eliminating the need for additional sensing modalities or specialized instrumentation, thereby enhancing practical applicability. A multi-signal comparative analysis is presented for optimal sensor selection. The framework is applied to all six IMU channels (three-axis linear acceleration and angular velocity: AccX/Y/Z and GyrX/Y/Z), enabling a systematic comparison to identify which signals yield optimal performance for detection, identification, and quantification tasks.

    \item \textit{Cross-segment parameter pooling for mission-level models.} A pooled FP-AR identification strategy is introduced that jointly estimates the projection coefficients from multiple data segments within a mission, yielding a single mission-level model that captures the shared dynamics, reduces the parameter variance, and minimizes the sensitivity to the segment selection.

    \item \textit{Unified statistical damage diagnosis framework.} A complete damage diagnosis framework is established that integrates damage detection via hypothesis testing on the model-based residuals, damage mode identification (isolation) through model validation criteria, and damage magnitude estimation via inverse and Bayesian quantification schemes, providing a statistically rigorous methodology.

    \item \textit{Novel Bayesian damage quantification with posterior uncertainty estimation.} A Bayesian approach is introduced for damage magnitude estimation that provides posterior damage size distributions based on properly defined priors, without invoking the asymptotic normality assumptions underlying conventional estimator-based confidence intervals. The method is systematically compared against the standard quantification technique, demonstrating superior uncertainty characterization.
    
\end{enumerate}

\section{Data generation} \label{sec:Data-generation}

\subsection{The hexacopter}
\begin{figure}[t] 
\centering
    \hspace{-0.6cm}\includegraphics[width=0.35\textwidth]{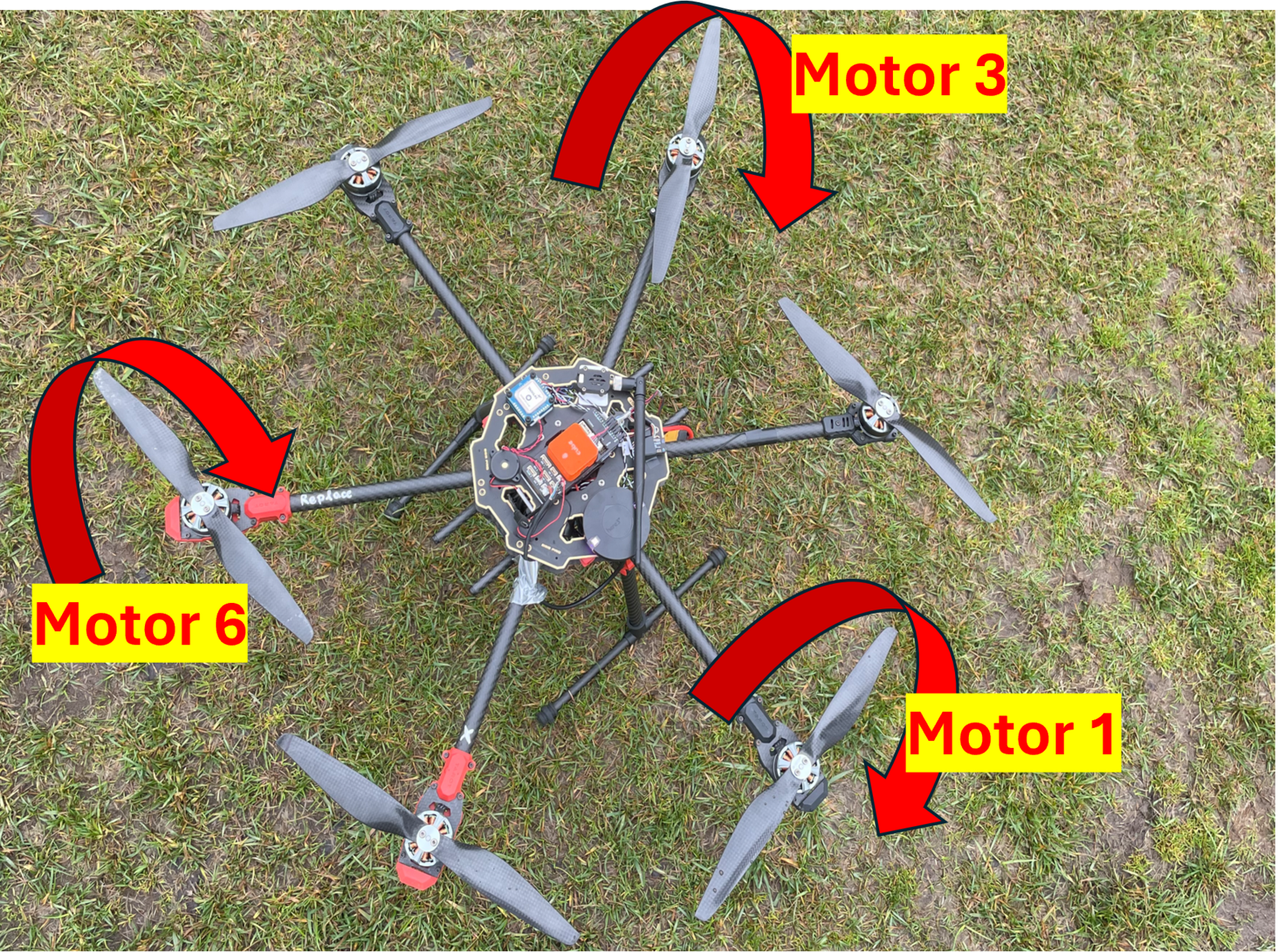}
    \hspace{-0.6cm}\includegraphics[width=0.35\textwidth]{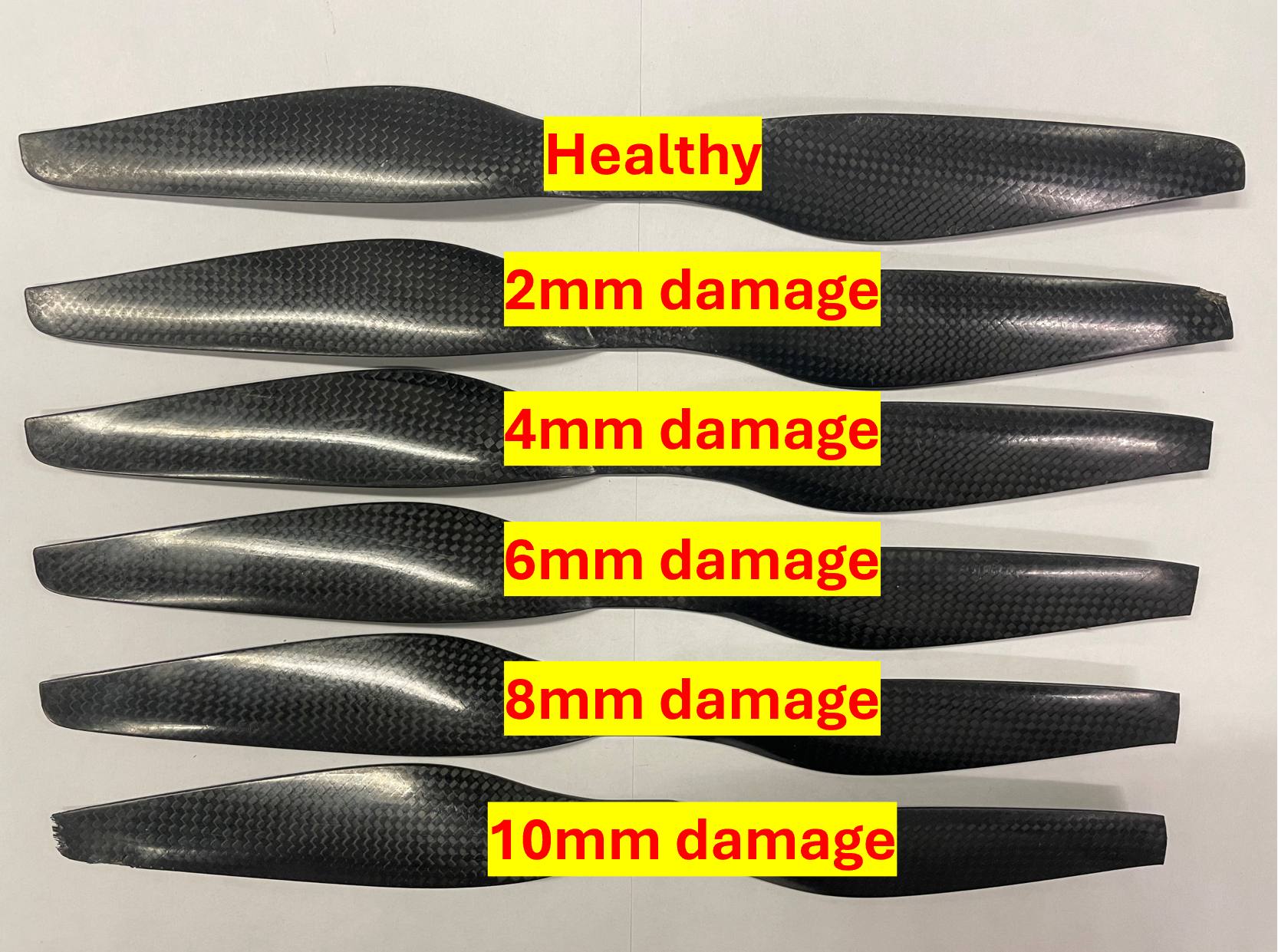}\includegraphics[width=0.35\textwidth]{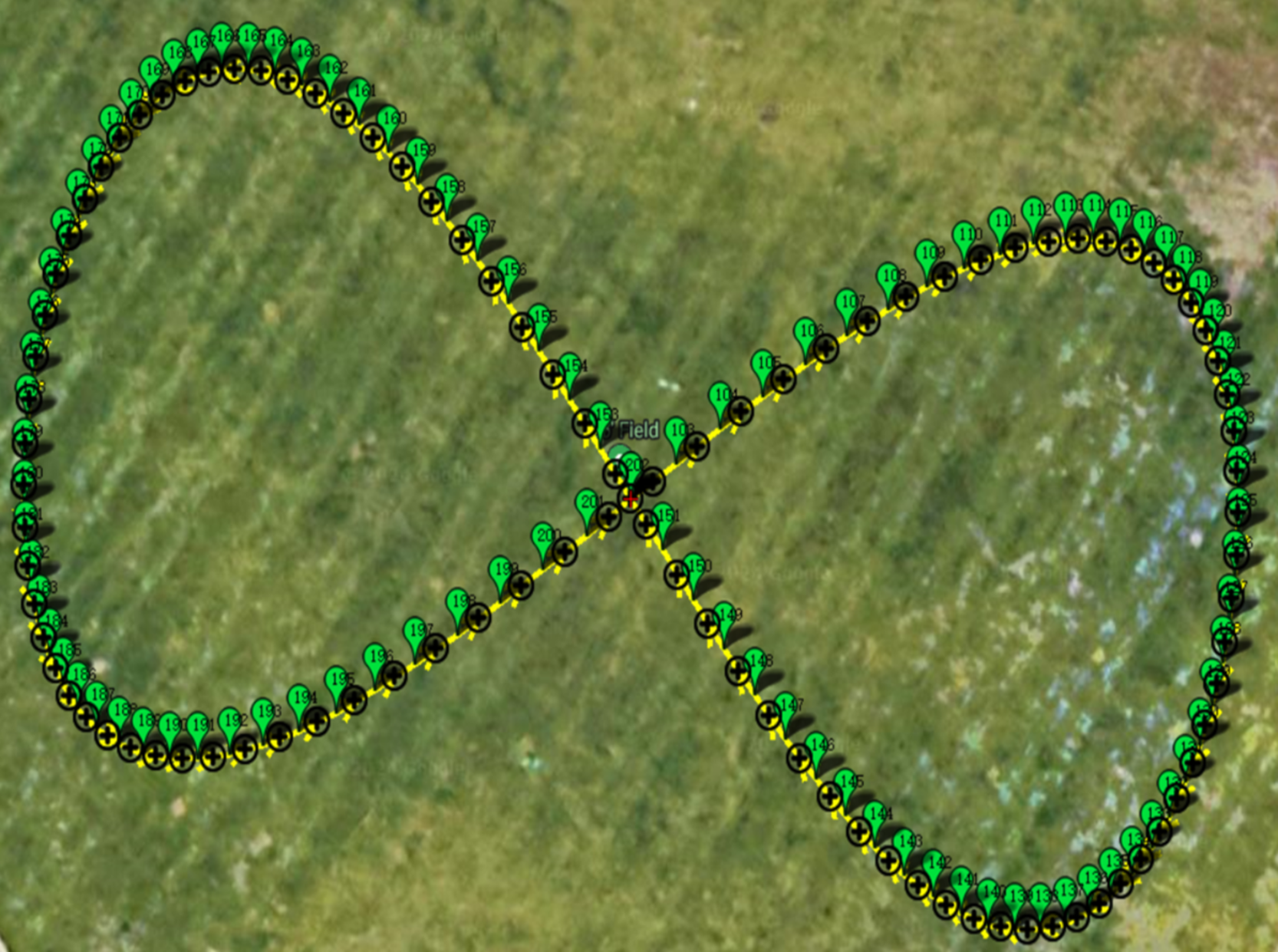}
  \caption{(Left) The custom-built hexacopter used for the flight tests. (Middle) Healthy and damaged propellers, with tips clipped by 2 to 10~mm in 2~mm increments. (Right) The programmed waypoints approximating a figure-eight flight path.}
  \label{fig:Drone}%
\end{figure}

A custom-built hexacopter, shown in Fig.~\ref{fig:Drone} (left), was used for the flight test campaign; its main physical and operational specifications are summarized in Table~\ref{tab:exp-params}. The platform is equipped with six T-Motor MN3110 (470~KV) motors driving 300~mm carbon-fiber propellers, and is powered by a 20{,}000~mAh battery, achieving stable hover at approximately 40\% throttle. Flight control is provided by a Pixhawk Cube Orange autopilot running the ArduPilot firmware. The autopilot integrates three 6-DoF IMUs for redundant inertial sensing, while its onboard logger records all essential flight signals, including the radio control (RC) commands and the electronic speed controller (ESC) telemetry.

\begin{table}[b]
  \centering
  \begin{threeparttable}
    \caption{Experimental parameters of the hexacopter flight tests}
    \label{tab:exp-params}
    \small
    \begin{tabular*}{0.7\textwidth}{@{\extracolsep{\fill}} l l @{}}
      \toprule
      Experimental parameters        & Value \\
      \midrule
      Hexacopter weight  & 8.65 kg \\
      Frame diameter     & 695 mm \\
      Propeller diameter & 300 mm \\
      Flight controller  & Pixhawk Cube Orange with ArduPilot \\
      Flight speed       & 2 m/s \\
      Flight altitude    & 9 m \\
      Sampling frequency & 1 kHz \\
      Damage levels      & 0, 2, 4, 6, 8, and 10 mm \\
      Tested motors      & Motor 1, Motor 3, and Motor 6 \\
      Flights per motor--damage case & 1 flight ($\approx 250$ s)\\
      Total number of flights & 18 \\
      Total recording duration       & $\approx 4{,}500$ s \\
      \bottomrule
    \end{tabular*}
  \end{threeparttable}
\end{table}

\subsection{Damage topology and experiment scenarios}

In order to emulate various levels of propeller tip damage, a series of modified propellers was prepared, with tip trimming ranging from 0 mm (healthy) to 10 mm in 2 mm increments, corresponding to relative damage levels between 0\% and 3.3\% of the propeller diameter, as illustrated in Fig.~\ref{fig:Drone} (middle). The considered damage states and damage modes are summarized in Table~\ref{tab:damage-conditions}. In order to assess motor-specific effects, the complete series of experiments was repeated on three motors sharing the same direction of rotation, namely Motors 1, 3, and 6 (Fig.~\ref{fig:Drone}).

Data acquisition was carried out while the hexacopter performed low-altitude flights under actual flight conditions (wind, gusts) in an outdoor environment. The flight path was designed as a figure-eight pattern, implemented by programming 101 waypoints into the mission planning software (Fig.~\ref{fig:Drone}, right). Upon reaching the designated takeoff location, the vehicle climbed to a target altitude of 9 m and subsequently traversed the waypoints at a constant speed of 2 m/s, completing one loop upon arrival at the final waypoint. Each test flight comprised two full loops followed by an automated landing. The complete series of flight experiments was conducted over several days spanning one week, under naturally varying ambient conditions (temperature, wind, and gusts), thus introducing realistic day-to-day environmental and operational variability into the collected data set.

Throughout the data collection process, the aircraft operated in fully autonomous mode. In this mode, the onboard flight controller processed the measurements of its internal IMUs and barometer through an EKF in order to estimate the key flight states, such as attitude (pitch/roll/yaw), relative altitude and velocity. Based on these estimated states, the control loops computed the appropriate pulse-width modulation (PWM) commands for the ESCs.





\begin{table}[b]
  \centering
  \begin{threeparttable}
    \caption{Summary of damage conditions (states and modes)}
    \label{tab:damage-conditions}
    \small
    \begin{tabular}{@{} l l l l @{}} 
      \toprule
      \textbf{Type} & \textbf{Label} & \textbf{Applies to} & \textbf{Description} \\
      \midrule

      \multirow{6}{*}{\textbf{Damage states}}
        & Healthy   & Motors 1, 3, 6 & No trimming (baseline condition) \\
        & 2\,mm     & Motors 1, 3, 6 & 2\,mm trimming of propeller tip \\
        & 4\,mm     & Motors 1, 3, 6 & 4\,mm trimming of propeller tip \\
        & 6\,mm     & Motors 1, 3, 6 & 6\,mm trimming of propeller tip \\
        & 8\,mm     & Motors 1, 3, 6 & 8\,mm trimming of propeller tip \\
        & 10\,mm    & Motors 1, 3, 6 & 10\,mm trimming of propeller tip \\
      \addlinespace

      \multirow{3}{*}{\textbf{Damage modes}}
        & Motor 1   & Motor 1        & Propeller tip trimming on Motor 1 \\
        & Motor 3   & Motor 3        & Propeller tip trimming on Motor 3 \\
        & Motor 6   & Motor 6        & Propeller tip trimming on Motor 6 \\
      \bottomrule
    \end{tabular}
  \end{threeparttable}
\end{table}

\subsection{The signals}

\begin{figure}[t!]
    \centering
    \begin{subfigure}[t]{0.48\textwidth}
        \includegraphics[width=\textwidth]{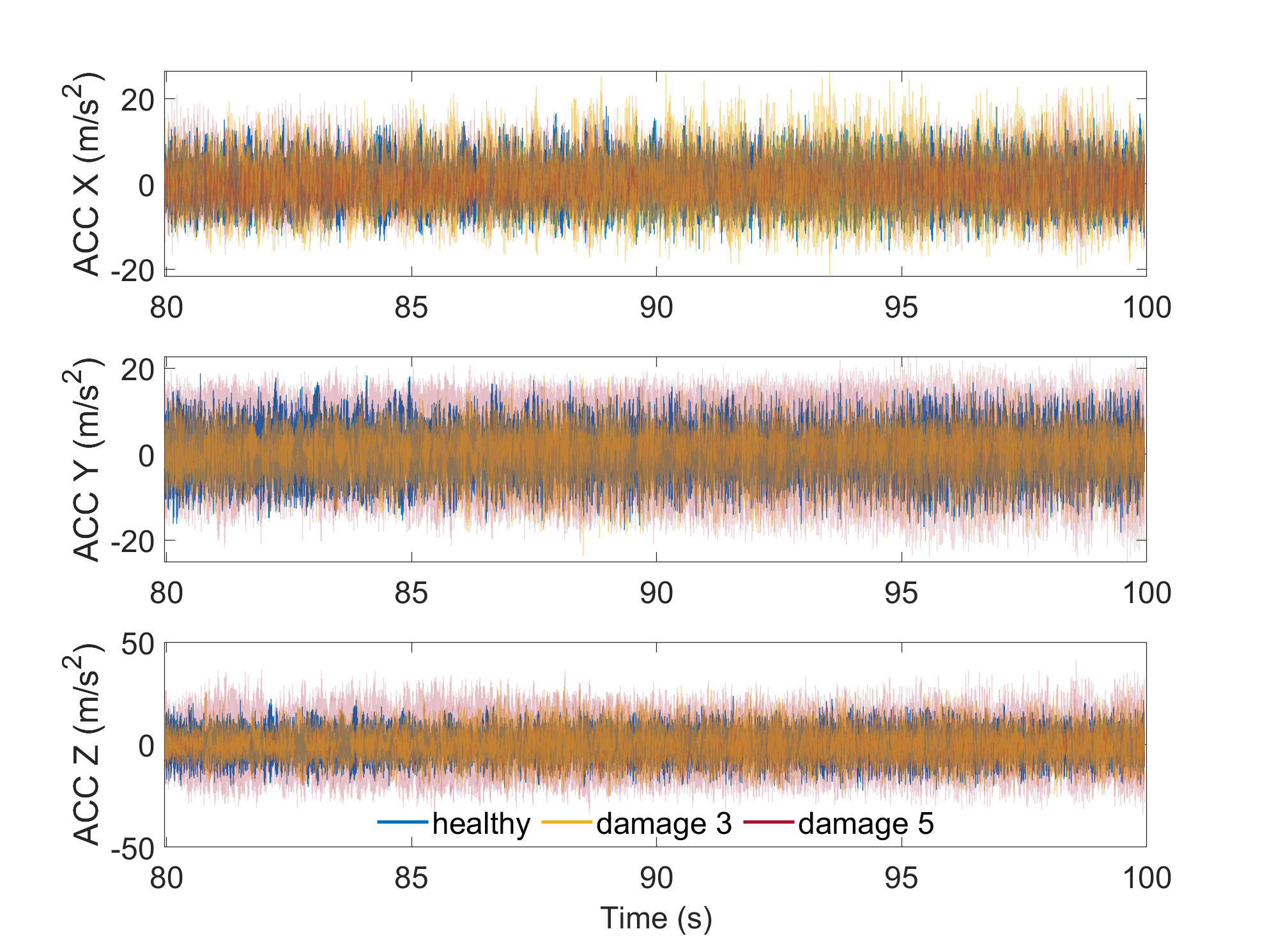}
        \caption{Linear acceleration (AccX) signals for three damage states: healthy, 6 mm (damage state 3), and 10 mm (damage state 5).}
        \label{fig:acc}
    \end{subfigure}
    \hfill
    \begin{subfigure}[t]{0.48\textwidth}
        \includegraphics[width=\textwidth]{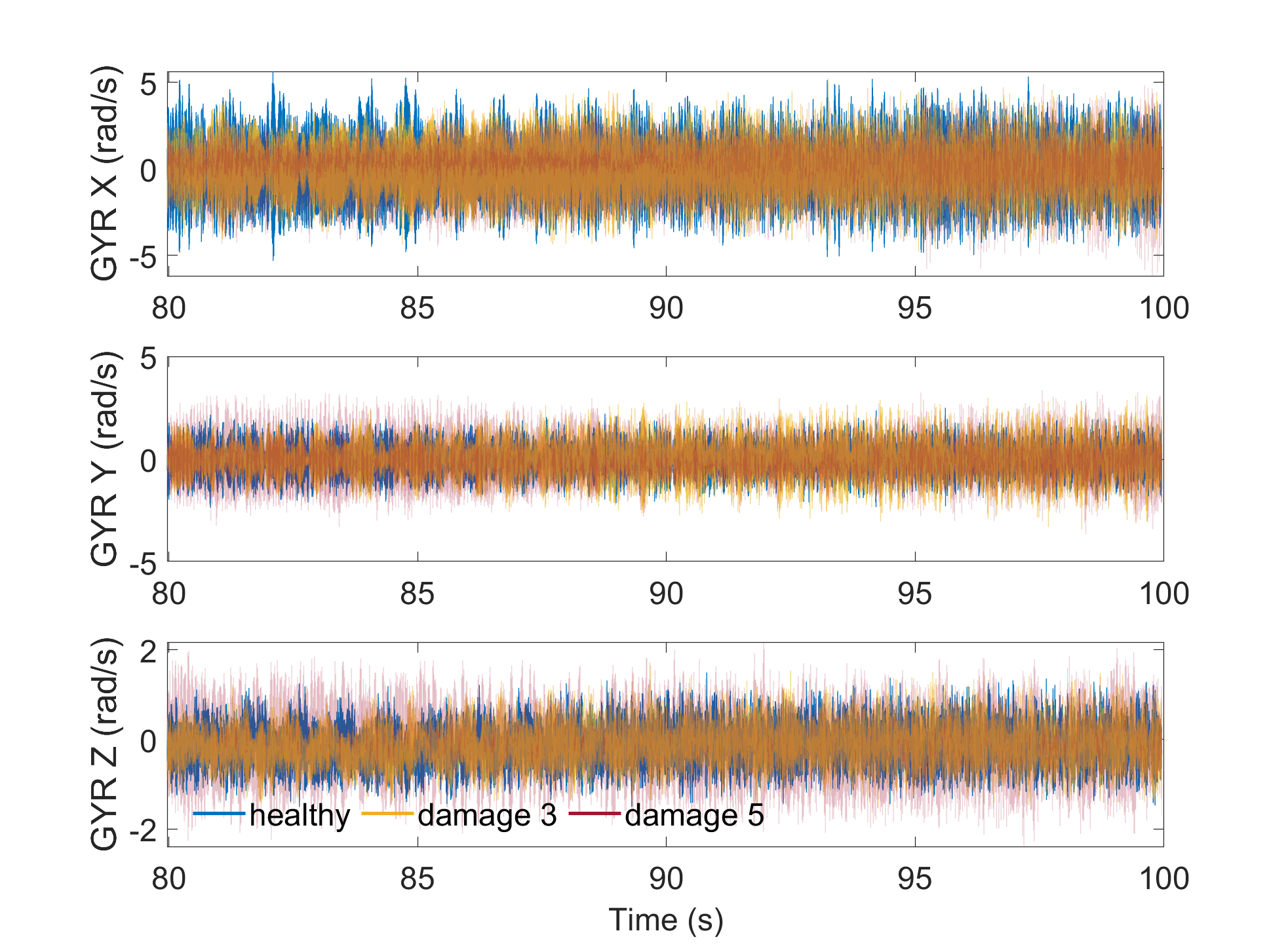}
        \caption{Angular velocity (GyrX) signals for three damage states: healthy, 6 mm (damage state 3), and 10 mm (damage state 5).}
        \label{fig:gyro}
    \end{subfigure}
    \caption{Indicative linear acceleration and angular velocity signals collected during the flight tests.}
    \label{fig:signals}
\end{figure}

With the flight altitude and speed kept constant, each of the three selected motors was tested across all defined damage levels. Six IMU signals, i.e., the three-axis linear acceleration (AccX/Y/Z) and the three-axis angular velocity (GyrX/Y/Z), were recorded at a sampling frequency of 1 kHz, while the altitude information was obtained from the onboard barometer. Fig.~\ref{fig:signals} presents indicative signal comparisons between the healthy state and two selected damage states (6 mm and 10 mm) for the AccX and GyrX signals of Motor 6 over the 80 to 100 s time window.

The resulting data set captures the dynamic behavior of the hexacopter under varying levels of propeller damage and real-world environmental disturbances. The presence of atmospheric turbulence introduces realistic variability, thus providing a challenging test environment for the assessment of the postulated FP-AR model-based framework. The recorded signals exhibit clear severity-dependent changes in amplitude and variance, furnishing the empirical basis for the model identification and statistical inference procedures developed in the sequel.

\section{Methodology} \label{sec:problem}
\label{sec:method}

\subsection{Baseline phase}

In the baseline phase, the dynamics of the hexacopter under the nominal (healthy) state are identified in order to determine the admissible time-domain model orders and initialize the subsequent functionally pooled estimation. Let $t$ designate the normalized discrete-time index ($t=1,2,\ldots$, with the corresponding physical time being $(t-1)T_s$, where $T_s$ is the sampling period) and $y[t]$ the measured response signal. A conventional AR model of order $n_a$ is employed~\cite{BoxJenkins2015,ljung1998system,hamilton2020time}:
\begin{equation}
\label{eq:healthy-ar}
y[t] + \sum_{i=1}^{n_a} a_i\,y[t-i] \;=\;  e[t],
\qquad e[t]\sim\mathcal N(0,\sigma^2),
\end{equation}
with $e[t]$ designating the one-step-ahead prediction error (innovation), assumed to be a zero-mean, white (serially uncorrelated) Gaussian sequence with variance $\sigma^2$.

The model is parameterized in terms of the parameter vector
$\bar {\boldsymbol{\theta}}=\big[ a_1,\ldots,a_{n_a};\sigma^2\big]^{\!\top}$,
which is estimated from the measured signals. Toward this end, the standard AR regressor vector
$\boldsymbol{\varphi}[t]=[-y[t-1],\ldots,-y[t-n_a]]^\top\!\in\!\mathbb{R}^{n_a}$ is defined and
stacked into the design matrix $\boldsymbol{\Phi}=[\boldsymbol{\varphi}[1]^\top;\ldots;\boldsymbol{\varphi}[N]^\top]\in\mathbb{R}^{N\times n_a}$, with $N$ designating the number of samples and
$\mathbf{y}=[y[1],\ldots,y[N]]^\top$ the corresponding response vector. The ordinary least squares (OLS) estimate of the AR parameter vector $\boldsymbol{\theta}=[a_1,\ldots,a_{n_a}]^\top$ is then
\begin{equation}
\widehat{\boldsymbol{\theta}}_{\!\text{OLS}}
=\arg\min_{\theta}\|\mathbf y-\boldsymbol{\Phi}\,\boldsymbol{\theta}\|_2^2
=(\boldsymbol{\Phi}^\top\boldsymbol{\Phi})^{-1}\boldsymbol{\Phi}^\top\mathbf y,
\label{eq:theta}
\end{equation}
with the corresponding estimated residuals
$\hat e[t]= y[t]- \boldsymbol \varphi[t]^\top\widehat{\boldsymbol{\theta}}_{\!\text{OLS}}$
and residual variance estimate
$\hat\sigma_e^{\,2}=\frac{1}{N-n_a}\sum_{t=1}^N {\hat e}[t]^2$.
When the residual sequence exhibits heteroscedasticity or serial correlation, a two-step weighted least squares (WLS)~\cite{Greene2018,carroll2017transformation} refinement is employed, whereby a consistent estimate of $\mathrm{cov}({\boldsymbol e})$ is constructed from the OLS residuals~\cite{seber2003linear,hamilton2020time} and the system
$(\boldsymbol\Phi^\top\mathrm{cov}(\boldsymbol e)^{-1} \boldsymbol\Phi)\widehat{\boldsymbol \theta}_{\!\text{WLS}}=\boldsymbol \Phi^\top \mathrm{cov}(\boldsymbol e)^{-1} \mathbf y$ is solved.

The baseline model order $n_a$ is selected by minimizing an appropriate information criterion, typically the Bayesian information criterion (BIC)~\cite{Schwarz1978}:
\begin{equation}
\mathrm{BIC}(n_a)
=
N\ln\!\left(\frac{\mathrm{RSS}(n_a)}{N}\right)
+
n_a\ln N,
\qquad
\mathrm{RSS}(n_a)
=
\sum_{t=1}^{N}\hat e[t]^2 ,
\label{eq:bic_rss}
\end{equation}
which may be supplemented by the residual sum of squares normalized by the signal sum of squares (RSS/SSS). Final model acceptance relies on residual diagnostics, namely whiteness tests, such as the Ljung--Box portmanteau test, verifying the absence of remaining serial correlation in the residual sequence. The retained baseline model order and the associated parameter estimates constitute the fixed time-domain structure for the subsequent functionally pooled identification.

\subsection{Functionally pooled AR modeling and damage mode representation}
\label{sec:fpar}

Two descriptors are used to characterize the damage condition: the damage level and the damage mode. In the present study, the damage level refers to the magnitude of the propeller tip damage, whereas the damage mode designates the index of the damaged motor. The two descriptors are treated differently within the modeling framework: the damage level is represented as a continuous variable, designated as $k$, while the damage mode is treated as a discrete index, designated as $m$. 

Functional pooling is performed along the continuous damage-level coordinate $k$, whereas each
damage mode $m$ corresponds to a separate identification problem. This allows the AR
model parameters to evolve smoothly with the damage level, while preserving the mode-specific dynamics~\cite{Kopsaftopoulos-etal-MSSP12}.
Hence, for a given damage mode $m$ and varying damage level $k$, the FP-AR model structure assumes the form:

\begin{equation}
\mathcal{M}^{(m)}\!\left(\boldsymbol{\theta},\sigma^{2}(k)\right):
\quad
y_{k}[t]
+
\sum_{i=1}^{n_a}
a_i(k)\,y_{k}[t-i]
=
e_{k}[t],\quad e_{k}[t] \sim \mathcal N\!\big(0,\sigma ^2(k)\big),
\label{eq:single_motor_fpar}
\end{equation}
with $y_k[t]$ designating the response signal, $e_k[t]$ the model innovations, and $\sigma^2(k)$ the corresponding, potentially level-dependent, innovations variance at damage level $k$, while $\boldsymbol{\theta}$ designates the model parameter vector defined in the sequel. The dependence of the model parameters on $k$ is captured through a low-dimensional functional subspace, with each AR model parameter expanded on a set of $r$ basis functions~\cite{RamsaySilverman2005,de1978practical,MagnusNeudecker2019}:
\begin{equation}\label{eq:coef-expansion}
a_i(k)=\sum_{j=1}^{r} a_{ij}\,G_j({k}),
\qquad i=1,\ldots,n_a,
\end{equation}
so that the identification problem reduces to the estimation of the projection coefficients $a_{ij}$, collected in the vector
$\boldsymbol{\theta}=[a_{11}\ \ldots\ a_{1r}\ \; \ldots \; \ a_{n_a 1}\ \ldots\ a_{n_a r}]^{\top}\in\mathbb{R}^{n_a r}$. 

Defining the regressor vector
$\boldsymbol \varphi_k[t]=[-y_k[t-1],\ldots,-y_k[t-n_a]]^\top$ and the functional feature vector
$\boldsymbol g({k})=[G_1({k}),\ldots,G_r({k})]^\top$,
Eqs.~\eqref{eq:single_motor_fpar}--\eqref{eq:coef-expansion} yield the linear regression form:
\begin{equation}\label{eq:sample}
y_k[t]=\big( \boldsymbol \varphi_k[t]\otimes \boldsymbol g({k}) \big)^{\!\top}\boldsymbol \theta + e_k[t],
\end{equation} 
where $\otimes$ designates the Kronecker product.
Pooling together the regression equations of the form of Eq.~\eqref{eq:sample} corresponding to all sampled damage levels $k\in\{k^{(1)},k^{(2)},\ldots,k^{(M)}\}$, with $M$ designating the number of sampled damage levels (cross-sections), yields, at each time instant $t$:

\begin{equation}
\begin{bmatrix}
y_{k^{(1)}}[t] \\
\vdots \\
y_{k^{(M)}}[t]
\end{bmatrix}
=
\begin{bmatrix}
\big(\boldsymbol{\varphi}_{k^{(1)}}[t]\otimes \boldsymbol g(k^{(1)})\big)^{\!\top} \\
\vdots \\
\big(\boldsymbol{\varphi}_{k^{(M)}}[t]\otimes \boldsymbol g(k^{(M)})\big)^{\!\top}
\end{bmatrix}
\boldsymbol{\theta}
+
\begin{bmatrix}
e_{k^{(1)}}[t] \\
\vdots \\
e_{k^{(M)}}[t]
\end{bmatrix},
\qquad
\mathbf{y}[t]
=
\boldsymbol{\Phi}[t]\,\boldsymbol{\theta}
+
\mathbf{e}[t].
\label{eq:stacked}
\end{equation}

Substituting the measured data for $t=1,\ldots,N$ into Eq.~\eqref{eq:stacked} and stacking the resulting equations over time yields the final regression equation, which no longer depends on the time index:
\begin{equation}
\mathbf{y}
=
\boldsymbol{\Phi}\,\boldsymbol{\theta}
+
\mathbf{e},
\qquad
\mathbf{y}=\begin{bmatrix}\mathbf{y}[1]\\ \vdots\\ \mathbf{y}[N]\end{bmatrix},
\quad
\boldsymbol{\Phi}=\begin{bmatrix}\boldsymbol{\Phi}[1]\\ \vdots\\ \boldsymbol{\Phi}[N]\end{bmatrix},
\quad
\mathbf{e}=\begin{bmatrix}\mathbf{e}[1]\\ \vdots\\ \mathbf{e}[N]\end{bmatrix},
\label{eq:stacked_full}
\end{equation}
with $\mathbf{y},\mathbf{e}\in\mathbb{R}^{NM}$ and $\boldsymbol{\Phi}\in\mathbb{R}^{NM\times n_a r}$. Weighted least squares estimation applied to Eq.~\eqref{eq:stacked_full} provides the estimate $\widehat{\boldsymbol \theta}_{\!\text{WLS}}$, from which the smooth, damage-level-dependent AR model parameters $a_i(k)$ are reconstructed via Eq.~\eqref{eq:coef-expansion}. Repeating the procedure for each damage mode $m$ yields a family of mode-specific FP-AR models.

The choice of the functional basis family $\{G_j\}$ and its dimensionality $r$ is critical. In this work, shifted Chebyshev polynomials of the second kind are employed as the functional basis; these are described in detail in Appendix A. The basis functions are evaluated on the normalized damage level $\tilde{k}=k/k_{\max}\in[0,1]$; for notational simplicity, this dependence is expressed in terms of $k$ throughout. Models are fitted for
$r=1,2,\ldots,r_{\max}$, and $r$ is selected by minimizing:
\begin{equation} \label{eq:bic-basis-dimension} \mathrm{BIC}(r) = N_p\ln\!\left(\frac{\mathrm{RSS}(r)}{N_p}\right) + p_{\mathrm{eff}}(r)\ln N_p, \qquad p_{\mathrm{eff}}(r)=n_a r,
\end{equation}
where $N_p=MN$ designates the total pooled sample count, ${\mathrm{RSS}}(r)$ is computed from the pooled fit of Eq.~\eqref{eq:stacked_full}, and
$p_{\mathrm{eff}}(r)$ designates the number of projection coefficients in $\boldsymbol \theta$. Final model acceptance is again based on whiteness tests of the residuals of Eq.~\eqref{eq:stacked_full}.

\subsection{Cross-segment parameter pooling}
\label{sec:pooling}

In the preceding modeling framework, each FP-AR model is identified from a
single response segment. Here, a segment refers to a short time window containing
$N_s$ consecutive samples extracted from the measured response, denoted as
$\mathbf{y}^{(s)}=\{y^{(s)}[t]\}_{t=1}^{N_s}$, where $s=1,\ldots,S$ indexes the
selected segments.

In practical deployments, however, single-segment FP-AR identification is found
to be sensitive to the selected time segment. Different segments collected under the
same operating condition and the same damage mode $m$ may
yield different AR orders, functional basis dimensions, and projection coefficient estimates.
This variability can be attributed to spectral fluctuations, mild
nonstationarity, and occasional segment-level anomalies.

Therefore, the objective is to obtain a more representative FP-AR model that
captures the segment-invariant dynamics while retaining sensitivity to
short-window information. To this end, parameter pooling is introduced. By
combining parameter estimates from multiple short, approximately stationary
segments, the pooled model reduces the estimation variance, attenuates
segment-specific idiosyncrasies, and mitigates the bias that may arise from
fitting a single long identification window. In this sense, the pooled FP-AR model constitutes an improved representation over its single-segment counterpart: by exploiting several data segments instead of a single one, it better captures the experimental uncertainty and variability inherent in the flight data, thus leading to more robust damage diagnosis.

To enable direct parameter pooling, the underlying dynamics are assumed to admit
a shared parameter dimension across the selected segments, namely a common AR
order and a common functional basis. Under this assumption, the common order and
basis are first identified, after which the segment-wise parameter estimates are
combined by averaging.


\subsubsection{Model structure selection for the pooled model}

Let $\{y^{(s)}[t]\}_{t=1}^{N_s}$, $s=1,\ldots,S$, designate the segments available for a given damage mode $m$. To enforce a common parameter dimension prior to averaging, a common AR order and a common functional basis are selected using the segment-wise BIC.

\begin{figure}[!t]
    \centering 
    
    \begin{subfigure}{.48\textwidth} 
        \centering
        \includegraphics[width=\textwidth]{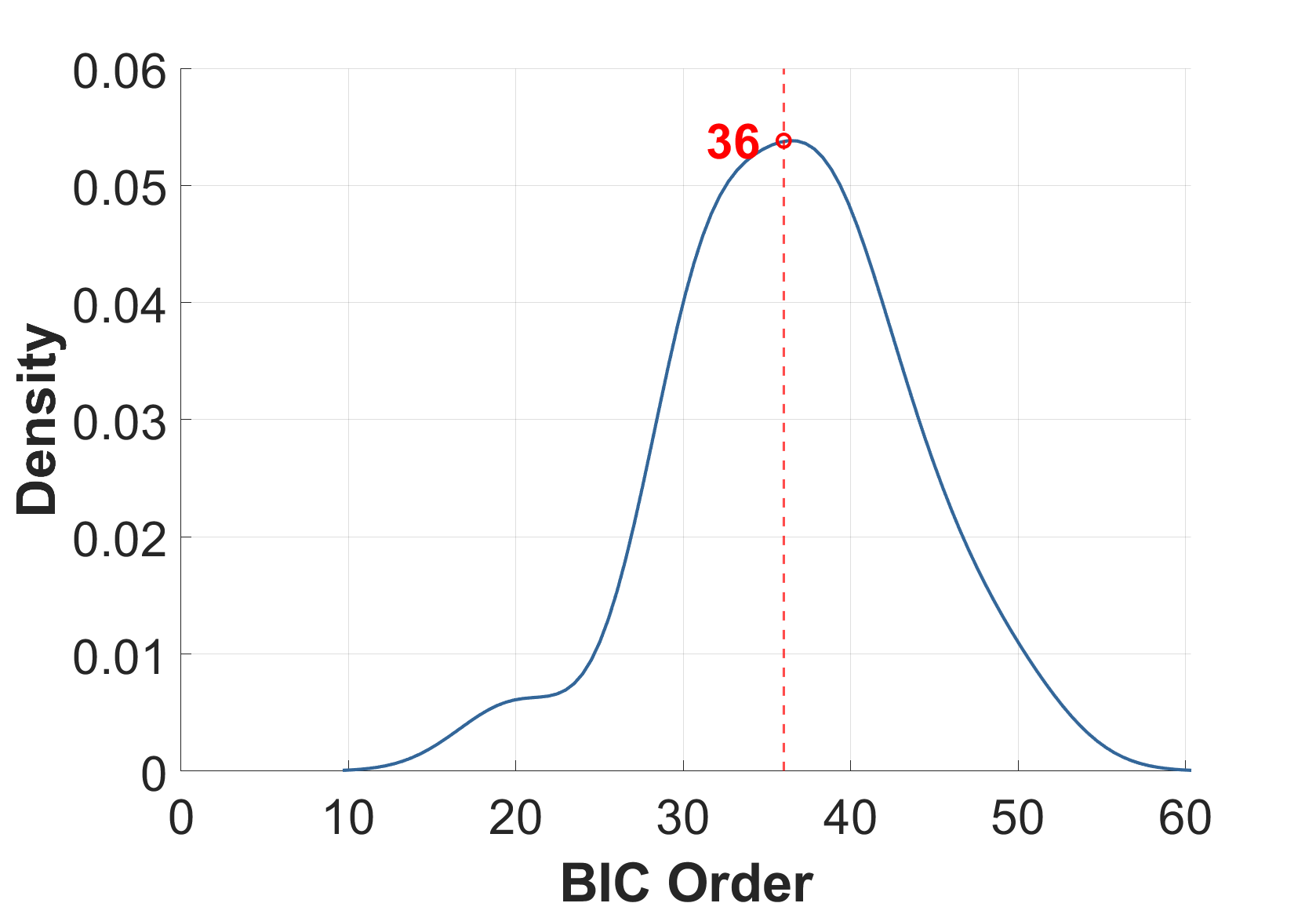}
        \caption{Order selection}
    \end{subfigure}
    \hfill 
    \begin{subfigure}{.48\textwidth}
        \centering
        \includegraphics[width=\textwidth]{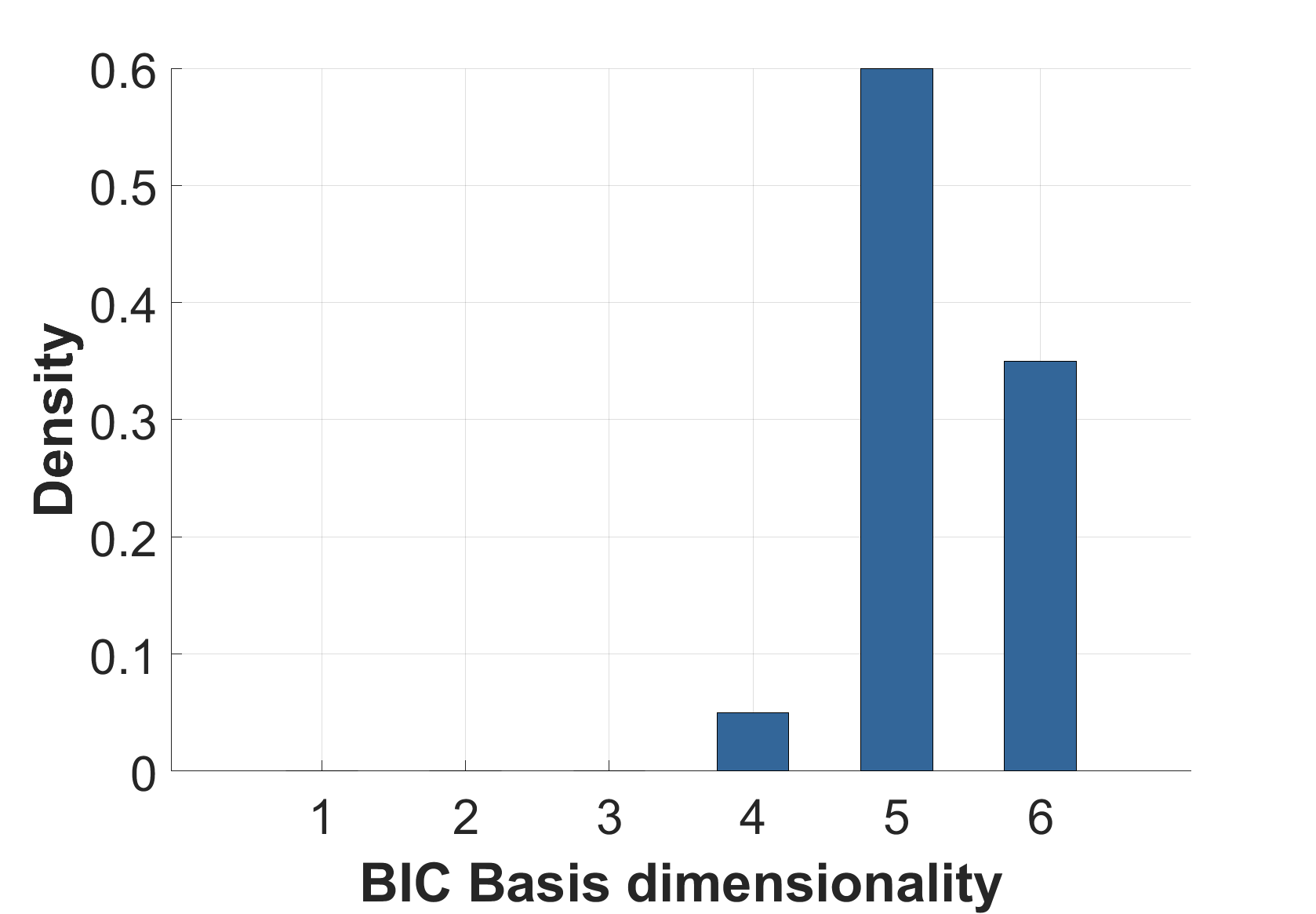}
        \caption{Basis selection}
    \end{subfigure}
   \caption{Order and basis selection for the Motor 1 AccX signal. (a) Gaussian KDE of the BIC-optimal AR orders across segments; the dominant peak (red dashed line) yields the pooled order (36). (b) Histogram of the BIC-optimal basis dimensionalities; the most frequent dimensionality is selected for the pooled FP-AR model.}
    \label{fig:order&basis density M1}
\end{figure}

For order selection, the same BIC criterion defined in Eq.~\eqref{eq:bic_rss} is
applied independently to each segment. Specifically, for the $s$th segment, AR
models with candidate orders $n\in\mathcal{N}$ are fitted, and the segment-wise
BIC value is denoted by $\mathrm{BIC}_s(n)$. The BIC-optimal order for segment
$s$ is then obtained as
\begin{equation}
\label{eq:segmentwise-bic-order}
\widehat n_a^{(s)}
=
\arg\min_{n\in\mathcal{N}}
\mathrm{BIC}_s(n).
\end{equation}
The pooled AR order is determined from the distribution of the segment-wise
BIC-optimal orders $\{\widehat n_a^{(s)}\}_{s=1}^{S}$. A Gaussian kernel density
estimate (KDE) is used to smooth this empirical distribution, and the integer location
of the dominant peak is adopted as the pooled AR order:
\begin{equation}
\label{eq:pooled-order}
n_a^{*}
=
\mathrm{round}
\left[
\arg\max_x \widehat f_n(x)
\right],
\end{equation}
where $\widehat f_n(x)$ designates the KDE of the segment-wise BIC-optimal orders. For illustration, Fig.~\ref{fig:order&basis density M1}(a) shows the KDE of the segment-wise BIC-optimal orders for the Motor 1 AccX signal, with the red dashed line indicating the dominant peak adopted as $n_a^*$.

With the pooled AR order $n_a^{*}$ fixed, the basis dimensionality is selected using
the BIC criterion defined in Eq.~\eqref{eq:bic-basis-dimension}. For each segment
$s$, candidate basis dimensionalities $r\in\mathcal R$ are evaluated with the AR order
fixed at $n_a^{*}$, and the segment-wise BIC-optimal basis dimensionality is obtained
as
\begin{equation}
\label{eq:segmentwise-basis}
\widehat r^{(s)}
=
\arg\min_{r\in\mathcal R}
\mathrm{BIC}_s(r;n_a^{*}).
\end{equation}
The pooled basis dimensionality is then chosen as the most frequently selected value
among the segment-wise BIC-optimal dimensionalities:
\begin{equation}
\label{eq:pooled-basis}
r^{*}
=
\mathrm{mode}
\left(
\widehat r^{(1)},\widehat r^{(2)},\ldots,\widehat r^{(S)}
\right).
\end{equation}
For illustration, Fig.~\ref{fig:order&basis density M1}(b) presents the segment-wise BIC-optimal basis dimensionalities via histograms for the Motor 1 AccX signal, with the most frequent count (tie-broken by the pooled BIC) yielding $r^*$. With $(n_a^*,r^*)$ determined, all segments share the same parameter dimension $p=n_a^* r^*$, enabling parameter pooling by simple averaging in the subsequent step.

\subsubsection{Parameter pooling}

With the common model structure $(n_a^{*},r^{*})$ fixed, each segment
$s=1,\ldots,S$ is fitted independently to obtain a segment-wise projection
coefficient vector
$\boldsymbol{\theta}^{(s)}\in\mathbb{R}^{p}$, where
$p=n_a^{*}r^{*}$. Following the least squares formulation of
Eq.~\eqref{eq:theta}, the segment-wise estimate is obtained as
\begin{equation}
\label{eq:segmentwise-theta}
\widehat{\boldsymbol{\theta}}^{(s)}
=
\arg\min_{\boldsymbol{\theta}\in\mathbb{R}^{p}}
\left\|
\mathbf{y}^{(s)}
-
\boldsymbol{\Phi}^{(s)}\boldsymbol{\theta}
\right\|_2^2 ,
\end{equation}
where $\boldsymbol{\Phi}^{(s)}$ is the segment-wise design matrix constructed
from the AR regressors and the selected functional basis. The corresponding
segment-wise residual variance is estimated as
\begin{equation}
\label{eq:segmentwise-variance}
\widehat{\sigma}_{e,s}^{2}
=
\frac{1}{N_s-n_a^{*}}
\left\|
\mathbf{y}^{(s)}
-
\boldsymbol{\Phi}^{(s)}
\widehat{\boldsymbol{\theta}}^{(s)}
\right\|_2^2 .
\end{equation}

The pooled projection coefficient vector is then obtained by averaging the
segment-wise estimates:
\begin{equation}
\label{eq:theta-pooling}
\widehat{\boldsymbol{\theta}}_{p}
=
\frac{1}{S}
\sum_{s=1}^{S}
\widehat{\boldsymbol{\theta}}^{(s)} .
\end{equation}
This pooled vector represents the segment-invariant projection coefficients under
the common order and basis dimensionality.

An empirical covariance of the pooled coefficient vector can be estimated by
combining the within-segment estimation uncertainty and the between-segment
variability:
\begin{equation}
\label{eq:theta-pooling-cov}
\widehat{\mathrm{Cov}}(\widehat{\boldsymbol{\theta}}_{p})
=
\frac{1}{S^2}
\sum_{s=1}^{S}
\widehat{\boldsymbol{\Sigma}}^{(s)}
+
\frac{1}{S(S-1)}
\sum_{s=1}^{S}
\left(
\widehat{\boldsymbol{\theta}}^{(s)}
-
\widehat{\boldsymbol{\theta}}_{p}
\right)
\left(
\widehat{\boldsymbol{\theta}}^{(s)}
-
\widehat{\boldsymbol{\theta}}_{p}
\right)^{\!\top},
\end{equation}
where
\begin{equation}
\label{eq:segmentwise-theta-cov}
\widehat{\boldsymbol{\Sigma}}^{(s)}
=
\widehat{\sigma}_{e,s}^{2}
\left[
\left(
\boldsymbol{\Phi}^{(s)}
\right)^{\!\top}
\boldsymbol{\Phi}^{(s)}
\right]^{-1}.
\end{equation}

The pooled model parameter functions are recovered from
$\widehat{\boldsymbol{\theta}}_{p}$ as
\begin{equation}
\label{eq:pooled-coefficient-field}
\widehat a_i(k)
=
\sum_{j=1}^{r^{*}}
\widehat\theta_{p,ij}\,
G_j^{*}(k),
\qquad
i=1,\ldots,n_a^{*},
\end{equation}
where $\widehat\theta_{p,ij}$ designates the entry of $\widehat{\boldsymbol{\theta}}_{p}$
corresponding to the $j$th projection coefficient of the $i$th AR lag, and
$\{G_j^{*}\}_{j=1}^{r^{*}}$ designates the selected functional basis.
The pooled estimation residuals are computed, in direct analogy with the single-segment FP-AR case, as the difference between the measured response and the corresponding model-based one-step-ahead prediction under $\widehat{\boldsymbol{\theta}}_{p}$ across the cross-sections:
\begin{equation}
\label{eq:pooled-residual}
e^{(s)}[t;\widehat{\boldsymbol{\theta}}_{p}]
=
y^{(s)}[t]
+
\sum_{i=1}^{n_a^{*}}
\widehat a_i(k)\,y^{(s)}[t-i],
\end{equation}
with $\widehat a_i(k)$ evaluated at the damage level of the cross-section to which segment $s$ belongs. The pooled residual variance is then estimated by
\begin{equation}
\label{eq:pooled-variance}
\widehat{\sigma}_{e,p}^{2}
=
\frac{
\displaystyle
\sum_{s=1}^{S}
\sum_{t=1}^{N_s}
\left(
e^{(s)}[t;\widehat{\boldsymbol{\theta}}_{p}]
\right)^2
}{
\displaystyle
\sum_{s=1}^{S}
\left(N_s-n_a^{*}\right)
}.
\end{equation}

\subsection{Inspection phase}
Let $y[t]$, $t=1,\ldots,N$, designate the response currently observed from the system in an \emph{unknown} health state. Damage detection, identification, and quantification are based on the mode-specific FP-AR models identified in the baseline phase. In the present formulation, the damage condition is described by two distinct descriptors: the continuous damage level $k$ and the discrete damage mode $m$, with $m$ designating the damaged motor index. For each candidate motor, a separate FP-AR model is available, with its AR model parameters expressed as smooth functions of the damage level $k$. The current response signal is evaluated using these mode-specific models, and the corresponding residual series is obtained for each candidate pair $(k, m)$. Subsequently, these residuals are used for damage detection, identification, and quantification within a statistical decision-making framework. It is noted that the inspection phase described below may be implemented on either model type, that is, it applies identically to both the single-segment FP-AR models of Section~\ref{sec:fpar} and the pooled FP-AR models of Section~\ref{sec:pooling}.

\subsubsection{Step 1: damage detection}

In the detection stage, the damage level $k$ and the residual variance 
$\sigma_e^2(k)$ are treated as unknown quantities to be estimated, while the model associated with each candidate damage mode $m$ is considered separately. Toward this end, the FP-AR model of Eq.~\eqref{eq:single_motor_fpar} is re-parameterized in terms of $k$:

\begin{equation}
\mathcal{M}^{(m)}\!\left(k,\sigma_{e}^{2}(k)\right):
\quad
y_{k}[t]
+
\sum_{i=1}^{n_a}
a_i(k)\,y_{k}[t-i]
=
e_{k}[t],\quad e_{k}[t] \sim \mathcal N\!\big(0,\sigma_e ^2(k)\big),
\label{eq:inspection_fpar}
\end{equation}
where the model parameters $a_i(k)=\sum_{j=1}^{r} \hat a_{ij}\,G_j(k)$ are expanded in the fixed functional basis, with the projection coefficient estimates $\hat a_{ij}$ carried over from the baseline identification.
Given a candidate mode $m$, the damage level is obtained by minimizing the pooled one-step-ahead prediction error via the nonlinear least squares (NLS) criterion
\begin{equation}
\label{eq:det-nls}
\widehat k=\arg\min_{k}\;
\sum_{t=n_a+1}^{N} e_{k}[t]^2,
\qquad
e_{k}[t]=y[t]+\sum_{i=1}^{n_a} a_i(k)\,y[t-i],
\end{equation}
and the associated residual variance estimate is
\begin{equation}
\label{eq:det-var}
\widehat\sigma_e^2(k)=\frac{1}{N-n_a}\sum_{t=n_a+1}^{N} e_{k}[t]^2
\Big|_{k=\widehat k}.
\end{equation}
In practice, a coarse global search on $k$ via a genetic algorithm (GA), followed by a local constrained refinement via sequential quadratic programming (SQP), offers reliable convergence to the minimizer of Eq.~\eqref{eq:det-nls}.

Under mild regularity assumptions, and provided that the true condition belongs to the considered candidate mode $m$ or the healthy case, the first-stage estimator of the damage level is asymptotically Gaussian. In the general case of an operating parameter vector $\mathbf{k}$, as $N\to\infty$ the estimator $\widehat{\mathbf{k}}$ is approximately distributed around the true value $\mathbf{k}$, with covariance given by a Cram\'er--Rao-type lower bound (see Appendix B):
\begin{equation}
\label{eq:det-cov}
\widehat{\mathbf{k}}\ \dot\sim\ \mathcal N\!\big(\mathbf{k},\ \boldsymbol{\Sigma}_{\mathbf{k}}\big),
\qquad
\boldsymbol{\Sigma}_{\mathbf{k}}=\sigma^2(\mathbf{k})\,
\Bigg[\sum_{t=1}^{N}
\boldsymbol{\epsilon}[t,\mathbf{k}]\,
\boldsymbol{\epsilon}[t,\mathbf{k}]^{\!\top}
\Bigg]^{-1}_{\mathbf{k}=\widehat{\mathbf{k}}},
\qquad
\boldsymbol{\epsilon}[t,\mathbf{k}]=\frac{\partial e_{\mathbf{k}}[t]}{\partial \mathbf{k}},
\end{equation}
where $\boldsymbol{\epsilon}[t,\mathbf{k}]$ designates the sensitivity of the model residual with respect to the operating parameters. In the present case, the operating parameter reduces to the scalar damage level, $\mathbf{k}\equiv k$, and Eq.~\eqref{eq:det-cov} takes the simplified form:
\begin{equation}
\label{eq:det-cov-scalar}
\widehat k\ \dot\sim\ \mathcal N\!\big(k,\ \Sigma_{k}\big),
\qquad
\Sigma_{k}=\sigma^2(k)\,
\Bigg[\sum_{t=1}^{N}
\Big(\big(\boldsymbol\varphi[t]\otimes \tfrac{\partial \boldsymbol g( k)}{\partial k}\big)^{\!\top}\widehat{\boldsymbol\theta}\Big)^{2}
\Bigg]^{-1}_{k=\widehat k},
\end{equation}
where $\boldsymbol\varphi[t]$ and $\boldsymbol g(k)$ are defined as in Eq.~\eqref{eq:sample}, while $\widehat{\boldsymbol{\theta}}$ designates the projection coefficient estimate of the corresponding mode-specific FP-AR model.

Since the healthy state corresponds to zero damage magnitude, detection is based on the hypothesis test
\begin{equation}
\label{eq:det-hyp}
H_0:\ k=0 \quad\text{(healthy)} \qquad \text{vs}\qquad
H_1:\ k\neq 0 \quad\text{(damaged)} .
\end{equation}
Under the null hypothesis $H_0$, the statistic
\begin{equation}
\label{eq:det-t}
t=\frac{\widehat k}{\widehat \sigma_{k}}
\end{equation}
follows a $t$-distribution with $N-n_a$ degrees of freedom (accounting for the fitted AR order), with $\widehat\sigma_{k}$ being the positive square root of $\widehat \Sigma_{k}$, that is, the estimated standard deviation of $\widehat k$. At the risk level $\alpha$, the null hypothesis is rejected by a given model whenever $|t| > t_{1-\alpha/2}(N-n_a)$. In practice, Eqs.~\eqref{eq:det-nls}--\eqref{eq:det-t} are evaluated for each candidate damage mode $m$, yielding one test statistic per mode-specific FP-AR model. Damage is declared when at least one of the resulting statistics exceeds the critical threshold, that is, when the null hypothesis $H_0$ is rejected by any candidate model; the system is deemed healthy only if $H_0$ is accepted by all candidate models. It is noted that performing multiple simultaneous tests increases the family-wise false alarm rate above the nominal per-test level $\alpha$; if strict control of the overall false alarm rate is desired, a Bonferroni-type adjustment of the per-test risk level (e.g., $\alpha/|\mathcal{M}|$) may be employed. It is stressed that Step~1 addresses solely the damage detection task; the identification of the damaged motor and the estimation of the damage magnitude are addressed in Steps~2 and~3, respectively.

\subsubsection{Step 2: damage mode identification}

Step~2 addresses solely the damage mode identification task, that is, the determination of which propeller has been damaged. Once damage occurrence has been detected in Step~1, the damage mode is identified by validating the candidate FP-AR models associated with the different admissible damage modes. For each candidate mode $m\in\mathcal{M}$, with $\mathcal{M}$ designating the set of admissible modes, the first-stage estimate of the damage level, denoted by $\widehat{k}^{(m)}$, is substituted into the corresponding FP-AR model, and the residual sequence $\widehat e_m[t]$ is obtained.

The current damage mode is then determined by examining the residual whiteness of each candidate model. Specifically, the model corresponding to the actually damaged propeller is expected to produce an uncorrelated (white) residual sequence, whereas models associated with incorrect damage modes tend to leave correlated dynamics in the residuals. This is evaluated through the following hypothesis test:
\begin{equation}
\begin{aligned}
H_0 &: \rho_m[\tau]=0, \quad \tau=1,\ldots,\tau_{\max}, \
\\
H_1 &: \rho_m[\tau]\neq 0 \quad \text{for some } \tau ,
\end{aligned}
\label{eq:whiteness_hypothesis}
\end{equation}
\begin{equation}
\widehat{\rho}_{m}[\tau]
=
\frac{
\sum_{t=\tau+1}^{N}
\widehat e_{m}[t]\widehat e_{m}[t-\tau]
}{
\sum_{t=1}^{N}
\widehat e_{m}^{2}[t]
},
\qquad
\tau=1,2,\ldots,\tau_{\max} ,
\label{eq:residual_autocorrelation_mode}
\end{equation}
where $\rho_m[\tau]$ designates the normalized residual autocorrelation at lag $\tau$ for candidate mode $m$, and $\tau_{\max}$ is the maximum lag considered.
For a prescribed maximum lag $\tau_{\max}$, the Ljung--Box statistic
\begin{equation}
\label{eq:Qstat}
Q(m)=N(N+2)\sum_{\tau=1}^{\tau_{\max}}\frac{\widehat\rho_{m}[\tau]^2}{N-\tau}
\end{equation}
approximately follows, under $H_0$, a $\chi^2$ distribution with $\tau_{\max}$ degrees of freedom,
and $H_0$ is accepted at risk level $\alpha$ whenever
\begin{equation}
\label{eq:LBtest}
Q(m)\le \chi^2_{1-\alpha}(\tau_{\max}).
\end{equation}

The current damage mode is declared as the mode $m$ satisfying the
whiteness test of Eq.~\eqref{eq:LBtest}. If more than one mode is accepted, the tie is broken
by selecting the mode attaining the minimal residual variance
\[
\widehat\sigma_e^2(m)=\frac{1}{N-n_a}\sum_{t=n_a+1}^{N}\widehat e_{m}[t]^2 .
\]
If no mode satisfies Eq.~\eqref{eq:LBtest}, the mode minimizing $Q(m)$ is retained while
flagging a model-mismatch condition; re-baselining or an expanded model set may then be
warranted. The mode estimate $\,\widehat m\,$ obtained here, together with the
associated level estimate $\,\widehat k\,$ from Step~1, is carried forward to
the magnitude estimation stage.

\subsubsection{Step 3: damage magnitude estimation (quantification)}

Step~3 addresses the damage magnitude (quantification) task: once the damage mode has been determined as $m=\widehat m$ in Step~2, the final estimation of the damage level $k$, along with the corresponding uncertainty characterization, is carried out. Two alternative routes are considered toward this end.

\paragraph{Nonlinear inverse optimization.}
With the damage mode fixed to the outcome of Step~2, $m=\widehat m$, the corresponding FP-AR (or pooled FP-AR) model, re-parameterized in terms of $k$, is employed, and the damage
magnitude is quantified by assigning uncertainty to the level estimate
$ k\equiv \widehat k^{(\widehat m)}$ obtained in Step~1 via the NLS criterion of Eq.~\eqref{eq:det-nls}.
A $(1-\alpha)$ two-sided confidence interval for the damage magnitude is therefore
\begin{equation}
\label{eq:ci-k1-only}
k \in \Big[\ \widehat k - t_{1-\alpha/2}(N-n_a)\,\widehat \sigma_{k},\ 
               \widehat k + t_{1-\alpha/2}(N-n_a)\,\widehat \sigma_{k}\ \Big],
\end{equation}
where $t_{1-\alpha/2}(N-n_a)$ is the critical value of the $t$-distribution with
$N-n_a$ degrees of freedom. 

\paragraph{Bayesian damage magnitude estimation.}

Alternatively, a Bayesian inverse formulation is employed to quantify the
 magnitude of the damage with explicit uncertainty characterization. After the damage
mode is identified in Step~2, the mode index is fixed as $m=\widehat m$, and
the corresponding FP-AR (or pooled FP-AR) model is used as the forward model. For notational
simplicity, the fixed mode index $\widehat{m}$ is omitted in the following
formulation, with $k$ designating the candidate damage magnitude under the identified mode.
The unknowns in the inverse problem are the damage magnitude $k$ and the
residual variance $\sigma_e^2$. Assuming prior independence, their joint prior is
written as
\begin{equation}
\label{eq:bayes-joint-prior}
p(k,\sigma_e^2)=p(k)p(\sigma_e^2).
\end{equation}

\begin{figure}[t!]
    \centering
    \begin{subfigure}[t]{0.48\textwidth}
        \includegraphics[width=\textwidth]{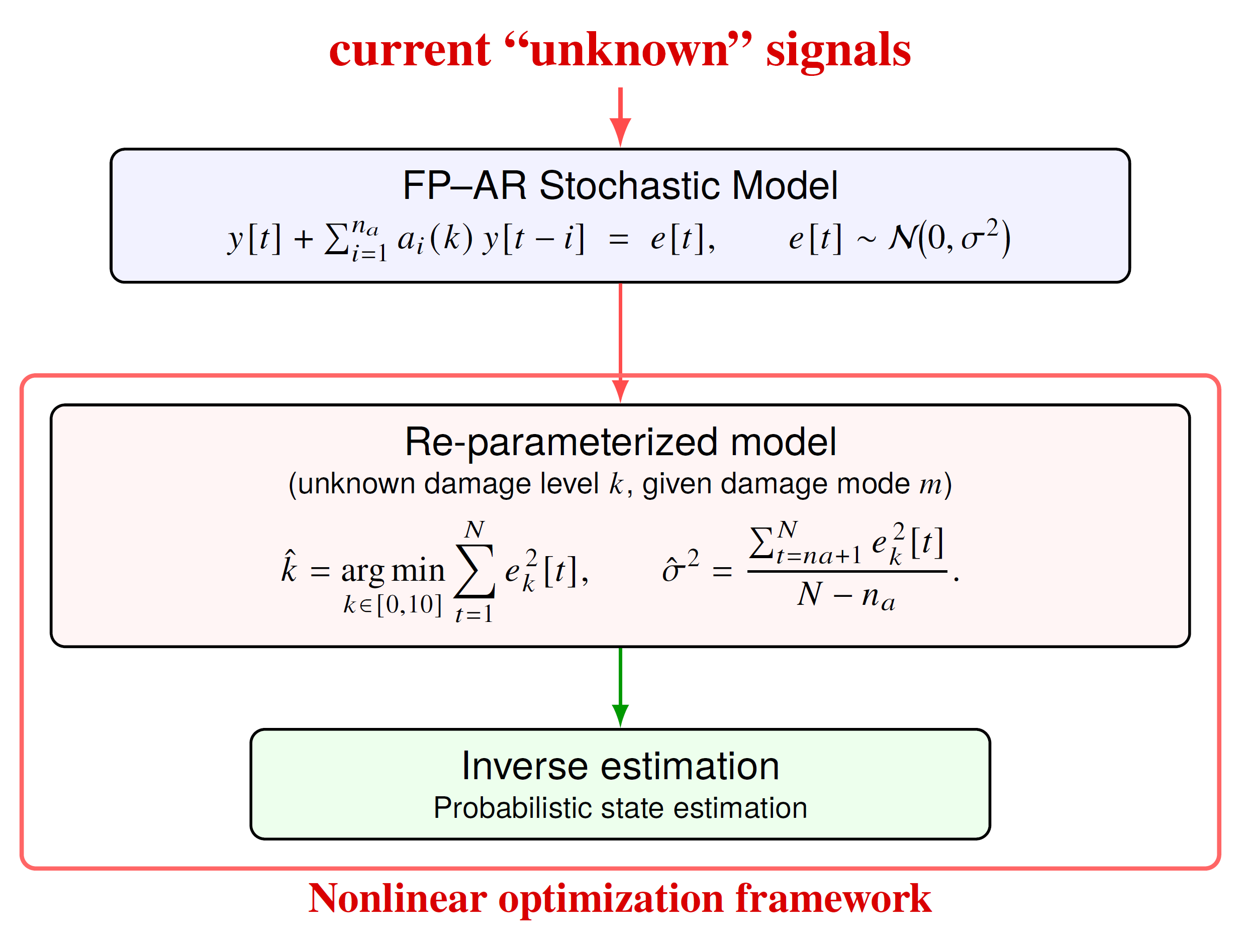}
        \caption{Inverse estimation}
        \label{fig:inverse}
    \end{subfigure}
    \begin{subfigure}[t]{0.48\textwidth}
        \includegraphics[width=\textwidth]{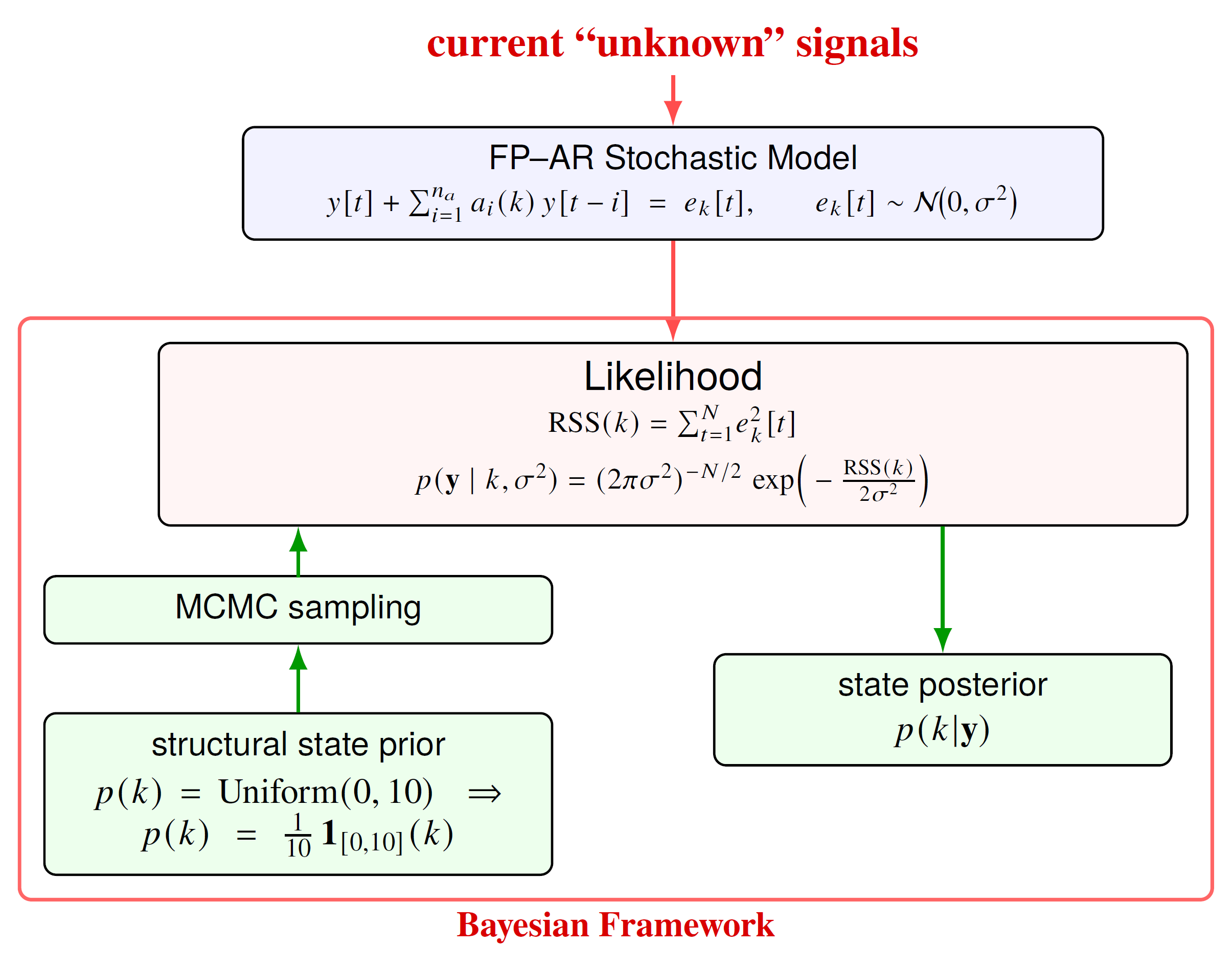}
        \caption{Bayesian estimation}
        \label{fig:bayesian}
    \end{subfigure}
    \caption{Block-diagram comparison of the two FP-AR damage quantification pipelines fed by the same flight segment. (a) Nonlinear inverse optimization: the FP-AR stochastic model is re-parameterized under fixed damage mode $m$; the damage level $k$ is obtained by minimizing the prediction-error sum of squares (RSS), yielding the point estimate $\hat{k}$ and the residual variance estimate $\hat{\sigma}_e^2$. (b) Bayesian estimation: the likelihood is built from the RSS and combined with a prior $p(k)$; MCMC sampling produces the posterior distribution for probabilistic inference.}
\label{fig:fp-ar-pipelines}
\end{figure}

The prior distribution of the damage magnitude is defined over the admissible
physical range as
\begin{equation}
\label{eq:k-prior-general}
p(k)=\pi_k(k),
\end{equation}
where $\pi_k(k)$ represents the selected prior model. In this study, two admissible
prior choices---a broad uniform prior and a sequential state-informed prior---are considered for $k$ in order to examine the sensitivity of the Bayesian
inversion to prior information. The residual variance $\sigma_e^2$ is assigned a
positive-support prior,
\begin{equation}
\label{eq:sigma-prior-general}
p(\sigma_e^2)=\pi_{\sigma}(\sigma_e^2),
\end{equation}
which is calibrated from baseline residual-variance statistics.

For a measured response segment
$\mathbf{y}=\{y[t]\}_{t=1}^{N}$, the FP-AR one-step-ahead residual under
a candidate $k$ is
\begin{equation}
\label{eq:bayes-residual}
e_{k}[t]
=
y[t]
+
\sum_{i=1}^{n_a}
a_i(k)y[t-i],
\end{equation}
with the corresponding residual sum of squares
\begin{equation}
\label{eq:bayes-rss}
\mathrm{RSS}(k)
=
\sum_{t=1}^{N} e_{k}^{2}[t].
\end{equation}
Assuming independent Gaussian residuals, the likelihood is defined as
\begin{equation}
\label{eq:bayes-likelihood}
p(\mathbf{y}\mid k,\sigma_e^2)
=
(2\pi\sigma_e^2)^{-N/2}
\exp\left[
-\frac{\mathrm{RSS}(k)}{2\sigma_e^2}
\right].
\end{equation}

According to Bayes' theorem, the joint posterior distribution is
\begin{equation}
\label{eq:bayes-joint-posterior}
p(k,\sigma_e^2\mid \mathbf{y})
=
\frac{
p(\mathbf{y}\mid k,\sigma_e^2)
p(k)p(\sigma_e^2)
}{
p(\mathbf{y})
}.
\end{equation}
Since the damage magnitude is the quantity of interest, the marginal posterior
of $k$ is obtained by integrating out $\sigma_e^2$:
\begin{equation}
\label{eq:bayes-k1-posterior}
p(k\mid \mathbf{y})
=
\int
p(k,\sigma_e^2\mid \mathbf{y})
\,\mathrm{d}\sigma_e^2 .
\end{equation}

In practice, the posterior is sampled using Markov chain Monte Carlo (MCMC) with an adaptive Metropolis
strategy. The sampled marginal posterior of $k$ is then used to compute the
maximum a posteriori (MAP) estimate, posterior mean, and credible interval:
\begin{equation}
\label{eq:bayes-estimates}
k^{\mathrm{MAP}}
=
\arg\max_{k\in[0,k_{\max}]}p(k\mid\mathbf{y}),
\qquad
\bar{k}
=
\mathbb{E}[k\mid\mathbf{y}],
\end{equation}
\begin{equation}
\label{eq:bayes-ci}
\mathrm{CI}_{1-\alpha}
=
\left[
F^{-1}_{k\mid\mathbf{y}}\left(\frac{\alpha}{2}\right),
F^{-1}_{k\mid\mathbf{y}}\left(1-\frac{\alpha}{2}\right)
\right],
\end{equation}
where $F^{-1}_{k\mid\mathbf{y}}(\cdot)$ designates the inverse cumulative distribution function (CDF)
of the marginal posterior distribution. In this study, $\alpha=0.05$ is used for
the 95\% credible interval.

When multiple test segments are available, the segment-wise posterior
distributions can be fused as
\begin{equation}
\label{eq:posterior-combination}
p_{\mathrm{comb}}(k)
\propto
\prod_{s=1}^{S} p_s(k\mid\mathbf{y}^{(s)}),
\end{equation}
followed by normalization over $k\in[0,k_{\max}]$. This Bayesian route provides a
full probabilistic characterization of the damage magnitude, complementing the
RSS-based inverse estimate with posterior uncertainty and credible bounds.
 
\begin{figure}[t!]
\centering
\includegraphics[width=0.9\columnwidth]{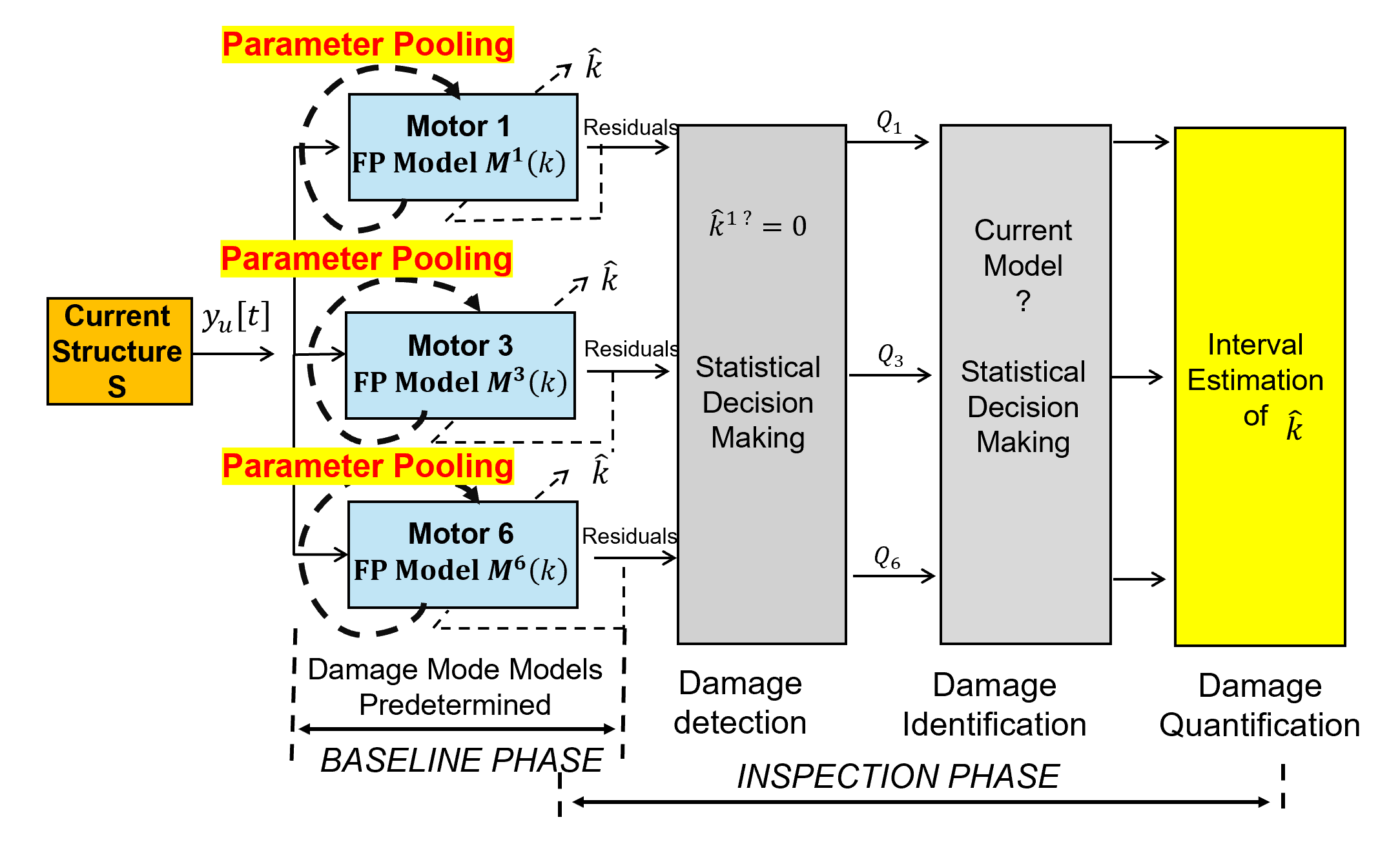}
\caption{Flowchart representation of the postulated FP-AR model-based damage diagnosis framework.}
\label{fig:framework_flowchart}
\end{figure}

The two FP-AR damage quantification pipelines are summarized in Fig.~\ref{fig:fp-ar-pipelines}, highlighting the inverse (RSS minimization) and Bayesian (likelihood combined with MCMC) routes. The complete FP-AR model-based damage diagnosis framework, integrating the baseline, functionally pooled identification, and inspection phases, is summarized in Fig.~\ref{fig:framework_flowchart}.

\section{Stochastic identification results}

\subsection{Non-parametric identification}

A non-parametric approach is first adopted in order to preliminarily evaluate how propeller damage influences the dynamic response of the aircraft, without relying on predefined model assumptions. This approach functions both as a diagnostic tool for detecting potential damage and as a stepping stone toward the subsequent parametric modeling. The primary focus lies in the analysis of the frequency-domain behavior through power spectral density (PSD) estimation via Welch's method~\cite{ljung1998system}, aiming to differentiate the signal patterns between the healthy and damaged conditions.

\begin{figure}[t!]
    \centering
    \begin{subfigure}[t]{0.32\textwidth}
        \includegraphics[width=\textwidth]{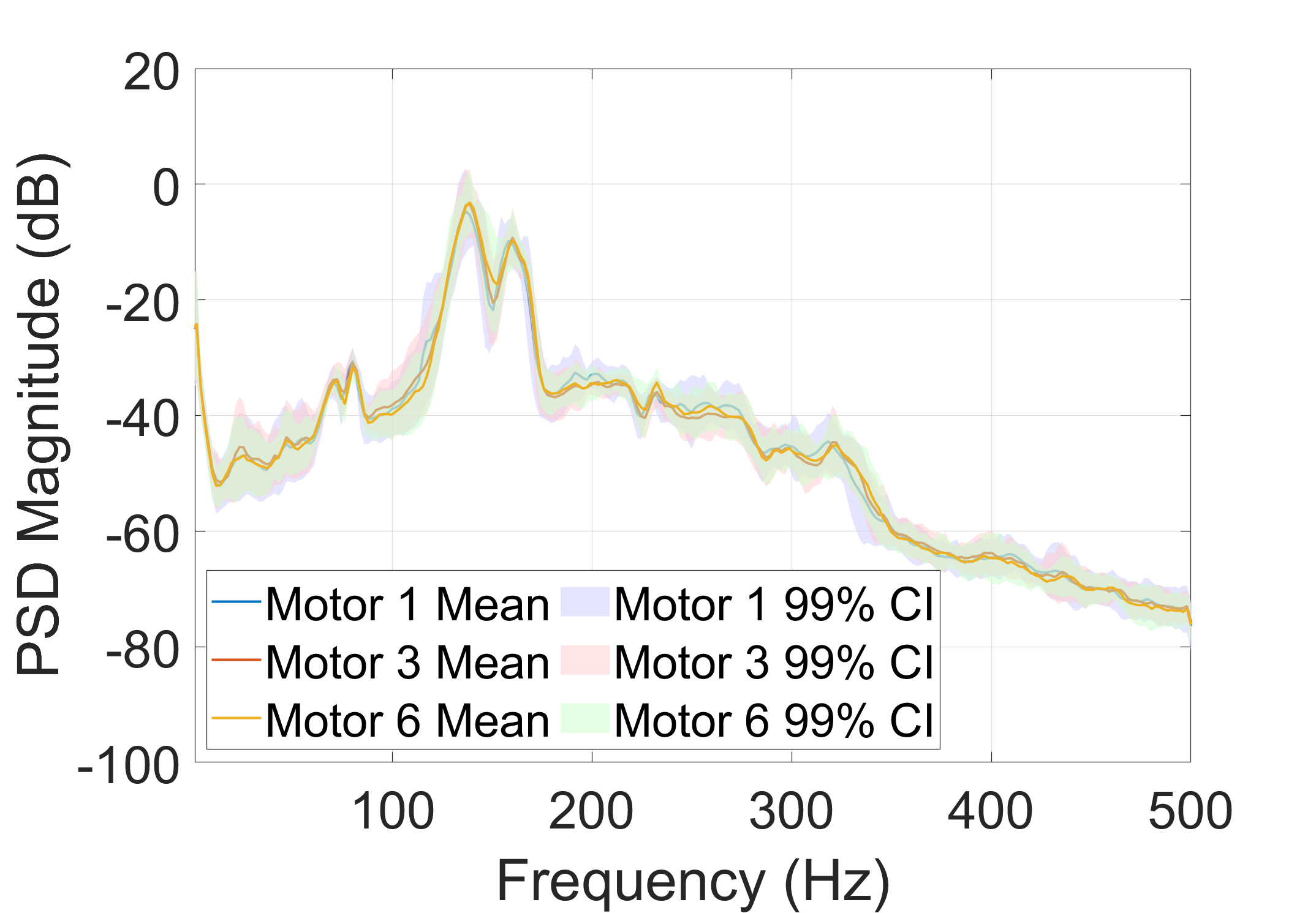}
        \caption{Healthy state}
        \label{fig:psd-healthy}
    \end{subfigure}
    \begin{subfigure}[t]{0.32\textwidth}
        \includegraphics[width=\textwidth]{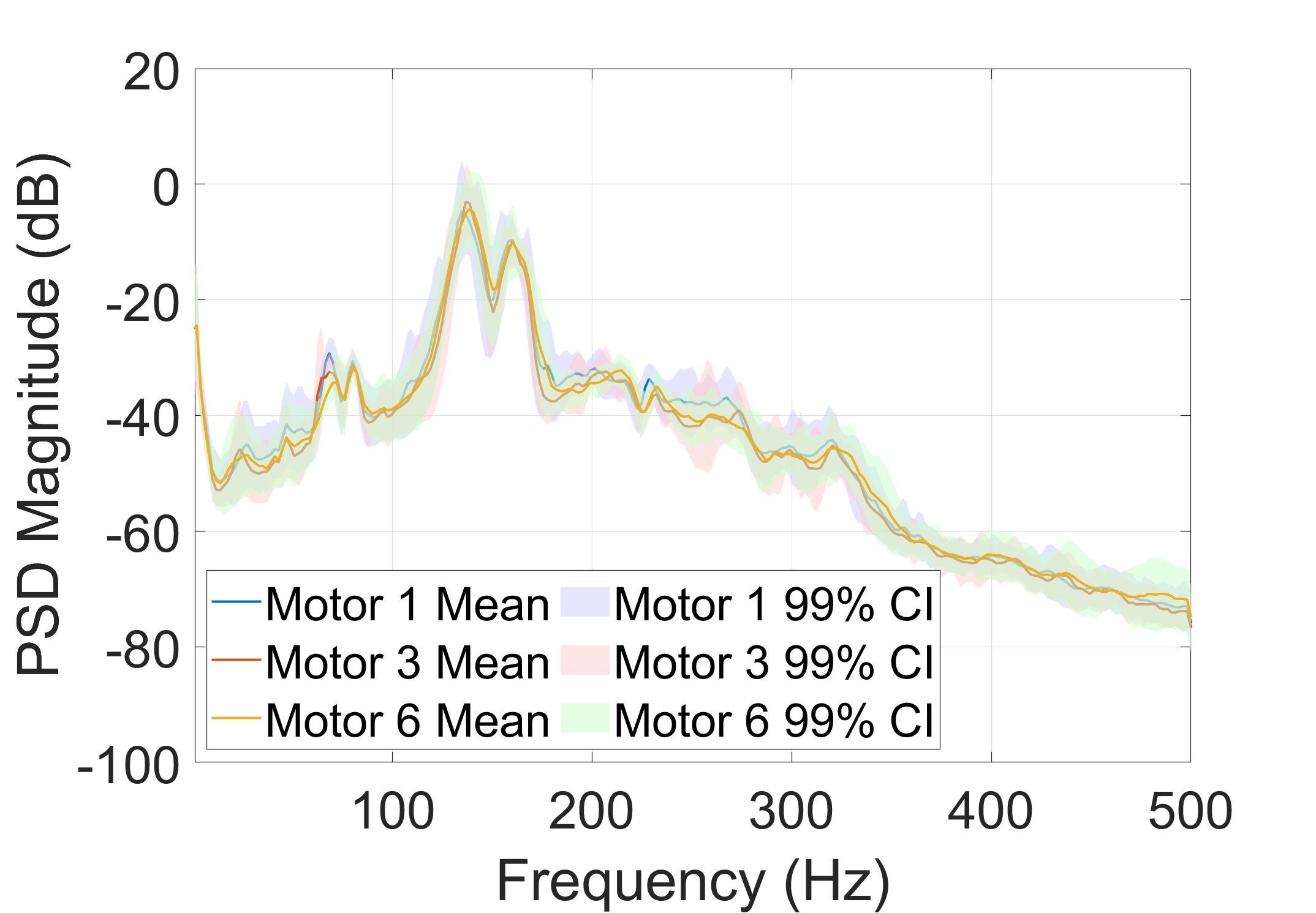}
        \caption{2 mm damage}
        \label{fig:psd-2}
    \end{subfigure}
    \begin{subfigure}[t]{0.32\textwidth}
        \includegraphics[width=\textwidth]{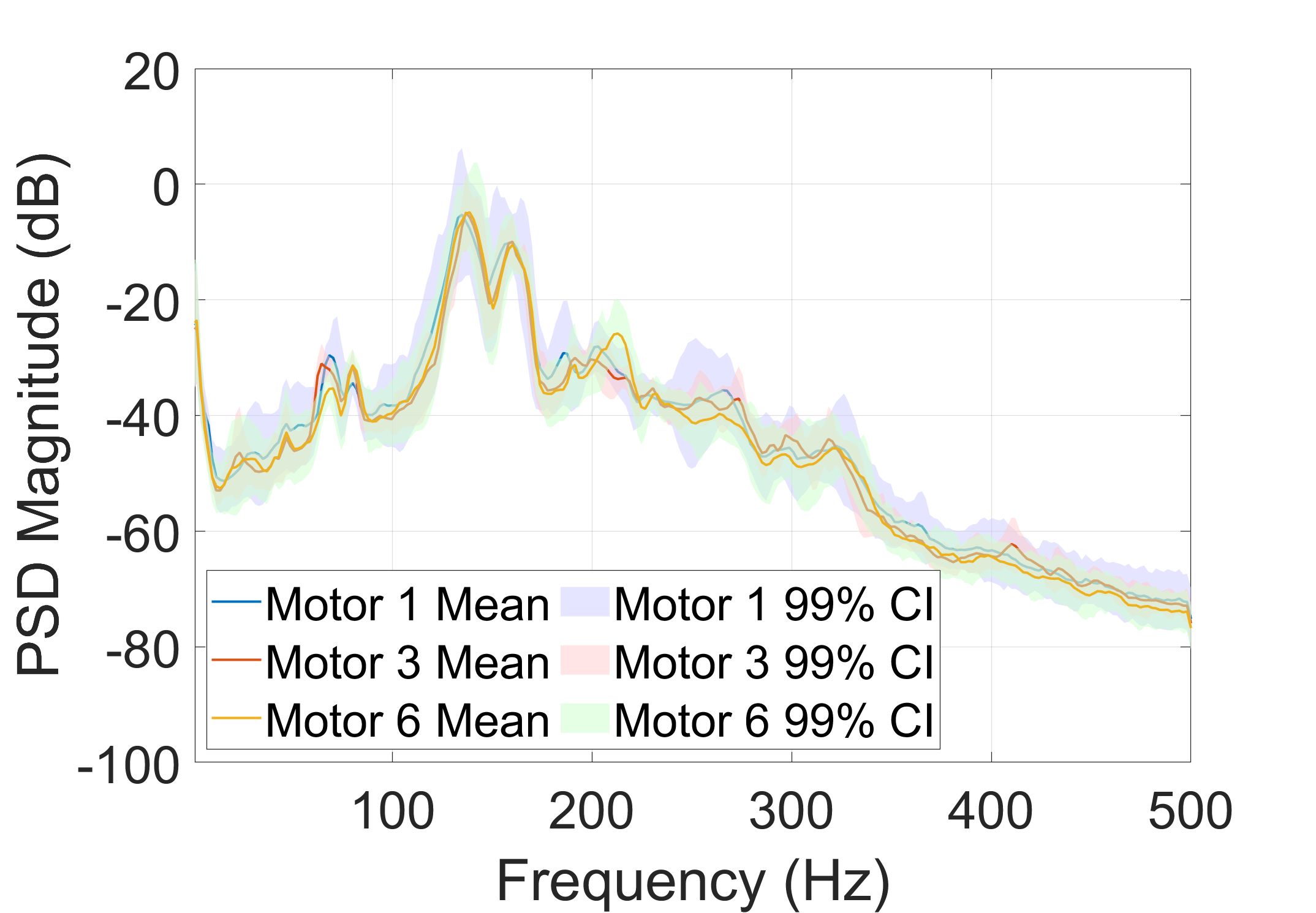}
        \caption{4 mm damage}
        \label{fig:psd-4}
    \end{subfigure}
        \begin{subfigure}[t]{0.32\textwidth}
        \includegraphics[width=\textwidth]{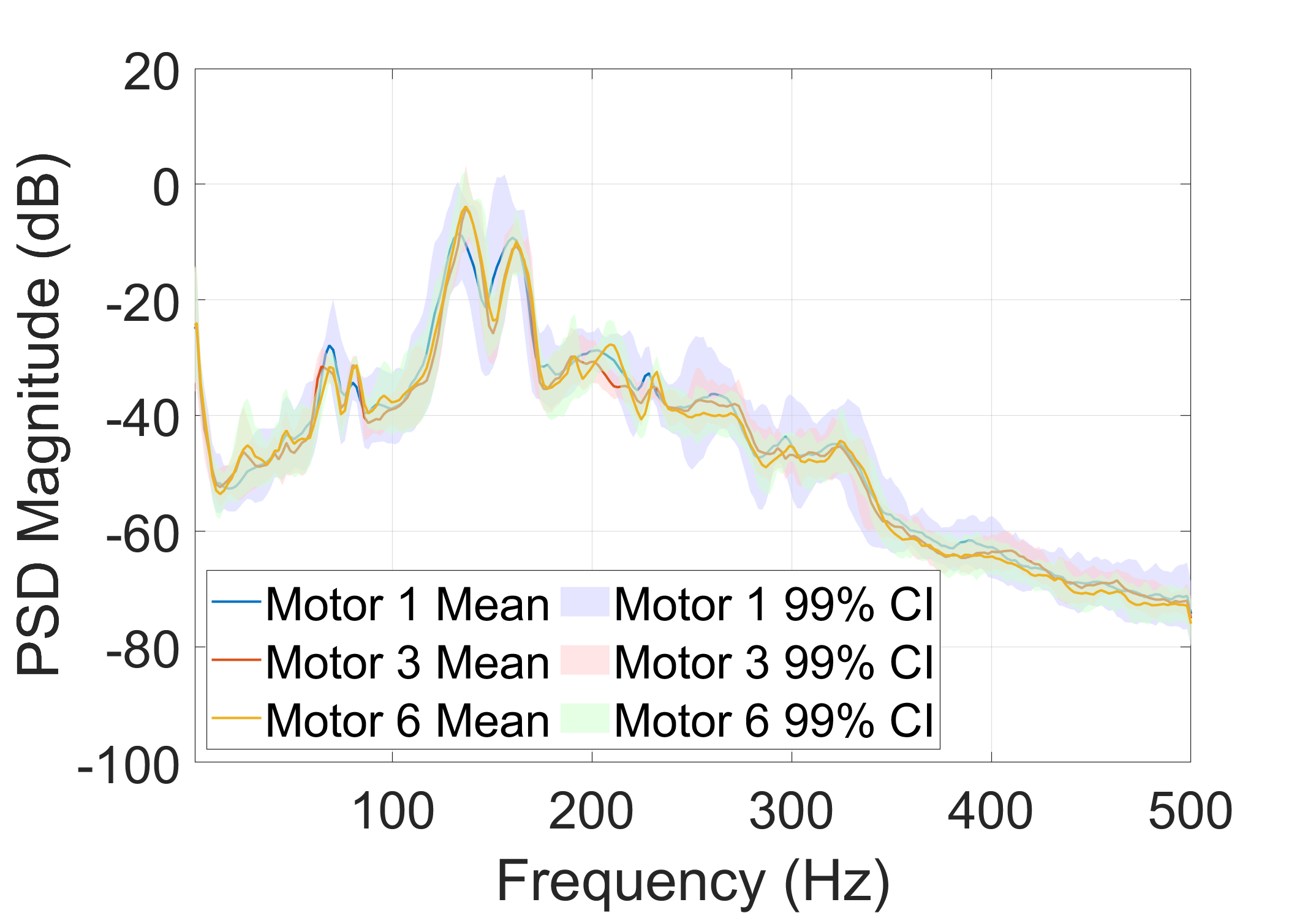}
        \caption{6 mm damage}
        \label{fig:psd-6}
    \end{subfigure}
    \begin{subfigure}[t]{0.32\textwidth}
        \includegraphics[width=\textwidth]{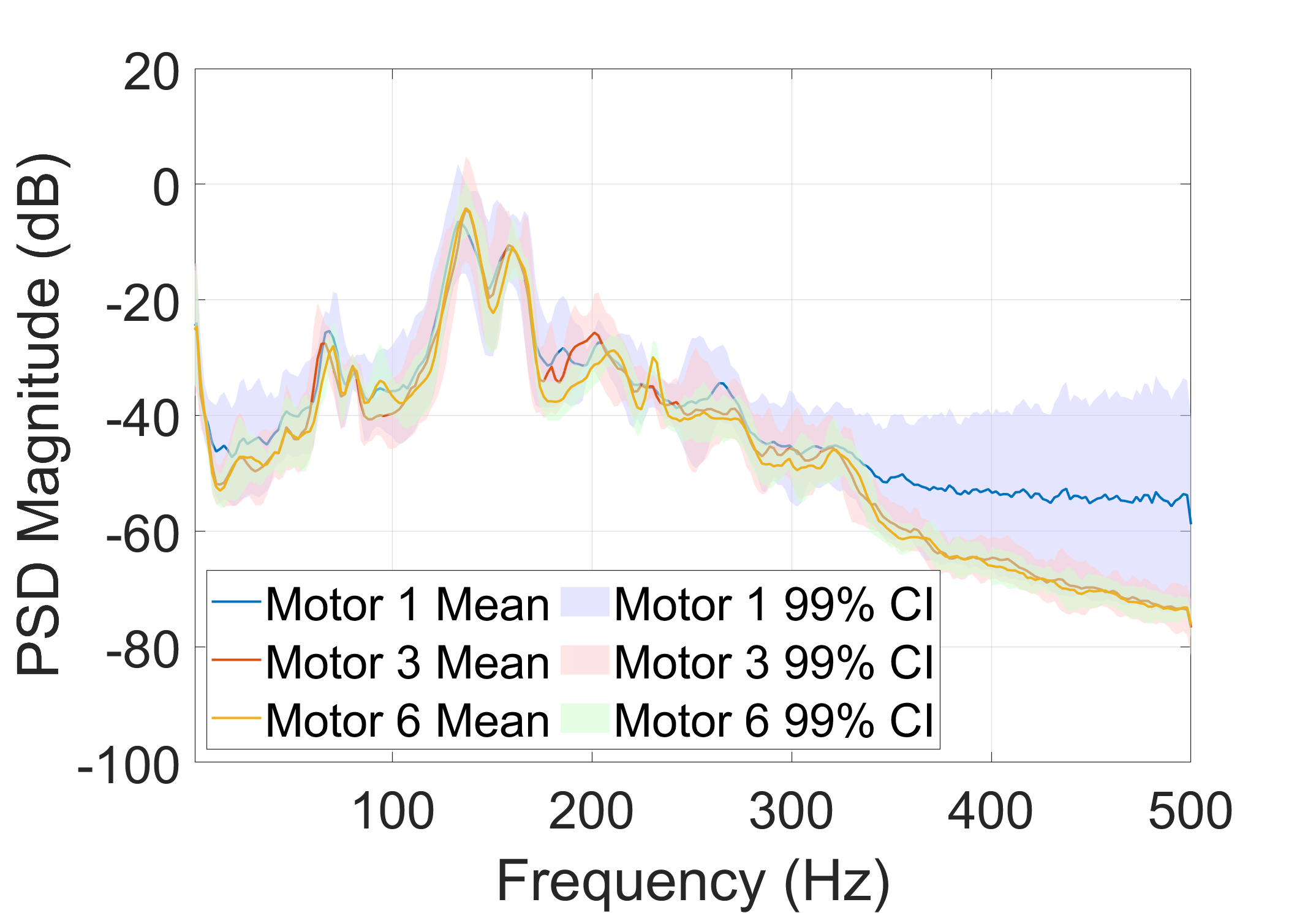}
        \caption{8 mm damage}
        \label{fig:psd-8}
    \end{subfigure}
    \begin{subfigure}[t]{0.32\textwidth}
        \includegraphics[width=\textwidth]{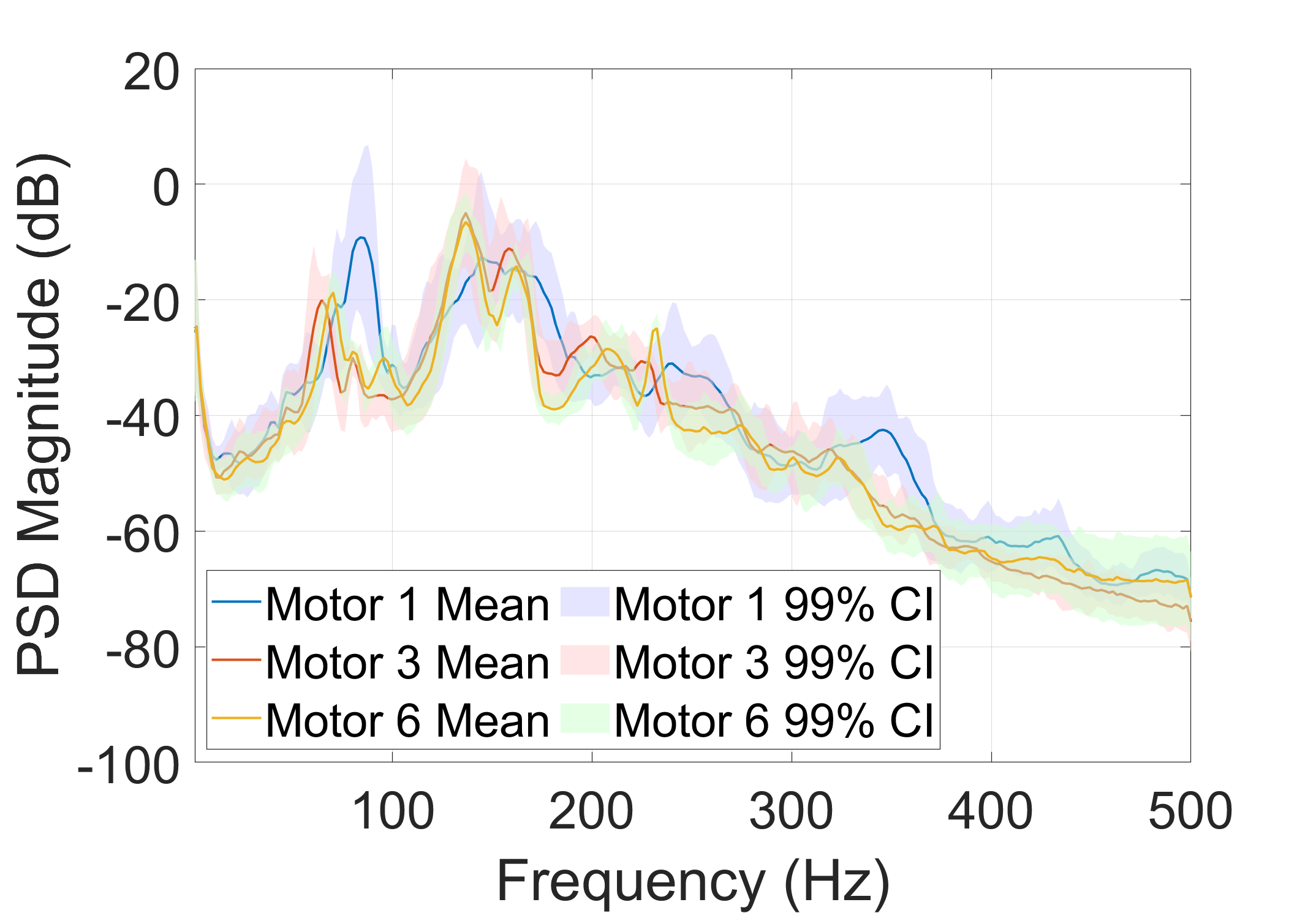}
        \caption{10 mm damage}
        \label{fig:psd-10}
    \end{subfigure}
    \caption{Welch-based PSD estimates of the GyrX (angular velocity) signals of the three motors for all considered damage states, from healthy to 10 mm. The 99\% confidence intervals of the estimated PSDs are shown as shaded areas.}
    \label{fig:PSD}
\end{figure}

Figure~\ref{fig:PSD} illustrates the PSD estimates of the GyrX signals of the three motors under all considered damage conditions, ranging from the healthy state to 10 mm damage. The spectral estimation settings are summarized in Table~\ref{tab:psd-params}. For each condition, 250 s of data were partitioned into 25 non-overlapping 10 s segments, a PSD estimate was computed per segment, and the figure reports the segment-wise mean along with the corresponding 99\% confidence intervals, shown as shaded areas. 

\begin{table}[b!]
  \centering
  \begin{threeparttable}
    \caption{Non-parametric (Welch-based) spectral estimation details}
    \label{tab:psd-params}
    \small
    \begin{tabular*}{0.7\textwidth}{@{\extracolsep{\fill}} l l @{}}
      \toprule
      Parameter          & Value \\
      \midrule
      Method             & Welch PSD (\texttt{pwelch}) \\
      Sampling frequency & 1 kHz \\
      Record duration    & 250 s \\
      Window type        & Hamming \\
      Window length      & 1024 samples \\
      Overlap            & 50\% \\
      FFT length         & 1024 points \\
      Segment partitioning & 25 $\times$ 10 s (non-overlapping) \\
      Confidence level   & 99\% \\
      \bottomrule
    \end{tabular*}
  \end{threeparttable}
\end{table}

The results exhibit frequency-dependent variations that correlate with the higher damage levels. For damage levels of 2 mm and 4 mm the PSDs do not exhibit clear differences with the healthy PSD. As the damage severity increases, distinct changes in the spectral content emerge most notably within frequency bands associated with the rotor blade harmonics and structural resonances. These shifts indicate altered system dynamics and airflow interactions caused by the tip damage, suggesting that higher damage levels affect the frequency-domain characteristics. It is worth noting that all three motors present similar PSDs for all structural states, except from the case of 10 mm and motor 1, which shows evident shifts in the PSD peaks compared to motors 3 and 6. 

\begin{figure}[t!]
    \centering
    \begin{subfigure}[t]{0.45\textwidth}
        \includegraphics[width=\textwidth]{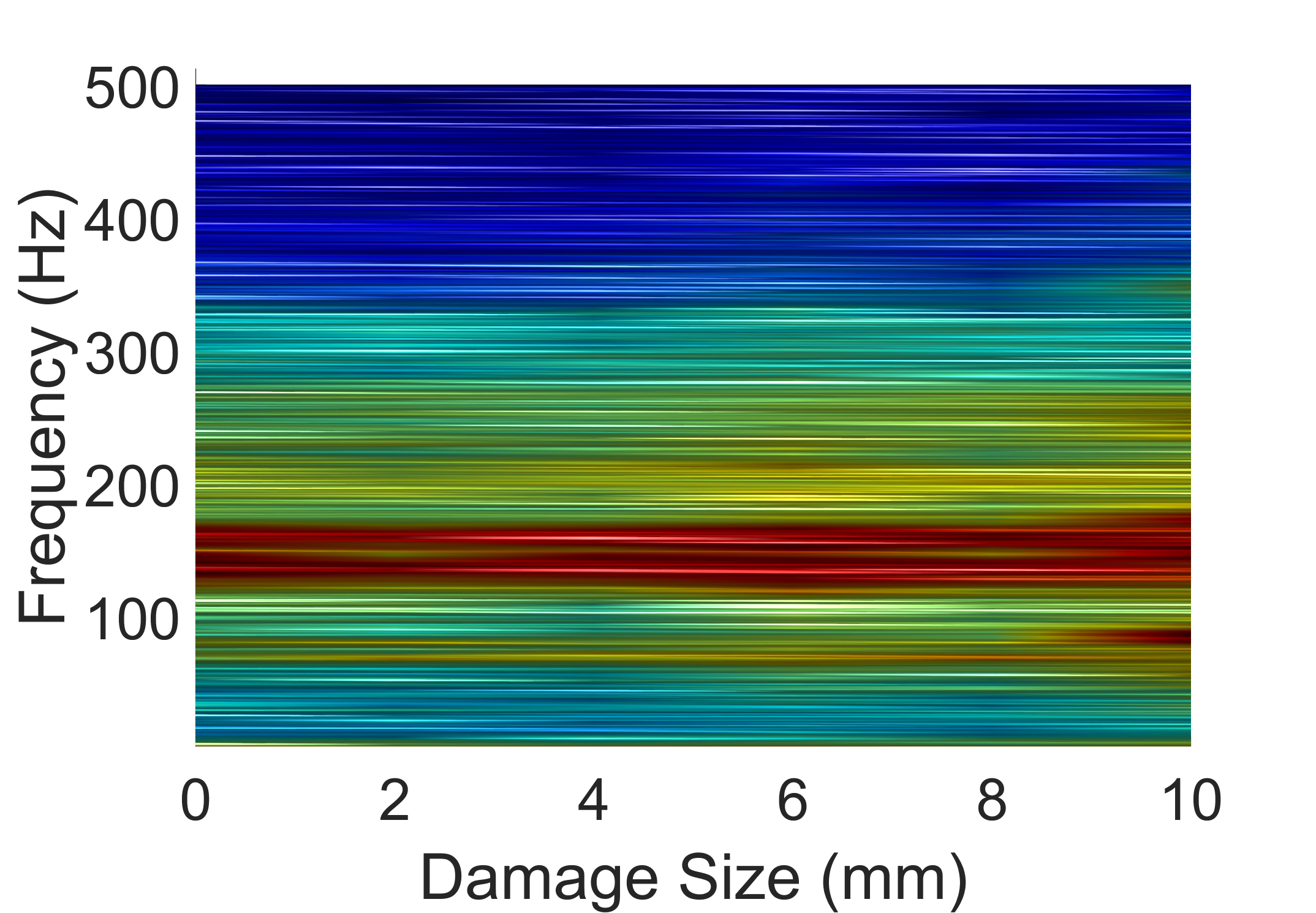}
        \caption{Motor 1 PSD }
        \label{fig:3d-psd-M1}
    \end{subfigure}
    \begin{subfigure}[t]{0.45\textwidth}
        \includegraphics[width=\textwidth]{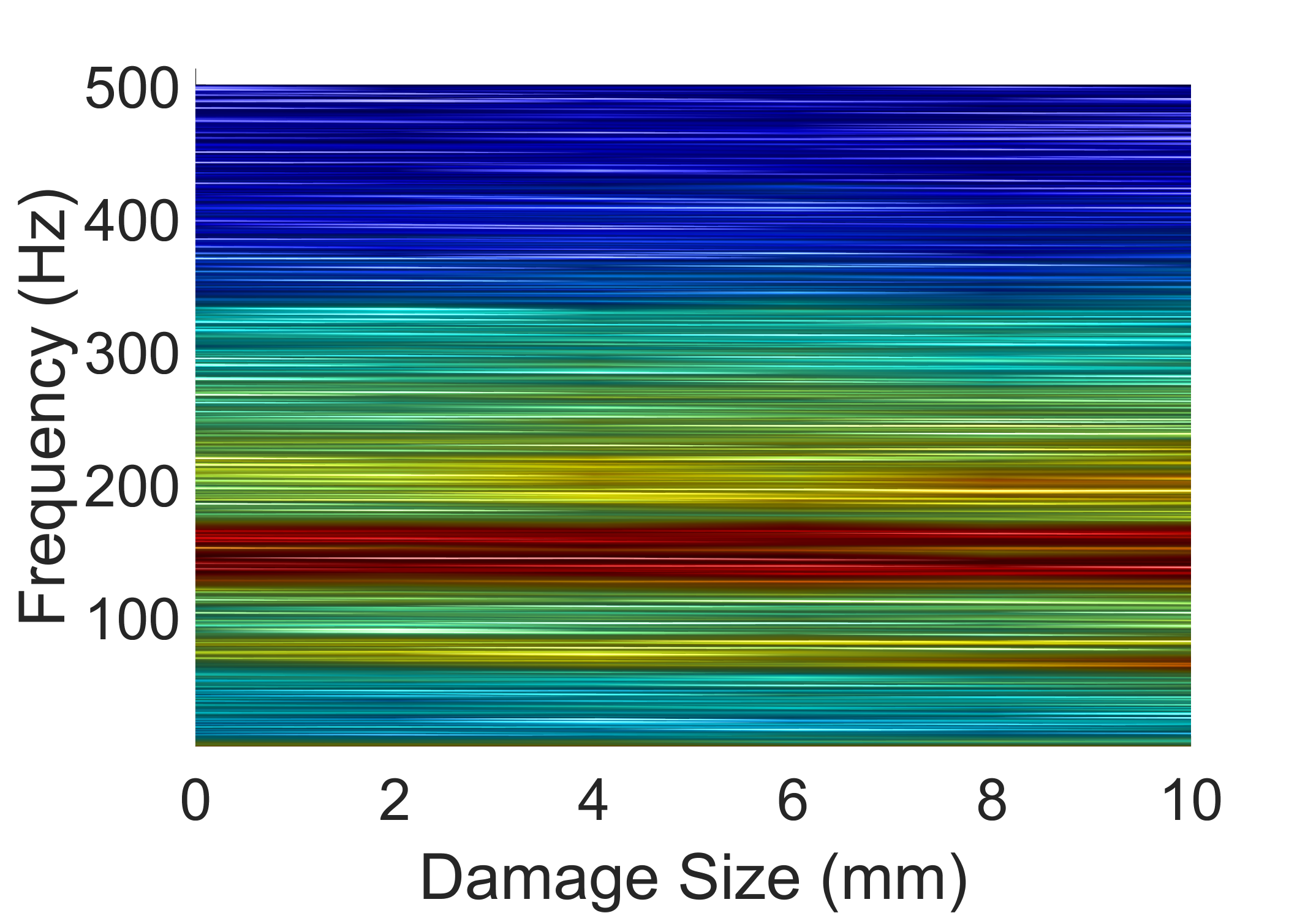}
        \caption{Motor 3 PSD}
        \label{fig:3d-psd-M3}
    \end{subfigure}
    \begin{subfigure}[t]{0.45\textwidth}
        \includegraphics[width=\textwidth]{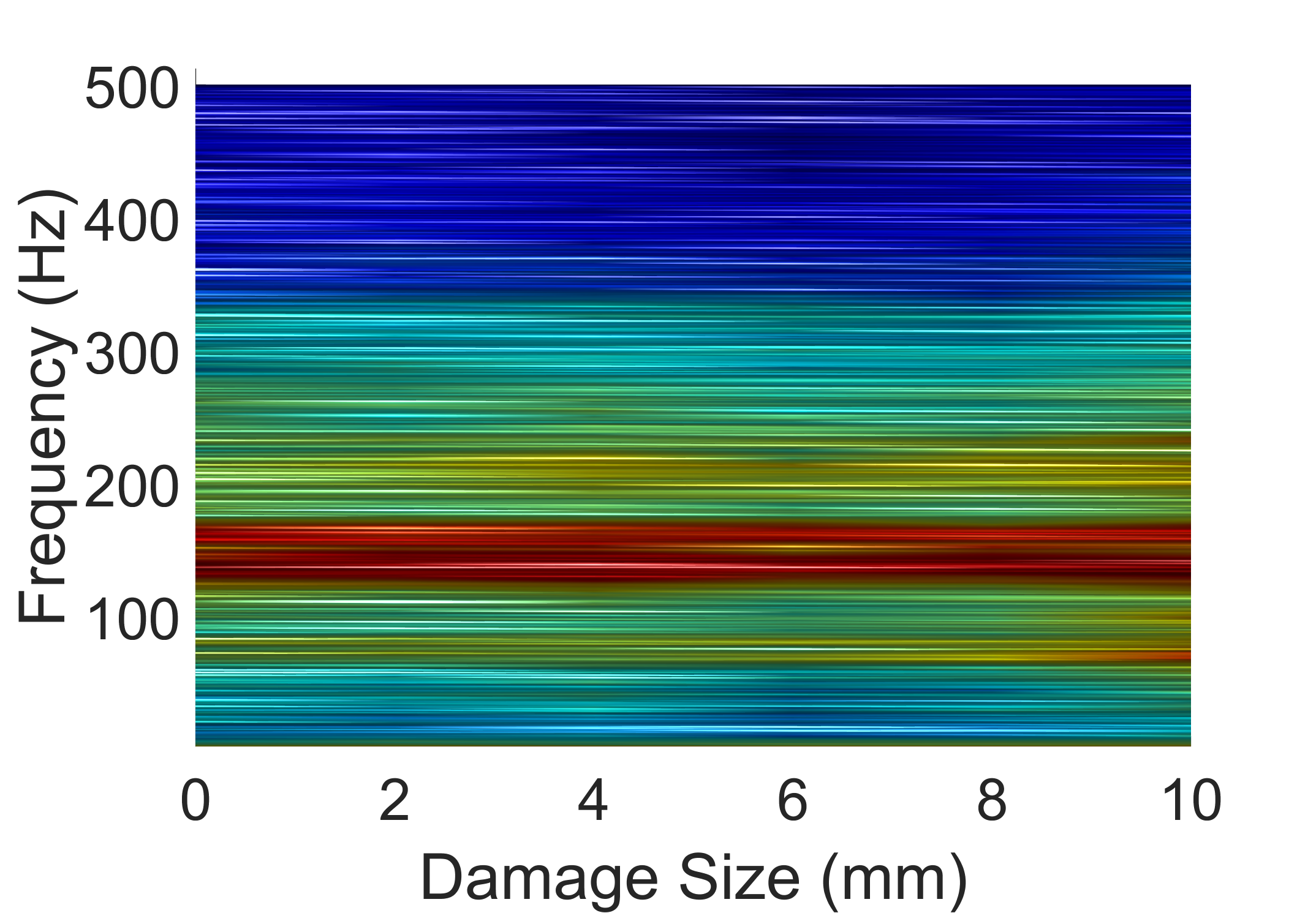}
        \caption{Motor 6 PSD}
        \label{fig:3d-psd-M6}
    \end{subfigure}
    \caption{Non-parametric damage topology representation: Welch-based PSD magnitude of the GyrX (angular velocity) signal versus frequency and damage size for Motors 1, 3, and 6.}
    \label{fig:3D-PSD}
\end{figure}

Fig.~\ref{fig:3D-PSD} presents the non-parametric damage topology representations for the three tested motors, showing the PSD variation across damage levels from 0 mm to 10 mm. Each plot depicts the evolution of the GyrX spectral content as the damage increases, revealing consistent trends in spectral amplitude and distribution. The analysis identifies frequency regions sensitive to the damage progression, particularly those aligned with the structural modes and rotor harmonics. The effect of damage becomes more evident with increasing level, especially after 6 mm.

These results lead to three key findings: (i) high-magnitude damage produces statistically significant spectral shifts beyond the baseline confidence bounds; (ii) the motors exhibit similar spectral evolution pattern that may challenge damage mode identification; and (iii) the damage severity correlates with systematic frequency-domain variations, establishing a strong basis for the subsequent model-based analysis.


\subsection{Parametric FP-model identification}

While the non-parametric analysis allows a preliminary detectability assessment of damage and reveals informative spectral patterns, it lacks the modeling precision, robustness, and compactness required for quantitative assessment. To address this limitation, FP-AR models are identified that capture the damage-dependent dynamics within a unified, statistically framework.

Conventional AR models treat each damage condition independently, resulting in multiple isolated models that fail to reflect the systematic evolution of the system dynamics across varying damage levels. In contrast, FP-AR models incorporate the damage magnitude directly into the model structure by representing the AR model parameters as continuous functions of the damage level. This formulation enables a single model to span the full range of the observed damage states.

The FP-AR approach offers three key advantages: (i) compactness, as it eliminates the need for multiple condition-specific models by encapsulating all damage states within a single representation; (ii) continuity, allowing the prediction of the dynamic behavior at any damage level within the observed range, including unseen intermediate values; and (iii) statistical efficiency, achieved by pooling information across conditions, which enhances the robustness and accuracy of the parameter estimation.


A single-segment (local) FP-AR model is first constructed using 4 s data segments (4{,}000 samples) extracted from a representative portion of a 250 s flight. This segment-based approach aims to capture the short-duration dynamics under consistent operating conditions. The GyrX signal of Motor 1 is selected as a representative example for model construction, using the 106--110 s interval.


\begin{figure}[t!]
    \centering
    \begin{subfigure}[t]{0.48\textwidth}
        \includegraphics[width=\textwidth]{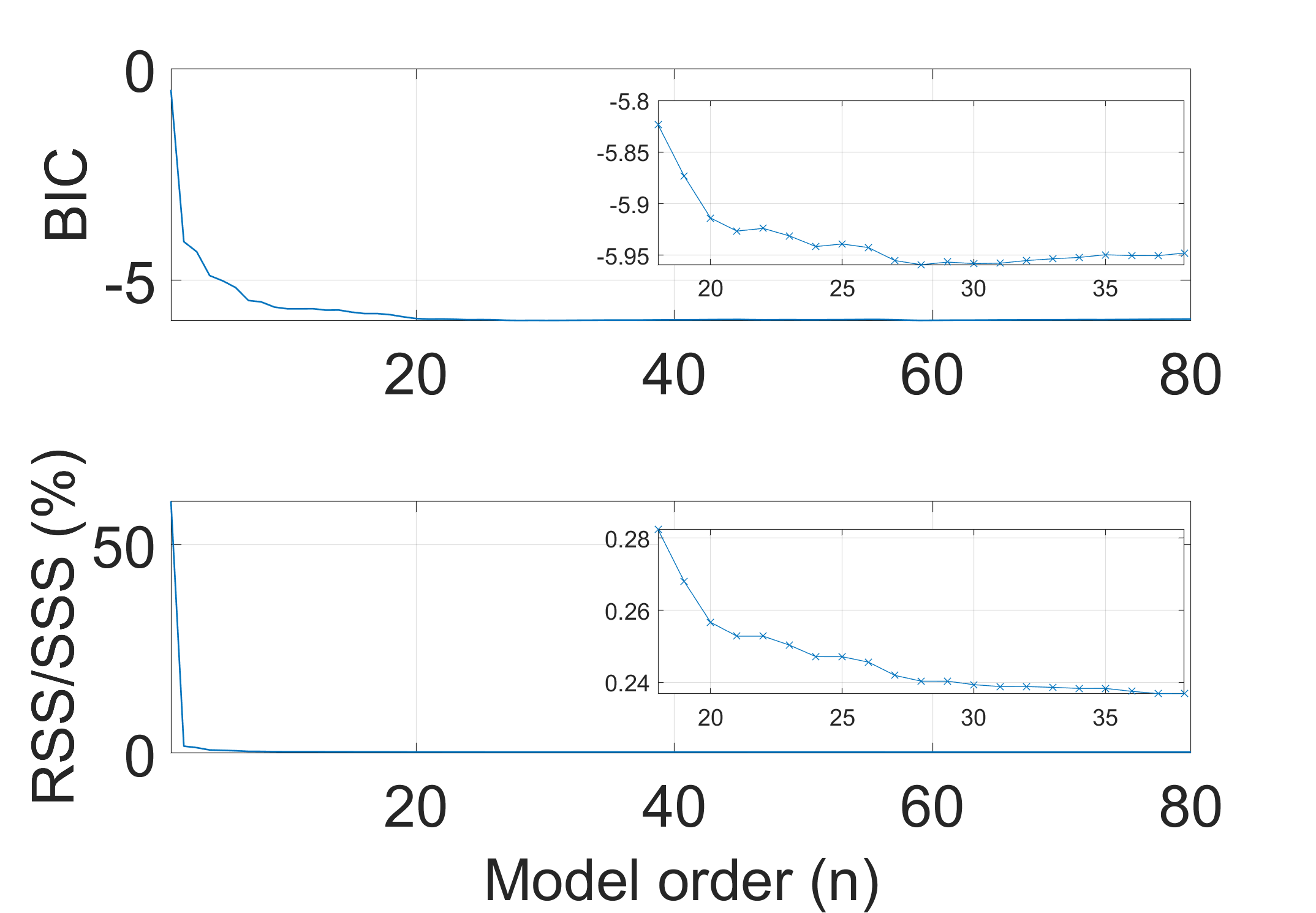}
        \caption{FP-AR model order selection}
        \label{fig:Order selection}
    \end{subfigure}
    \hfill
    \begin{subfigure}[t]{0.48\textwidth}
        \includegraphics[width=\textwidth]{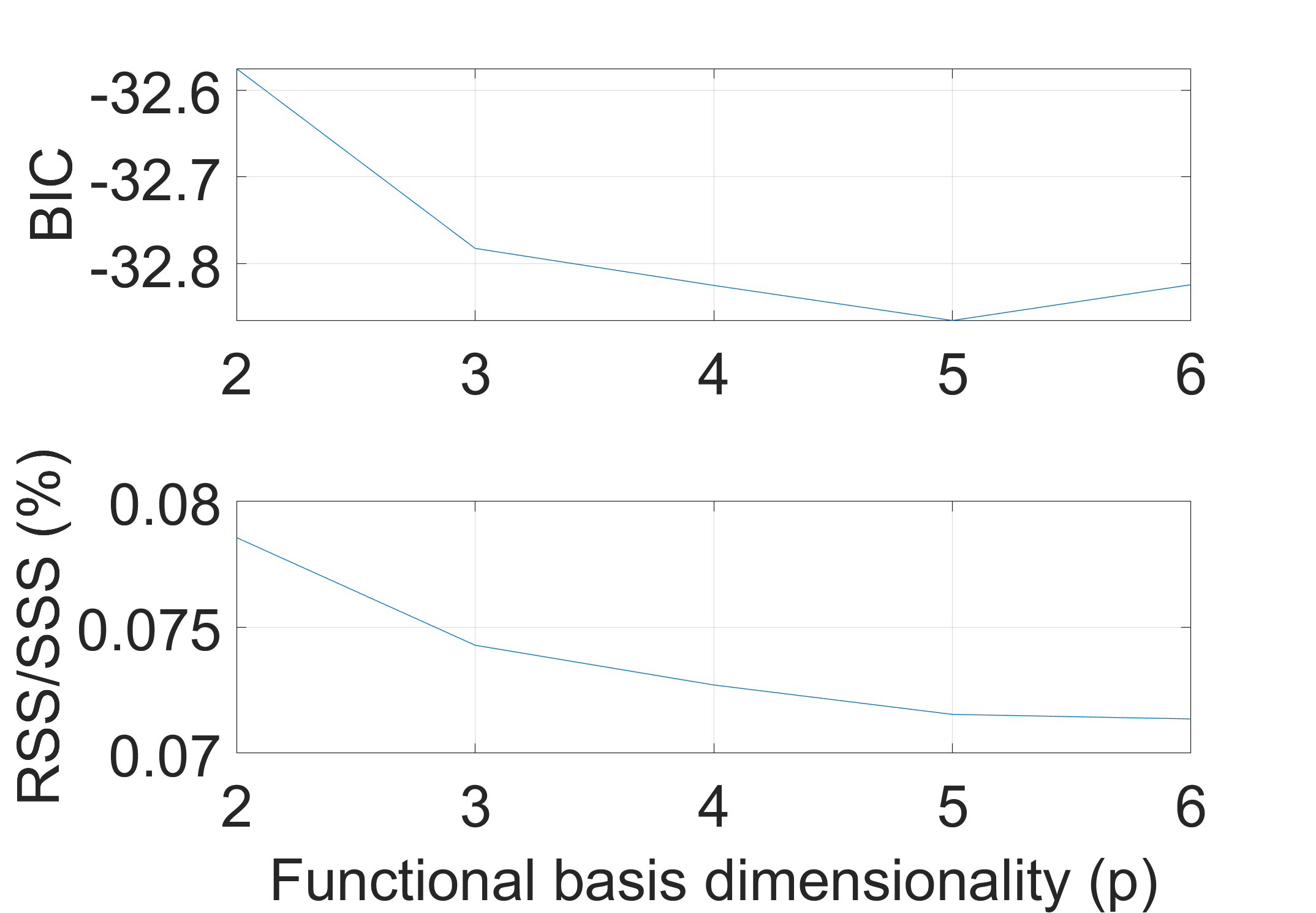}
        \caption{FP-AR model basis dimensionality selection}
        \label{fig:Basis selection}
    \end{subfigure}
    \caption{Selection of the model order and functional basis dimensionality for the FP-AR model constructed from the GyrX (angular velocity) signal of Motor 1.}
    \label{fig:Unpooled Order & Basis selection}
\end{figure}

\begin{figure}[t!]
\centering
\includegraphics[width=0.9\columnwidth]{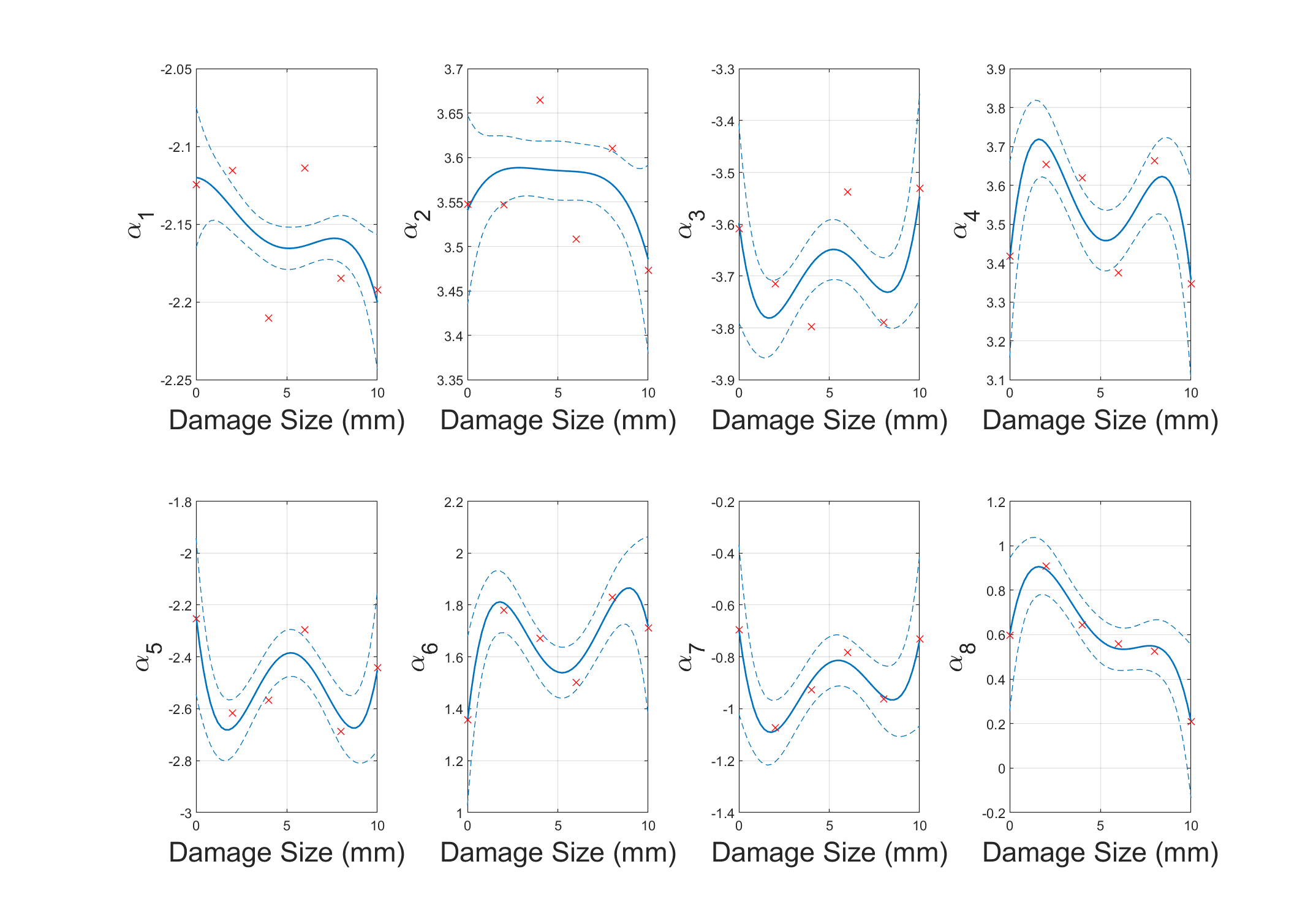}
\caption{FP-AR model parameters versus damage size. The 99\% confidence intervals of the estimated parameters are shown as dashed lines.}
\label{fig:FP parameters}
\end{figure}

Initial AR models are identified separately for each damage state using standard identification procedures. These serve as a reference for selecting the model order and for informing the structure of the FP-AR model. A total of $M = 6$ cross-sections (experimental data records), each corresponding to a distinct damage level from 0 mm to 10 mm, are used to span the full damage range. The AR order is selected as $n_a = 34$ based on the minimum BIC, as shown in Fig.~\ref{fig:Order selection}. The functional basis dimensionality is determined by evaluating up to six Chebyshev polynomials of the second kind. The optimal basis dimensionality, $r = 5$, is selected according to the lowest BIC and RSS/SSS ratio (see Fig.~\ref{fig:Basis selection}). Figure~\ref{fig:FP parameters} displays the first eight FP-AR model parameters, $a_1(k)$ to $a_8(k)$, as functions of the damage size. The red dots represent the parameter estimates at the $M=6$ discrete damage levels based on independent AR models. 

\begin{figure*}[t!]
	\centering
	\begin{subfigure}{.3\textwidth}
		\centering
		\includegraphics[width=\textwidth]{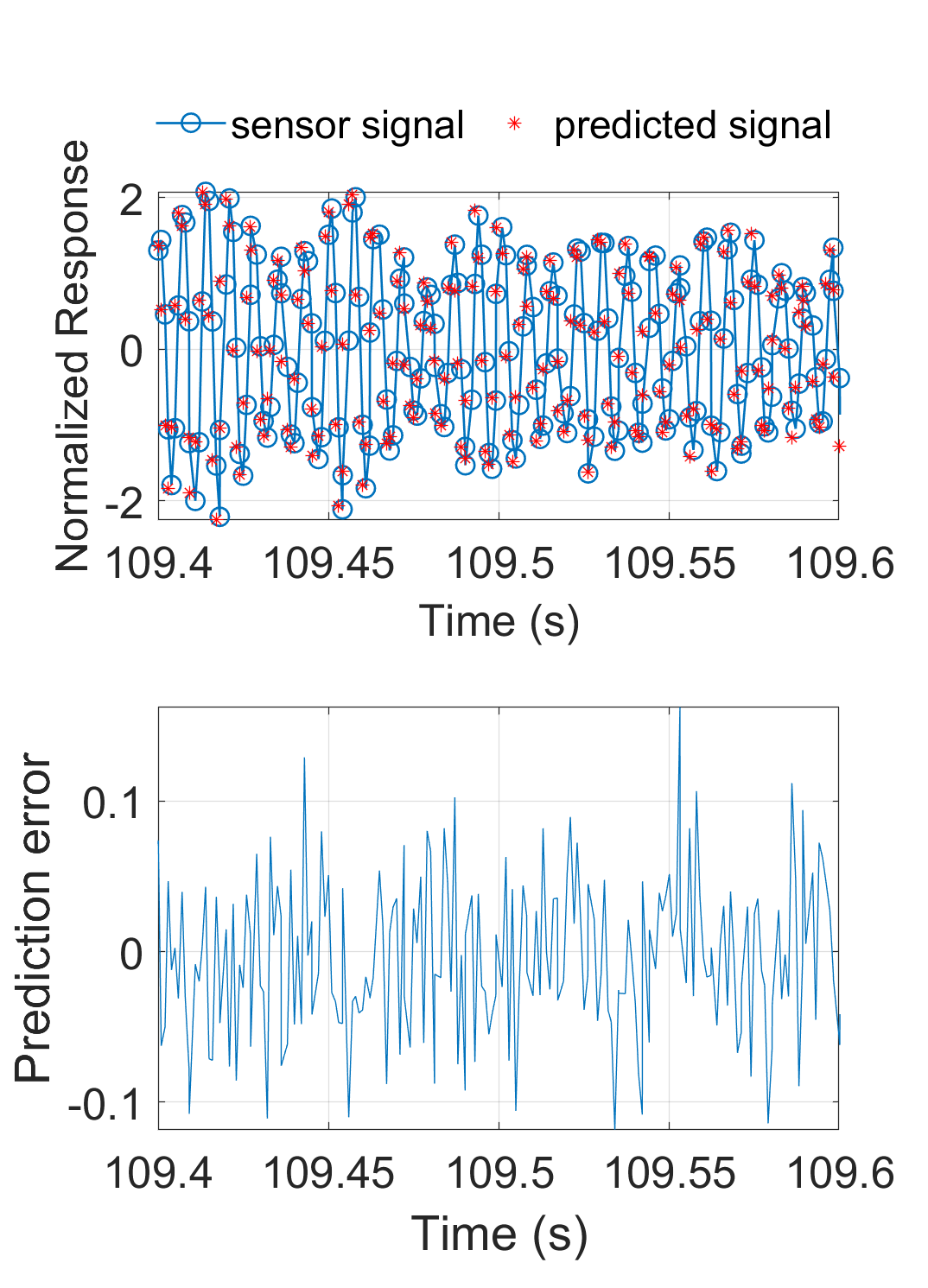}
		\caption{}
		\label{fig:damage 1 pre}
	\end{subfigure}
	\begin{subfigure}{.3\textwidth}
		\centering
		\includegraphics[width=\textwidth]{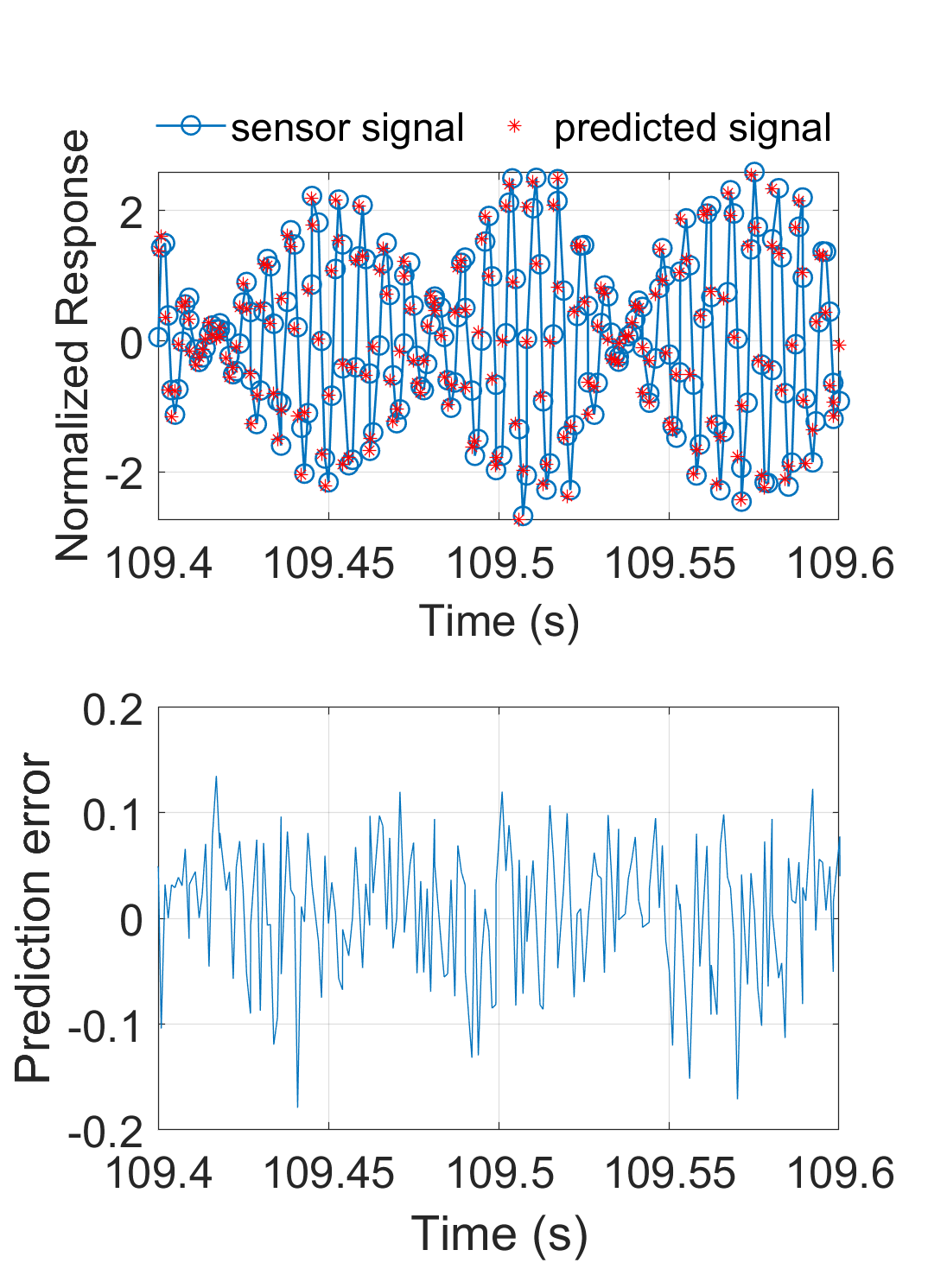}
		\caption{}
		\label{fig:damage 3 pre}
	\end{subfigure}
    \begin{subfigure}{.3\textwidth}
		\centering
		\includegraphics[width=\textwidth]{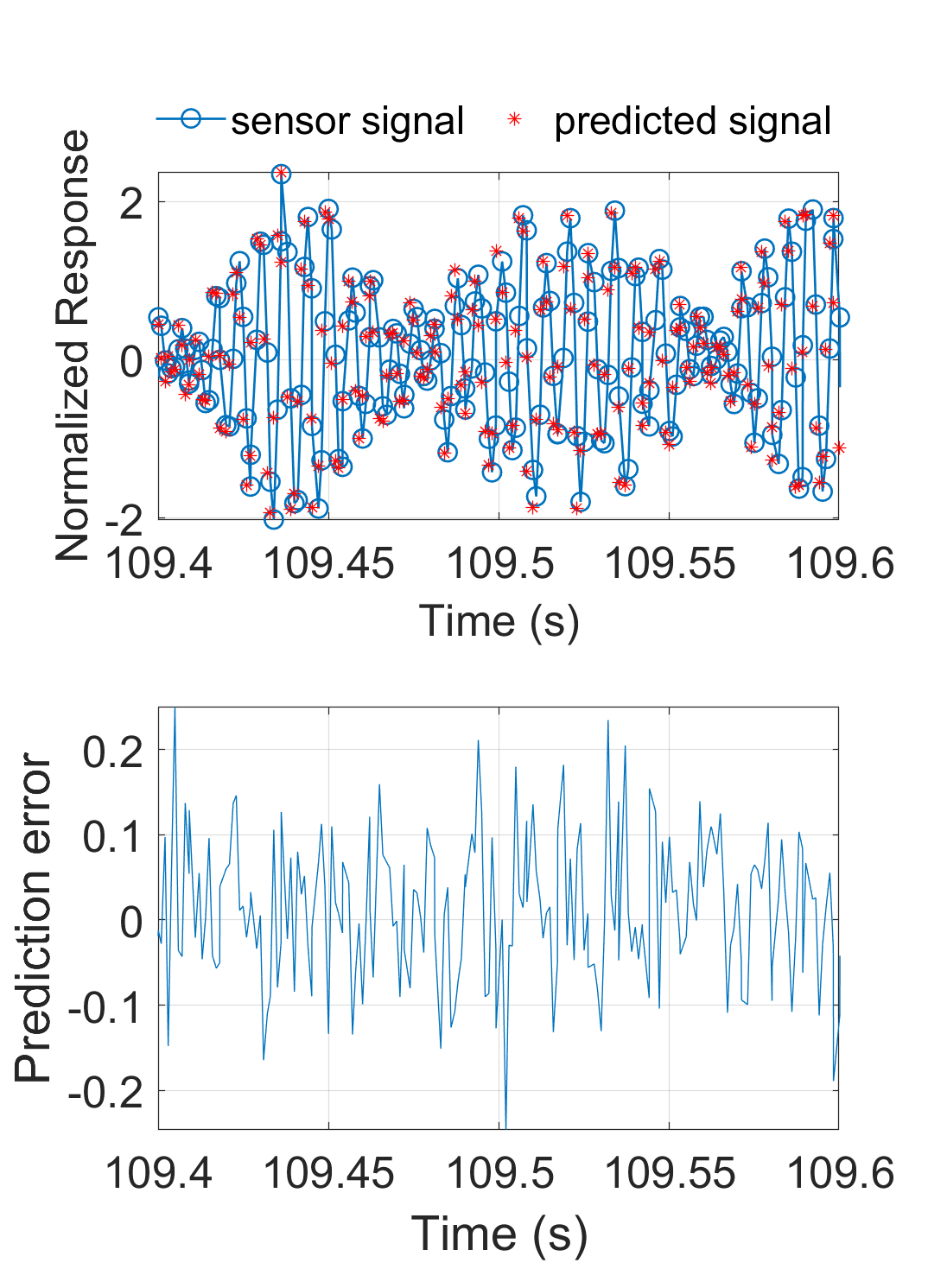}
		\caption{}
		\label{fig:damage 5 pre}
	\end{subfigure}
	\caption{Signal prediction over a 0.2 s window based on the identified FP-AR model: (a) healthy state; (b) 6 mm damage; (c) 10 mm damage.}
    \label{fig:damage all pre}
\end{figure*}

Following the FP-AR model identification, signal predictions are generated as illustrated in Fig.~\ref{fig:damage all pre}. The plots depict the model-based one-step-ahead predictions for the healthy state and damage states 3 (6 mm) and 5 (10 mm) over a 0.2 s window of the training segment (109.4--109.6 s), corresponding to 200 samples of the 4{,}000-sample record. The predicted signals closely follow the measured data, demonstrating the model's ability to reconstruct the short-term dynamics based on the selected order and functional basis dimensionality. This consistency confirms the effectiveness of the identified parameters in representing the system behavior across the different damage conditions.

\begin{figure}[htbp]
  \centering

  \begin{subfigure}[t]{0.95\textwidth}
    \centering
    \begin{minipage}{0.48\textwidth}
      \centering
      \includegraphics[width=\linewidth]{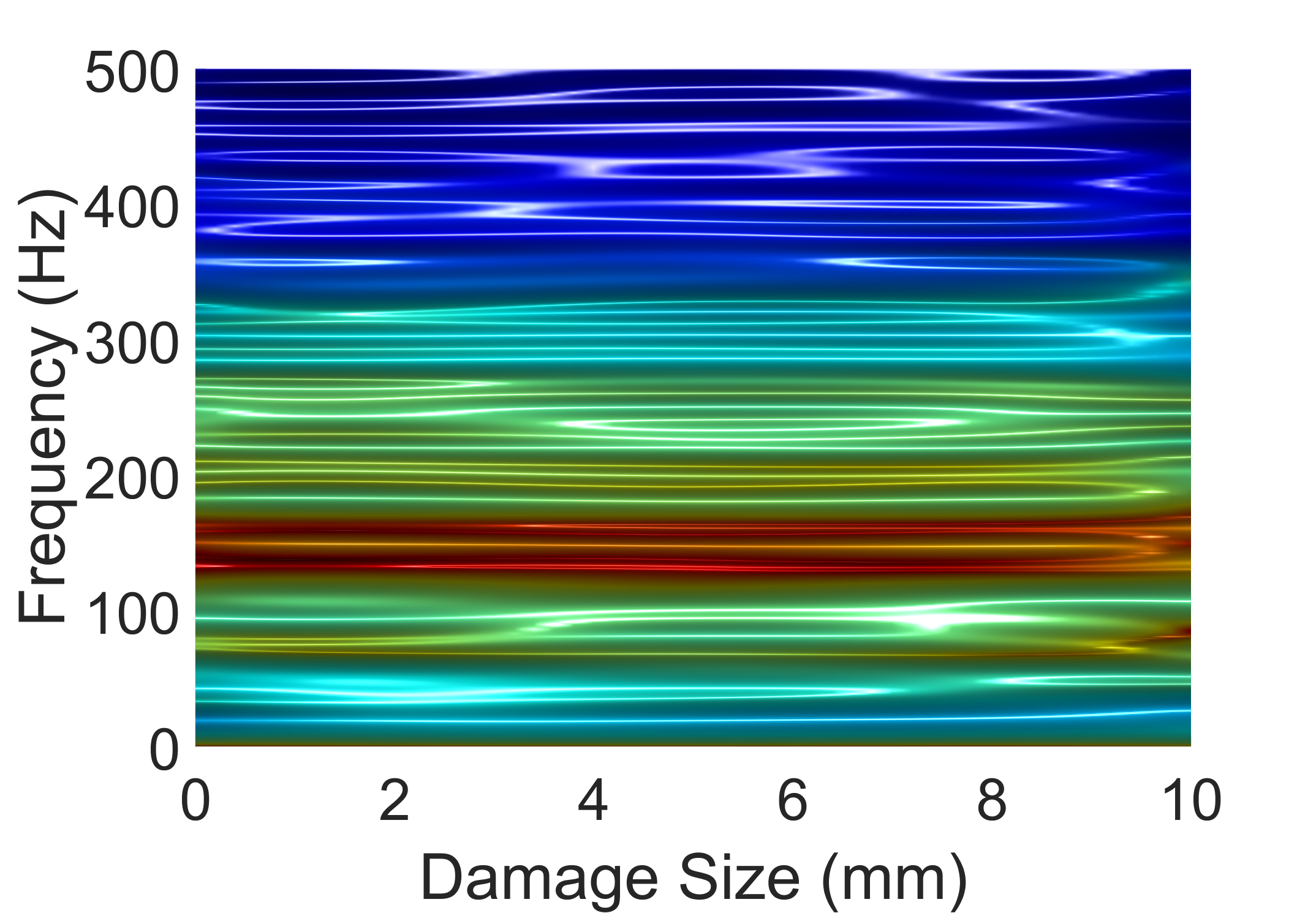}
    \end{minipage}\hfill
    \begin{minipage}{0.48\textwidth}
      \centering
      \includegraphics[width=\linewidth]{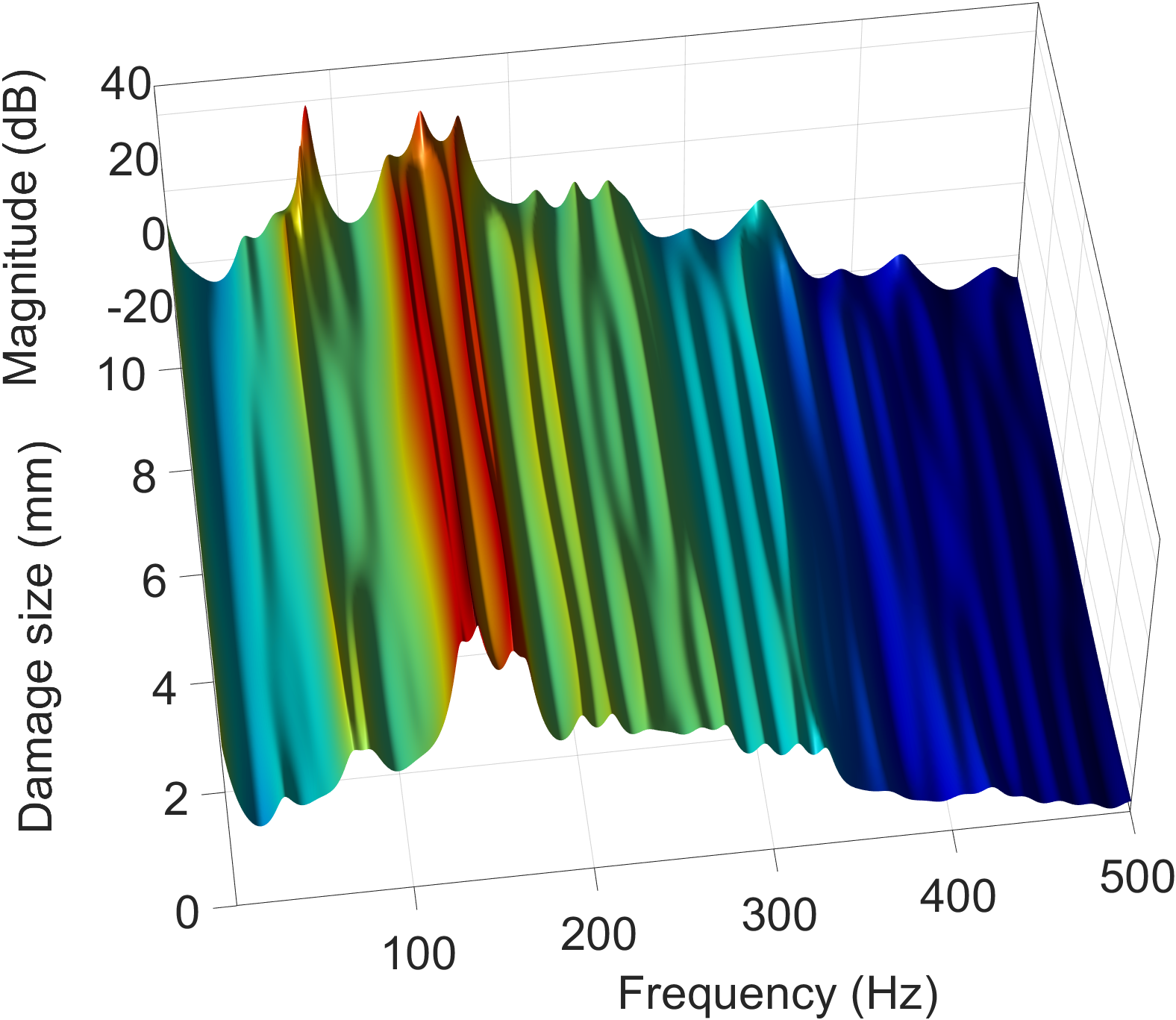}
    \end{minipage}
    \caption{Motor 1 FP-AR model-based spectral estimation (left: 2D; right: 3D)}
    \label{fig:m1_pair}
  \end{subfigure}

  \vspace{0.8em}

  \begin{subfigure}[t]{0.95\textwidth}
    \centering
    \begin{minipage}{0.48\textwidth}
      \centering
      \includegraphics[width=\linewidth]{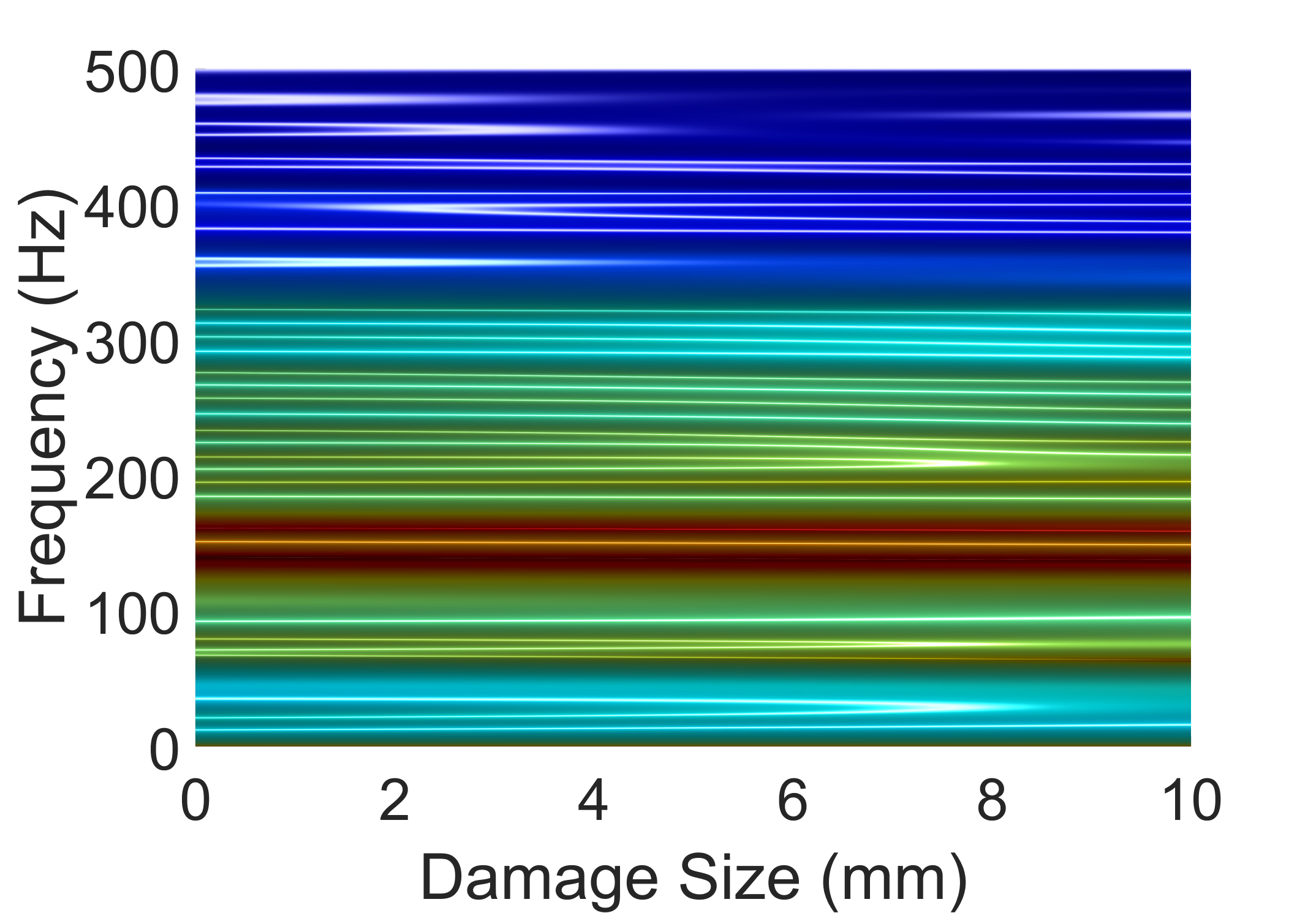}
    \end{minipage}\hfill
    \begin{minipage}{0.48\textwidth}
      \centering
      \includegraphics[width=\linewidth]{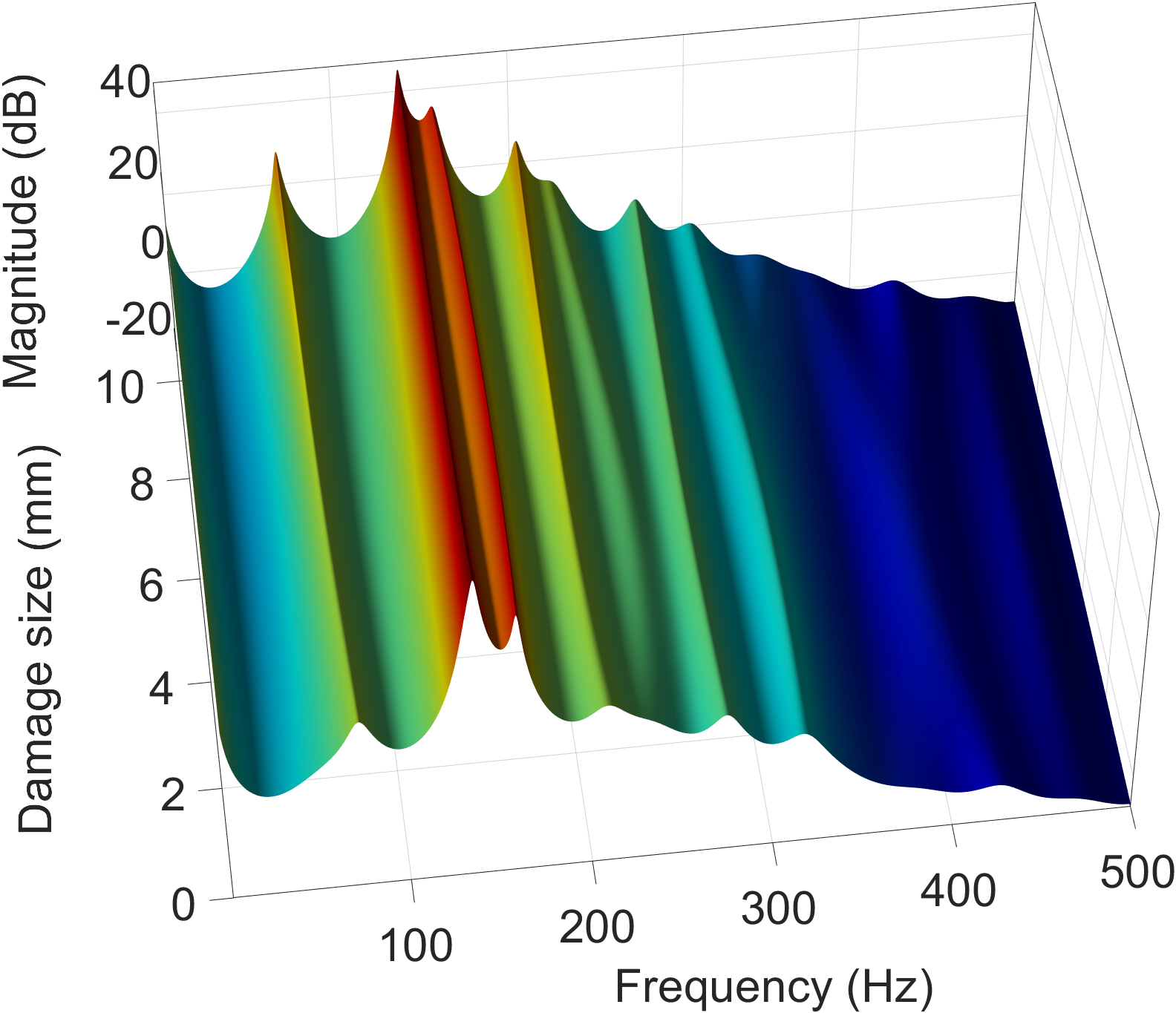}
    \end{minipage}
    \caption{Motor 3 FP-AR model-based spectral estimation (left: 2D; right: 3D)}
    \label{fig:m3_pair}
  \end{subfigure}

  \vspace{0.8em}

  \begin{subfigure}[t]{0.95\textwidth}
    \centering
    \begin{minipage}{0.48\textwidth}
      \centering
      \includegraphics[width=\linewidth]{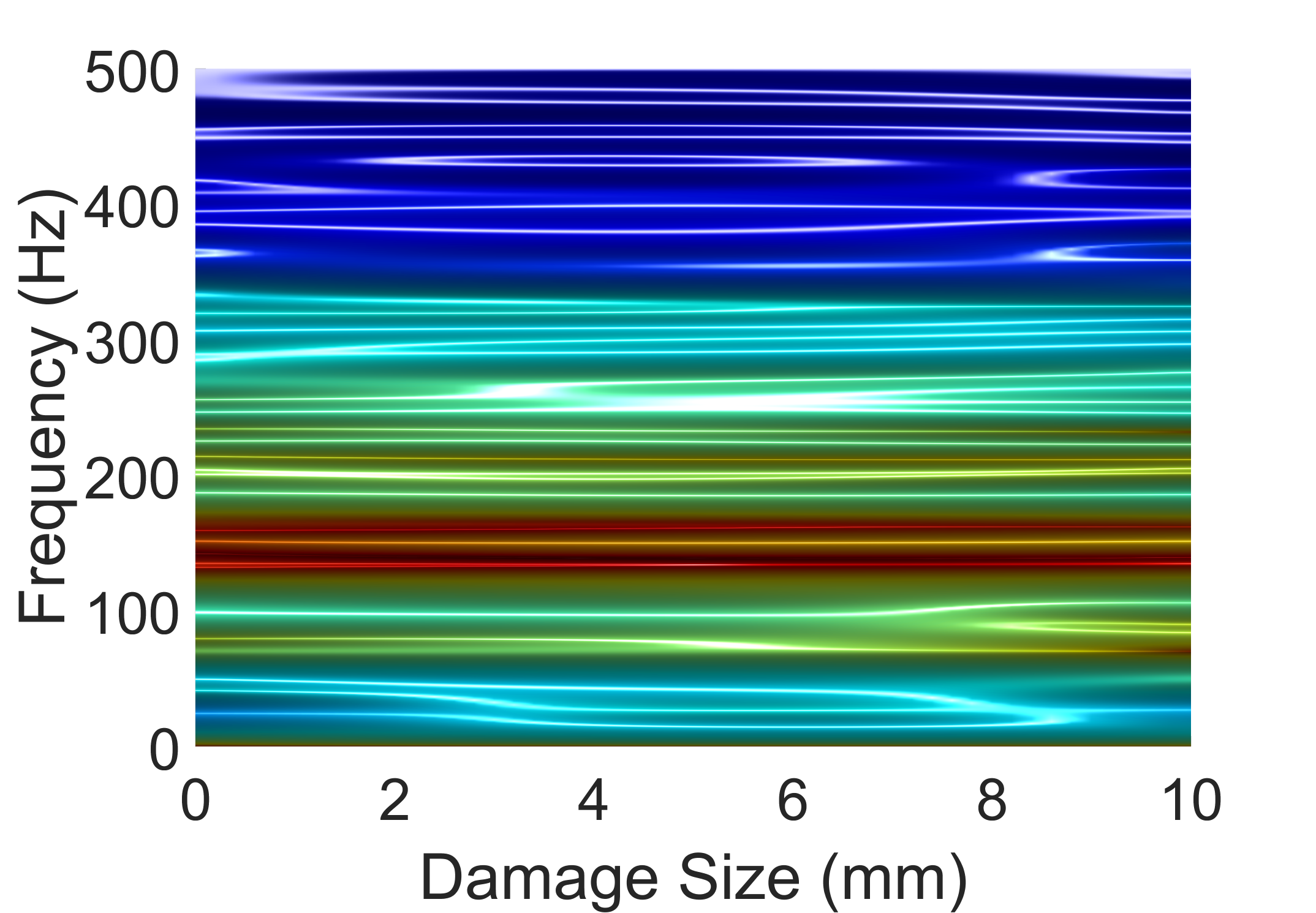}
    \end{minipage}\hfill
    \begin{minipage}{0.48\textwidth}
      \centering
      \includegraphics[width=\linewidth]{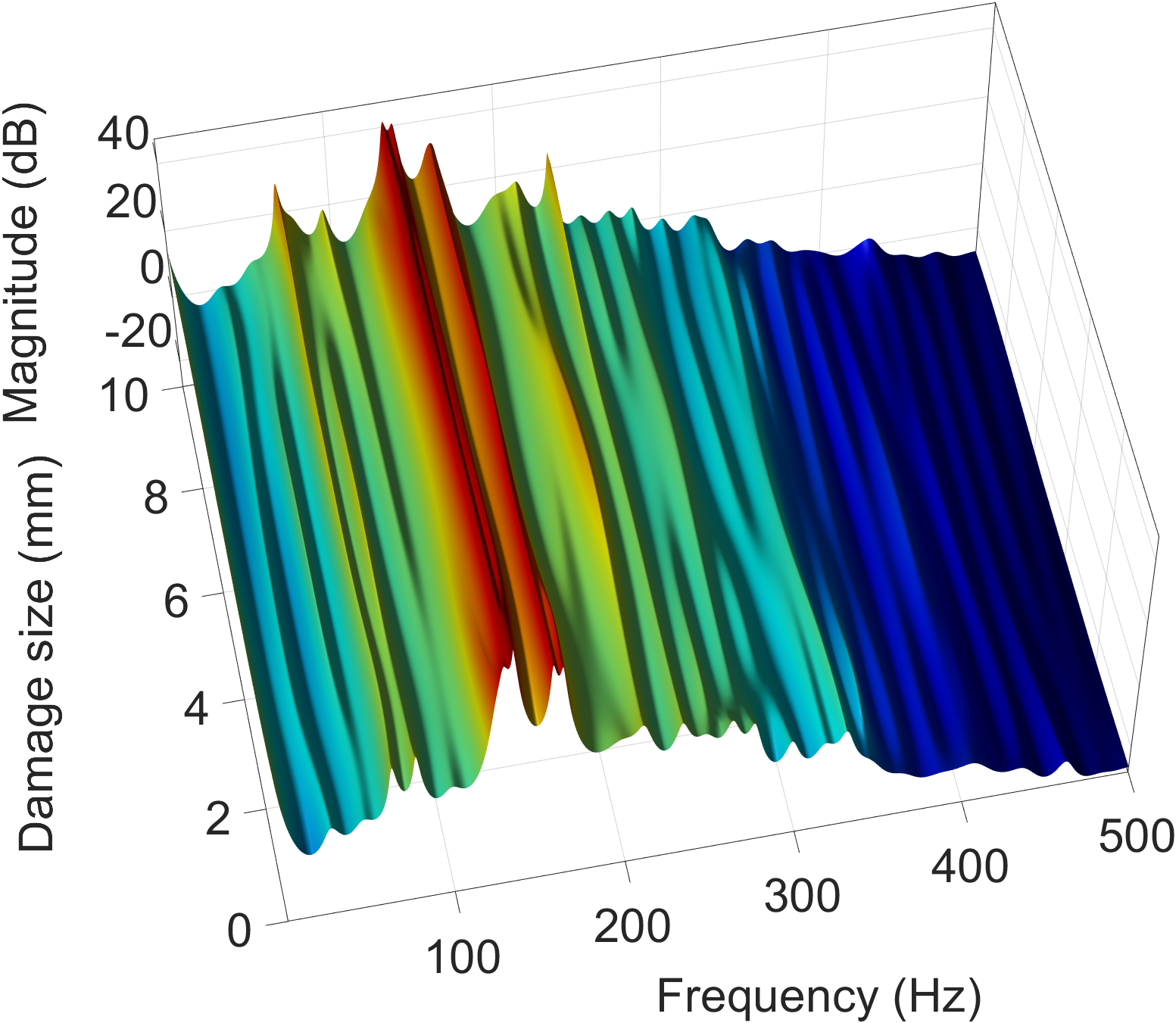}
    \end{minipage}
    \caption{Motor 6 FP-AR model-based spectral estimation (left: 2D; right: 3D)}
    \label{fig:m6_pair}
  \end{subfigure}

  \caption{FP-AR model-based spectral estimation for the GyrX signal: each row corresponds to one motor (Motors 1, 3, and 6), with the left panel reporting the 2D parametric spectrum (top view) and the right panel the corresponding 3D parametric spectrum.}
  \label{fig:ThreeStages_PSD}
\end{figure}

FP-AR models are subsequently constructed individually for each motor using 4 s data segments (4{,}000 samples) extracted from the GyrX signal recorded during flight. For Motor 1, the 106--110 s segment is used, while for Motors 3 and 6, the 90--94 s interval is selected. 
The selected FP-AR model configurations are FP-AR$(33)_5$ for Motor 1, FP-AR$(27)_3$ for Motor 3, and FP-AR$(62)_5$ for Motor 6, with FP-AR$(n_a)_r$ designating an FP-AR model of order $n_a$ and functional basis dimensionality $r$. Fig.~\ref{fig:ThreeStages_PSD} presents the resulting parametric damage topology representations, illustrating the spectral magnitude as a function of both frequency and damage size for each motor. In comparison to the non-parametric results of Fig.~\ref{fig:3D-PSD}, the FP-AR models produce significantly cleaner representations with reduced noise influence, allowing for a more detailed and reliable characterization of the damage-dependent spectral evolution.

These models effectively capture the systematic progression of the spectral features as the damage increases, laying the foundation for quantitative damage assessment. By modeling the dynamic response as a continuous function of the damage magnitude, rather than as a set of discrete states, the approach enables finer resolution in damage quantification and supports the derivation of confidence intervals for statistical reliability, as further discussed.

\section{Damage diagnosis results} \label{sec:diagnosis}

\subsection{Local FP-AR model-based diagnosis}

Damage detection, identification, and quantification are carried out for propeller damage of varying severity (0--10 mm) occurring independently at any one of the three monitored motor locations. All test cases involve one of the six predefined damage levels (0, 2, 4, 6, 8, and 10 mm) applied to a single motor at a time. The analysis is repeated across all three motor positions (Motors 1, 3, and 6), using the respective AccX signal as the response measurement channel. 

For damage detection, particular attention is placed on distinguishing the healthy case (0 mm) from all damaged conditions, as accurate separation of the healthy case is essential for reliable damage onset detection. Once damage is detected, the identification and quantification tasks focus on determining the affected motor and differentiating among the five damaged states (2--10 mm). The 0 mm condition is excluded from this phase, as the objective shifts from detecting the presence of damage to characterizing its location and severity.

All detection, identification, and quantification tasks are based on single-channel AccX measurements, with no cross-signal information used; the same procedure applies to any of the six available signals, and the AccX signal is selected here for demonstration purposes. The inherent similarity of the spectral content across certain cases, especially at the lower damage levels, poses challenges for diagnosis, underscoring the need for high-fidelity models and statistically robust decision metrics. The summary results for all signals are presented in Table \ref{tab:local-fpar}.

\begin{table}[b!]
  \centering
  \begin{threeparttable}
    \caption{Single-segment (local) FP-AR model structures}
    \label{tab:local-fpar}
    \small
    \begin{tabular}{@{} l l l r r r @{}} 
      \toprule
      Signal & Motor & Segment used (s) & Order $n_a$ & Basis $r$ & \#Params \\
      \midrule
      AccX & M1 & 74--78  & 33 & 5 & 165 \\
            & M3 & 74--78  & 35 & 2 & 70  \\
            & M6 & 74--78  & 36 & 4 & 144 \\
      AccY & M1 & 106--110 & 34 & 6 & 204 \\
            & M3 & 122--126 & 35 & 5 & 175 \\
            & M6 & 74--78   & 54 & 3 & 162 \\
      AccZ & M1 & 106--110 & 57 & 5 & 285 \\
            & M3 & 110--114 & 58 & 2 & 116 \\
            & M6 & 130--134 & 53 & 3 & 159 \\
      GyrX & M1 & 106--110 & 57 & 5 & 285 \\
            & M3 & 90--94 & 27 & 3 & 81 \\
            & M6 & 90--94 & 62 & 5 & 310 \\
      GyrY & M1 & 106--110 & 33 & 6 & 198 \\
            & M3 & 110--114 & 40 & 5 & 200 \\
            & M6 & 122--126 & 39 & 6 & 234 \\
      GyrZ & M1 & 106--110 & 27 & 6 & 162 \\
            & M3 & 90--94 & 38 & 2 & 76  \\
            & M6 & 90--94  & 38 & 3 & 114 \\
      \bottomrule
    \end{tabular}
   \begin{tablenotes}\footnotesize
      \item \textit{Note:} All local FP-AR models are constructed using 4 s of flight data. 
            The number of projection coefficients is computed as \(\#\mathrm{Params}= n_a \times r\).
    \end{tablenotes}
  \end{threeparttable}
\end{table}

\paragraph{Damage detection}

\begin{figure}[t!]
\centering

\begin{subfigure}[t]{0.92\columnwidth}
  \centering
  \includegraphics[width=\linewidth]{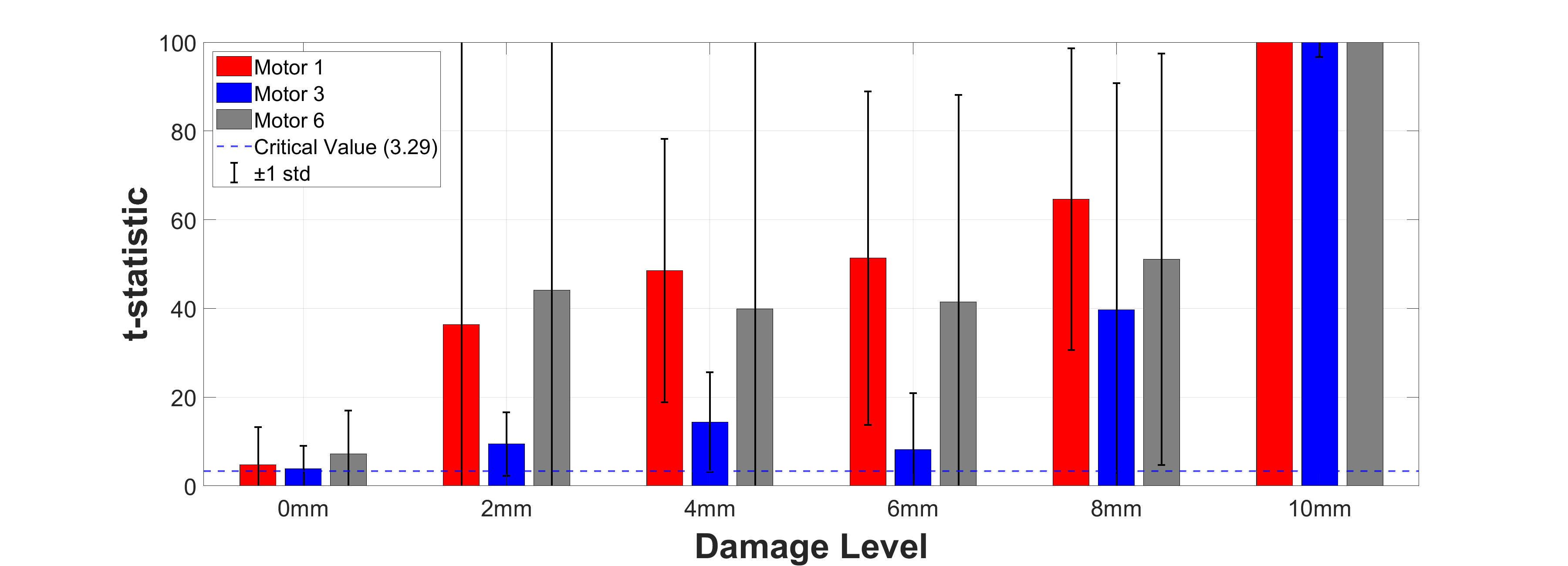}
  \caption{Motor~1, AccX (single segment 90--94\,s)}
  \label{fig:unpooled_dd_m1}
\end{subfigure}
\vspace{0.6em}

\begin{subfigure}[t]{0.92\columnwidth}
  \centering
  \includegraphics[width=\linewidth]{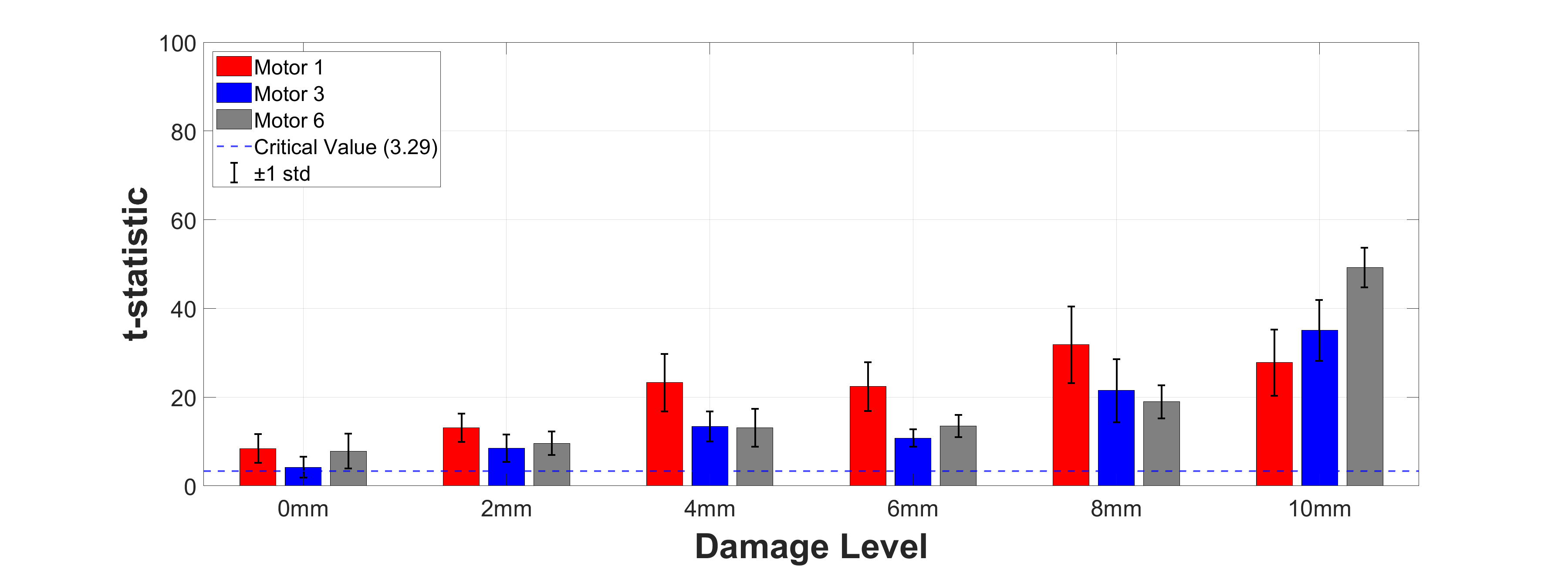}
  \caption{Motor~3, AccX (single segment 90--94\,s)}
  \label{fig:unpooled_dd_m3}
\end{subfigure}
\vspace{0.6em}

\begin{subfigure}[t]{0.92\columnwidth}
  \centering
  \includegraphics[width=\linewidth]{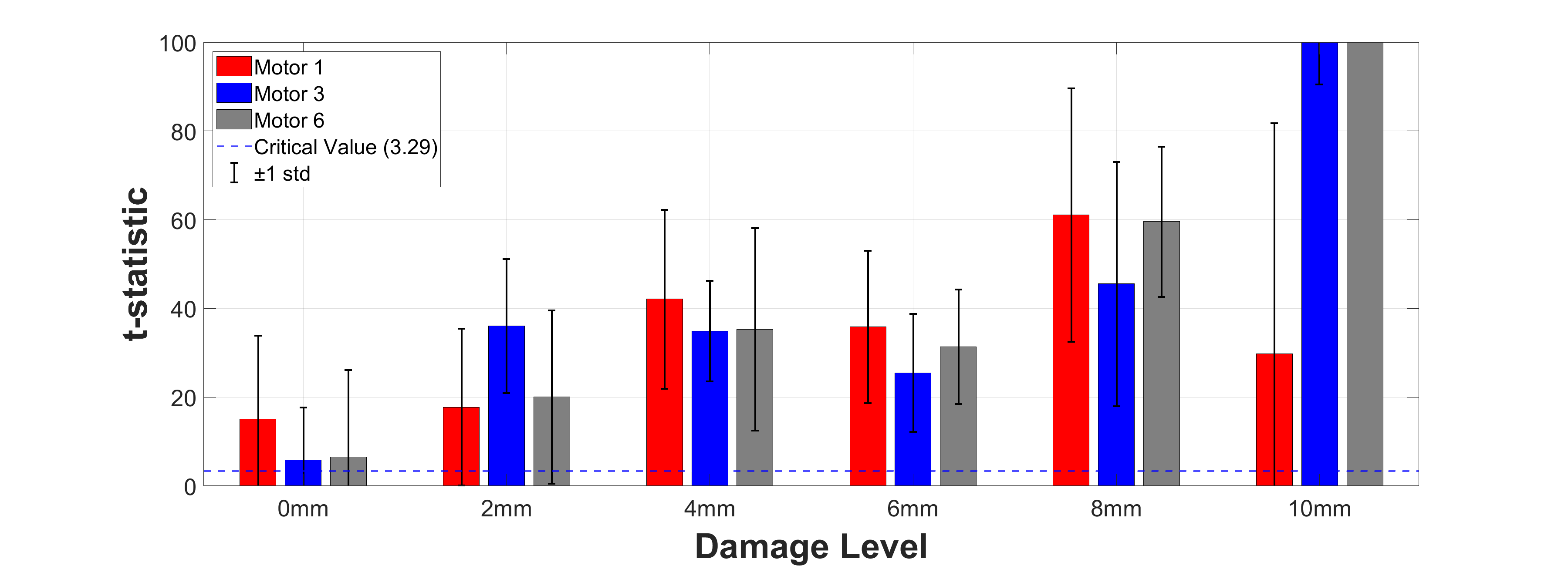}
  \caption{Motor~6, AccX (single segment 50--54\,s)}
  \label{fig:unpooled_dd_m6}
\end{subfigure}

\caption{Local (unpooled) FP-AR damage detection using $t$-statistics. For each motor, bars show the mean $t$-statistic across 21 test sets from 58--142\,s; error bars denote $\pm 1$ standard deviation. The dashed line indicates the critical point at risk level $\alpha = 0.001$. Damage is declared when the $t$-statistic exceeds the critical threshold.}
\label{fig:unpooled_dd_all}
\end{figure}

Damage detection results using the AccX signals of the three motors are presented in the sequel. As illustrated in Fig.~\ref{fig:unpooled_dd_all}, a local FP-AR model is constructed individually for each motor using a 4 s segment of its AccX signal. The resulting model is then applied to test data drawn from all three motors, specifically across 21 independent segments (58--142 s), for computing the associated $t$-statistics. The figure presents the mean and standard deviation of the $t$-statistics obtained from these 21 test sets. A risk level of $\alpha = 0.001$ is selected to define the corresponding critical threshold for damage detection.

Fig.~\ref{fig:unpooled_dd_all}(c) demonstrates representative results based on the Motor 6 AccX data, clearly distinguishing the damaged conditions (4--10 mm) from the healthy case. In these scenarios, the corresponding $t$-statistics consistently exceed the critical threshold, indicating successful detection of the damage presence. For lower damage levels (e.g., 2 mm), statistical significance is maintained in most test cases, underscoring the method's sensitivity at early damage stages.

However, as also seen in Figs.~\ref{fig:unpooled_dd_all}(a) and~\ref{fig:unpooled_dd_all}(b), there are instances where the $t$-statistics for the healthy (0 mm) case slightly exceed the critical threshold, corresponding to occasional false alarms. These may stem not only from motor-to-motor variability, but also from the characteristics of the specific data segment used to build the reference model. Nevertheless, it is important to emphasize that the framework consistently detects the presence of damage when it exists, confirming its effectiveness and robustness.

\paragraph{Damage identification}

\begin{figure}[!t]
\centering

\begin{subfigure}[t]{0.92\columnwidth}
  \centering
  \includegraphics[width=\linewidth]{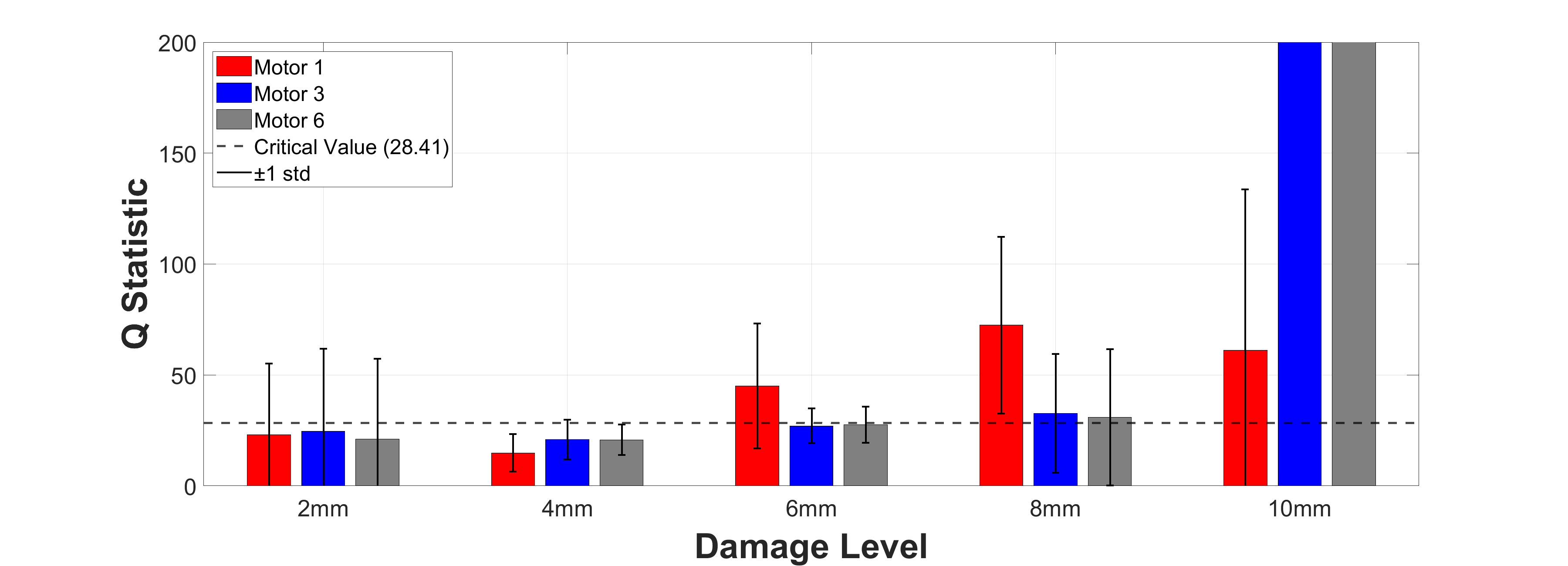}
  \caption{Motor~1, AccX (single segment 74--78\,s)}
  \label{fig:unpooled_loc_m1}
\end{subfigure}
\vspace{0.6em}

\begin{subfigure}[t]{0.92\columnwidth}
  \centering
  \includegraphics[width=\linewidth]{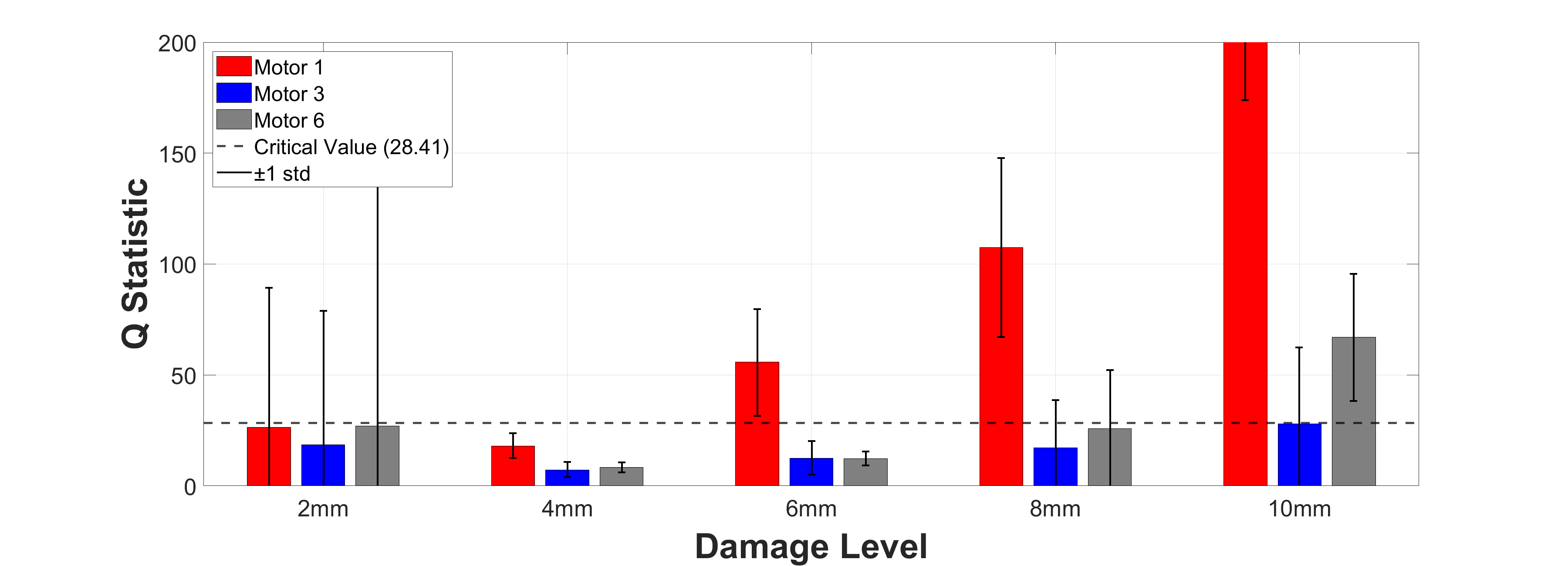}
  \caption{Motor~3, AccX (single segment 74--78\,s)}
  \label{fig:unpooled_loc_m3}
\end{subfigure}
\vspace{0.6em}

\begin{subfigure}[t]{0.92\columnwidth}
  \centering
  \includegraphics[width=\linewidth]{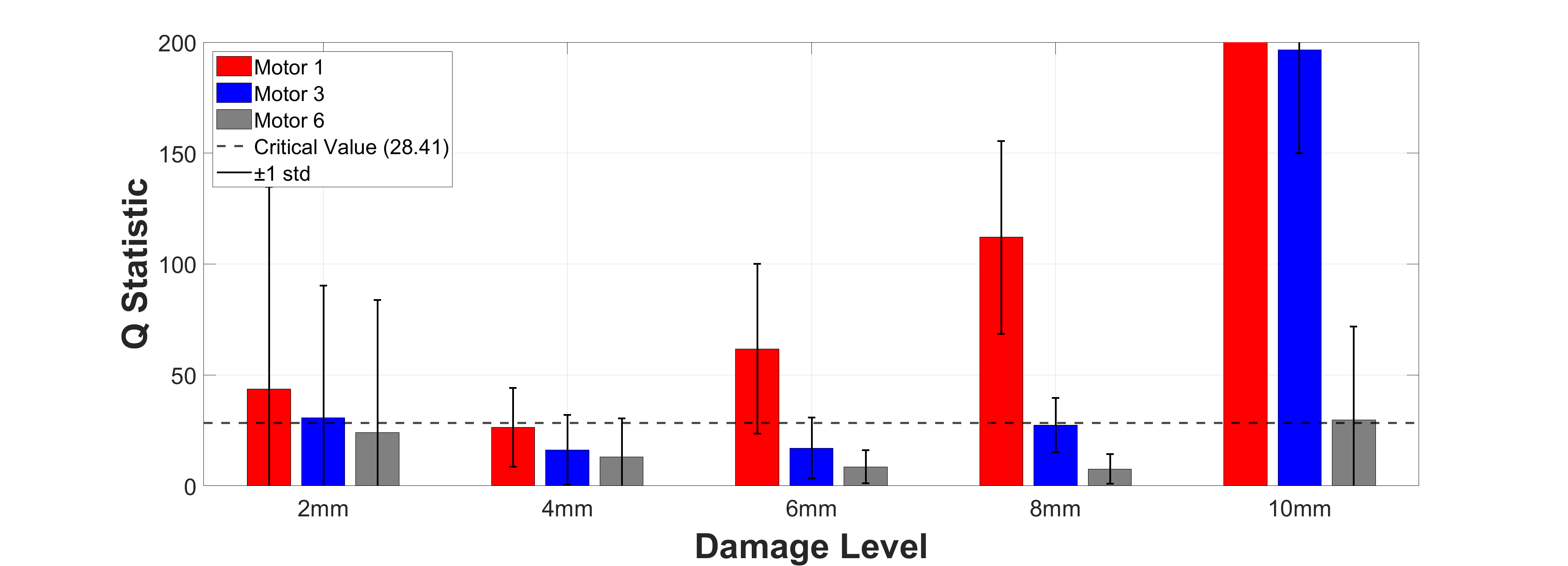}
  \caption{Motor~6, AccX (single segment 74--78\,s)}
  \label{fig:unpooled_loc_m6}
\end{subfigure}

\caption{Local FP-AR motor identification using $Q$-statistics. For each motor, bars show the mean $Q$-statistic across 21 test sets from 58--142\,s; error bars denote $\pm 1$ standard deviation. The dashed line marks the critical value at risk level $\alpha=0.1$. A candidate motor model (M1/M3/M6) is accepted for a given damage level when its $Q$-statistic lies below the critical threshold; otherwise, it is rejected.}
\label{fig:unpooled_loc_all}
\end{figure}

Damage identification, that is, the localization of the damaged propeller, results based on the AccX signals are presented next. The same local FP-AR models used for damage detection, each constructed from a 4 s AccX segment of Motors 1, 3, and 6, are applied here. As shown in Fig.~\ref{fig:unpooled_loc_all}, the $Q$-statistics are computed over 21 test segments (58--142 s) from all motors, and their means and standard deviations are presented. A risk level of $\alpha = 0.1$ defines the threshold for the damaged motor identification.

From the results, it can be observed that the test data originating from the same motor as the model consistently yield the lowest $Q$-statistic. This highlights the model's ability to correctly identify the corresponding motor, and thus localize the damage, despite being trained on a short 4 s segment and without utilizing any cross-motor information. However, it should also be noted that under certain damage conditions, particularly at the lower levels such as 2 mm and 4 mm, the differences in the $Q$-statistics between motors are relatively small, as it would be expected from the non-parametric analysis. Furthermore, due to the variability among models trained on different motors, it becomes challenging to define a unified threshold for damage identification based solely on the $Q$-statistic, especially when distinguishing between different damage levels within the same motor model.

\paragraph{Damage quantification}

\begin{figure*}[t!]
	\centering
	\begin{subfigure}{.32\textwidth}
		\centering
		\captionsetup{width=\linewidth}
		\includegraphics[width=\linewidth]{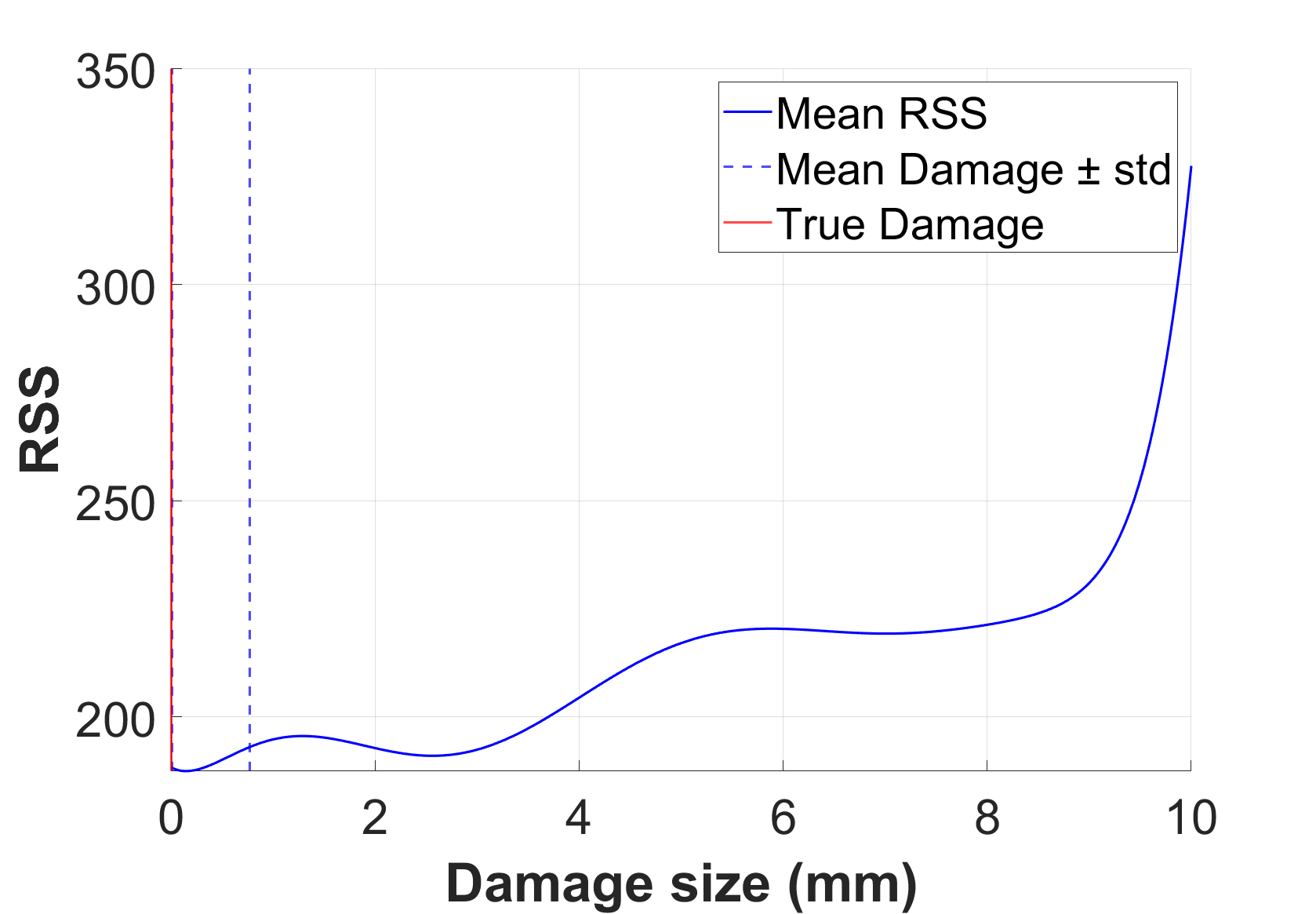}
		\caption{M1, 0 mm damage}
		
	\end{subfigure}
	\begin{subfigure}{.32\textwidth}
		\centering
		\captionsetup{width=\linewidth}
		\includegraphics[width=\linewidth]{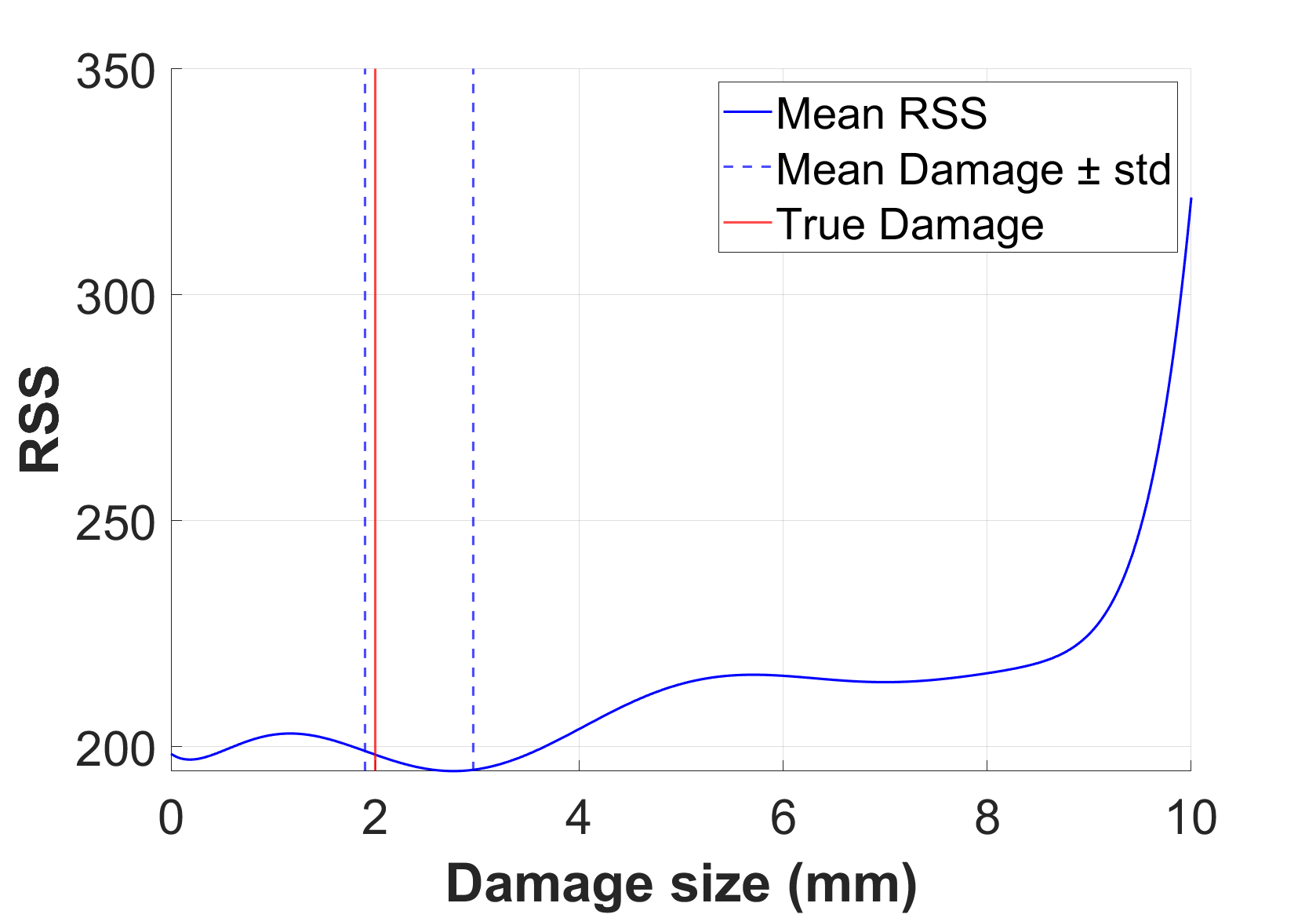}
		\caption{M1, 2 mm damage}
	\end{subfigure}
    \begin{subfigure}{.32\textwidth}
		\centering
		\captionsetup{width=\linewidth}
		\includegraphics[width=\linewidth]{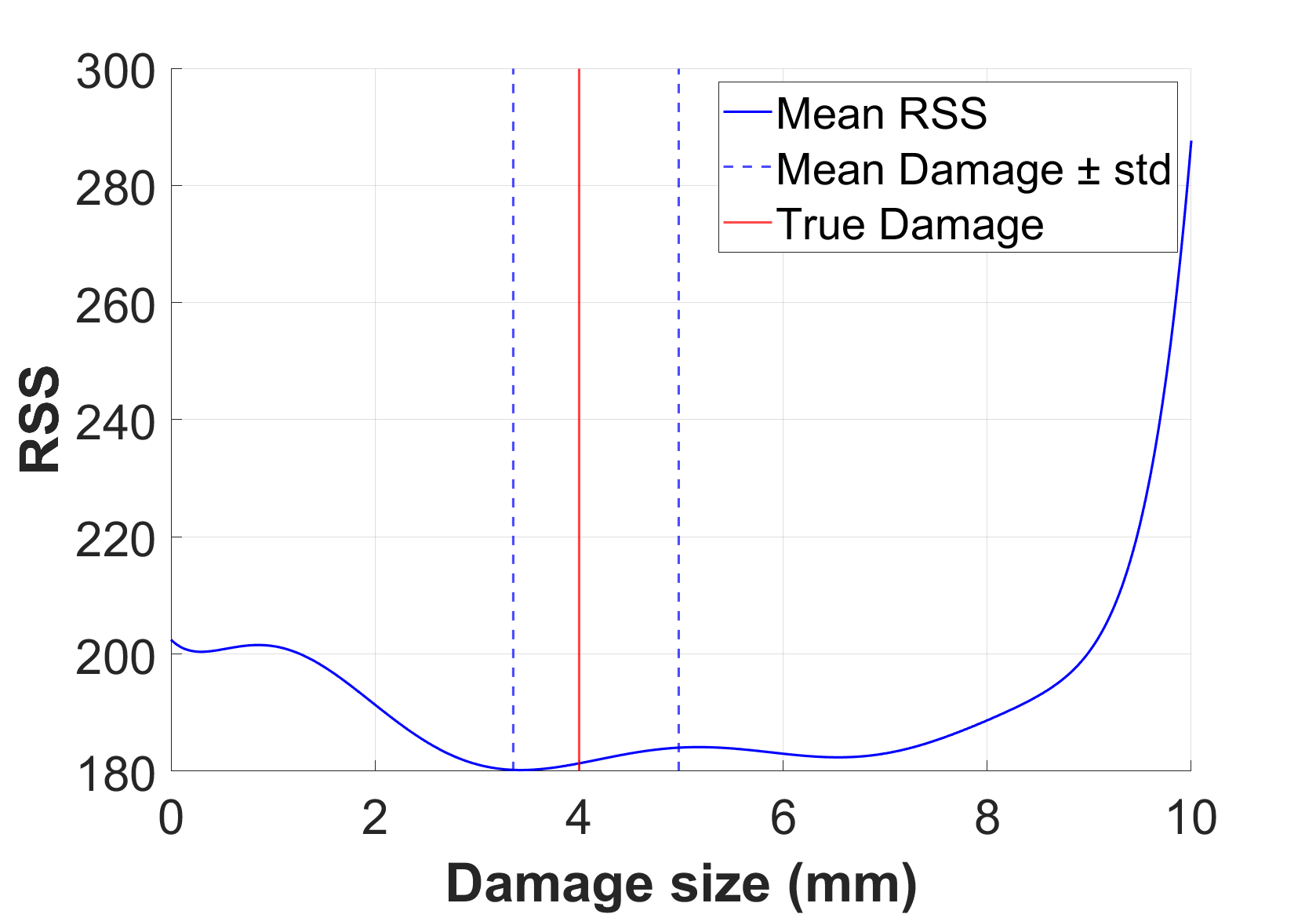}
		\caption{M1, 4 mm damage}
	\end{subfigure}
 
	\begin{subfigure}{.32\textwidth}
		\centering
		\captionsetup{width=\linewidth}
		\includegraphics[width=\linewidth]{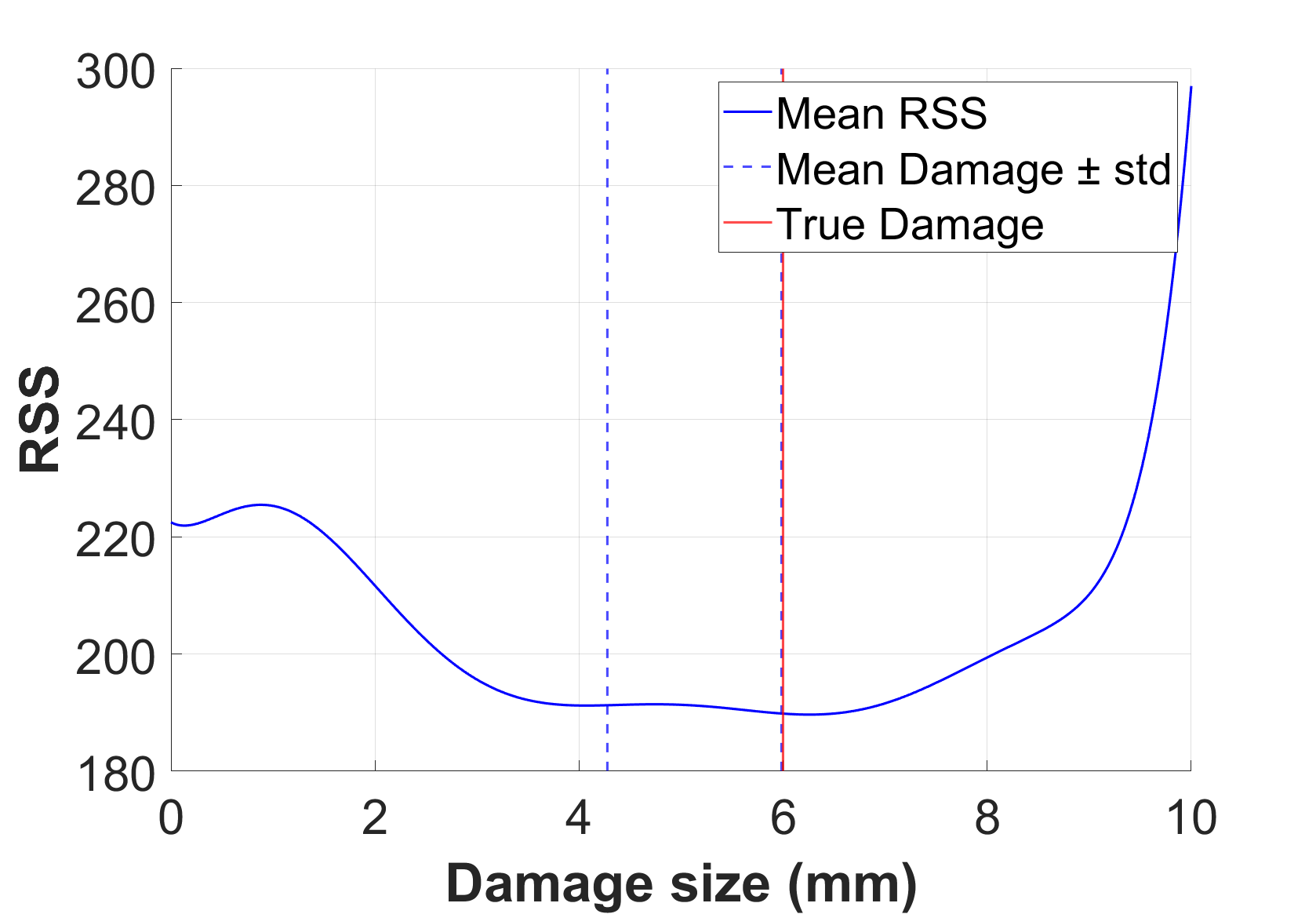}
		\caption{M1, 6 mm damage}
		
	\end{subfigure}
	\begin{subfigure}{.32\textwidth}
		\centering
		\captionsetup{width=\linewidth}
		\includegraphics[width=\linewidth]{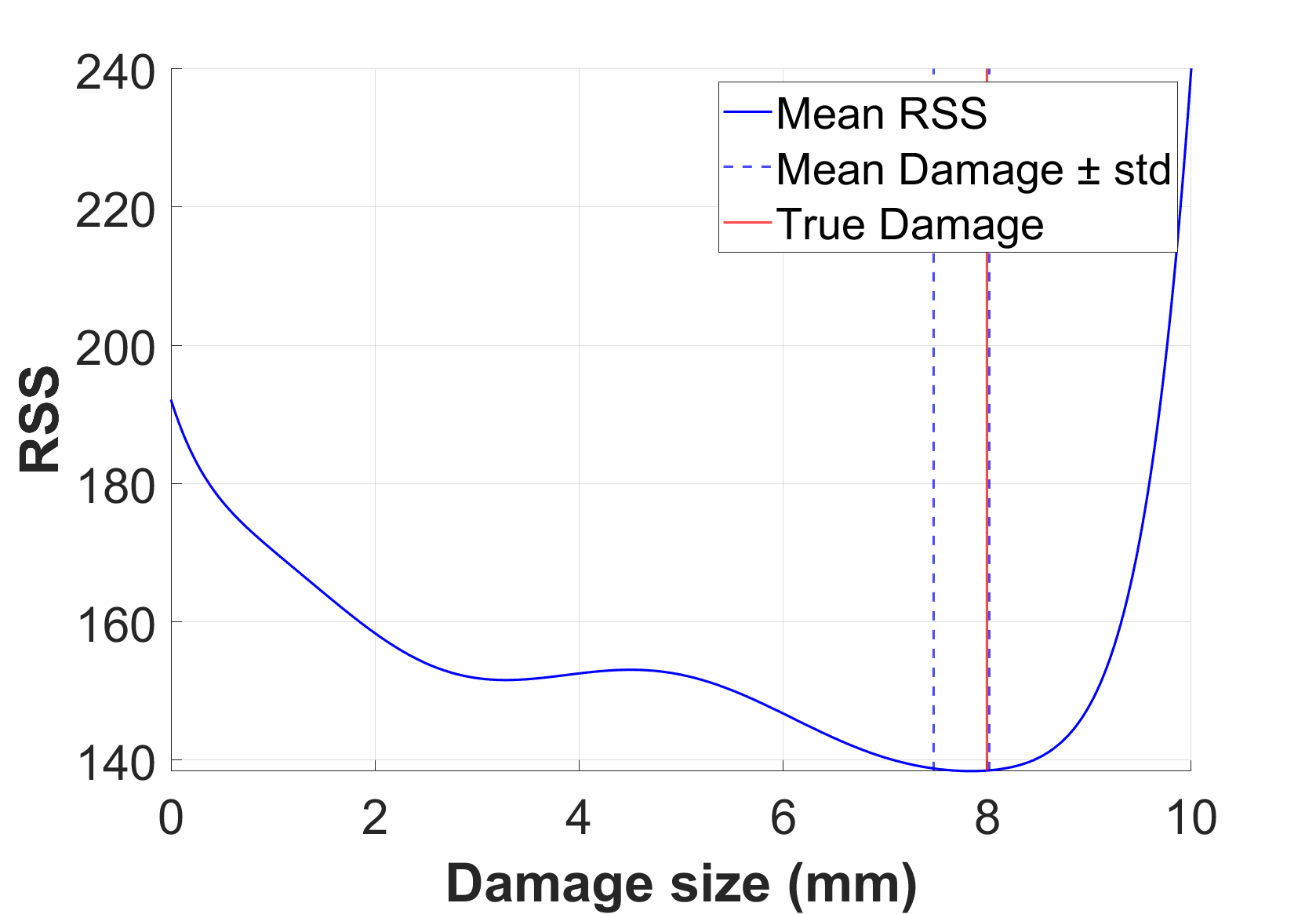}
		\caption{M1, 8 mm damage}
	\end{subfigure}
    \begin{subfigure}{.32\textwidth}
		\centering
		\captionsetup{width=\linewidth}
		\includegraphics[width=\linewidth]{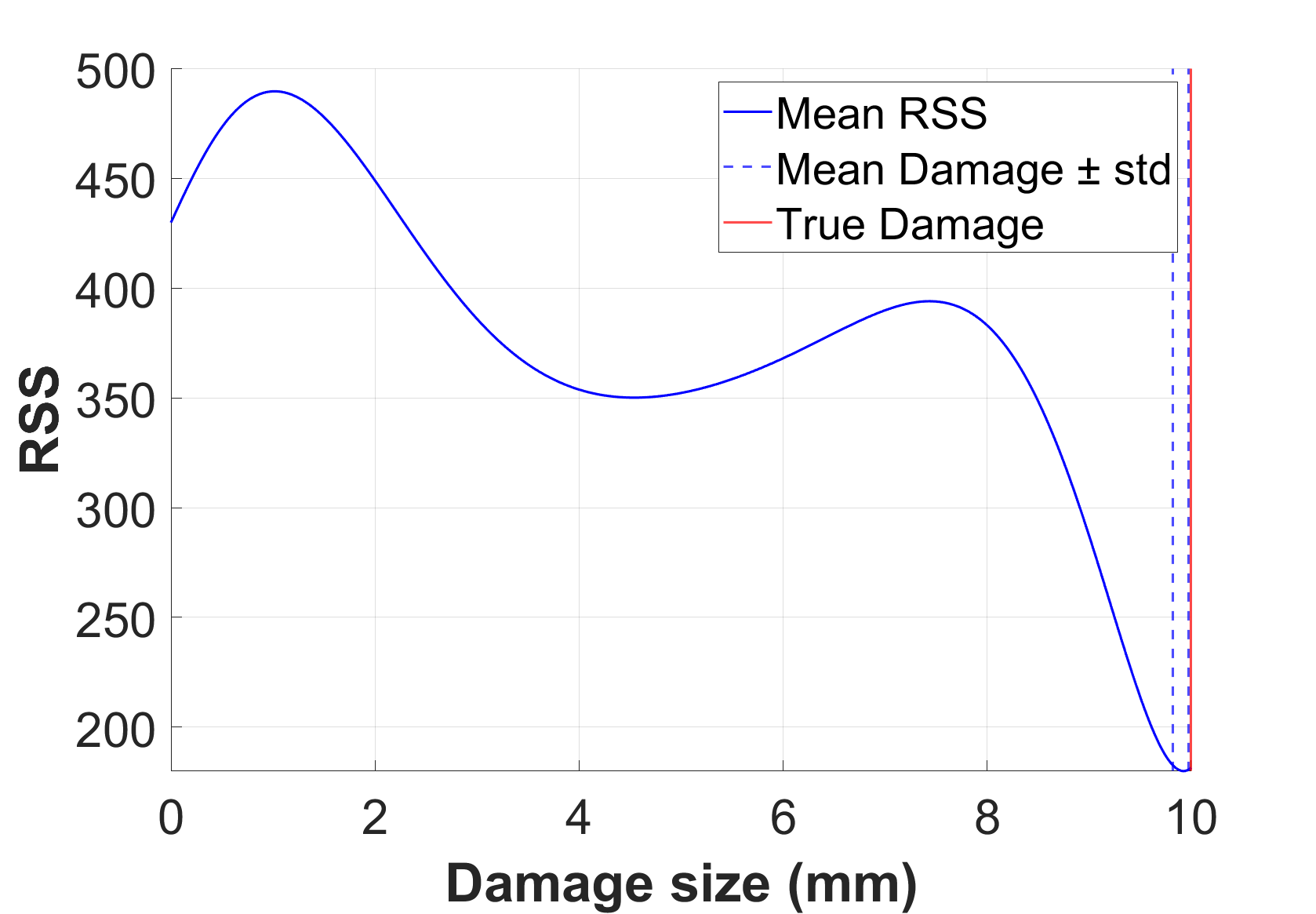}
		\caption{M1, 10 mm damage}
	\end{subfigure}
	\caption{Local (unpooled) FP-AR RSS-based damage size estimation using the AccX signal of Motor 1, across all damage levels (healthy to 10 mm).}
 \label{fig:Unpooled damage estimation M1}
\end{figure*}

\begin{figure*}[t!]
	\centering
	\begin{subfigure}{.32\textwidth}
		\centering
		\captionsetup{width=\linewidth}
		\includegraphics[width=\linewidth]{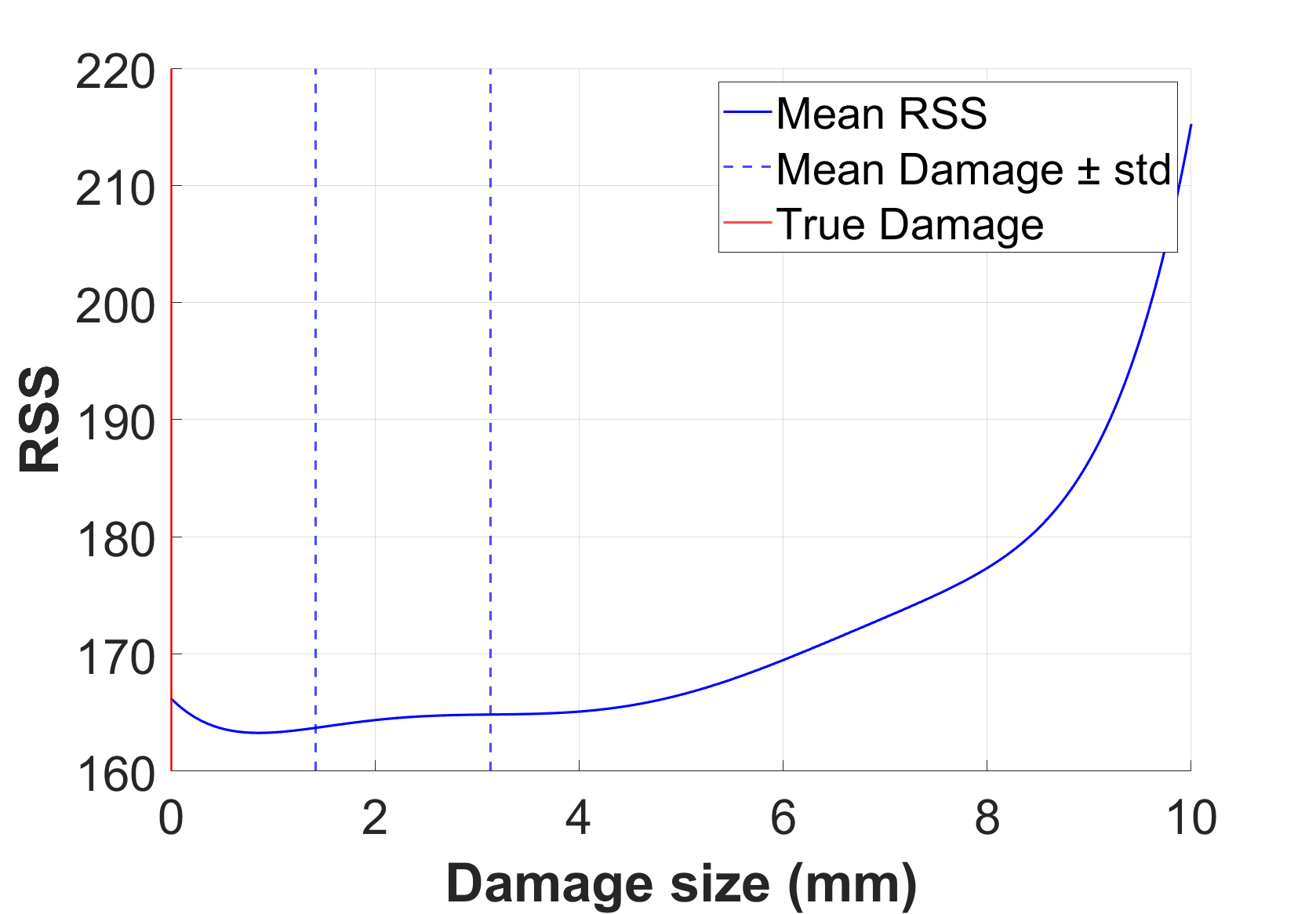}
		\caption{M6, 0 mm damage}
	\end{subfigure}
	\begin{subfigure}{.32\textwidth}
		\centering
		\captionsetup{width=\linewidth}
		\includegraphics[width=\linewidth]{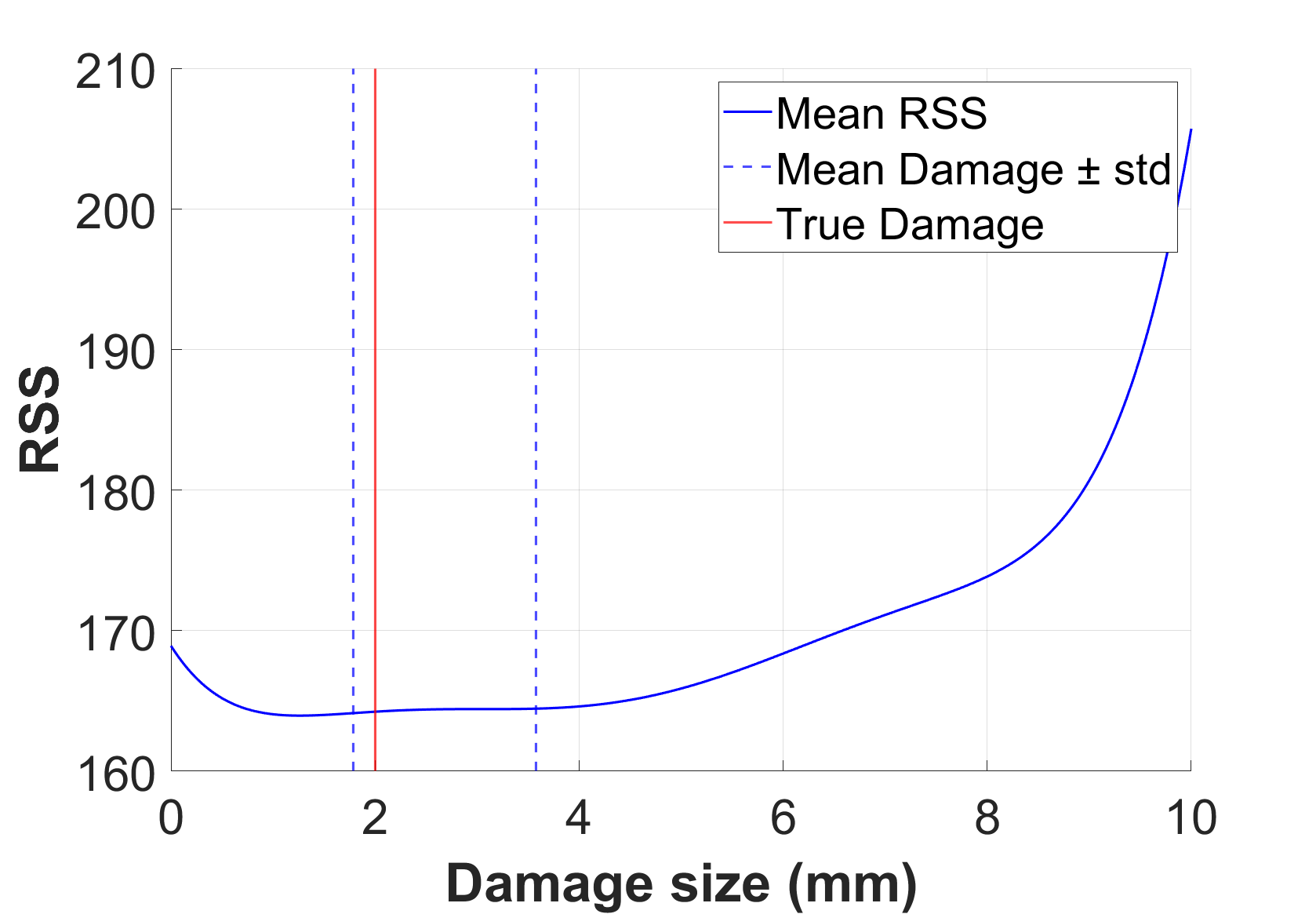}
		\caption{M6, 2 mm damage}
	\end{subfigure}
    \begin{subfigure}{.32\textwidth}
		\centering
		\captionsetup{width=\linewidth}
		\includegraphics[width=\linewidth]{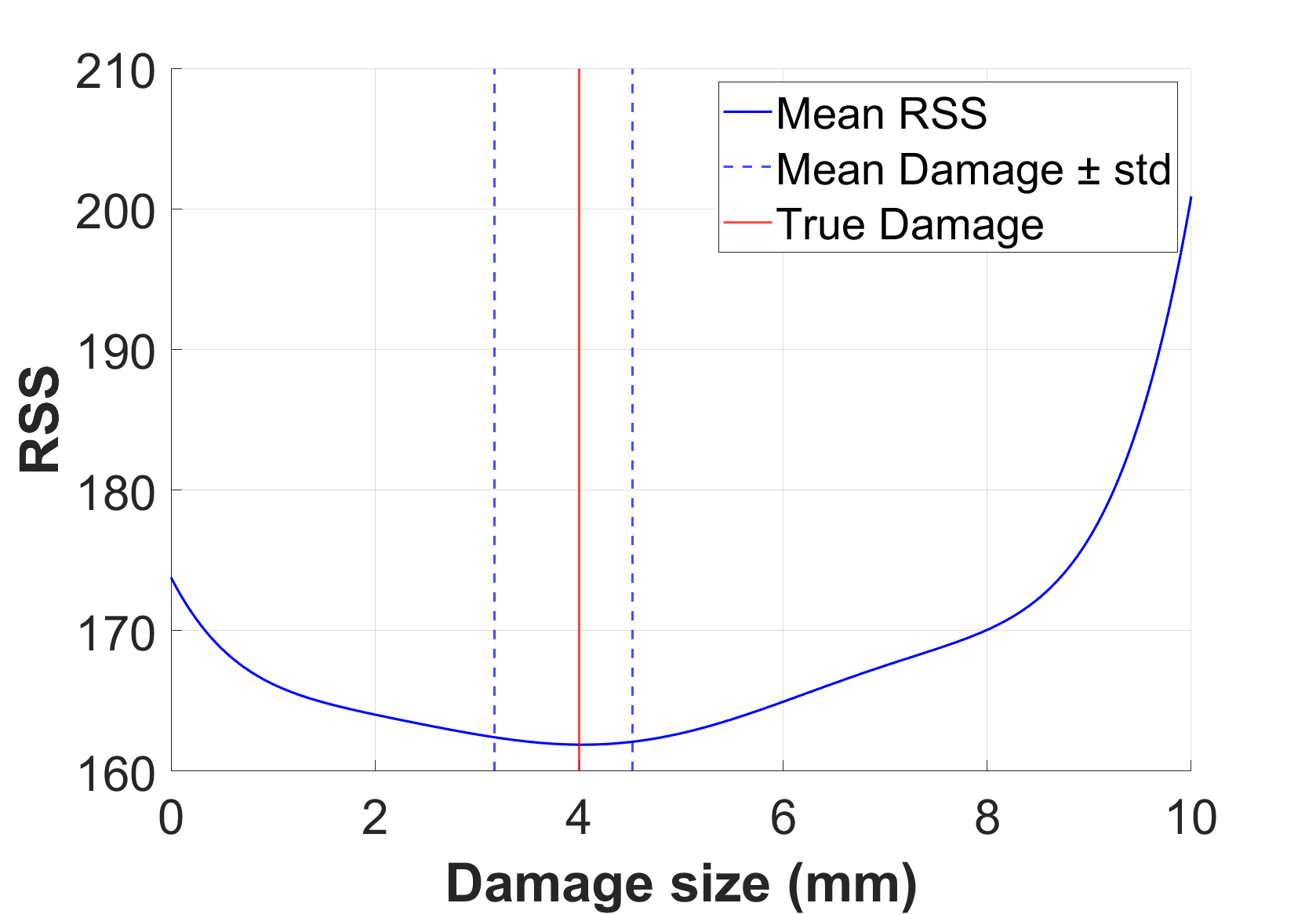}
		\caption{M6, 4 mm damage}
	\end{subfigure}
 
	\begin{subfigure}{.32\textwidth}
		\centering
		\captionsetup{width=\linewidth}
		\includegraphics[width=\linewidth]{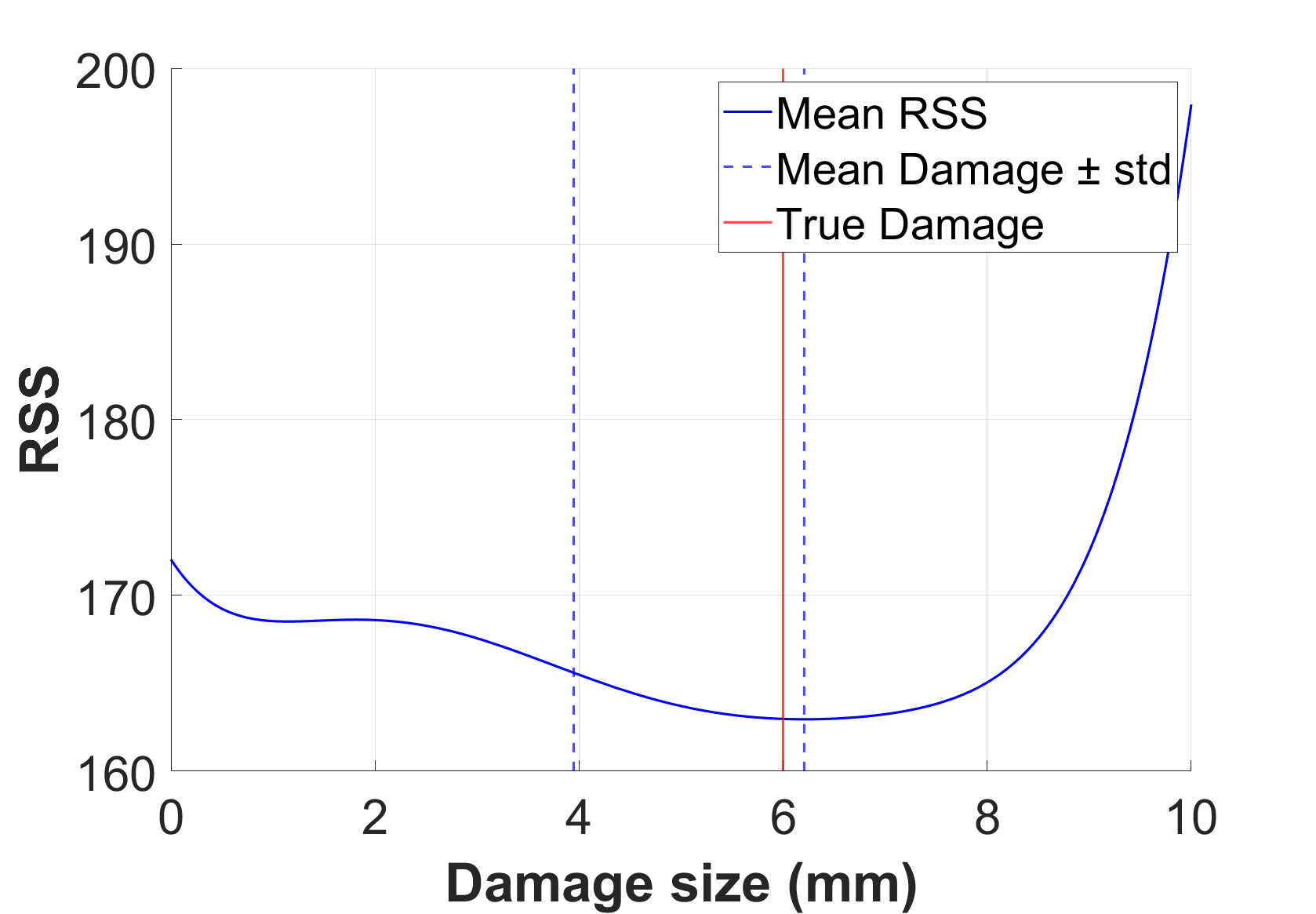}
		\caption{M6, 6 mm damage}
	\end{subfigure}
	\begin{subfigure}{.32\textwidth}
		\centering
		\captionsetup{width=\linewidth}
		\includegraphics[width=\linewidth]{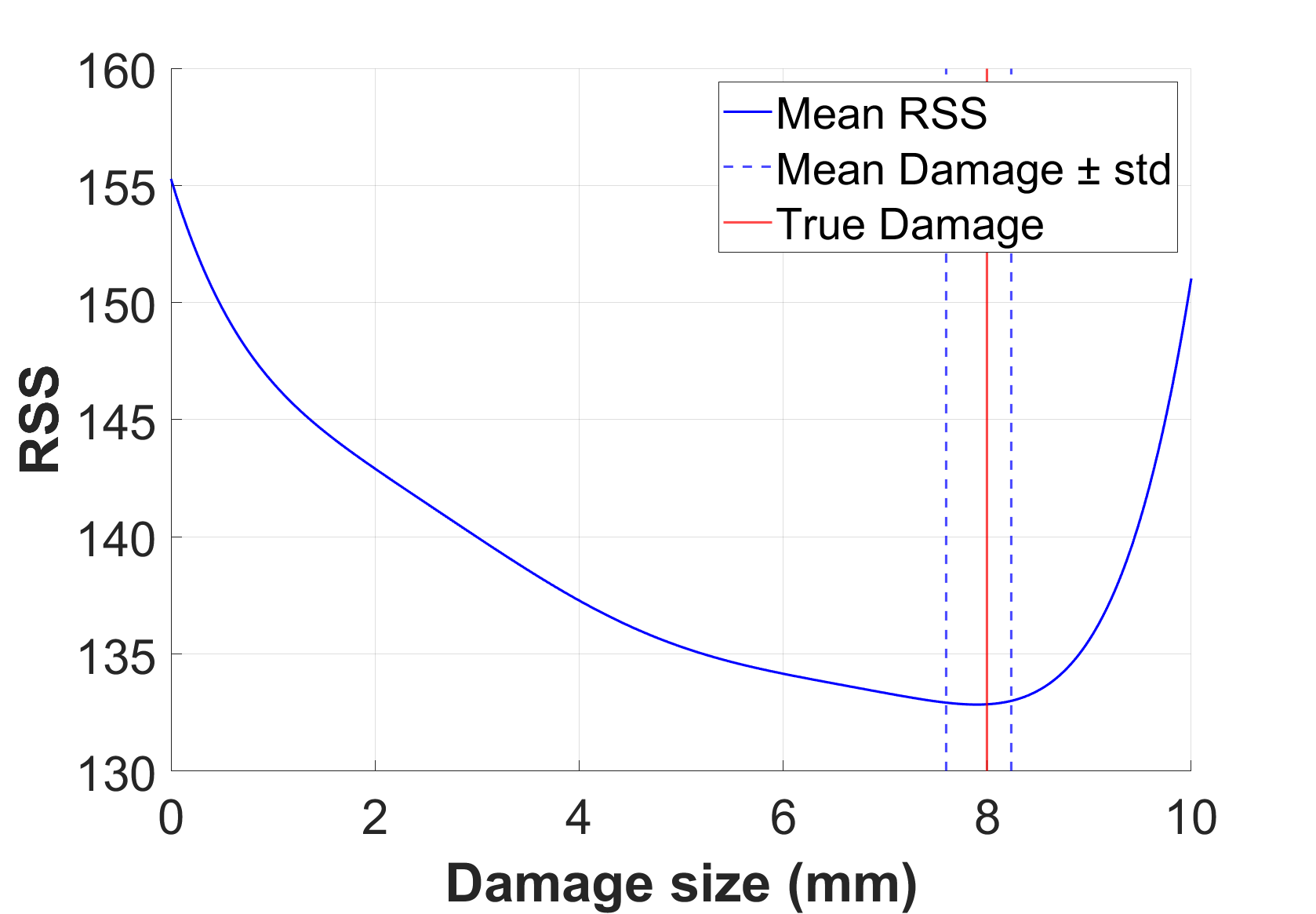}
		\caption{M6, 8 mm damage}
	\end{subfigure}
    \begin{subfigure}{.32\textwidth}
		\centering
		\captionsetup{width=\linewidth}
		\includegraphics[width=\linewidth]{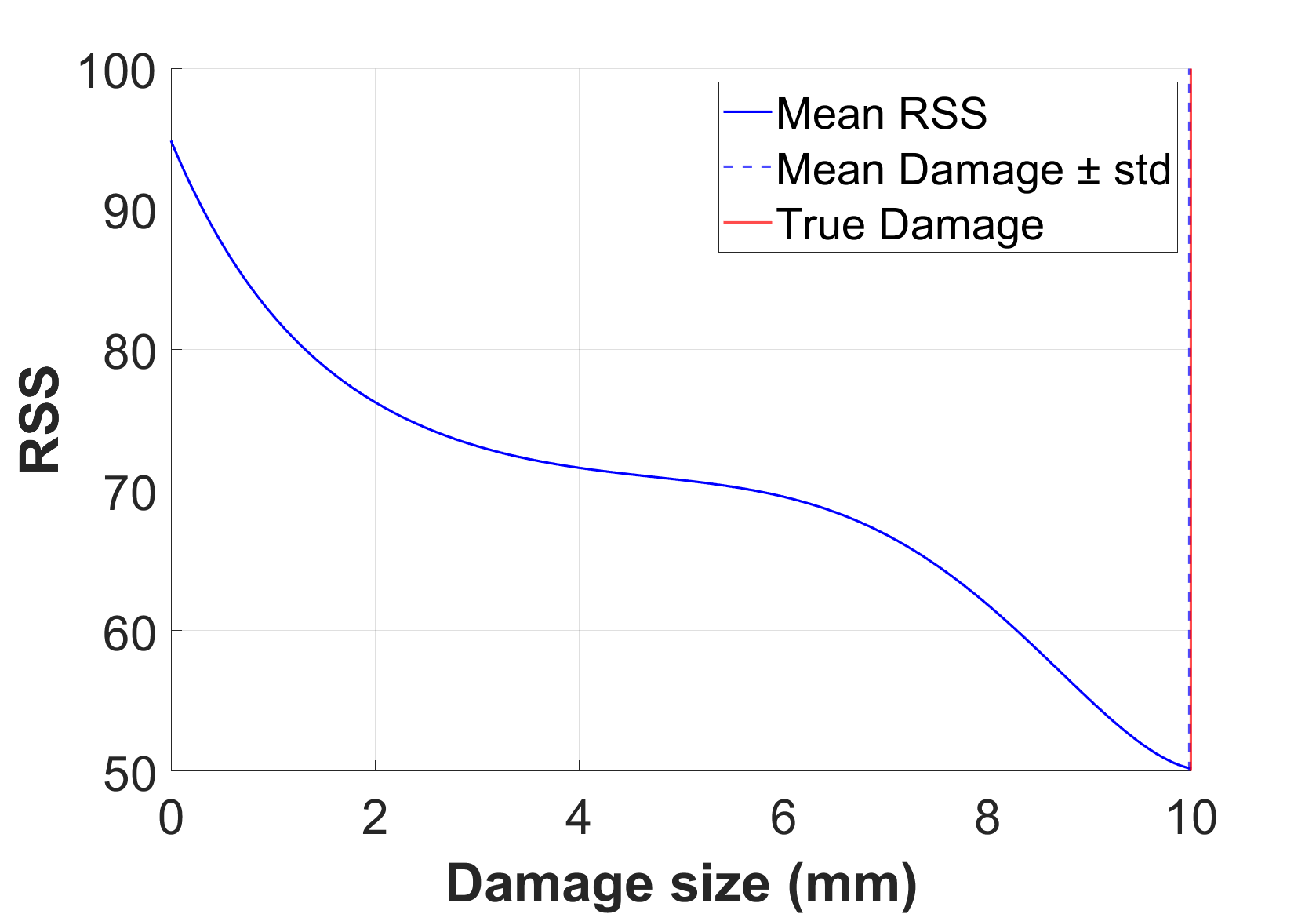}
		\caption{M6, 10 mm damage}
	\end{subfigure}
	\caption{Local (unpooled) FP-AR RSS-based damage size estimation using the AccX signal of Motor 6, across all damage levels (healthy to 10 mm).}
 \label{fig:Unpooled damage estimation M6}
\end{figure*}

Fig.~\ref{fig:Unpooled damage estimation M1} presents representative damage quantification results using the AccX signal over 21 independent test segments for Motor 1; the corresponding Motor 3 and Motor 6 results are provided in Appendix C (Figs.~\ref{fig:Unpooled damage estimation M3} and~\ref{fig:Unpooled damage estimation M6}). The same local FP-AR models used for damage detection and identification, each trained on a 4 s segment from Motors 1, 3, and 6, are employed here. For each test case, the RSS is computed across the candidate damage levels, and the estimated damage level is reported as the mean $\pm$ one standard deviation across the test segments (blue dashed lines), compared against the true damage level.

The results show that in most cases, especially for damage below 10 mm, the local FP-AR framework produces highly accurate estimates. This suggests the model's robustness in capturing damage-induced variations, even at small damage magnitudes such as 2 mm, especially considering that the flight controller partially compensates for the effects of damage and returns the aircraft to steady flight. Increased accuracy with damage severity is observed.

However, a slight overestimation is observed under the 0 mm condition, consistent with the earlier damage detection results in which the $t$-statistics occasionally exceeded the threshold. Furthermore, under specific configurations, such as the Motor 3 model with a relatively low basis dimensionality of $r=2$, notable deviations between the estimated and true damage levels occur at intermediate damage states (e.g., 4 mm and 6 mm), indicating the influence of the model complexity on the quantification accuracy.

\subsection{Pooled FP-AR model-based diagnosis}

While the local FP-AR model achieves damage detection, identification, and quantification using only a short segment (4 s out of a 250 s flight) of AccX signal data, certain limitations are observed. In particular, relying solely on a single short data segment may not provide sufficient robustness for accurate detection and identification across varying damage scenarios. Notably, a trade-off often emerges between the detection and identification performance: some models exhibit strong damage detection capabilities, yet perform suboptimally in identifying the damaged motor, whereas others demonstrate the opposite behavior. This performance inconsistency highlights the challenge of generalizability when models are constructed from limited and localized information.

\begin{figure}[!t]
    \centering 
    
    \begin{subfigure}{.3\textwidth} 
        \centering
        \includegraphics[width=\textwidth]{Figure/BIC_order_selection/BIC_Distribution_M1_ACC_X_shuffle.png}
        \caption{AccX}
    \end{subfigure}
    \hfill 
    \begin{subfigure}{.3\textwidth}
        \centering
        \includegraphics[width=\textwidth]{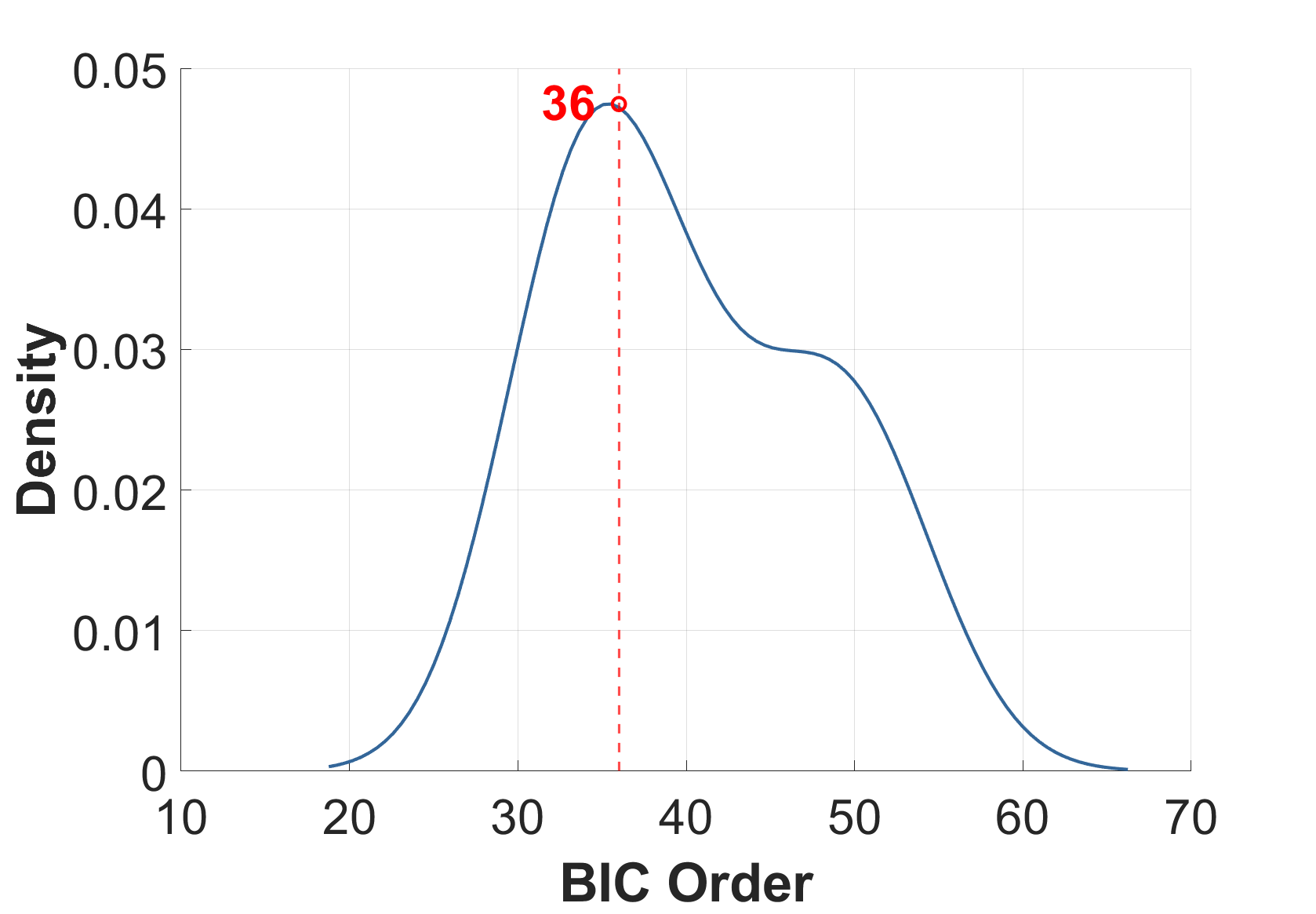}
        \caption{AccY}
    \end{subfigure}
    \hfill
    \begin{subfigure}{.3\textwidth}
        \centering
        \includegraphics[width=\textwidth]{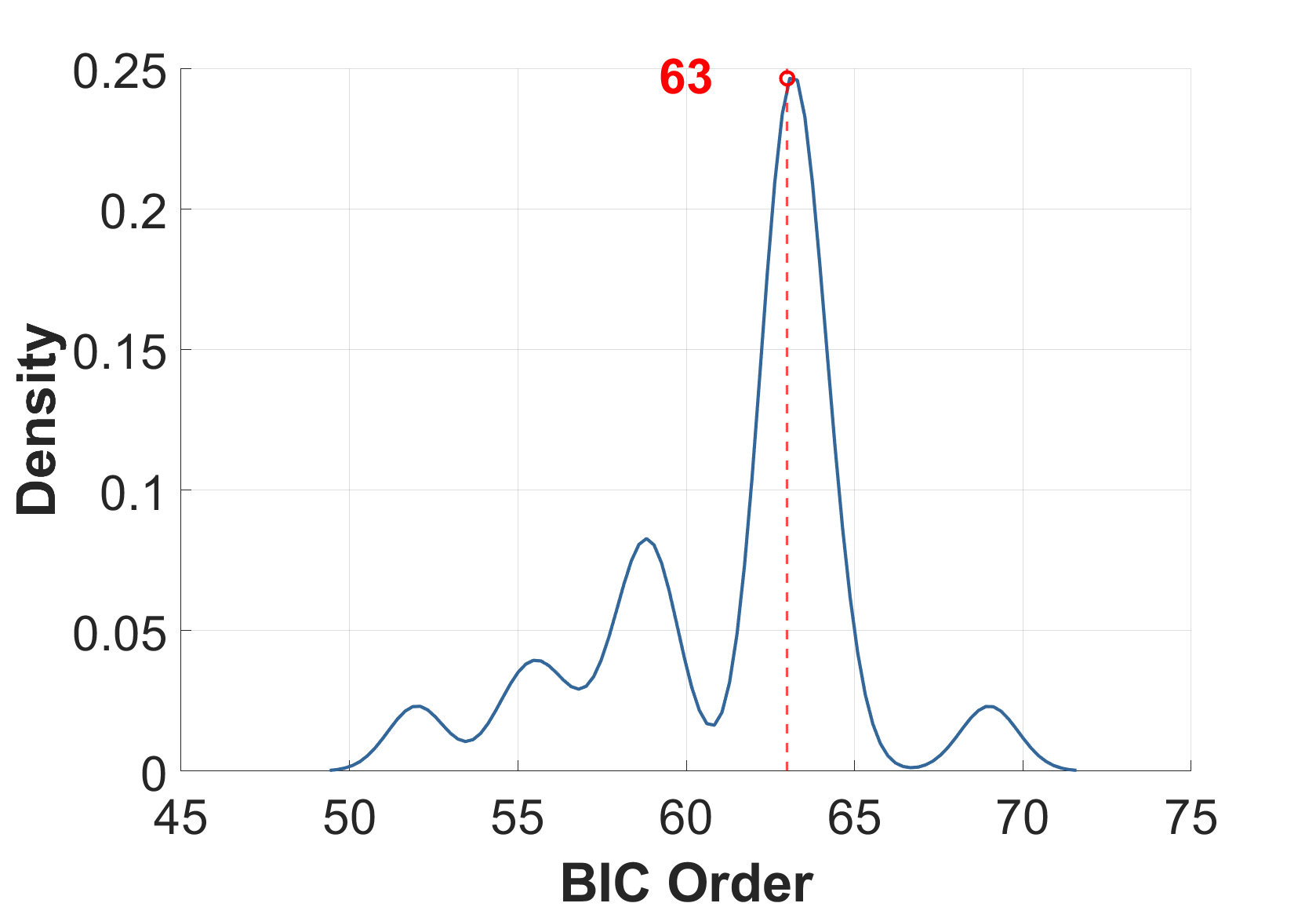}
        \caption{AccZ}
    \end{subfigure}
    \begin{subfigure}{.3\textwidth}
        \centering
        \includegraphics[width=\textwidth]{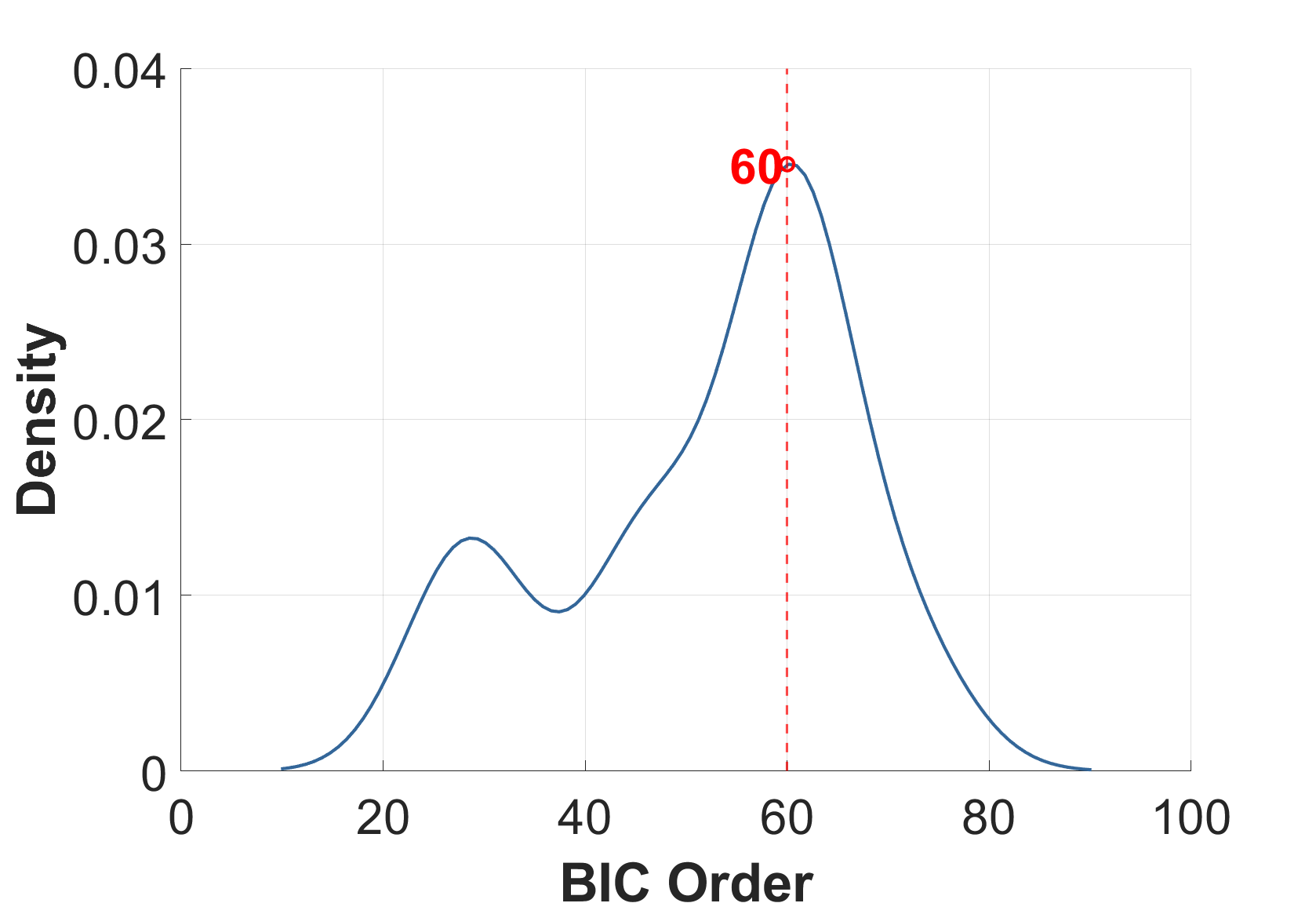}
        \caption{GyrX}
    \end{subfigure}
    \hfill
    \begin{subfigure}{.3\textwidth}
        \centering
        \includegraphics[width=\textwidth]{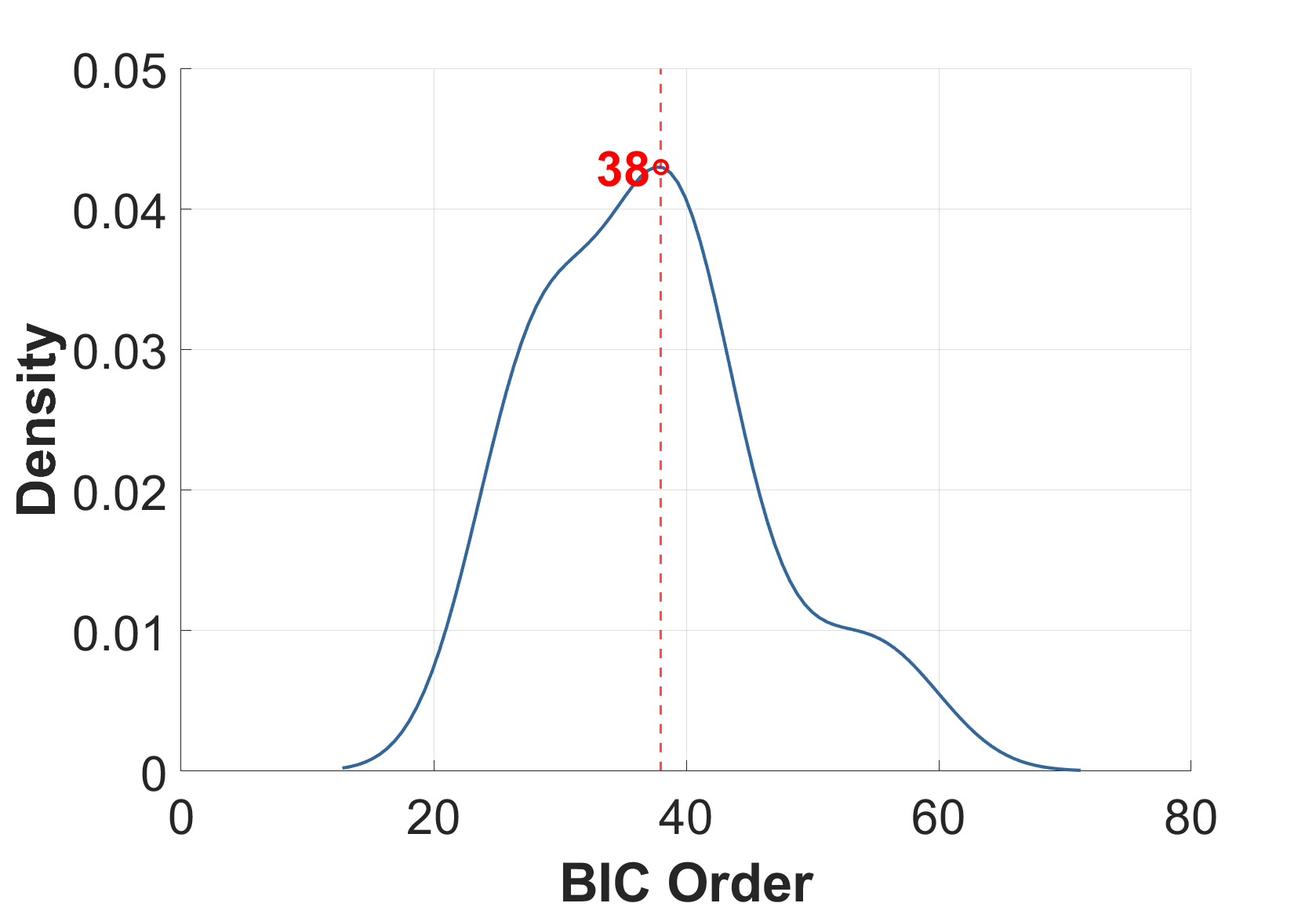}
        \caption{GyrY}
    \end{subfigure}
    \hfill
    \begin{subfigure}{.3\textwidth}
        \centering
        \includegraphics[width=\textwidth]{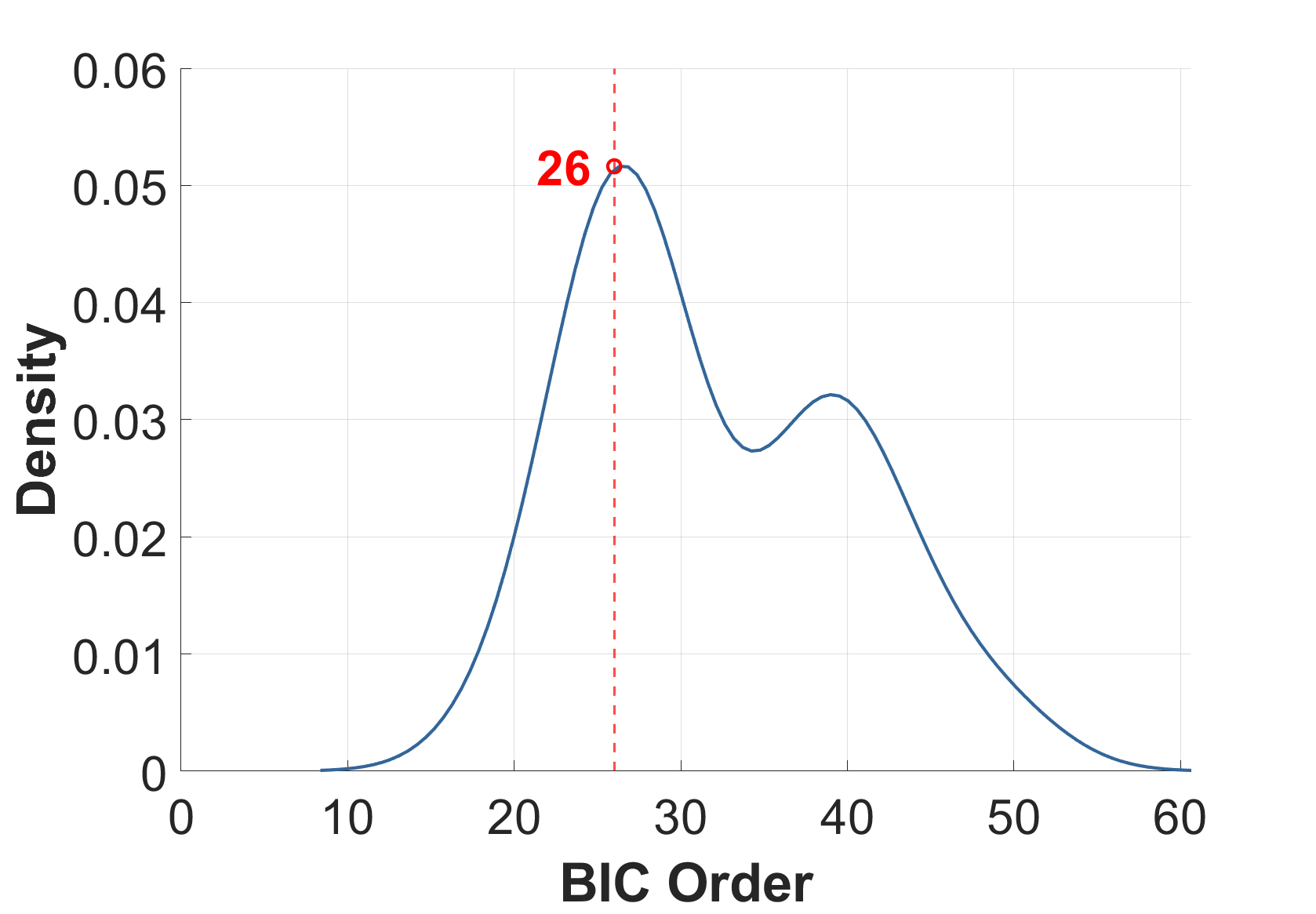}
        \caption{GyrZ}
    \end{subfigure}
    \caption{Gaussian KDEs of the BIC-optimal model orders for the six signals of Motor 1. The red vertical line indicates the peak of the density curve, representing the most probable model order selected for the pooled FP-AR model.}
    \label{fig:order density M1}
\end{figure}

To enhance the model robustness and generalizability, a pooled FP-AR model is constructed using a broader temporal range of the flight data. Specifically, 36 signal segments are extracted from the 30--174 s interval, among which 20 segments are used for training and the remaining 16 for testing. The model construction follows the cross-segment parameter pooling procedure described in Section~\ref{sec:pooling}, employing a statistically driven strategy to ensure a consistent structure across all segments. The pooled model order and basis dimensionality are determined by analyzing the distributions of the locally identified orders and basis dimensionalities across the 20 training segments, with the most probable values selected via the Gaussian KDE peak for the model order and the histogram mode for the basis dimensionality, as shown in Figs.~\ref{fig:order density M1} and~\ref{fig:basis density M1} (Motor 1 is used as an example; the results for Motors 3 and 6 can be found in Appendix C). Table~\ref{tab:global-fpar} summarizes the pooled FP-AR structure per signal--motor pair (selected order and basis dimensionality) and the resulting projection coefficient counts.

\begin{figure}[!t]
    \centering 
    
    \begin{subfigure}{.3\textwidth} 
        \centering
        \includegraphics[width=\textwidth]{Figure/BIC_basis_selection/BIC_Basis_Histogram_Motor_1_Acc_X.png}
        \caption{AccX}
    \end{subfigure}
    \hfill 
    \begin{subfigure}{.3\textwidth}
        \centering
        \includegraphics[width=\textwidth]{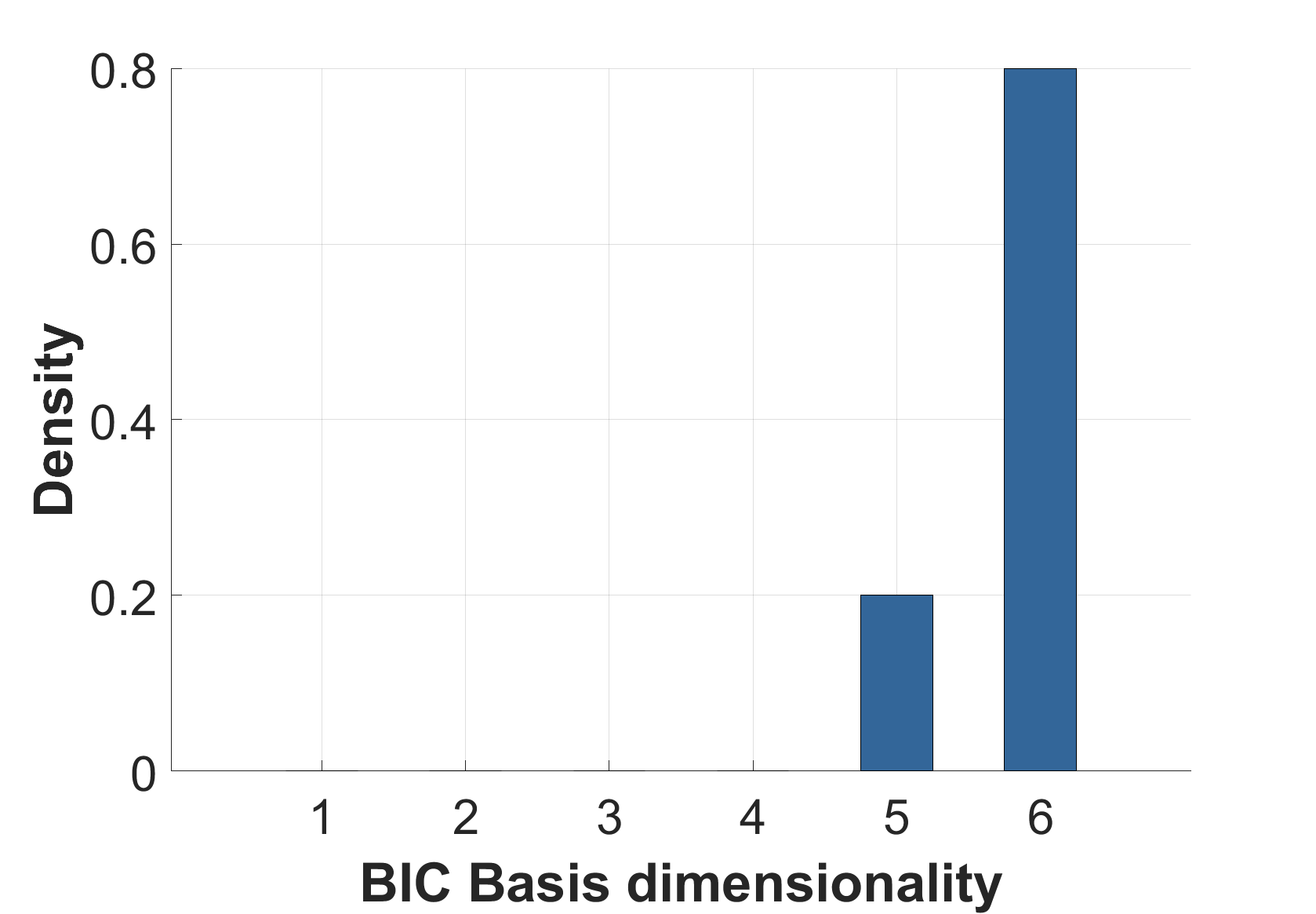}
        \caption{AccY}
    \end{subfigure}
    \hfill
    \begin{subfigure}{.3\textwidth}
        \centering
        \includegraphics[width=\textwidth]{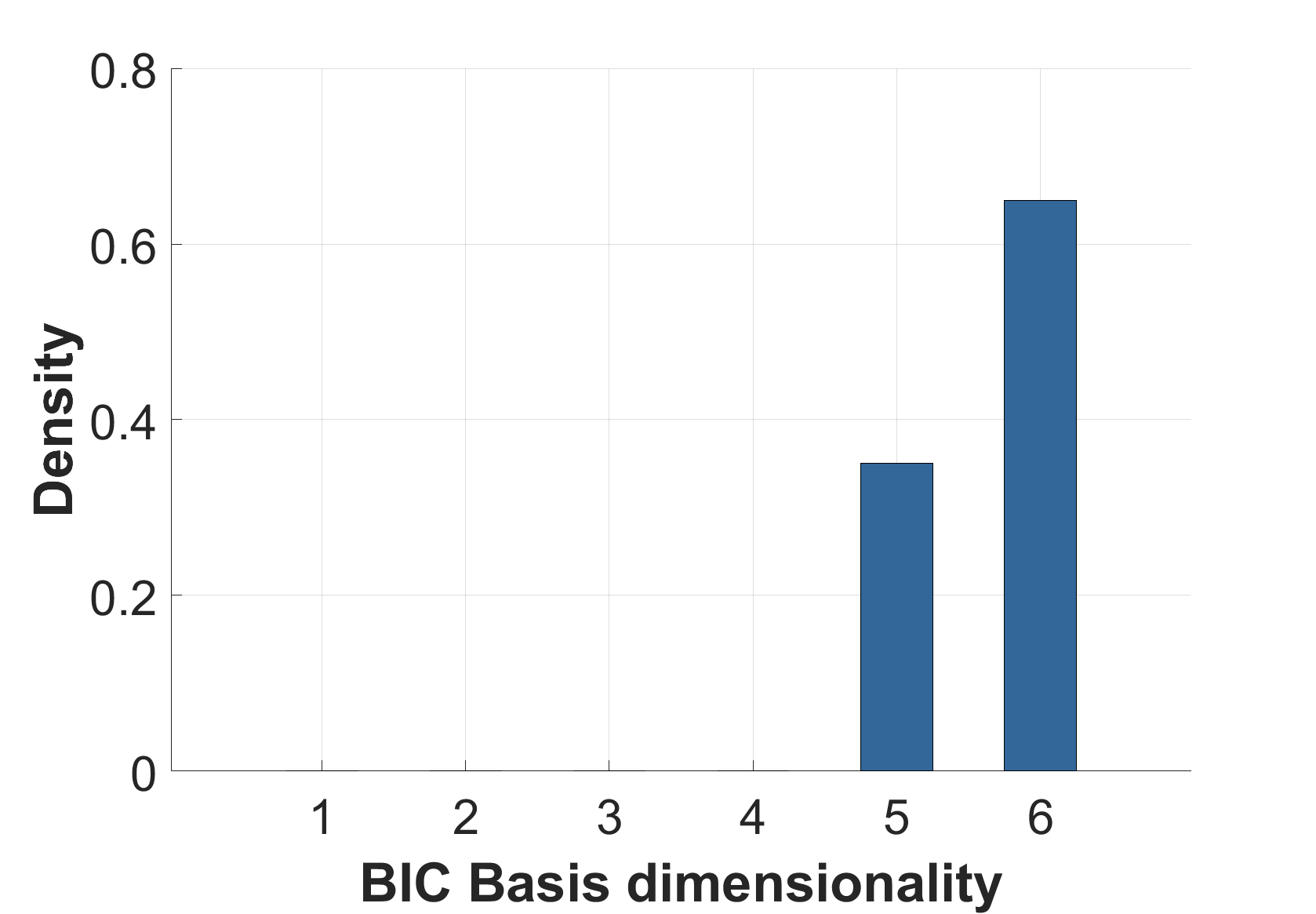}
        \caption{AccZ}
    \end{subfigure}

    \begin{subfigure}{.3\textwidth}
        \centering
        \includegraphics[width=\textwidth]{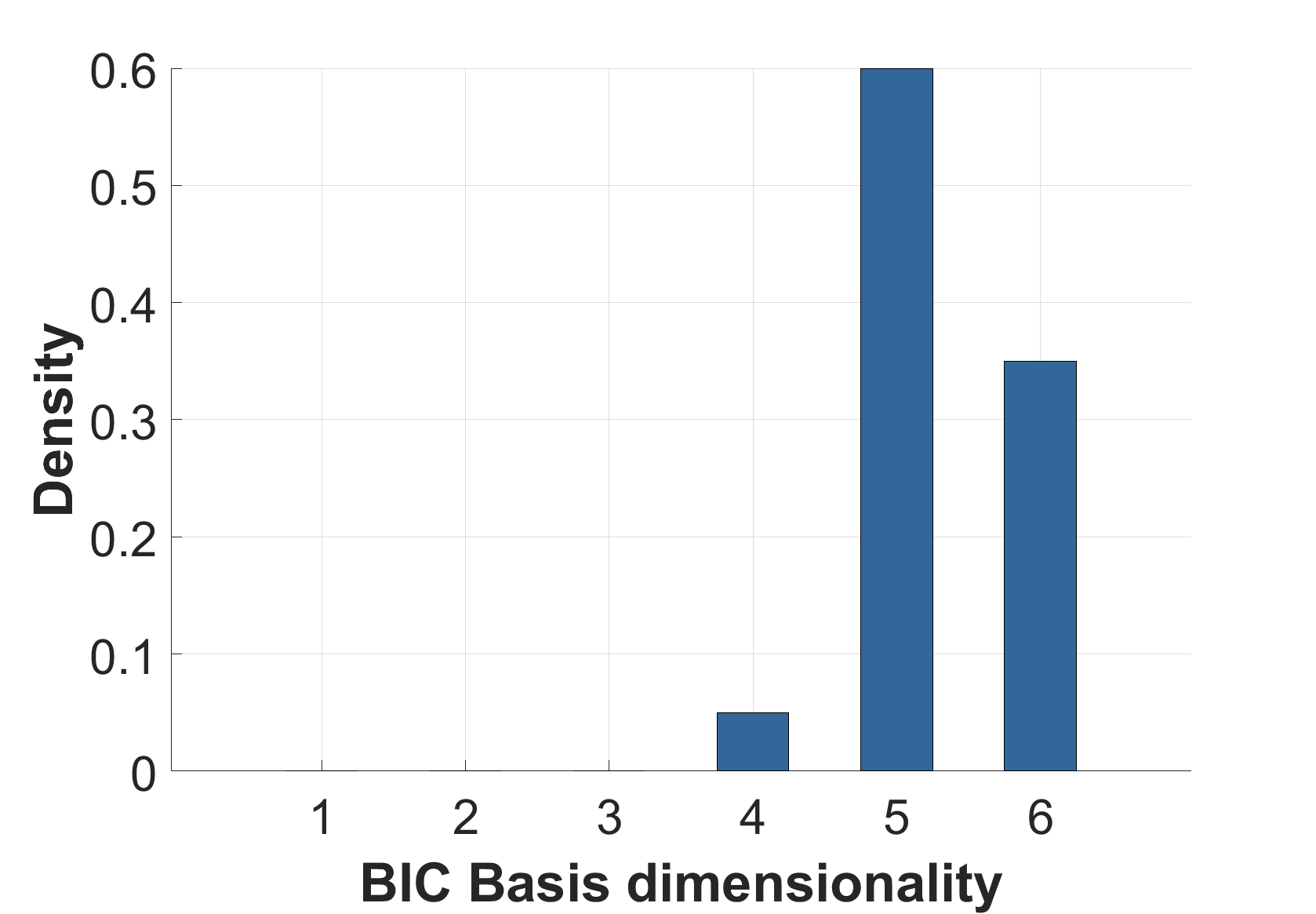}
        \caption{GyrX}
    \end{subfigure}
    \hfill
    \begin{subfigure}{.3\textwidth}
        \centering
        \includegraphics[width=\textwidth]{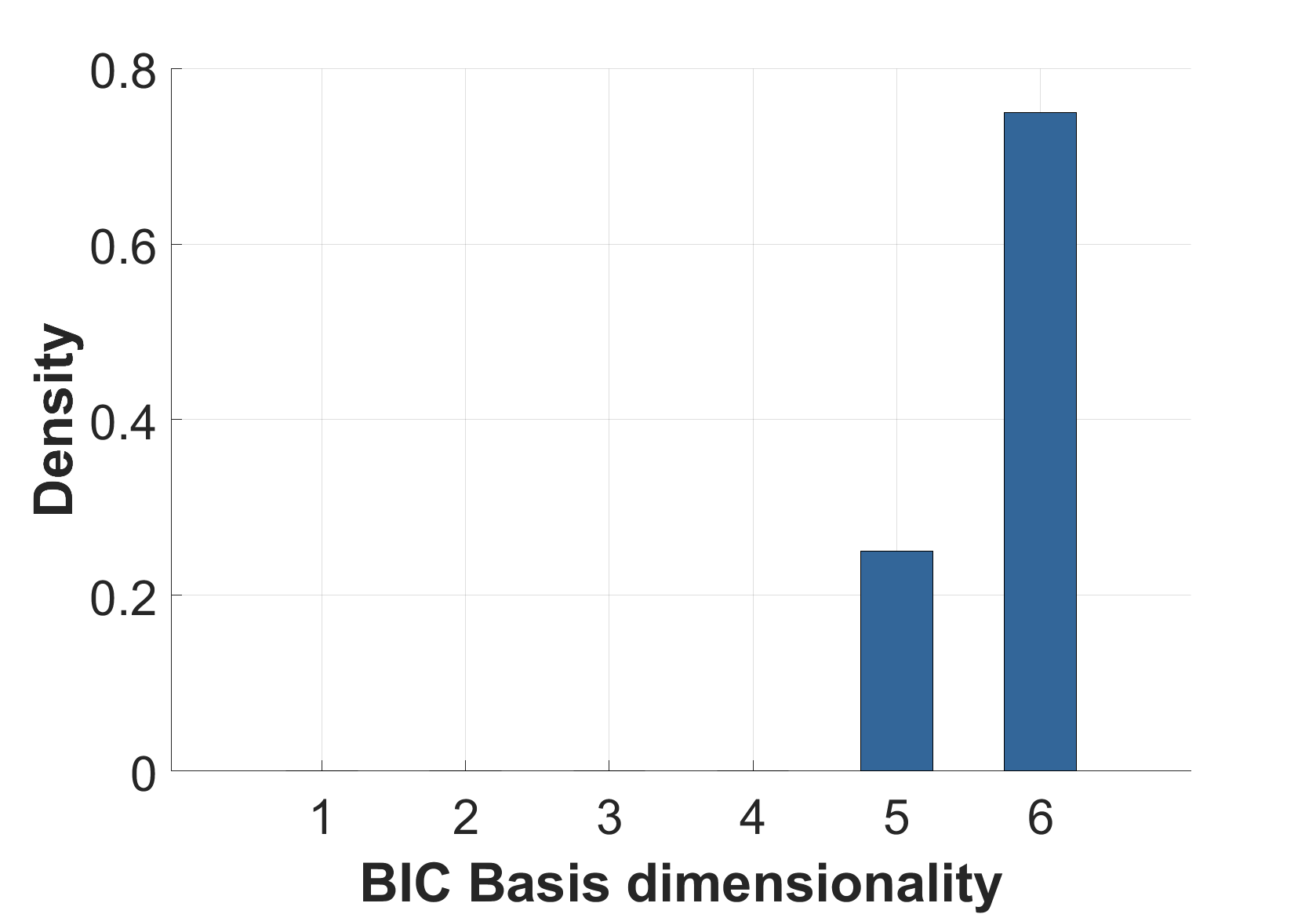}
        \caption{GyrY}
    \end{subfigure}
    \hfill
    \begin{subfigure}{.3\textwidth}
        \centering
        \includegraphics[width=\textwidth]{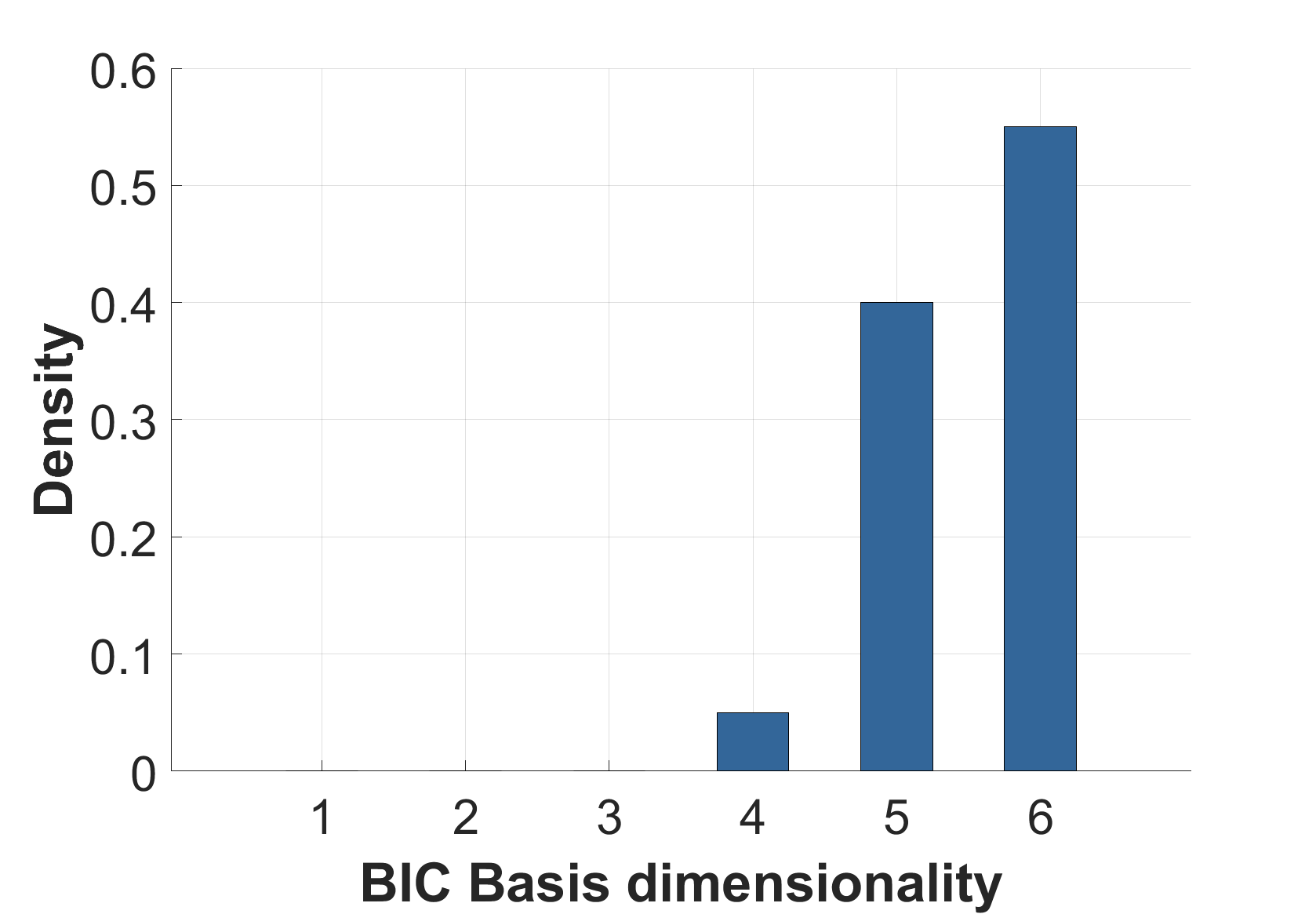}
        \caption{GyrZ}
    \end{subfigure}
    \caption{Histograms of the BIC-optimal basis dimensionalities for the six signals of Motor 1. The histograms show the frequency of the basis dimensionalities obtained from the training segments, with the most frequent value selected for the pooled FP-AR model.}
    \label{fig:basis density M1}
\end{figure}

Once the common model structure (order and basis dimensionality) is fixed, a new set of FP-AR models is fitted to the training segments using these common settings. The final pooled FP-AR model is obtained by averaging the estimated projection coefficients across all trained models. This parameter pooling effectively captures the representative system dynamics, improving the robustness to segment-to-segment variability and supporting reliable damage diagnosis under varying operating conditions.

\begin{table}[b!]
  \centering
  \begin{threeparttable}
    \caption{Pooled FP-AR model structures}
    \label{tab:global-fpar}
    \small
    \begin{tabular}{@{} l l l r r r @{}} 
      \toprule
      Signal & Motor & Segment source (s) & Order $n_a^*$ & Basis $r^*$ & \#Projection coefficients \\
      \midrule
      AccX & M1 & 30--174 & 36 & 5 & 3600 \\
            & M3 & 30--174 & 32 & 3 & 1920  \\
            & M6 & 30--174 & 36 & 3 & 3160 \\
      AccY & M1 & 30--174 & 36 & 6 & 4320 \\
            & M3 & 30--174 & 35 & 5 & 3500 \\
            & M6 & 30--174 & 33 & 3 & 1980 \\
      AccZ & M1 & 30--174 & 63 & 5 & 6300 \\
            & M3 & 30--174 & 63 & 2 & 2520 \\
            & M6 & 30--174 & 60 & 3 & 3600 \\
      GyrX & M1 & 30--174 & 60 & 5 & 6000 \\
            & M3 & 30--174 & 57 & 5 & 5900 \\
            & M6 & 30--174 & 49 & 3 & 2940 \\
      GyrY & M1 & 30--174 & 38 & 6 & 4560 \\
            & M3 & 30--174 & 41 & 5 & 4100 \\
            & M6 & 30--174 & 39 & 4 & 3120\\
      GyrZ & M1 & 30--174 & 26 & 5 & 2600 \\
            & M3 & 30--174 & 39 & 2 & 780  \\
            & M6 & 30--174 & 28 & 3 & 1680 \\
      \bottomrule
    \end{tabular}
    \begin{tablenotes}\footnotesize
      \item \textit{Note:} A total of 36 segments are extracted from the 30--174 s window; 
      twenty segments are used for training and sixteen for testing.  
      The total number of estimated projection coefficients is computed as \(\#\mathrm{Projection\ coefficients}= n_a^* \times r^* \times \#\mathrm{training\ segments}\).
    \end{tablenotes}
  \end{threeparttable}
\end{table}

\subsubsection{Damage detection}

Damage detection results based on the pooled FP-AR model using the AccX signals are presented in the sequel. As described earlier, the pooled model is constructed by aggregating information from 20 signal segments and averaging the projection coefficients under a common model order and basis dimensionality. This results in a statistically representative model for each motor and signal source.

\begin{figure}[t!]
\centering

\begin{subfigure}[t]{0.92\columnwidth}
  \centering
  \includegraphics[width=\linewidth]{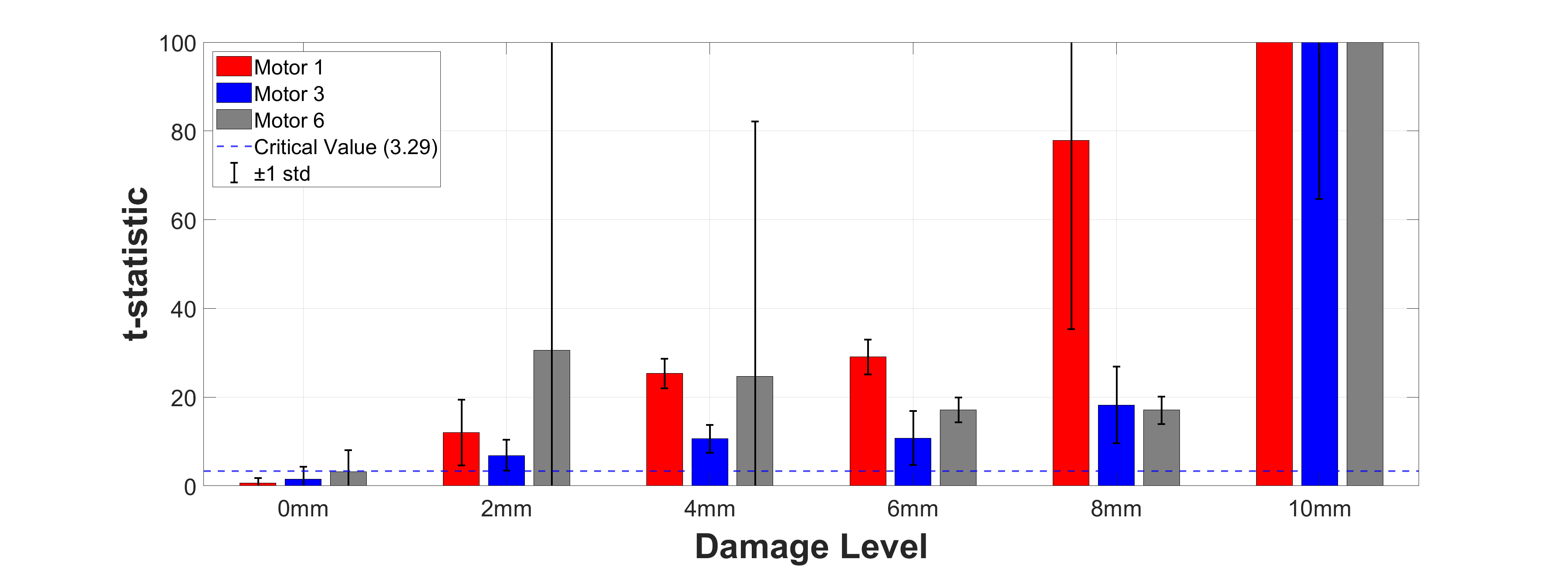}
  \caption{Motor~1, AccX (pooled)}
  \label{fig:pooled_health_m1}
\end{subfigure}
\vspace{0.6em}

\begin{subfigure}[t]{0.92\columnwidth}
  \centering
  \includegraphics[width=\linewidth]{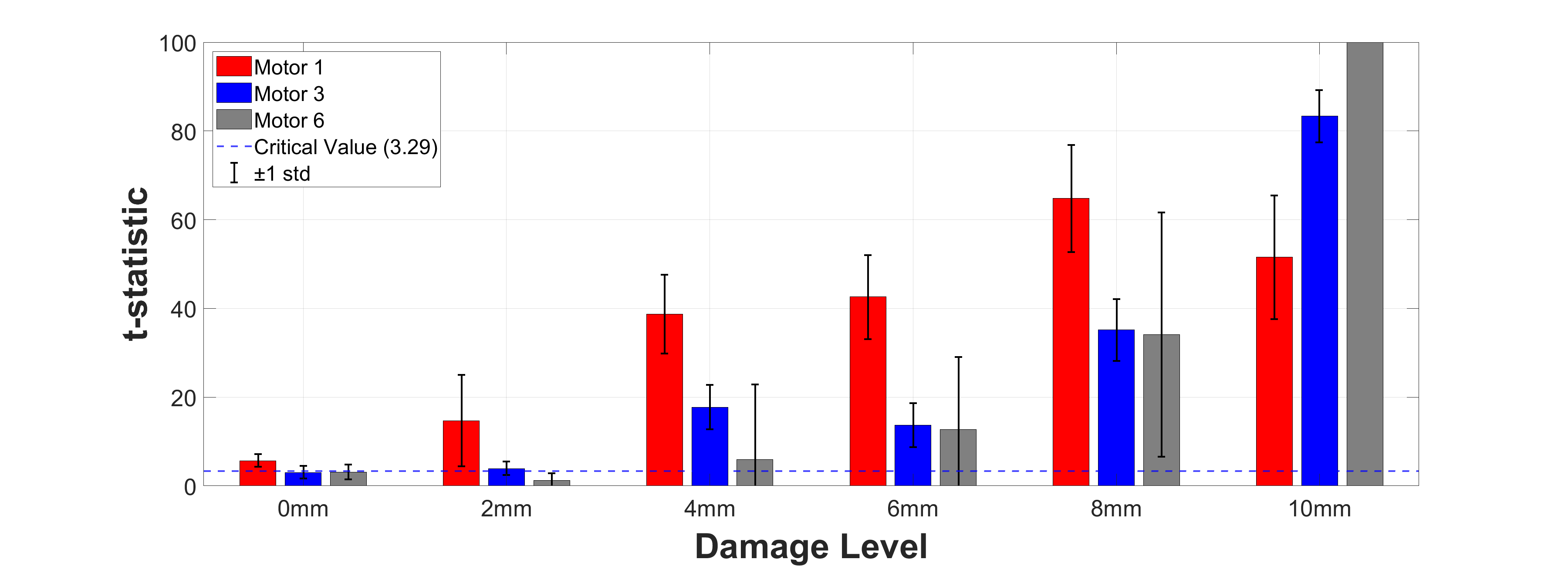}
  \caption{Motor~3, AccX (pooled)}
  \label{fig:pooled_health_m3}
\end{subfigure}
\vspace{0.6em}

\begin{subfigure}[t]{0.92\columnwidth}
  \centering
  \includegraphics[width=\linewidth]{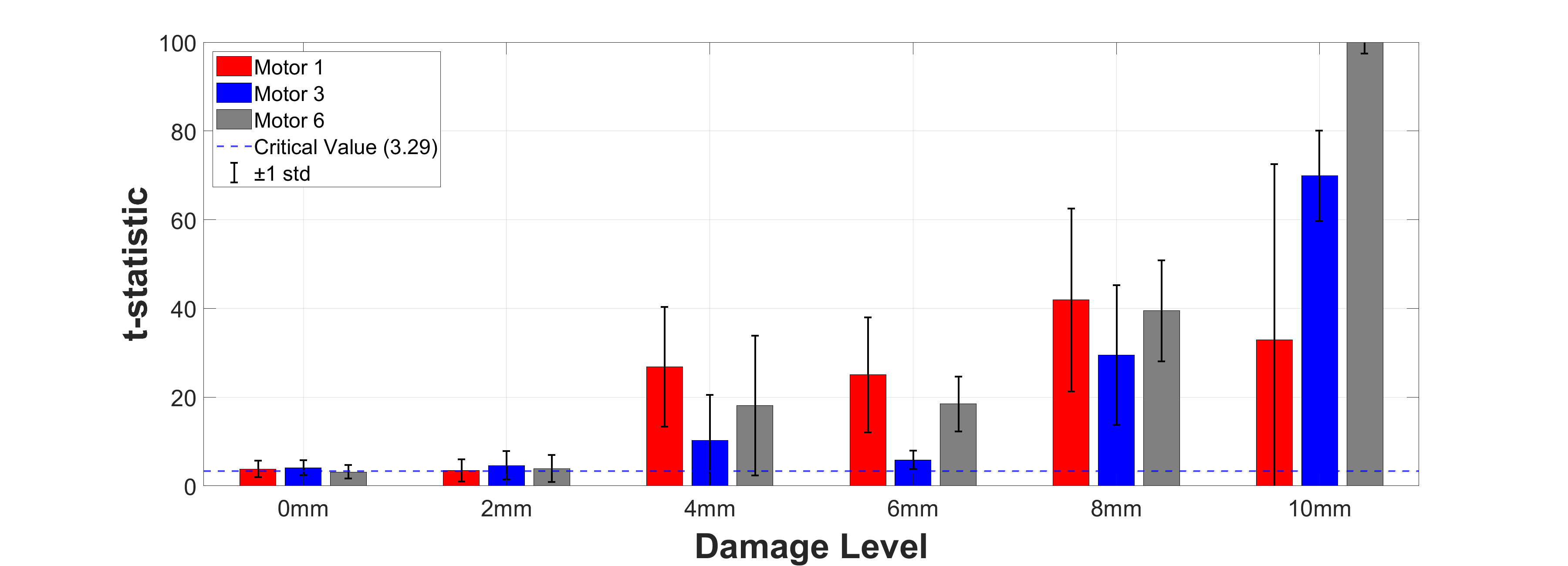}
  \caption{Motor~6, AccX (pooled)}
  \label{fig:pooled_health_m6}
\end{subfigure}

\caption{Pooled FP-AR damage detection using $t$-statistics. Models are trained on 20 randomly selected training segments from the 30--174\,s window and evaluated on 16 testing segments from the same window. For each motor, bars show the mean $t$-statistic across testing segments; error bars denote $\pm 1$ standard deviation. The dashed line indicates the critical value at risk level $\alpha=0.001$. Damage is declared when the $t$-statistic exceeds the critical threshold.}
\label{fig:pooled_health_all}
\end{figure}

Figure~\ref{fig:pooled_health_all} presents representative results based on the AccX data of Motors 1, 3, and 6. Compared to the local FP-AR approach, the pooled model exhibits improved performance: the $t$-statistics for nearly all damaged cases (2--10 mm) consistently exceed the critical threshold, while those for the healthy (0 mm) condition remain well below it in nearly all instances. This indicates enhanced discrimination capability and reduced false alarm rates. The improved reliability stems from the model's ability to generalize across multiple operating conditions, avoiding the overfitting and sensitivity that may arise from single-segment training. The results confirm that the pooled FP-AR framework not only preserves the sensitivity to small damage, but also offers superior robustness in characterizing the healthy state, marking a clear advantage over its local model counterpart.

\subsubsection{Damage identification}

\begin{figure}[!t]
\centering
\begin{subfigure}[t]{0.92\columnwidth}
  \centering
  \includegraphics[width=\linewidth]{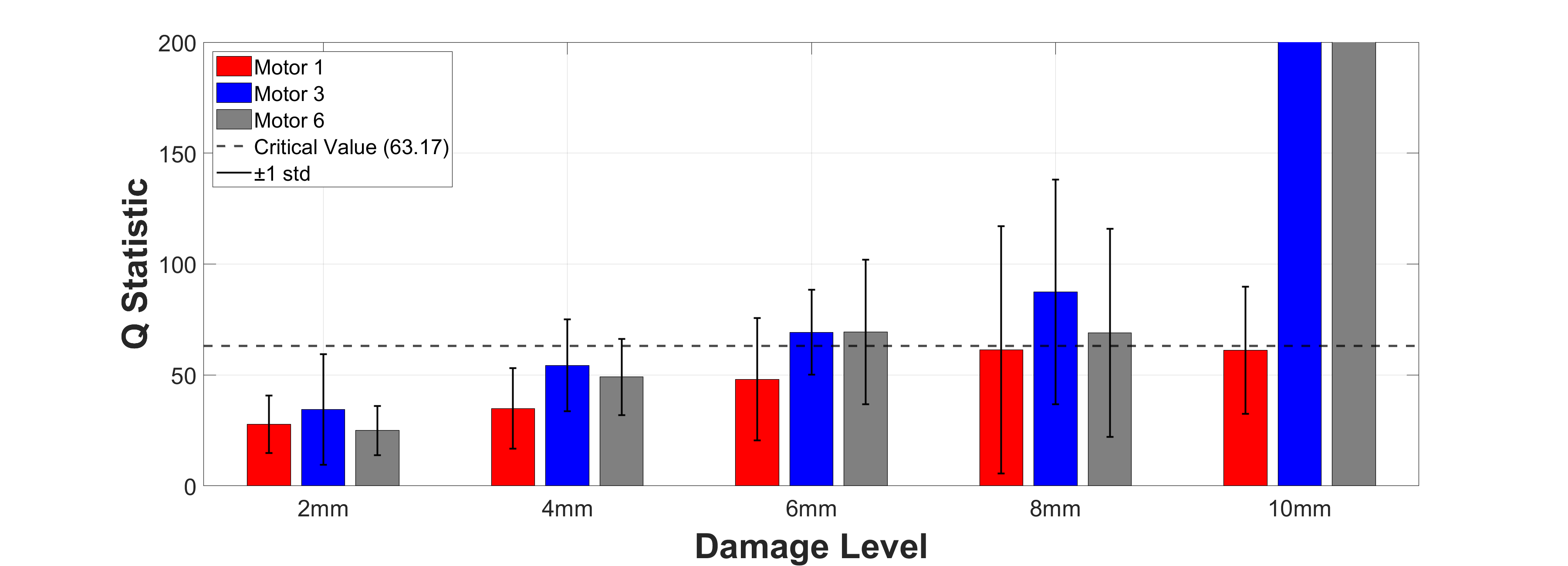}
  \caption{Motor 1}
  \label{fig:pooled_q_m1}
\end{subfigure}
\vspace{0.6em}

\begin{subfigure}[t]{0.92\columnwidth}
  \centering
  \includegraphics[width=\linewidth]{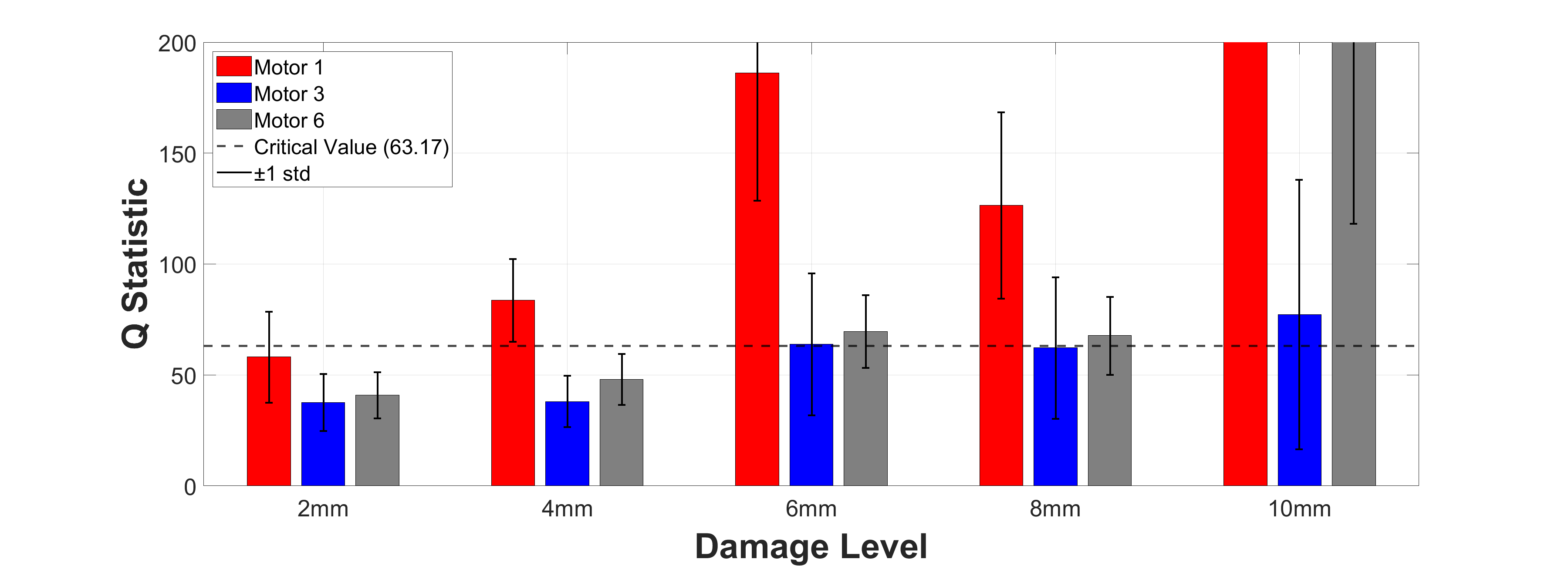}
  \caption{Motor 3}
  \label{fig:pooled_q_m3}
\end{subfigure}
\vspace{0.6em}

\begin{subfigure}[t]{0.92\columnwidth}
  \centering
  \includegraphics[width=\linewidth]{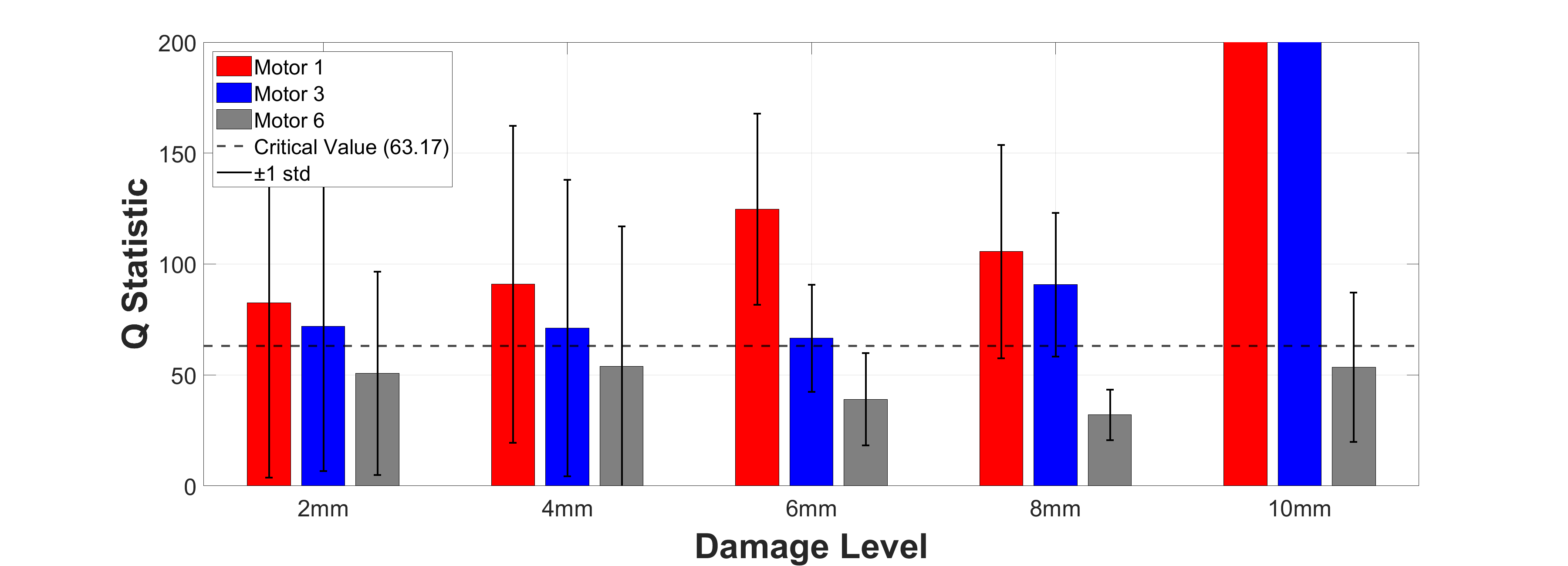}
  \caption{Motor 6}
  \label{fig:pooled_q_m6}
\end{subfigure}

\caption{Damaged motor identification using $Q$-statistics with pooled FP-AR models. Each model is trained on 20 randomly selected training segments from the 30--174\,s window and evaluated on 16 testing segments from the same window. Bars show the mean $Q$-statistic for each candidate motor model (M1, M3, M6); error bars denote $\pm 1$ standard deviation. The dashed line marks the critical value at risk level $\alpha=0.1$. For a given damage level, the corresponding damage topology (motor model) is accepted if its $Q$-statistic falls below the critical threshold; otherwise it is rejected.}
\label{fig:pooled_q_all}
\end{figure}

Damage identification results based on the pooled FP-AR model using the AccX signals are presented next. As shown in Fig.~\ref{fig:pooled_q_all}, the $Q$-statistics are computed across 16 test segments for each motor. The figure reports the mean and standard deviation of the $Q$-statistics, with a risk level of $\alpha = 0.1$ used to define the identification threshold. In most cases, the test data corresponding to the correct motor consistently yield the lowest $Q$-statistics, indicating successful identification of the damaged motor. Clear separation is observed across motors, including at the lower damage levels such as 2 mm and 4 mm, supporting the reliability of the pooled FP-AR framework for motor-level damage identification.

\subsubsection{Damage quantification}
\paragraph{Nonlinear inverse optimization.}
The first part of the damage quantification analysis is based on the nonlinear
RSS-based estimator derived from the pooled FP-AR models. For each signal--motor
configuration, the RSS criterion of Eq.~\eqref{eq:det-nls} is minimized continuously over the
admissible damage range, yielding the corresponding damage size estimate. The variability of the estimates is
characterized by the standard deviation across the test segments.

\begin{figure*}[t!]
	\centering
	\begin{subfigure}{.32\textwidth}
		\centering
		\captionsetup{width=\linewidth}
		\includegraphics[width=\linewidth]{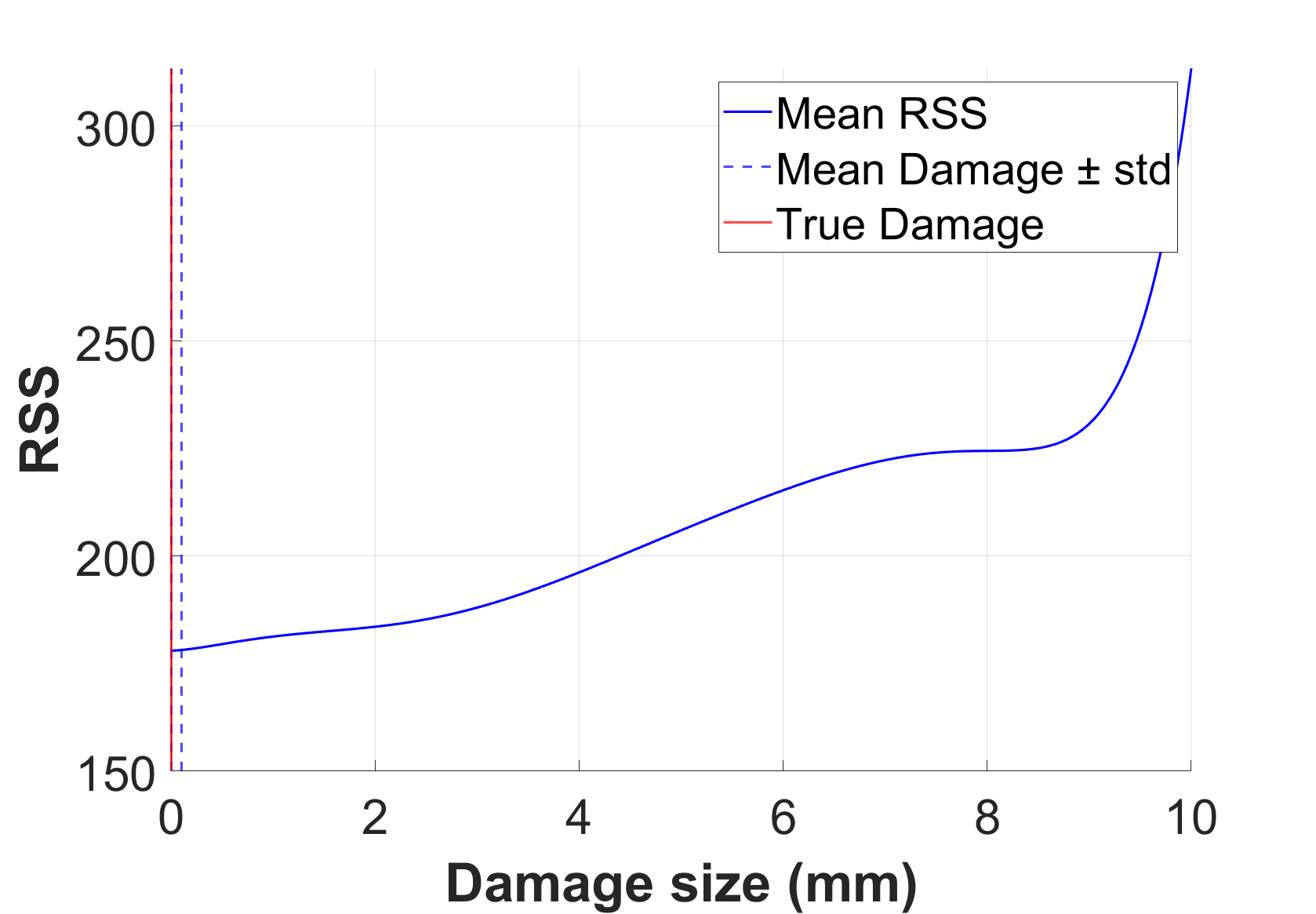}
		\caption{M1, 0 mm damage}
	\end{subfigure}
	\begin{subfigure}{.32\textwidth}
		\centering
		\captionsetup{width=\linewidth}
		\includegraphics[width=\linewidth]{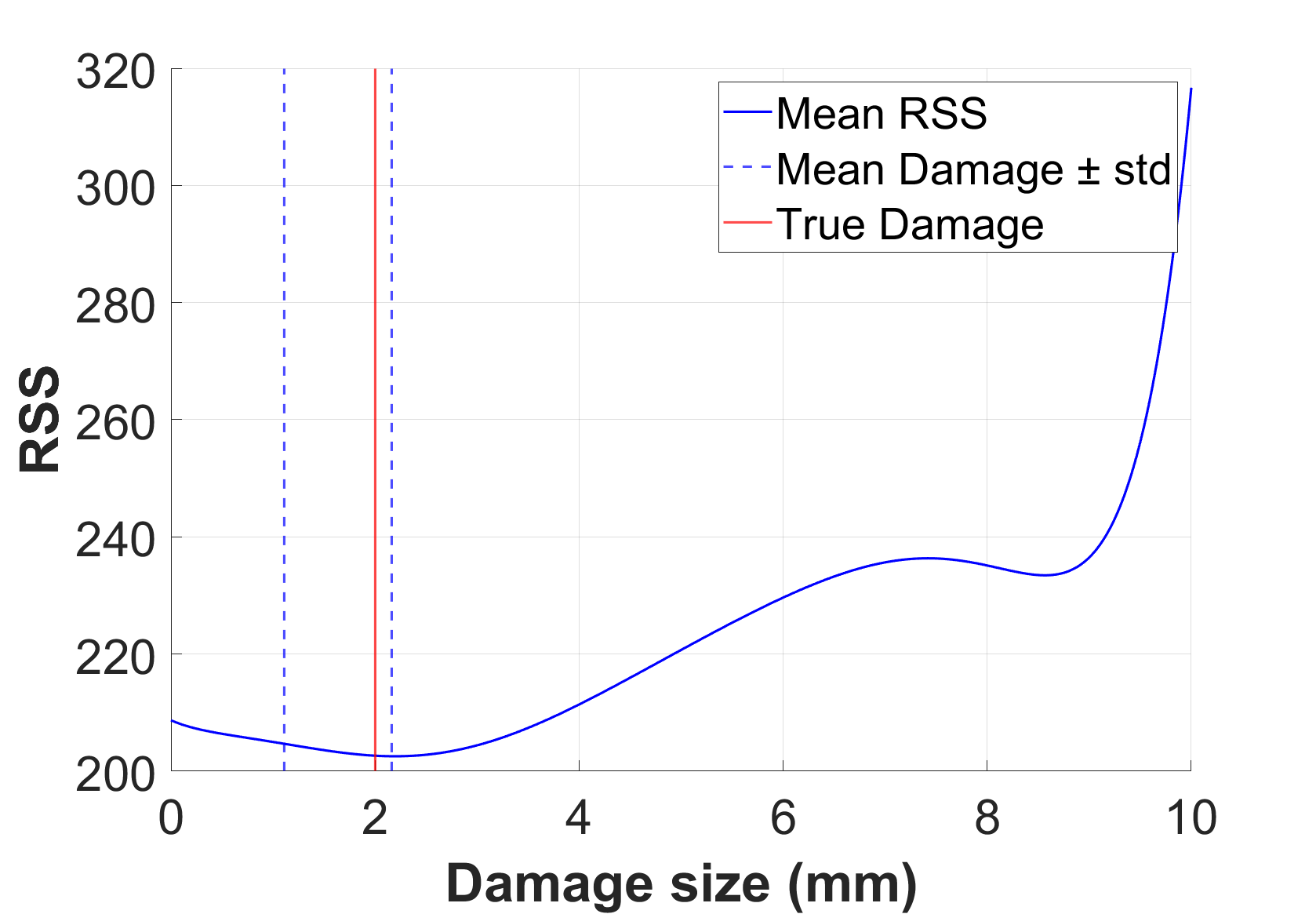}
		\caption{M1, 2 mm damage}
	\end{subfigure}
    \begin{subfigure}{.32\textwidth}
		\centering
		\captionsetup{width=\linewidth}
		\includegraphics[width=\linewidth]{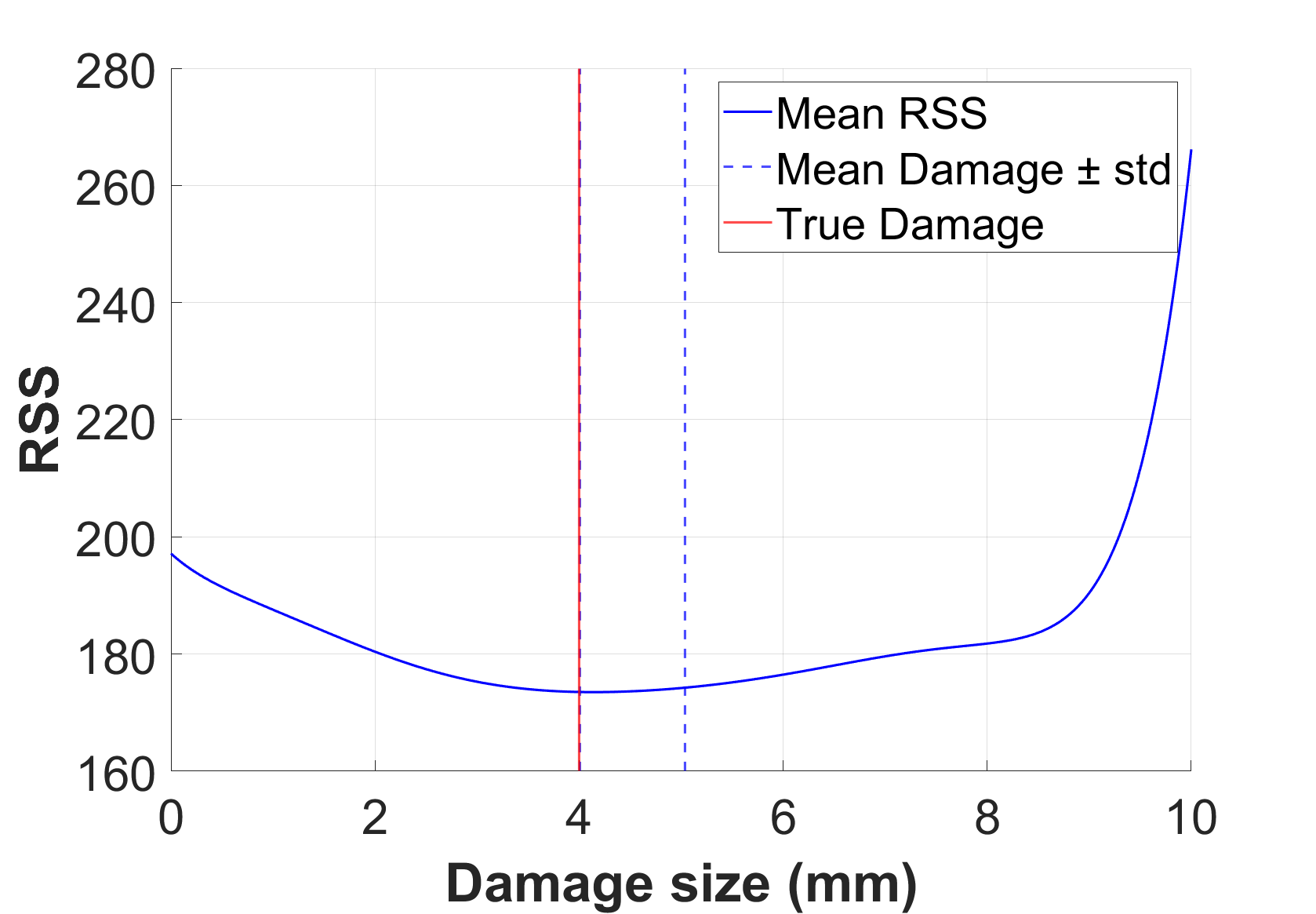}
		\caption{M1, 4 mm damage}
	\end{subfigure}
 
	\begin{subfigure}{.32\textwidth}
		\centering
		\captionsetup{width=\linewidth}
		\includegraphics[width=\linewidth]{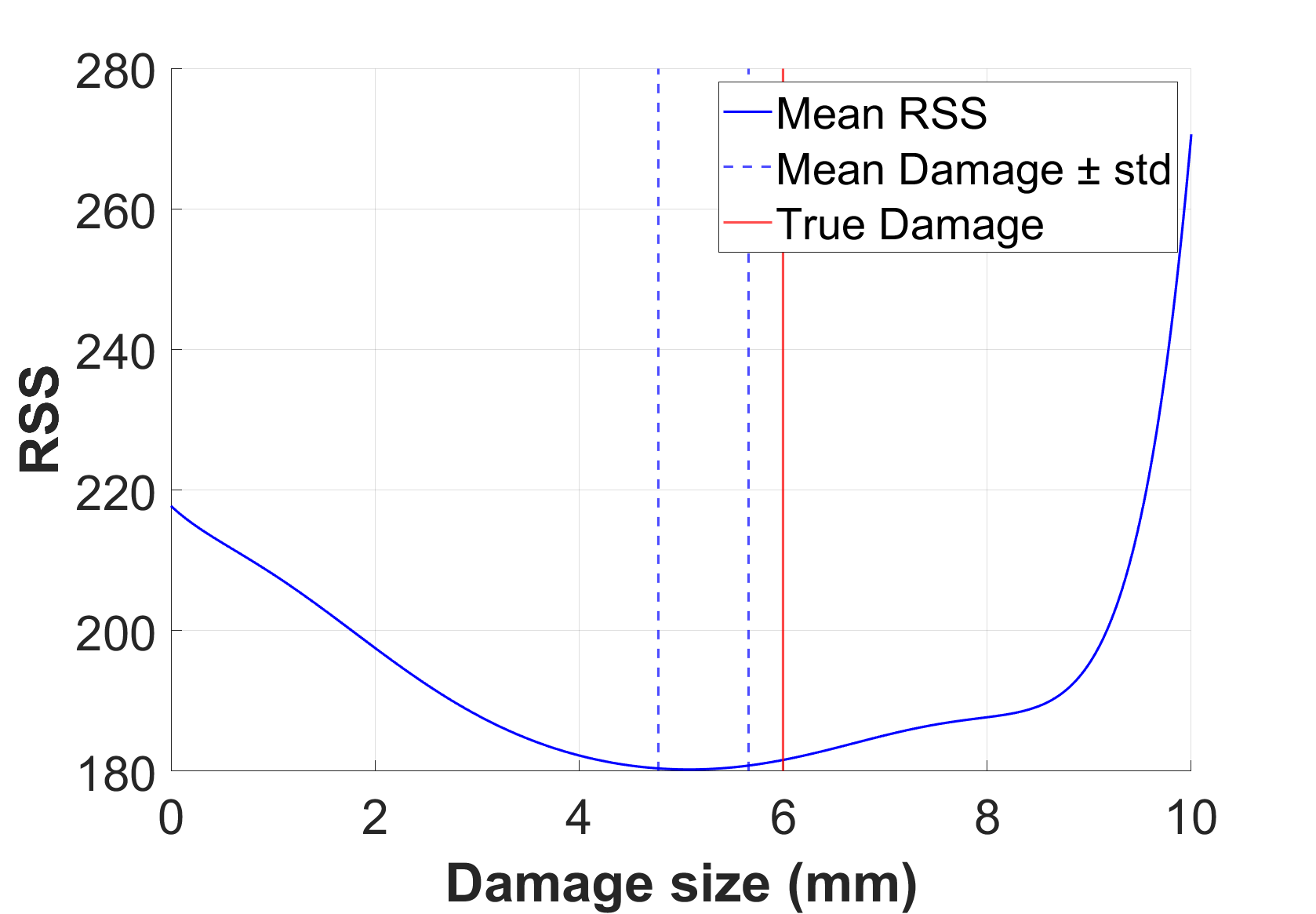}
		\caption{M1, 6 mm damage}
	\end{subfigure}
	\begin{subfigure}{.32\textwidth}
		\centering
		\captionsetup{width=\linewidth}
		\includegraphics[width=\linewidth]{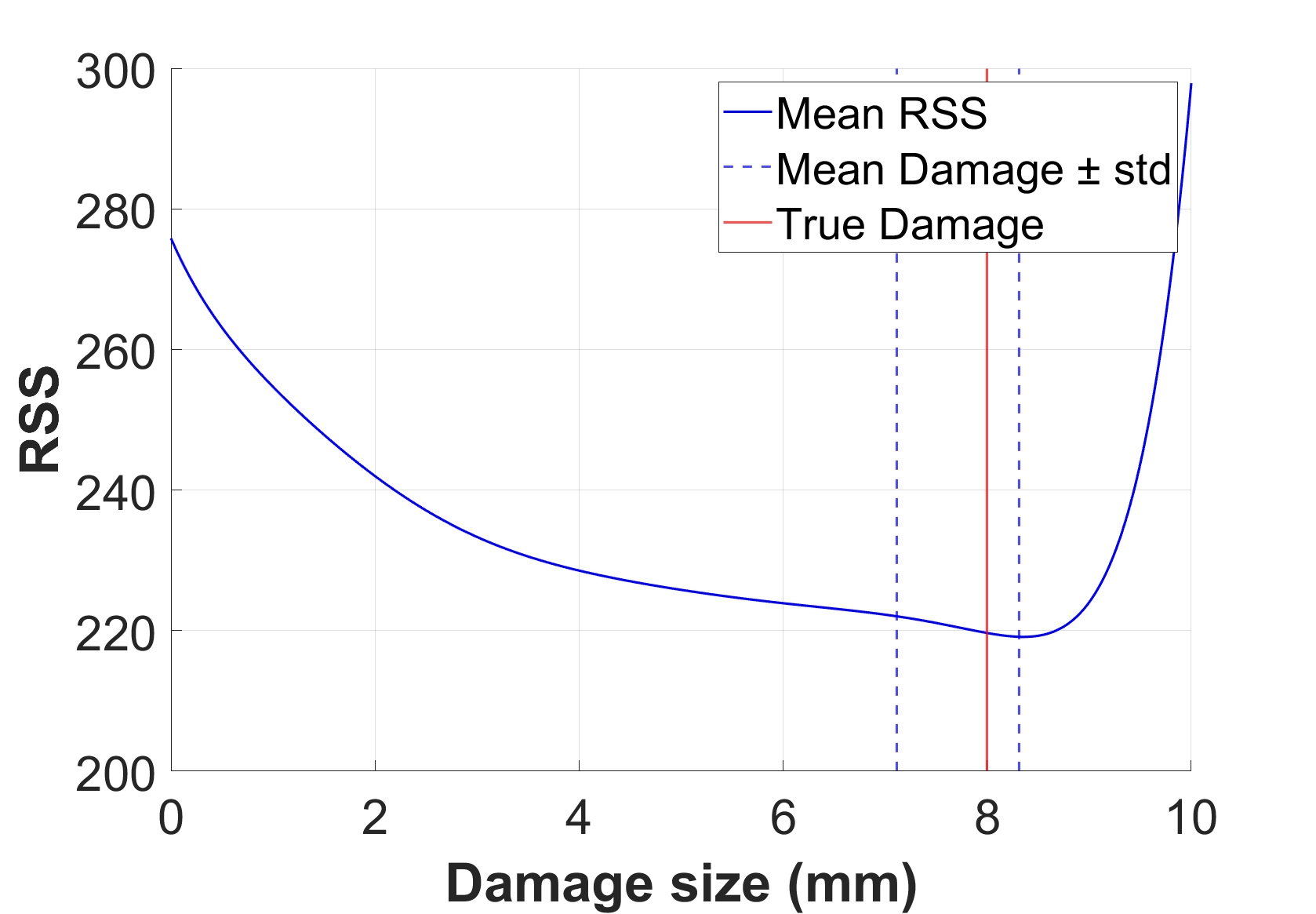}
		\caption{M1, 8 mm damage}
	\end{subfigure}
    \begin{subfigure}{.32\textwidth}
		\centering
		\captionsetup{width=\linewidth}
		\includegraphics[width=\linewidth]{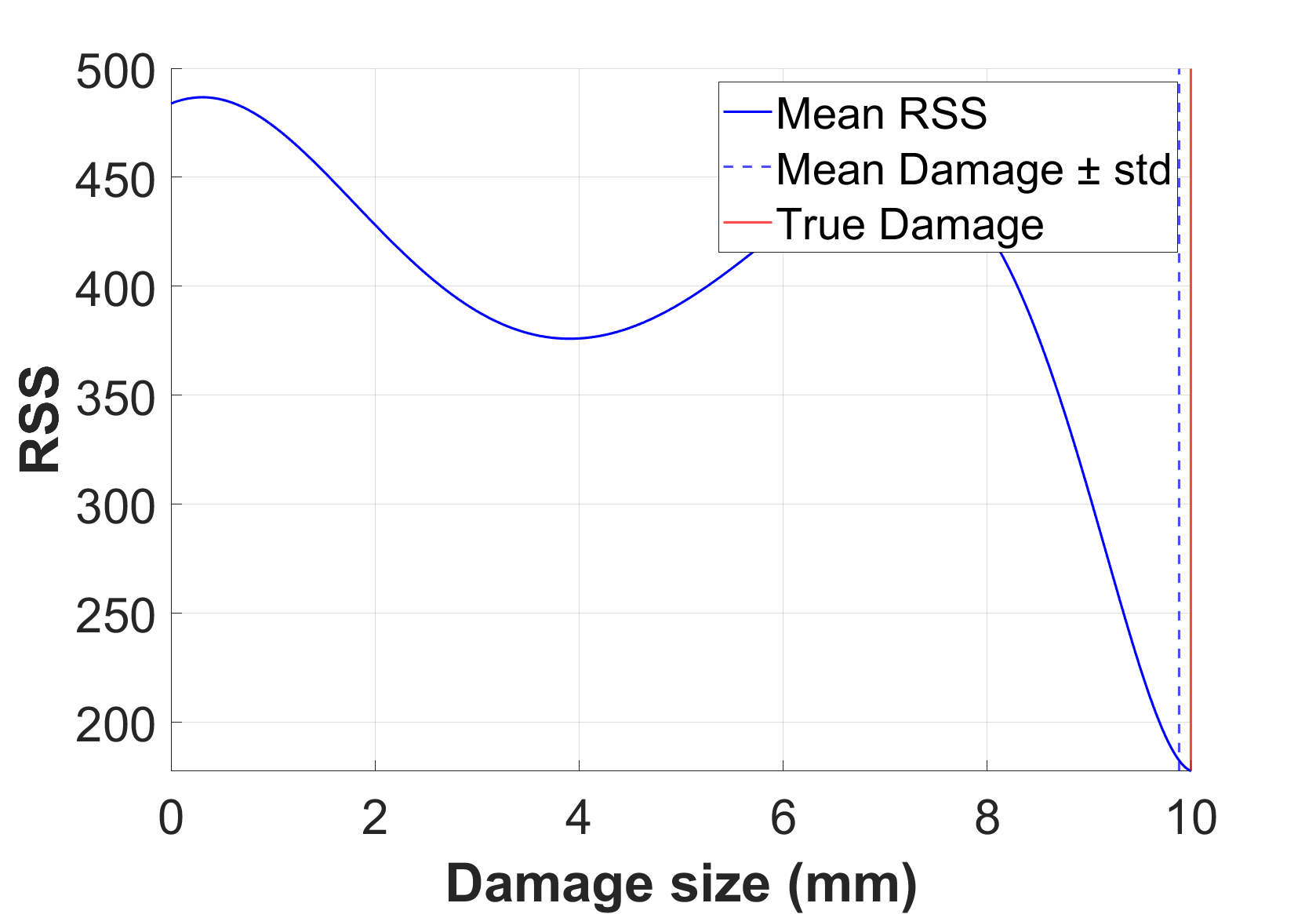}
		\caption{M1, 10 mm damage}
	\end{subfigure}
	\caption{Pooled FP-AR RSS-based damage size estimation using the AccX signal of Motor 1, across all damage levels (healthy to 10 mm).}
 \label{fig:Global damage estimation RSS M1}
\end{figure*}

\begin{figure*}[t!]
	\centering
	\begin{subfigure}{.32\textwidth}
		\centering
		\captionsetup{width=\linewidth}
		\includegraphics[width=\linewidth]{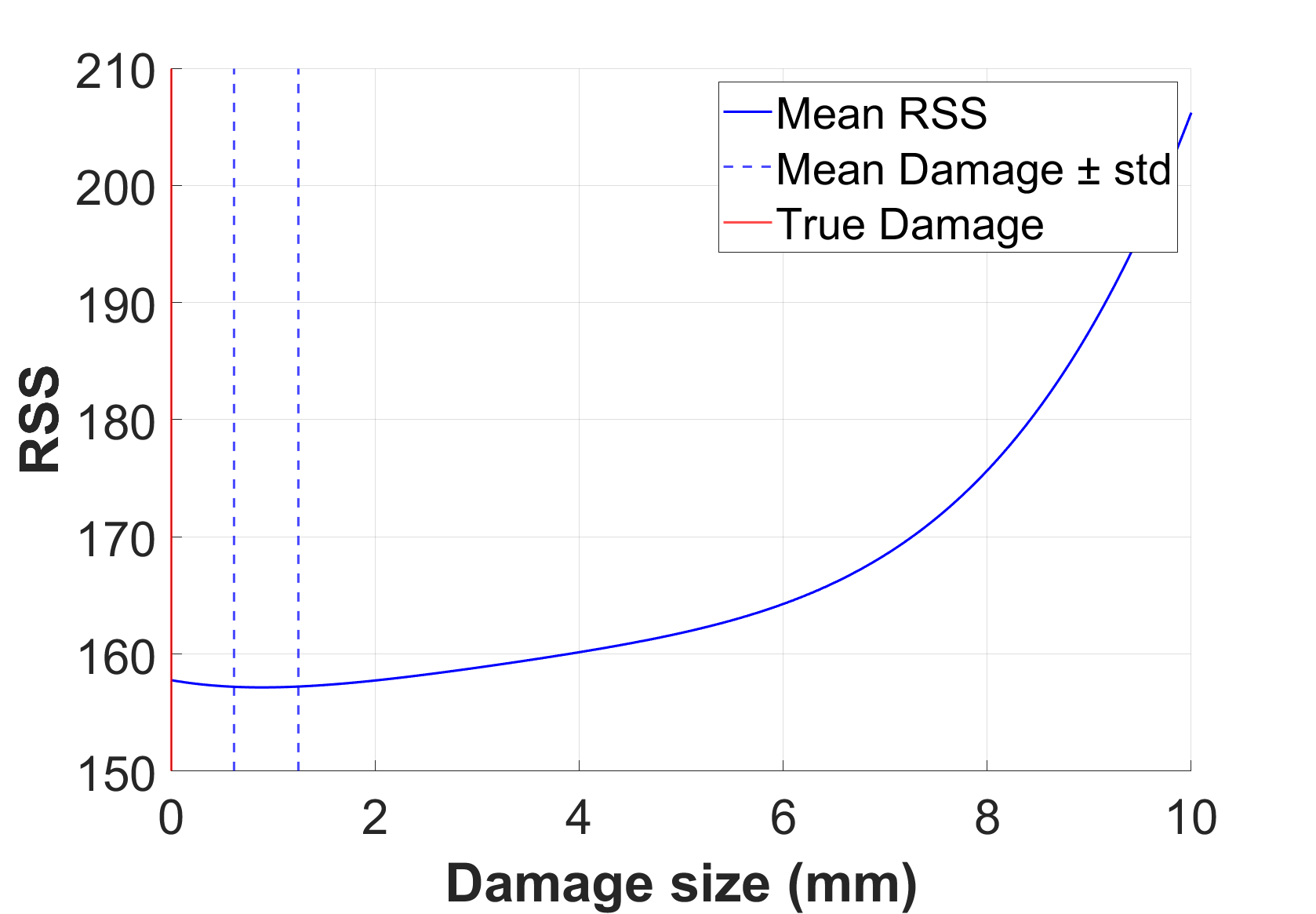}
		\caption{M6, 0 mm damage}
	\end{subfigure}
	\begin{subfigure}{.32\textwidth}
		\centering
		\captionsetup{width=\linewidth}
		\includegraphics[width=\linewidth]{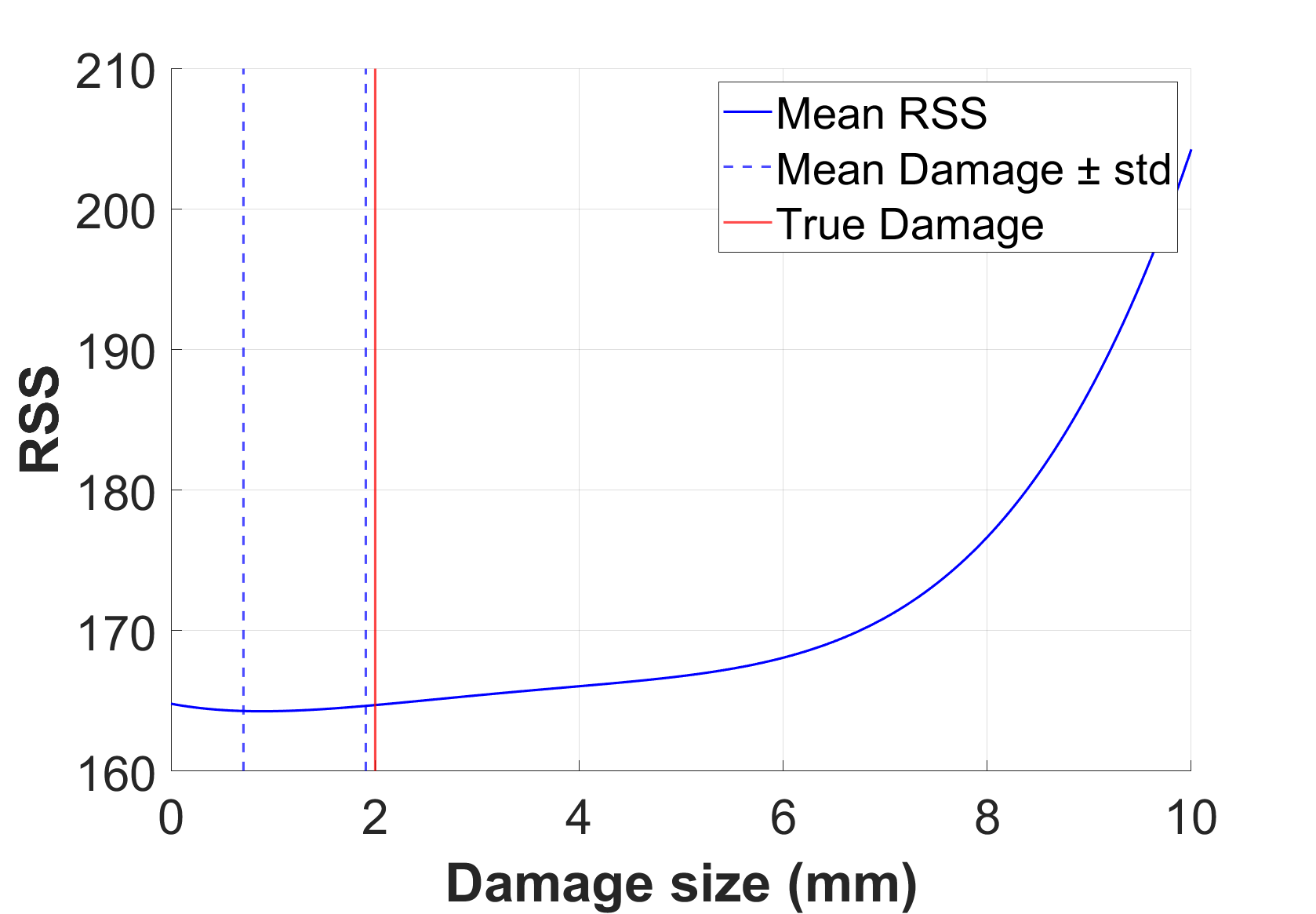}
		\caption{M6, 2 mm damage}
	\end{subfigure}
    \begin{subfigure}{.32\textwidth}
		\centering
		\captionsetup{width=\linewidth}
		\includegraphics[width=\linewidth]{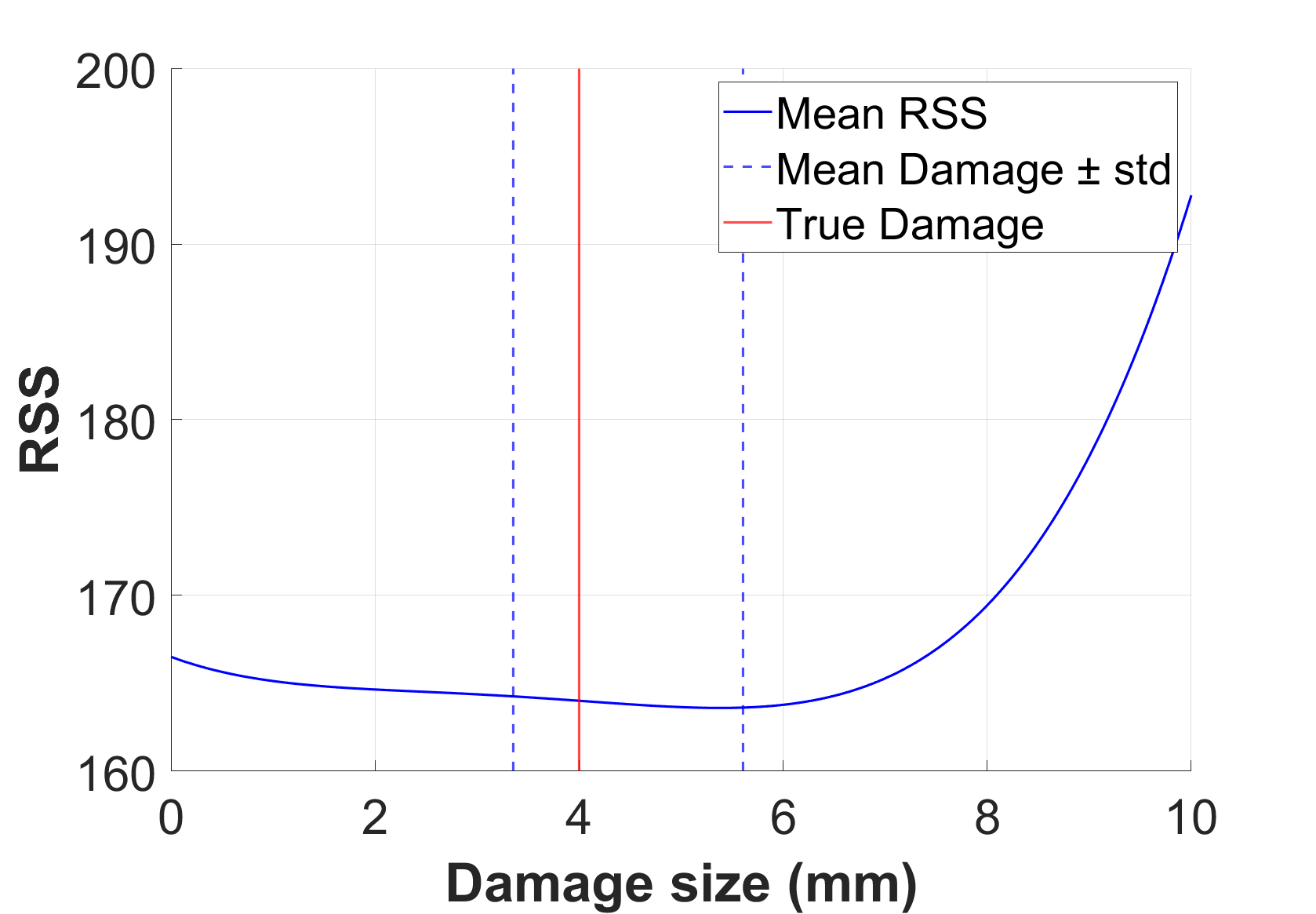}
		\caption{M6, 4 mm damage}
	\end{subfigure}
 
	\begin{subfigure}{.32\textwidth}
		\centering
		\captionsetup{width=\linewidth}
		\includegraphics[width=\linewidth]{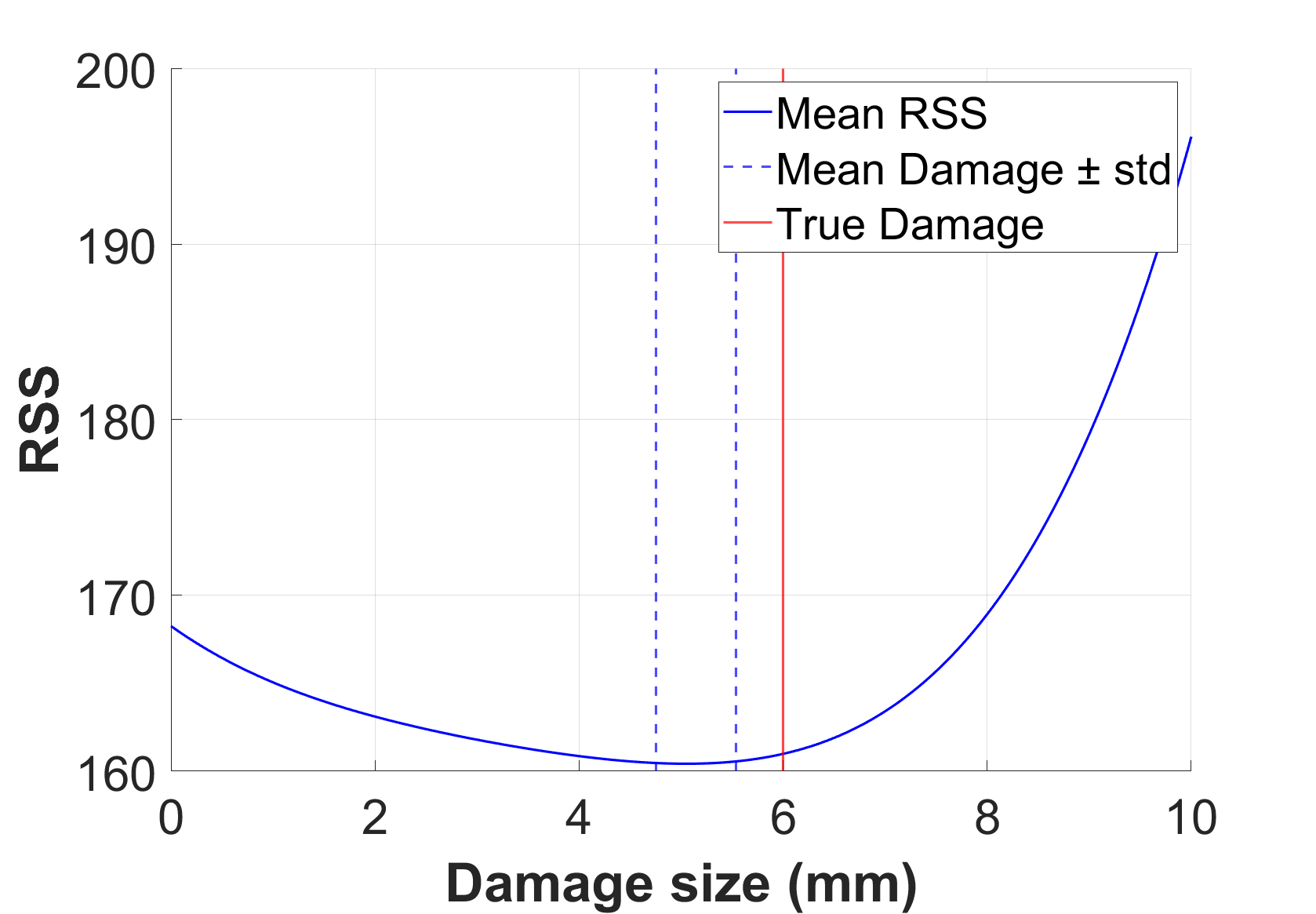}
		\caption{M6, 6 mm damage}
		
	\end{subfigure}
	\begin{subfigure}{.32\textwidth}
		\centering
		\captionsetup{width=\linewidth}
		\includegraphics[width=\linewidth]{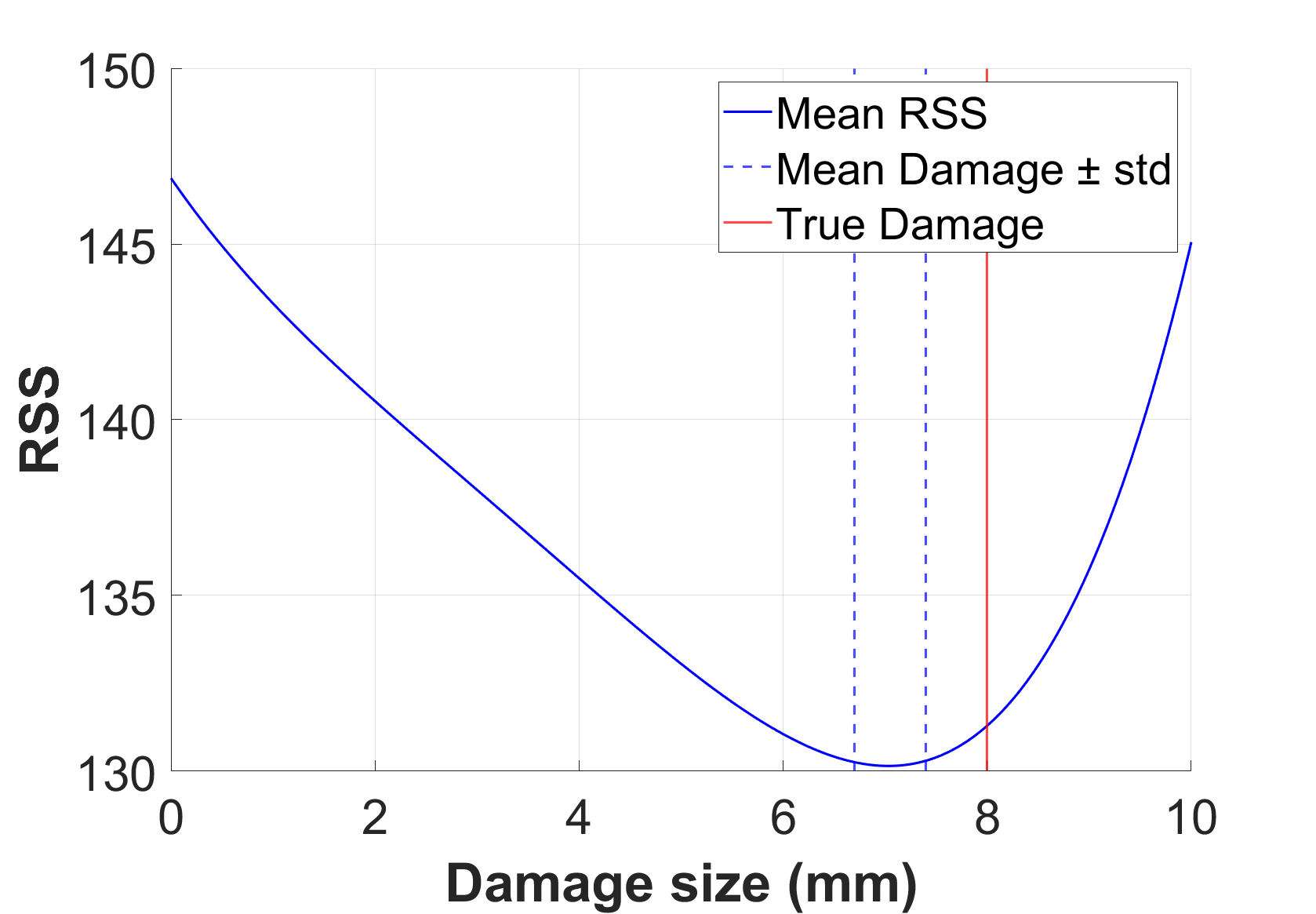}
		\caption{M6, 8 mm damage}
	\end{subfigure}
    \begin{subfigure}{.32\textwidth}
		\centering
		\captionsetup{width=\linewidth}
		\includegraphics[width=\linewidth]{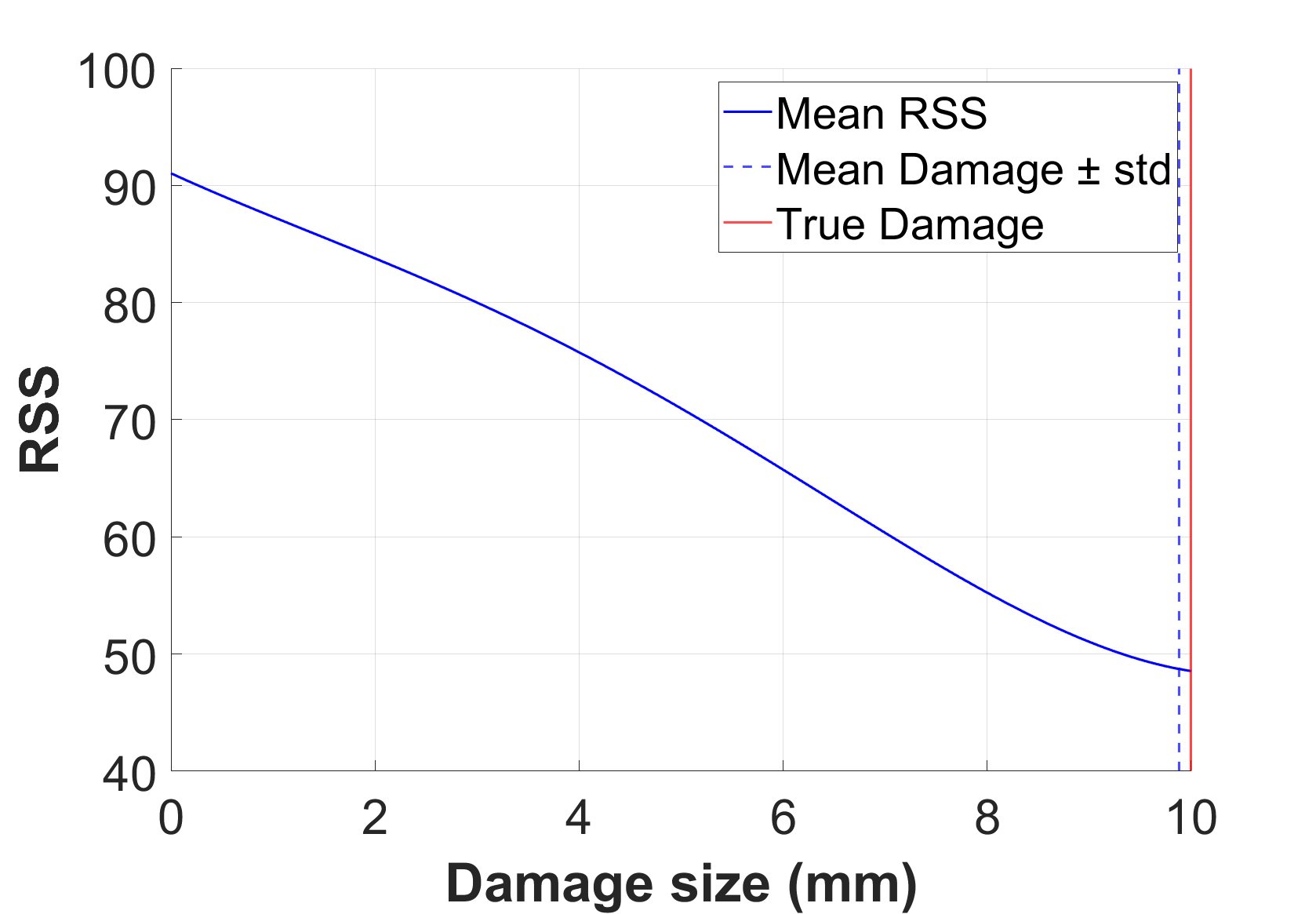}
		\caption{M6, 10 mm damage}
	\end{subfigure}
	\caption{Pooled FP-AR RSS-based damage size estimation using the AccX signal of Motor 6, across all damage levels (healthy to 10 mm).}
 \label{fig:Global damage estimation RSS M6}
\end{figure*}

Fig.~\ref{fig:Global damage estimation RSS M1} presents the damage quantification results for Motor 1; the corresponding Motor 3 and 6 results are provided in Appendix C (Figs.~\ref{fig:Global damage estimation RSS M3} and~\ref{fig:Global damage estimation RSS M6}). The results indicate that aggregating the parameter estimates across multiple segments yields a smoother RSS landscape and reduced between-segment variability, thereby improving the robustness to segment-to-segment fluctuations and operating-condition changes. Relative to the local (single-segment) FP-AR fits, the pooled model achieves comparable point accuracy in locating the RSS minimum across damage levels and motors, while offering more stable minima---most notably at the healthy and higher-damage regimes, where the pooled curves are less flat and the uncertainty bands tighten. In cases with limited basis dimensionality (e.g., Motor 3), the advantage manifests itself primarily as consistency rather than a systematic reduction of bias. Overall, pooling does not uniformly surpass the local model in absolute error, but provides a statistically steadier estimator and a more regularized objective surface, which are desirable for reliable decision-making under repeated flights.

\begin{figure}[!t]
	\centering
	\begin{subfigure}{.3\textwidth}
		\centering
		\includegraphics[width=\linewidth]{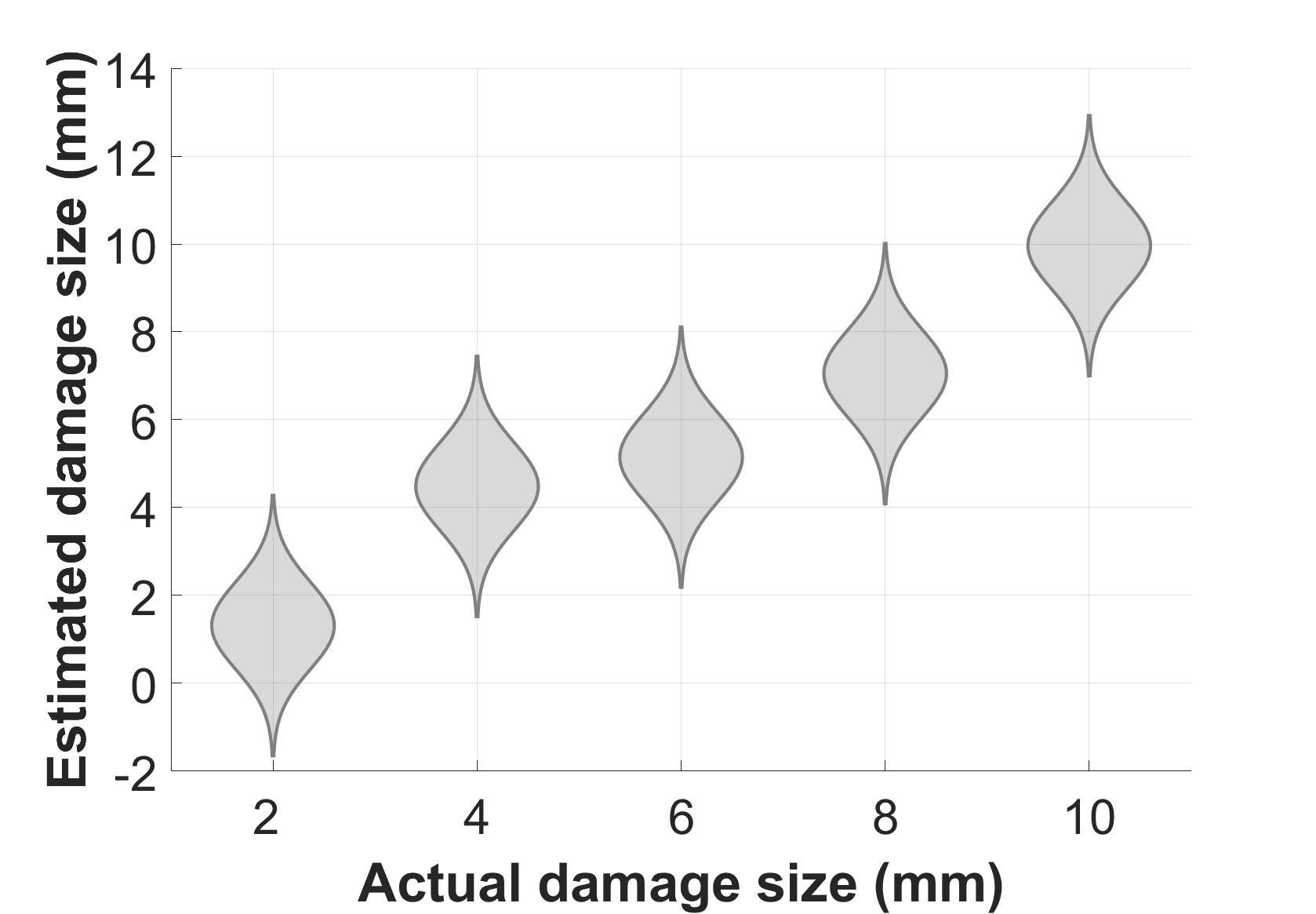}
		\caption{AccX}
		\label{fig:AccX violin}
	\end{subfigure}
    \hfill 
	\begin{subfigure}{.3\textwidth}
		\centering
		\includegraphics[width=\linewidth]{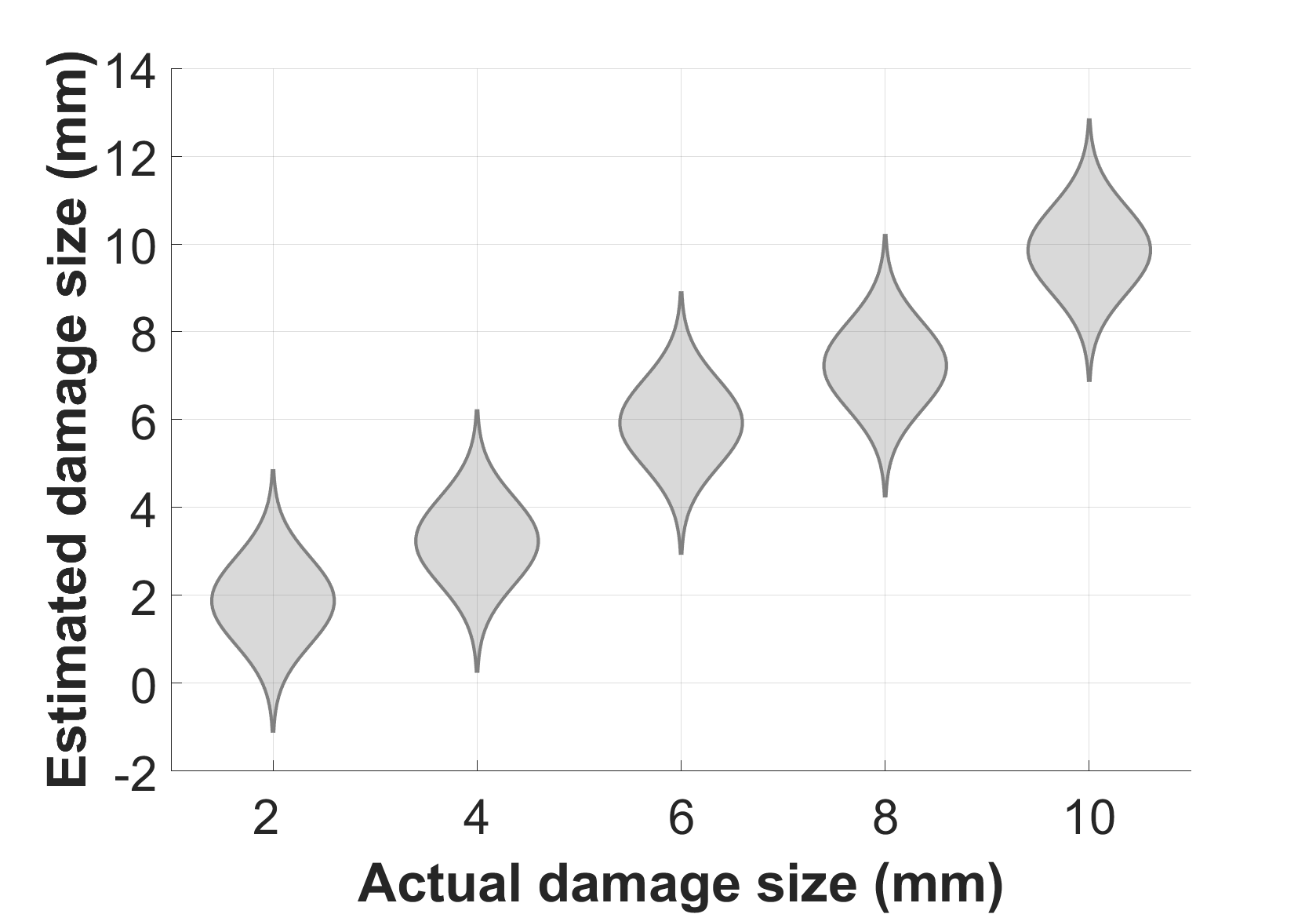}
		\caption{AccY}
		\label{fig:AccY violin}
	\end{subfigure}
    \hfill 
    \begin{subfigure}{.3\textwidth}
		\centering
		\includegraphics[width=\linewidth]{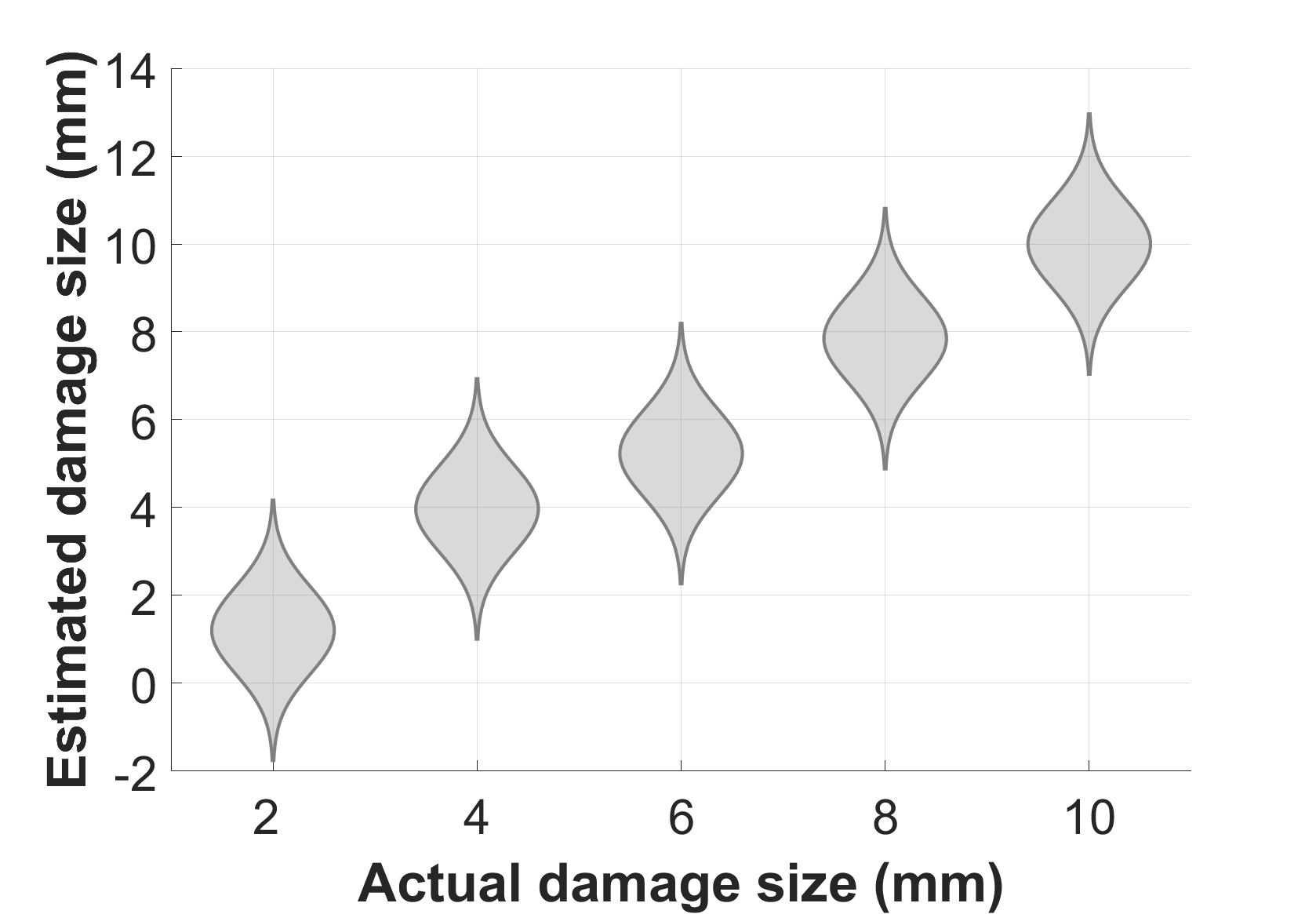}
		\caption{AccZ}
		\label{fig:AccZ violin}
	\end{subfigure}

	\begin{subfigure}{.3\textwidth}
		\centering
		\includegraphics[width=\linewidth]{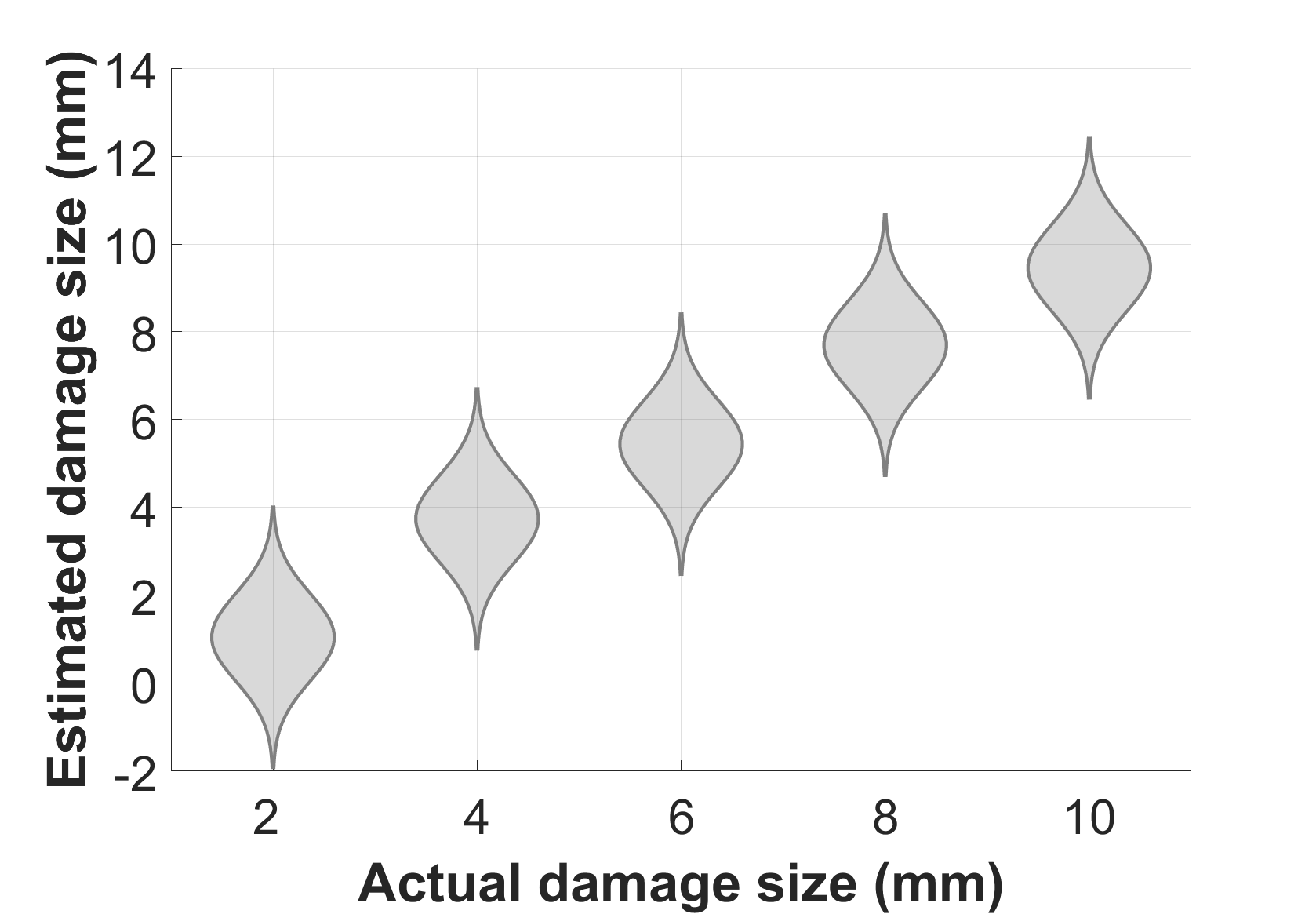}
		\caption{GyrX}
		\label{fig:GyrX violin}
	\end{subfigure}
    \hfill 
	\begin{subfigure}{.3\textwidth}
		\centering
		\includegraphics[width=\linewidth]{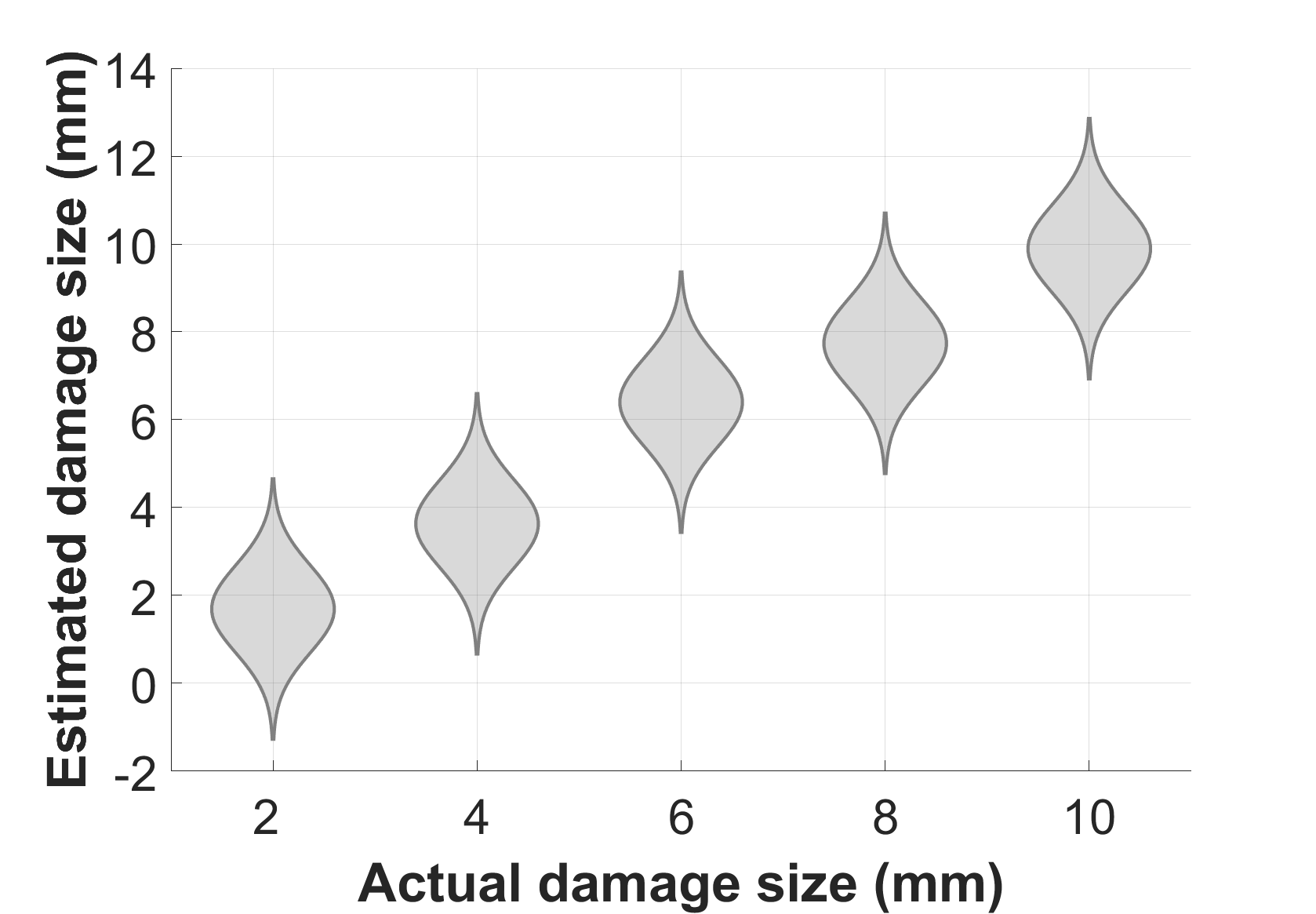}
		\caption{GyrY}
		\label{fig:GyrY violin}
	\end{subfigure}
    \hfill 
    \begin{subfigure}{.3\textwidth}
		\centering
		\includegraphics[width=\linewidth]{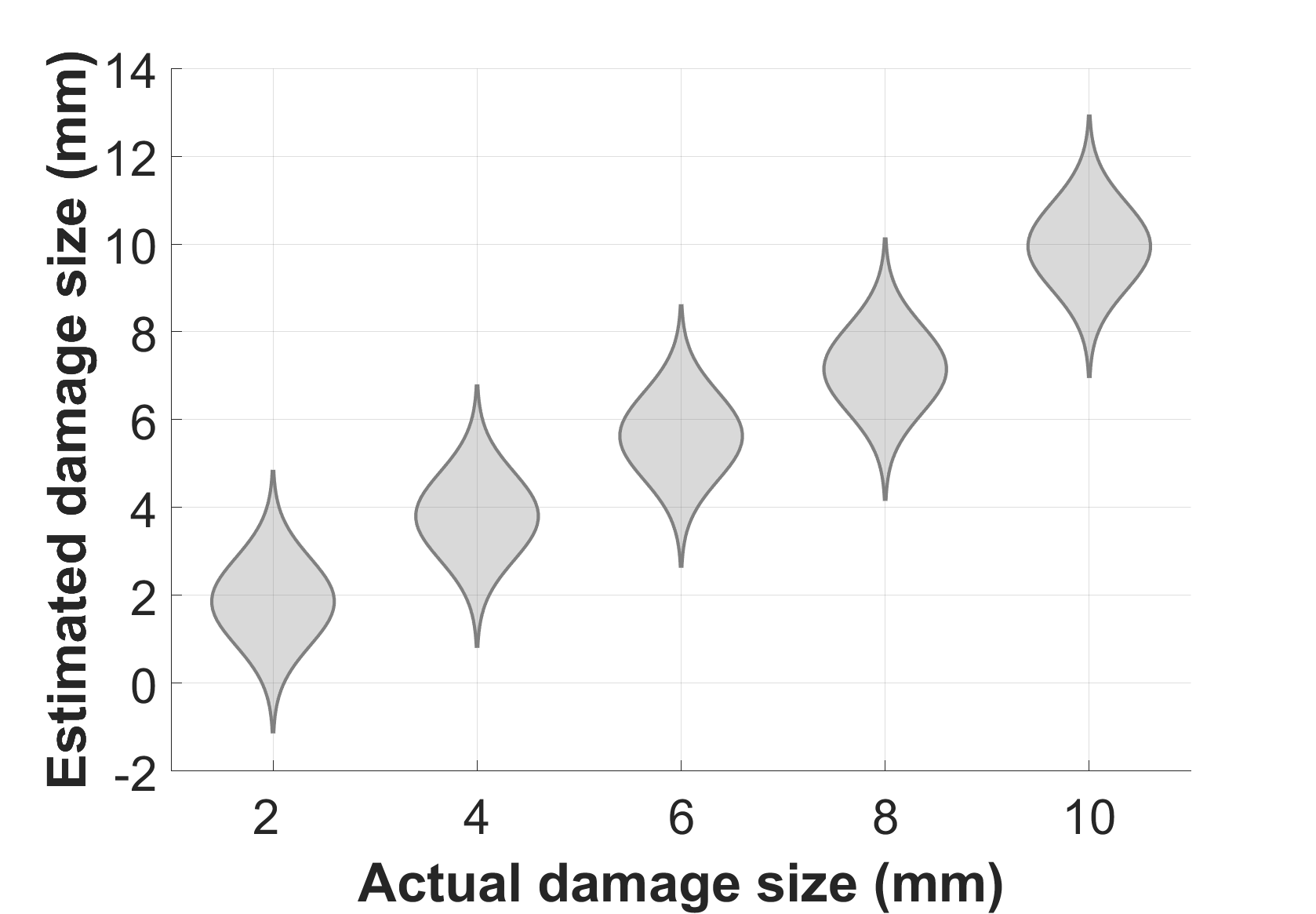}
		\caption{GyrZ}
		\label{fig:GyrZ violin}
	\end{subfigure}
    
	\caption{Violin plots of the estimated damage levels for the six signals (AccX/Y/Z and GyrX/Y/Z) of Motor 6 across the actual damage states (healthy to 10 mm).}
    \label{fig:M6 Violin}
\end{figure}

To assess the estimation performance across the different signal sources, Fig.~\ref{fig:M6 Violin} summarizes the final damage estimation results for the six signals using violin plots. Each distribution represents 400 damage estimates per level, with Gaussian KDE applied and normalized across damage levels for comparison. The shape of each violin reflects the spread and central tendency of the estimates. These results demonstrate varying levels of sensitivity depending on the signal source, with AccX and GyrY generally exhibiting tighter distributions and higher accuracy. In general, the estimation accuracy improves with increasing damage magnitude, as larger damage induces more pronounced changes in the system dynamics, making it easier to capture.

\begin{figure}[!t]
	\centering
	\begin{subfigure}{.3\textwidth}
		\centering
		\includegraphics[width=\linewidth]{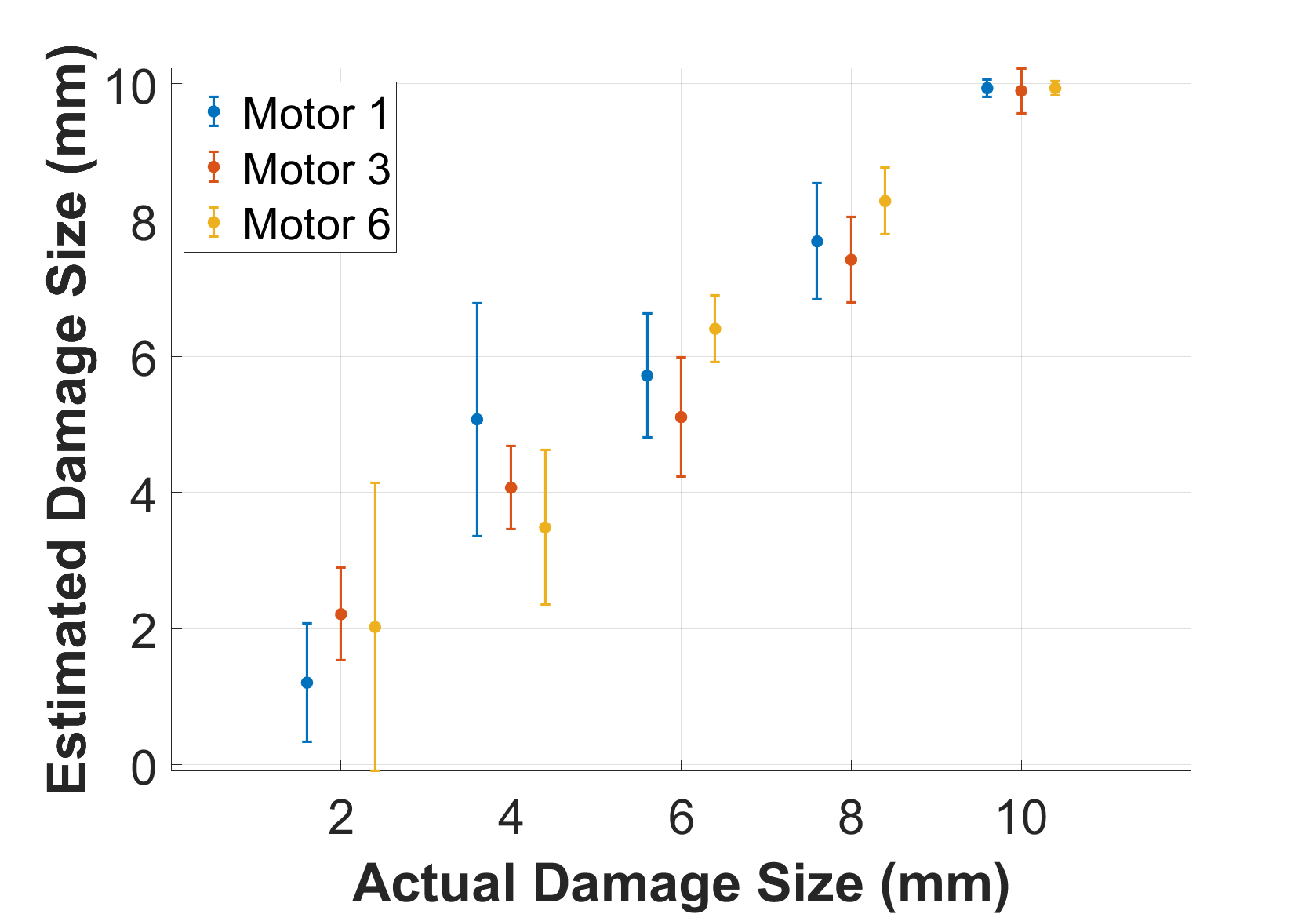}
		\caption{AccX}
	\end{subfigure}
    \hfill 
	\begin{subfigure}{.3\textwidth}
		\centering
		\includegraphics[width=\linewidth]{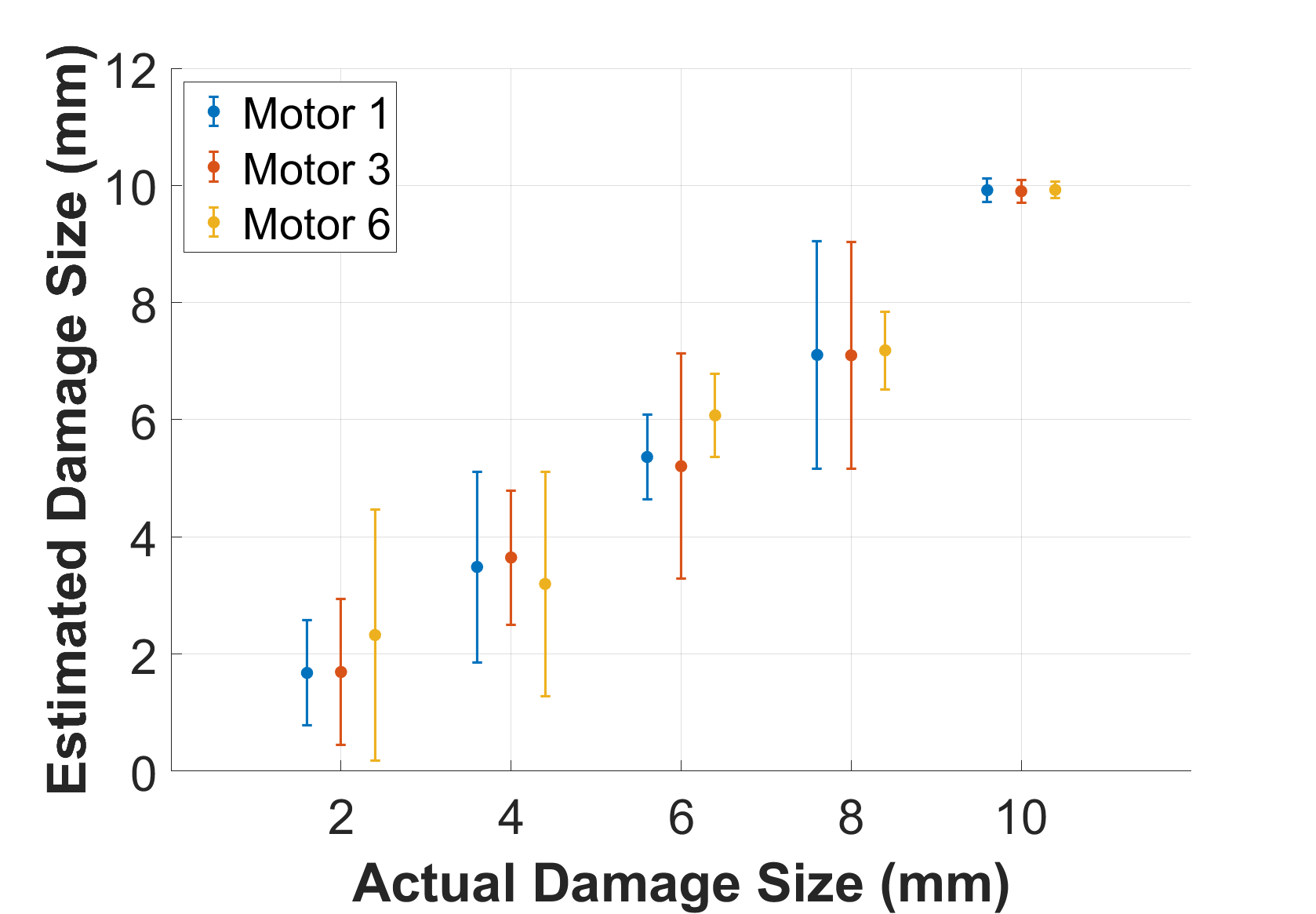}
		\caption{AccY}
	\end{subfigure}
    \hfill 
    \begin{subfigure}{.3\textwidth}
		\centering
		\includegraphics[width=\linewidth]{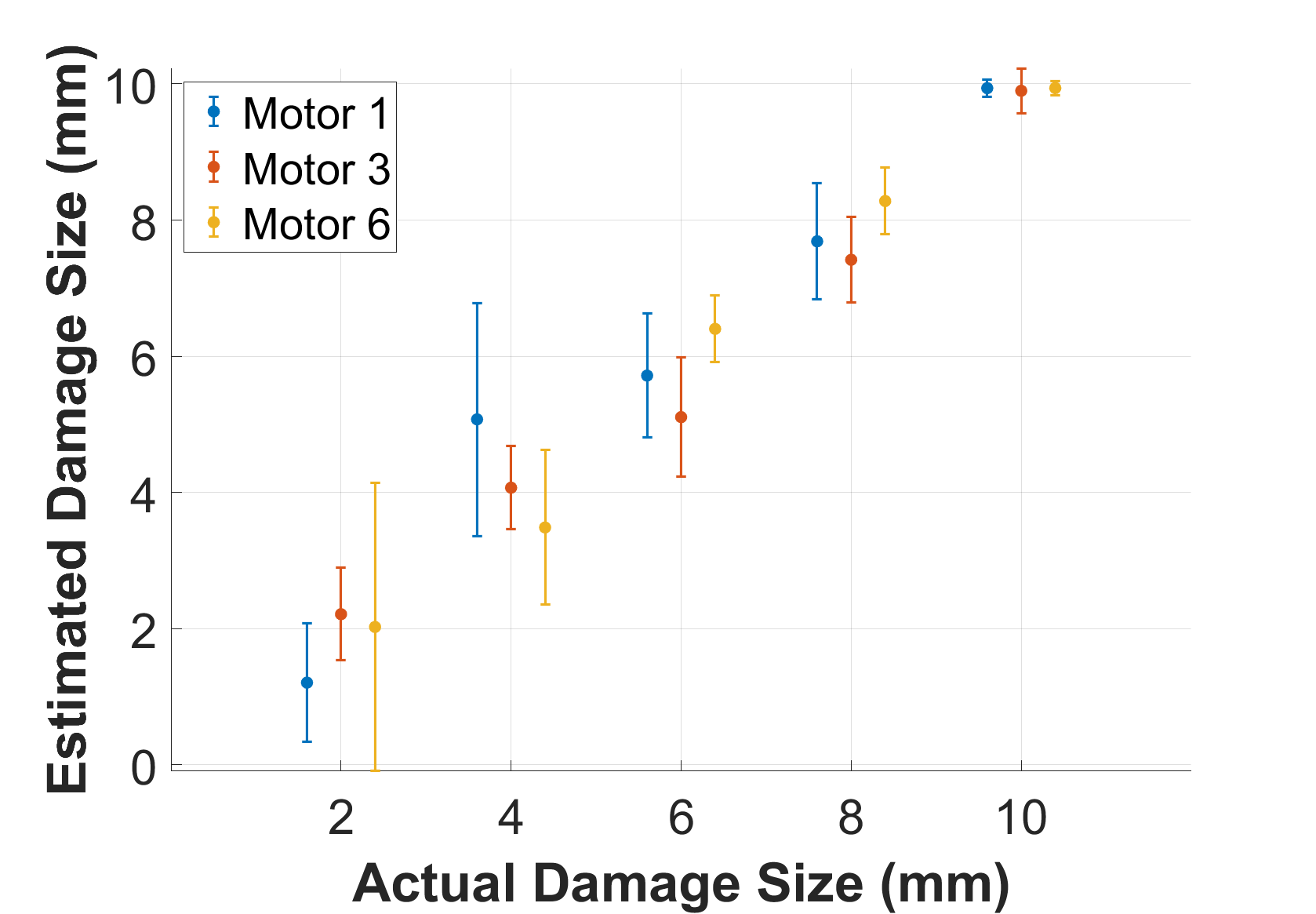}
		\caption{AccZ}
	\end{subfigure}

	\begin{subfigure}{.3\textwidth}
		\centering
		\includegraphics[width=\linewidth]{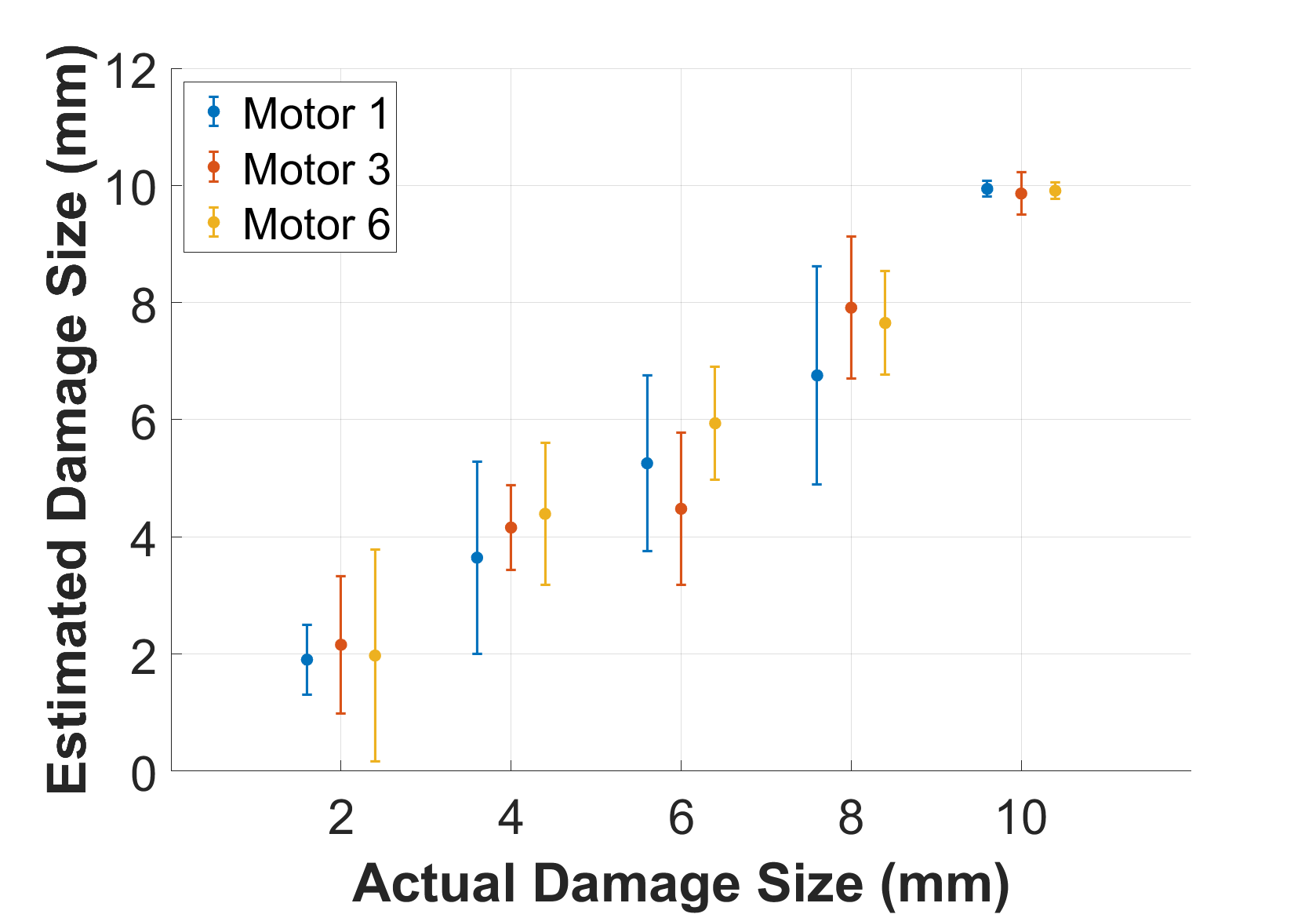}
		\caption{GyrX}
	\end{subfigure}
    \hfill 
	\begin{subfigure}{.3\textwidth}
		\centering
		\includegraphics[width=\linewidth]{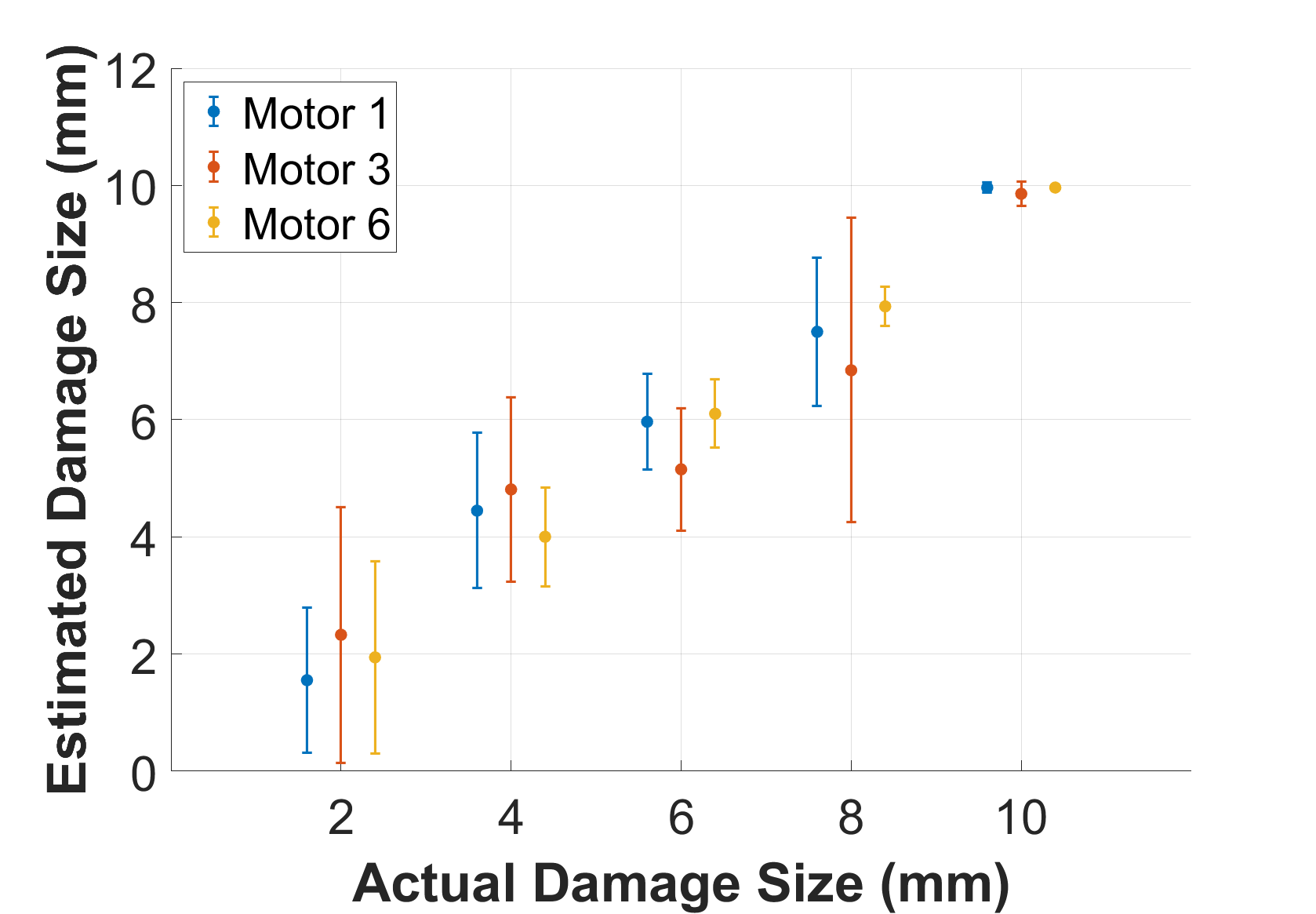}
		\caption{GyrY}
	\end{subfigure}
    \hfill 
    \begin{subfigure}{.3\textwidth}
		\centering
		\includegraphics[width=\linewidth]{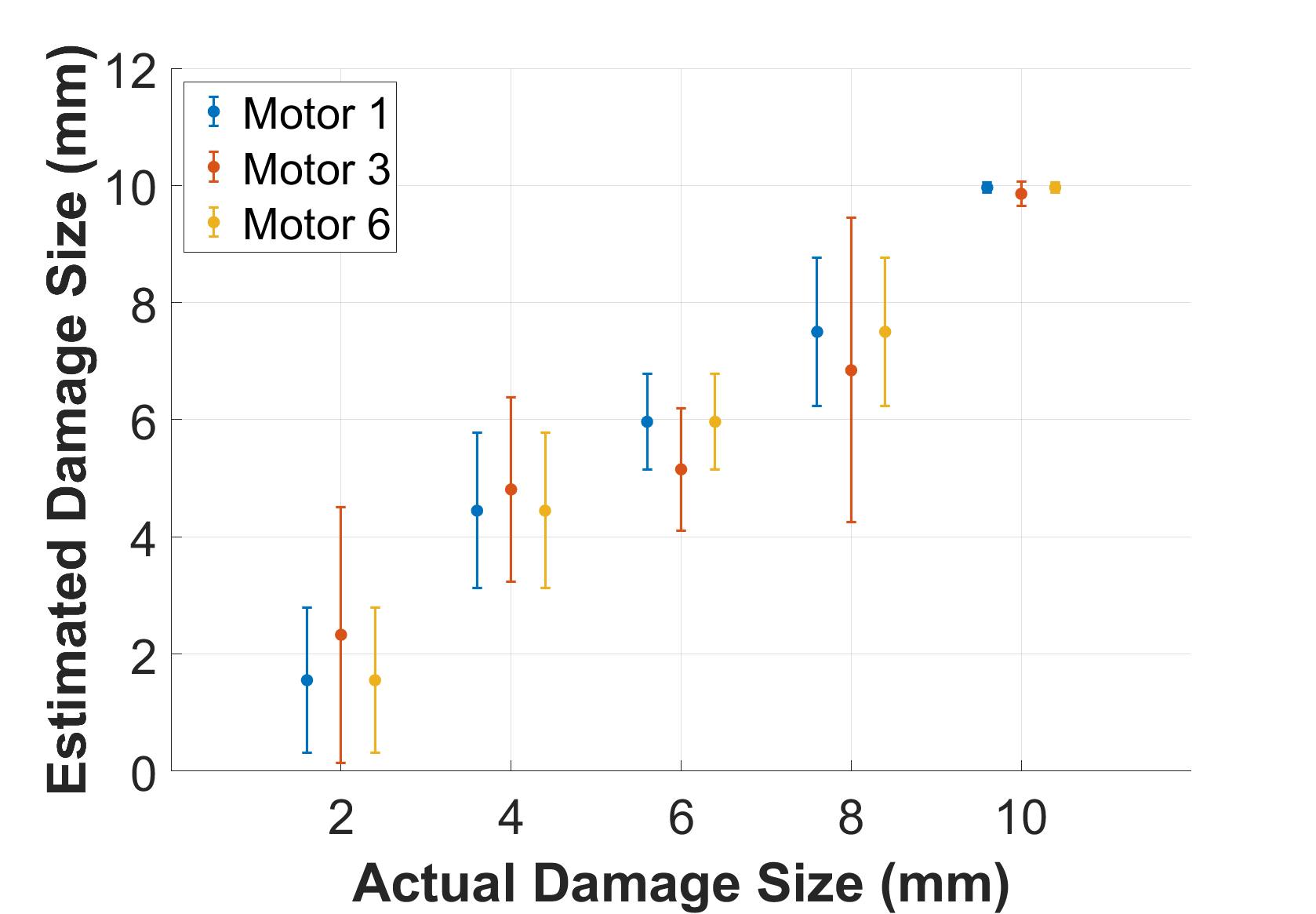}
		\caption{GyrZ}
	\end{subfigure}
    
	\caption{Summary plots with error bars of the estimated damage levels for the six signals (AccX/Y/Z and GyrX/Y/Z) of Motor 6 across the actual damage states (healthy to 10 mm).}
    \label{fig:Global damage estimation dotplot all}
\end{figure}

Furthermore, Fig.~\ref{fig:Global damage estimation dotplot all} displays the summary results of the estimated damage levels for all signal--motor combinations, providing a concise overview of the model performance across all configurations. The AccX signal of Motor 6 shows the highest overall accuracy and consistency, likely due to stronger signal-to-noise characteristics and a more informative dynamic response in this configuration. In contrast, signals such as AccZ or GyrZ of Motor 3 exhibit relatively higher variance, suggesting reduced sensitivity or model expressiveness under certain conditions. These plots allow a comparative interpretation of the quantification performance across motors and sensing modalities. Complementing the dot-plot visualization, Table~\ref{tab:dot-error-matrix-nohealthy} reports the numerical mean $\pm$ standard deviation of the estimated damage for every signal--motor pair at each damage level, enabling a quantitative comparison across configurations.


\begin{table*}[b!]
\centering
\footnotesize
\setlength{\tabcolsep}{4pt}
\caption{Summary of the damage size estimation results obtained using the nonlinear
inverse optimization approach for each signal--motor combination at damage levels of 2--10 mm.
The results are reported as mean $\pm$ standard deviation over 16 test data
sets.}
\label{tab:dot-error-matrix-nohealthy}
\begin{tabularx}{\textwidth}{l l *{5}{>{\centering\arraybackslash}X}}
\toprule
 & & \multicolumn{5}{c}{\textbf{Estimated damage size (mm)}} \\
\cmidrule(lr){3-7}
\textbf{Signal} & \textbf{Motor} & \textbf{2} & \textbf{4} & \textbf{6} & \textbf{8} & \textbf{10} \\
\midrule
  & M1 & 1.64 ± 1.05 & 4.52 ± 1.03 & 5.22 ± 0.89 & 7.71 ± 1.19 & 9.95 ± 0.14 \\
AccX & M3 & 1.23 ± 0.62 & 5.30 ± 0.61 & 4.48 ± 1.45 & 7.18 ± 0.62 & 9.91 ± 0.16 \\
  & M6 & 1.31 ± 1.20 & 4.48 ± 2.26 & 5.15 ± 0.78 & 7.05 ± 0.70 & 9.96 ± 0.15 \\
\midrule
  & M1 & 1.48 ± 1.00 & 3.69 ± 2.32 & 5.71 ± 1.34 & 6.70 ± 2.31 & 9.96 ± 0.11 \\
AccY & M3 & 1.16 ± 1.06 & 4.00 ± 2.45 & 5.95 ± 1.83 & 7.24 ± 1.56 & 9.88 ± 0.25 \\
  & M6 & 1.87 ± 2.02 & 3.24 ± 2.26 & 5.92 ± 0.55 & 7.23 ± 0.59 & 9.86 ± 0.25 \\
\midrule
  & M1 & 1.31 ± 1.13 & 5.21 ± 1.74 & 5.40 ± 0.76 & 7.81 ± 1.51 & 9.94 ± 0.14 \\
AccZ & M3 & 1.57 ± 0.56 & 4.00 ± 0.47 & 4.83 ± 0.77 & 7.20 ± 0.59 & 9.83 ± 0.46 \\
  & M6 & 1.21 ± 1.37 & 3.97 ± 1.65 & 5.23 ± 0.69 & 7.84 ± 0.40 & 10.00 ± 0.00 \\
\midrule
  & M1 & 1.46 ± 1.20 & 3.80 ± 2.44 & 5.56 ± 1.22 & 5.98 ± 2.29 & 9.95 ± 0.13 \\
GyrX & M3 & 1.03 ± 0.87 & 3.76 ± 0.92 & 4.73 ± 2.60 & 8.05 ± 0.42 & 9.93 ± 0.09 \\
  & M6 & 1.05 ± 1.49 & 3.74 ± 0.95 & 5.44 ± 1.14 & 7.70 ± 0.61 & 9.46 ± 1.76 \\
\midrule
  & M1 & 1.34 ± 1.45 & 3.83 ± 1.07 & 5.61 ± 0.78 & 6.38 ± 1.75 & 9.96 ± 0.10 \\
GyrY & M3 & 2.01 ± 1.68 & 4.41 ± 1.45 & 4.17 ± 2.01 & 6.04 ± 2.77 & 9.82 ± 0.22 \\
  & M6 & 1.69 ± 1.73 & 3.63 ± 0.97 & 6.40 ± 0.51 & 7.74 ± 0.46 & 9.89 ± 0.20 \\
\midrule
  & M1 & 0.82 ± 1.03 & 5.36 ± 1.62 & 5.16 ± 0.93 & 6.66 ± 1.35 & 9.95 ± 0.13 \\
GyrZ & M3 & 2.03 ± 0.99 & 3.98 ± 0.83 & 2.96 ± 1.13 & 5.92 ± 1.22 & 9.99 ± 0.02 \\
 & M6 & 1.86 ± 1.90 & 3.81 ± 1.71 & 5.63 ± 0.84 & 7.15 ± 0.53 & 9.95 ± 0.12 \\
\bottomrule
\end{tabularx}
\end{table*}

\paragraph{Bayesian damage magnitude estimation.}
To further enhance the damage quantification performance and provide uncertainty
information, Bayesian estimation is introduced as a probabilistic extension of
the preceding RSS-based estimator. For each candidate damage size, the pooled
FP-AR model is used to compute the prediction residuals, and the residual
variance is used to define the likelihood function. The posterior distribution
of the damage size is then obtained by combining this likelihood with a
prescribed prior distribution. In contrast to the deterministic RSS-based
estimator, which only selects the minimum-RSS candidate, the Bayesian approach
characterizes the full posterior probability distribution and, therefore, provides
both a point estimate and an uncertainty measure.

In this study, two different prior settings are considered.
The first is a broad uniform prior over the full admissible damage range,
$k \sim \mathcal{U}(0,10)$ mm, which is used as a weakly informative reference.
The second is a sequential state-informed prior, where the prior range is
restricted according to the previously known or previously estimated damage
state. Specifically, the 2, 4, 6, and 8 mm cases use prior ranges of 0--4,
2--6, 4--8, and 6--10 mm, respectively, while the 10 mm case also employs the 6--10 mm range. For each damage level, the Bayesian inversion is performed over 16 test segments.
For each test segment, posterior samples are generated using an adaptive
Metropolis MCMC algorithm with 10 chains and 500 samples per chain. The first
100 samples of each chain are discarded as burn-in, and the remaining samples
are used to compute the posterior mean and standard deviation of the estimated
damage size. These posterior statistics are then summarized across the 16 test
segments in the corresponding tables.

Table~\ref{tab:bayesian1} summarizes the Bayesian damage size
estimation results obtained using the broad uniform prior over 0--10 mm. Overall,
the estimates follow the increasing trend of the true damage levels, indicating
that the residual-based likelihood contains useful damage-severity
information. Several signal--motor combinations provide accurate estimates, such
as AccX of Motor 3 at 2 mm and 10 mm, AccZ of Motor 3 at 10 mm, and GyrZ of
Motor 3 at 10 mm. However, noticeable deviations are also observed in several
cases. For example, the damage level of 10 mm is substantially underestimated for
AccX-M1, AccY-M1, AccZ-M1, GyrX-M1, GyrX-M6, GyrY-M1, GyrY-M3, and GyrZ-M6.
Similarly, some 8 mm cases are underestimated, such as AccX-M1, AccY-M3,
AccZ-M1, and GyrY-M3. These deviations suggest that the broad prior does not
always lead to a posterior distribution concentrated around the correct damage
level. By examining the posterior distributions of the deviated cases, it is found
that the abnormal estimates are mainly associated with multi-peak posterior
densities. Instead of being concentrated around a single damage level, the
posterior probability mass is often spread over several competing modes. This
indicates that multiple candidate damage sizes can produce similar residual
energy in the RSS-based likelihood.

\begin{figure}[t]
\centering

\begin{subfigure}[t]{0.32\textwidth}
    \centering
    \includegraphics[width=\textwidth]{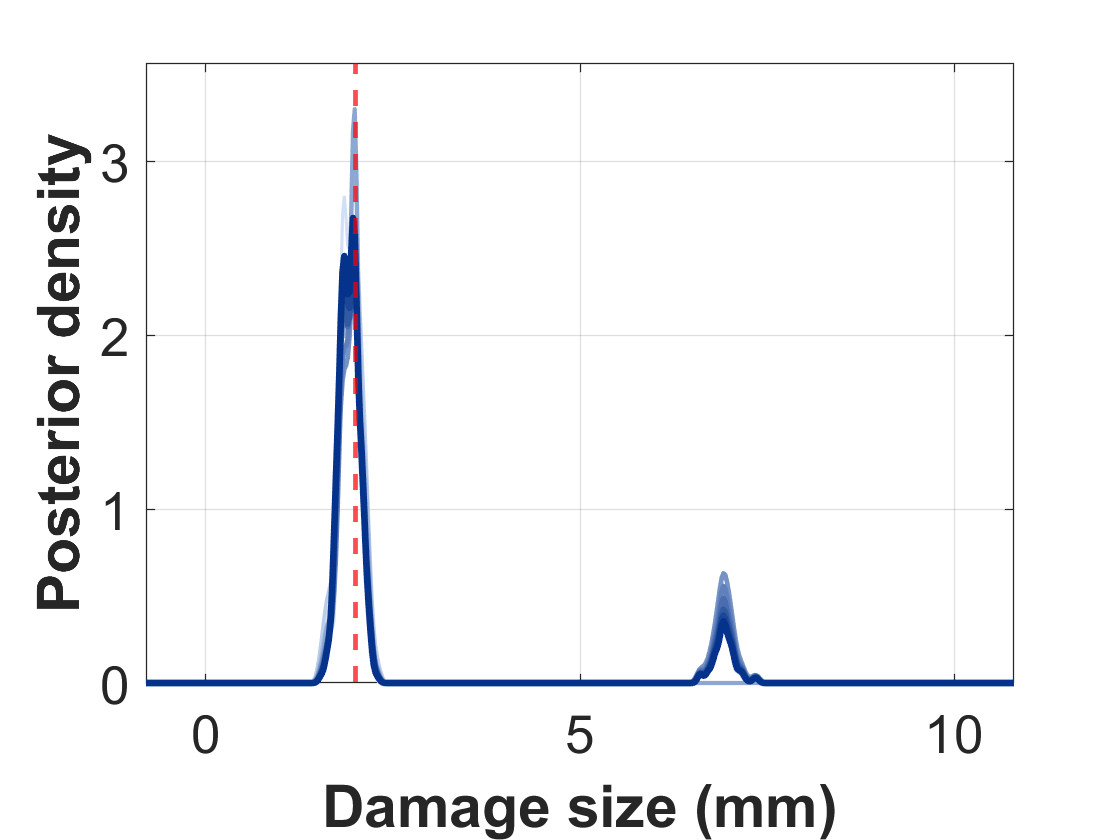}
    \caption{}
    \label{fig:bayes_multipeak_accx_m1_10mm}
\end{subfigure} 
\hfill
\begin{subfigure}[t]{0.32\textwidth}
    \centering
    \includegraphics[width=\textwidth]{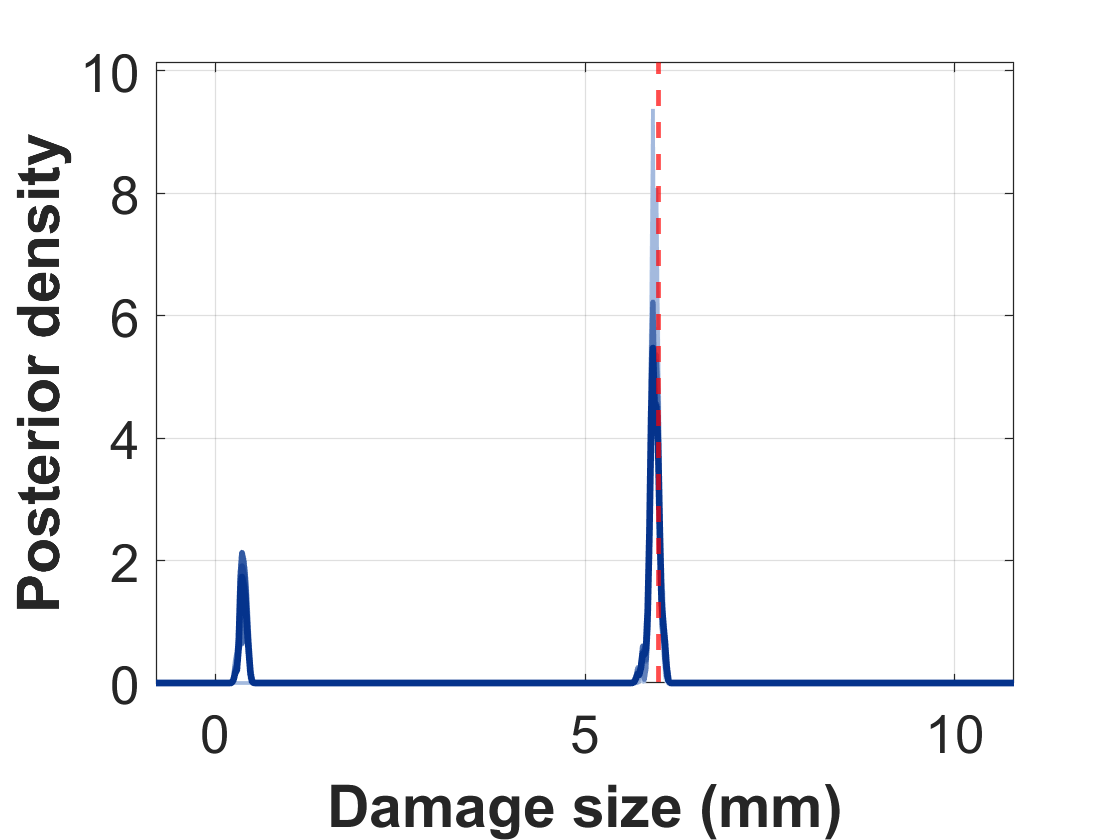}
    \caption{}
    \label{fig:bayes_multipeak_accy_m1_10mm}
\end{subfigure}
\hfill
\begin{subfigure}[t]{0.32\textwidth}
    \centering
    \includegraphics[width=\textwidth]{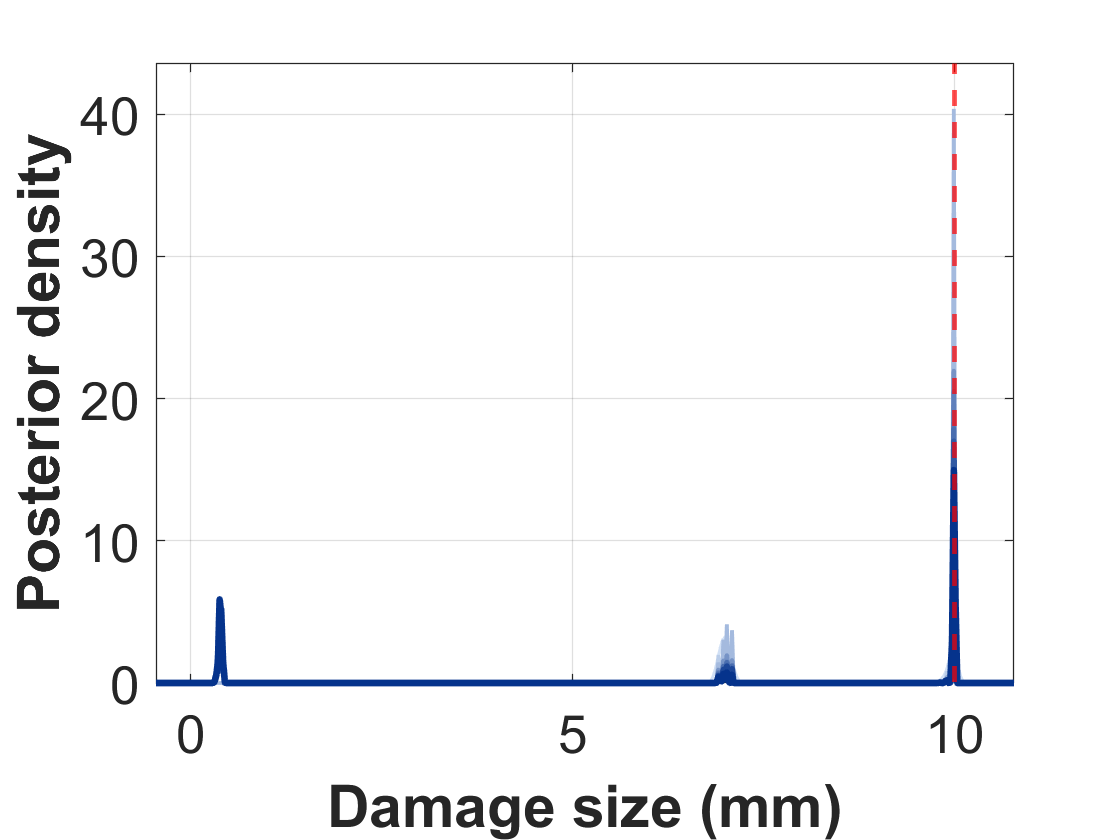}
    \caption{}
    \label{fig:bayes_multipeak_accz_m1_10mm}
\end{subfigure}

\vspace{0.8em}

\begin{subfigure}[t]{0.32\textwidth}
    \centering
    \includegraphics[width=\textwidth]{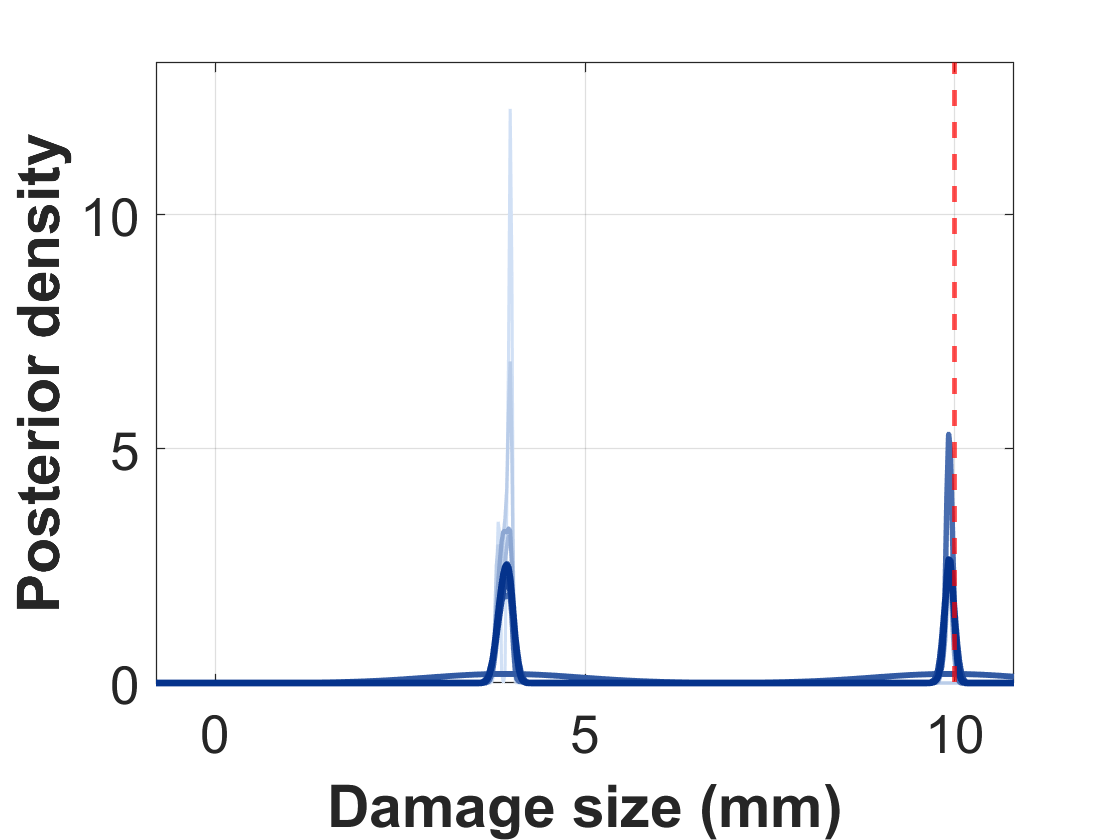}
    \caption{}
    \label{fig:bayes_multipeak_gyrx_m1_10mm}
\end{subfigure}
\hfill
\begin{subfigure}[t]{0.32\textwidth}
    \centering
    \includegraphics[width=\textwidth]{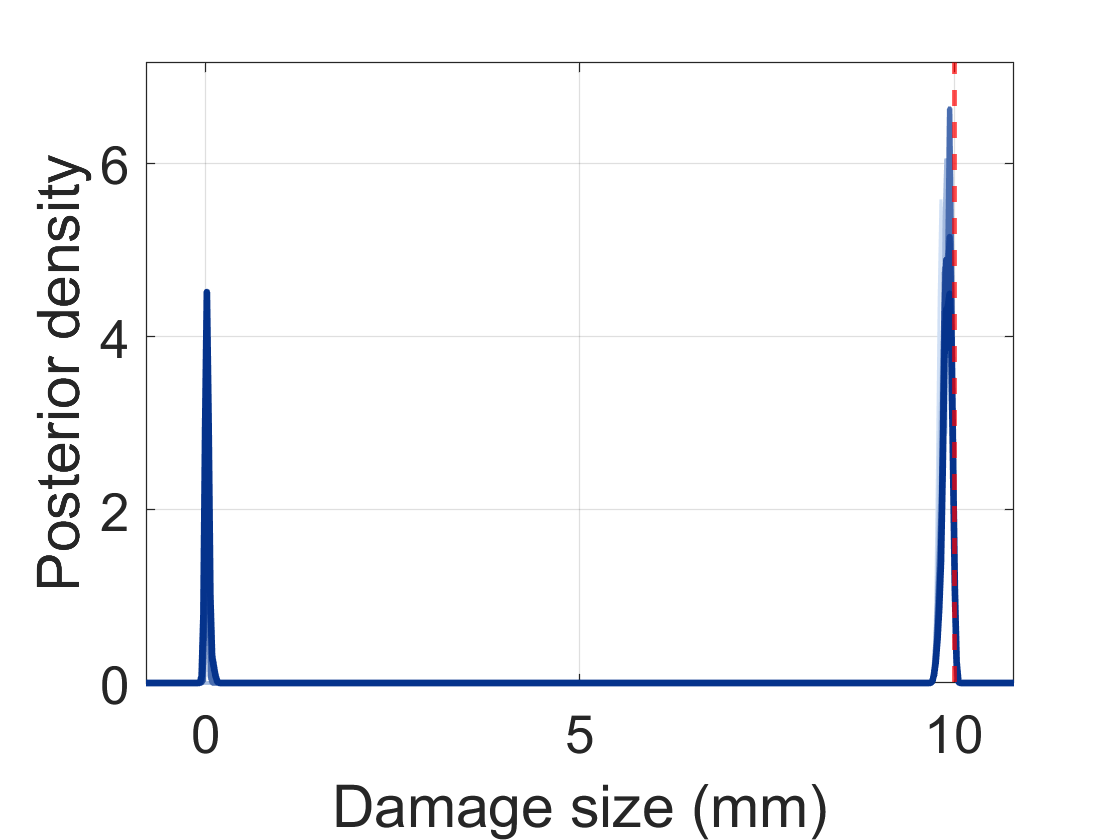}
    \caption{}
    \label{fig:bayes_multipeak_gyrx_m6_10mm}
\end{subfigure}
\hfill
\begin{subfigure}[t]{0.32\textwidth}
    \centering
    \includegraphics[width=\textwidth]{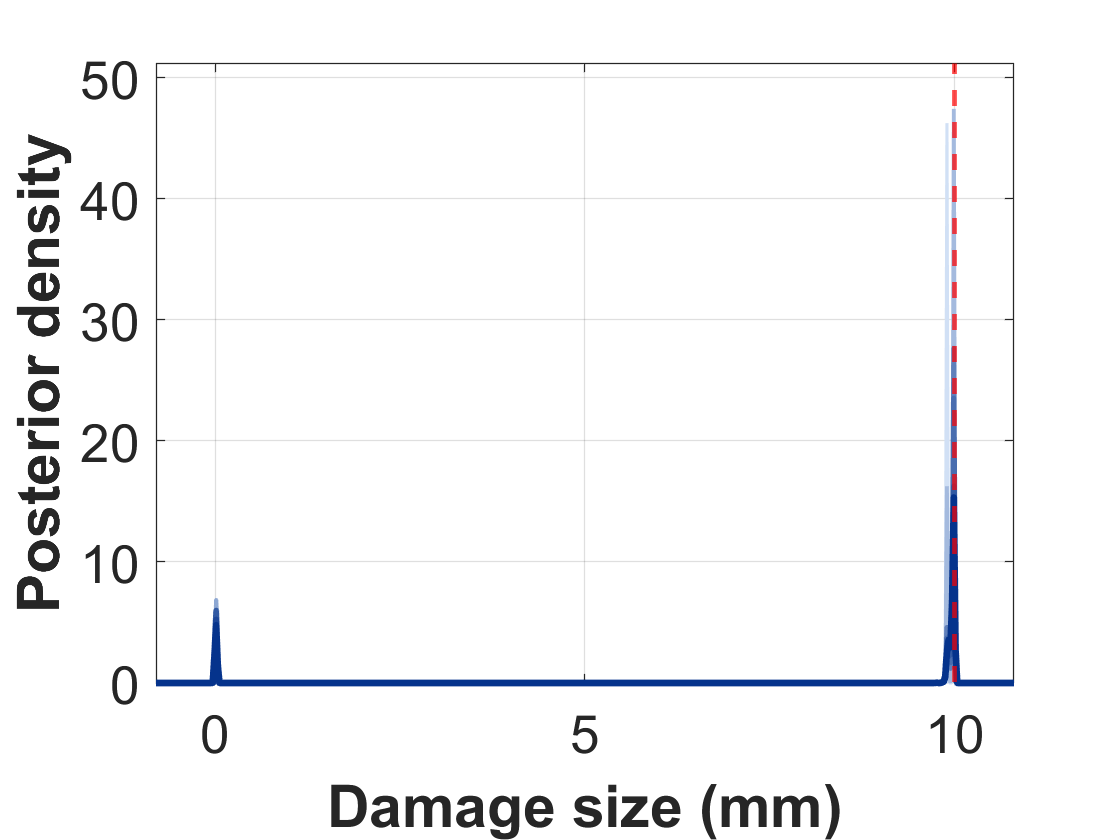}
    \caption{}
    \label{fig:bayes_multipeak_gyrz_m6_10mm}
\end{subfigure}

\caption{Posterior density (KDE) examples of damage magnitude showing
multi-peak behavior under the broad uniform prior over 0--10 mm. The red dashed
line indicates the actual damage level, while progressively darker curves
represent posterior densities accumulated over more MCMC chains. Panels:
(a) AccY-M3, $k=2$ mm; (b) AccY-M1, $k=6$ mm; (c) AccY-M1, $k=10$ mm;
(d) GyrX-M1, $k=10$ mm; (e) GyrX-M6, $k=10$ mm; (f) GyrZ-M6, $k=10$ mm.}
\label{fig:bayes_uniform_multipeak}
\end{figure}

Fig.~\ref{fig:bayes_uniform_multipeak} shows six representative posterior distributions selected from the deviated cases under the broad uniform prior. The red dashed line indicates the actual damage level, and the posterior density curves become progressively darker as more MCMC chains are accumulated. These examples demonstrate that the posterior mass is not always dominated by a single mode near the actual damage level. Instead, secondary modes can remain prominent and shift the posterior estimate away from the actual damage size.

Under the broad uniform prior, all damage levels within 0--10 mm are assigned equal prior probability; therefore, the posterior is strongly influenced by the shape of the RSS surface. If the RSS surface contains multiple comparable local minima, the posterior mean can be shifted away from the true value, even when one of the modes is close to the correct damage level.

\begin{table*}[b!]
\centering
\footnotesize
\setlength{\tabcolsep}{4pt}
\caption{Bayesian damage size estimation results obtained using the broad
uniform prior over 0--10 mm for each signal--motor combination at damage levels
of 2--10 mm. Values are reported as mean $\pm$ standard deviation over 16 test
data sets.}
\label{tab:bayesian1}
\begin{tabularx}{\textwidth}{l l *{5}{>{\centering\arraybackslash}X}}
\toprule
 & & \multicolumn{5}{c}{\textbf{Estimated damage size (mm)}} \\
\cmidrule(lr){3-7}
\textbf{Signal} & \textbf{Motor} & \textbf{2} & \textbf{4} & \textbf{6} & \textbf{8} & \textbf{10} \\
\midrule
      & M1 & 2.49 ± 1.34 &  4.55 ± 1.13  & 5.49 ± 0.56 & 6.31 ± 1.59 & 5.18 ± 0.44  \\
AccX & M3 & 1.98 ± 0.54  &  5.17 ± 0.62  & 4.44 ± 1.09 & 8.24 ± 0.78 & 9.84 ± 0.14 \\
  & M6 & 2.09 ± 0.71 &  4.18 ± 0.89  & 5.08 ± 0.63 & 7.11 ± 0.56 &  9.73 ± 0.79  \\
\midrule
  & M1 & 2.70 ± 0.78 &  3.94 ± 0.51  &  4.95 ± 0.81  & 6.17 ± 1.13 & 6.44 ± 1.13 \\
AccY & M3 & 2.81 ± 1.29 &  4.03 ± 0.68 & 5.22 ± 0.73  & 4.72 ± 0.62  & 9.12 ± 1.62 \\
  & M6 & 2.92 ± 0.88 & 3.91 ± 0.94 & 5.77 ± 0.89 & 7.08 ± 0.81 & 9.76 ± 0.30 \\
\midrule
  & M1 & 1.86 ± 0.98 & 4.81 ± 1.50 & 5.77 ± 0.73 & 5.79 ± 1.98 & 5.46 ± 0.45 \\
AccZ & M3  & 2.08 ± 1.12 & 3.32 ± 0.78 & 4.71 ± 0.81 & 7.62 ± 0.79 & 9.83 ± 0.17 \\
  & M6 & 1.97 ± 1.27 & 4.22 ± 1.07 & 5.41 ± 0.64 & 7.49 ± 0.45 & 9.69 ± 0.98 \\
\midrule
  & M1 & 3.55 ± 0.62 & 4.12 ± 0.82 & 5.92 ± 0.91 & 6.69 ± 0.82 & 5.82 ± 1.64 \\
GyrX & M3 & 1.86 ± 1.07 & 3.38 ± 1.53 & 3.12 ± 1.00 & 7.98 ± 1.79 & 9.74 ± 0.33 \\
  & M6 & 2.00 ± 0.64 & 4.76 ± 0.83 & 5.93 ± 1.42 & 7.36 ± 1.27 & 6.07 ± 0.54 \\
\midrule
  & M1 & 4.70 ± 0.59 & 5.82 ± 1.38 & 6.44 ± 0.80 & 7.30 ± 1.77 & 6.82 ± 1.12 \\
GyrY & M3 & 3.38 ± 0.78 & 4.47 ± 1.12 & 5.09 ± 0.88 & 4.55 ± 0.48  & 5.73 ± 1.00  \\
  & M6 & 3.21 ± 1.14 & 4.40 ± 0.71 & 5.64 ± 0.71 & 6.89 ± 1.31 & 9.94 ± 0.04 \\
\midrule
  & M1 & 2.84 ± 0.64 & 5.17 ± 1.62 & 5.35 ± 0.33 & 6.63 ± 1.69 & 9.87 ± 0.24 \\
GyrZ & M3 & 3.04 ± 1.52 & 1.71 ± 1.05 & 2.31 ± 1.43 & 6.67 ± 1.39 & 9.95 ± 0.02 \\
 & M6 & 2.89 ± 1.90 & 4.29 ± 1.33 & 5.75 ± 1.09 & 7.11 ± 0.46 & 7.39 ± 0.49 \\
\bottomrule
\end{tabularx}
\end{table*}

Motivated by the multi-peak posterior behavior observed under the broad uniform prior, the sequential state-informed prior is subsequently applied to restrict the Bayesian update to a locally admissible damage range informed by the preceding damage state. Table~\ref{tab:bayesian2} summarizes the corresponding damage estimation results. Compared with the broad-prior case, the estimates become substantially more stable and are generally closer to the actual damage levels. This improvement is especially clear for the 10 mm cases, where several broad-prior estimates were previously biased toward lower damage levels. For example, AccX-M1, AccY-M1, AccZ-M1, GyrX-M1, GyrX-M6, GyrY-M1, GyrY-M3, and GyrZ-M6 are all recovered to values close to 10 mm after applying the state-informed prior. Posterior standard deviations are also reduced in many cases, indicating a more concentrated posterior distribution.

\begin{figure}[t]
\centering

\begin{subfigure}[t]{0.32\textwidth}
    \centering
    \includegraphics[width=\textwidth]{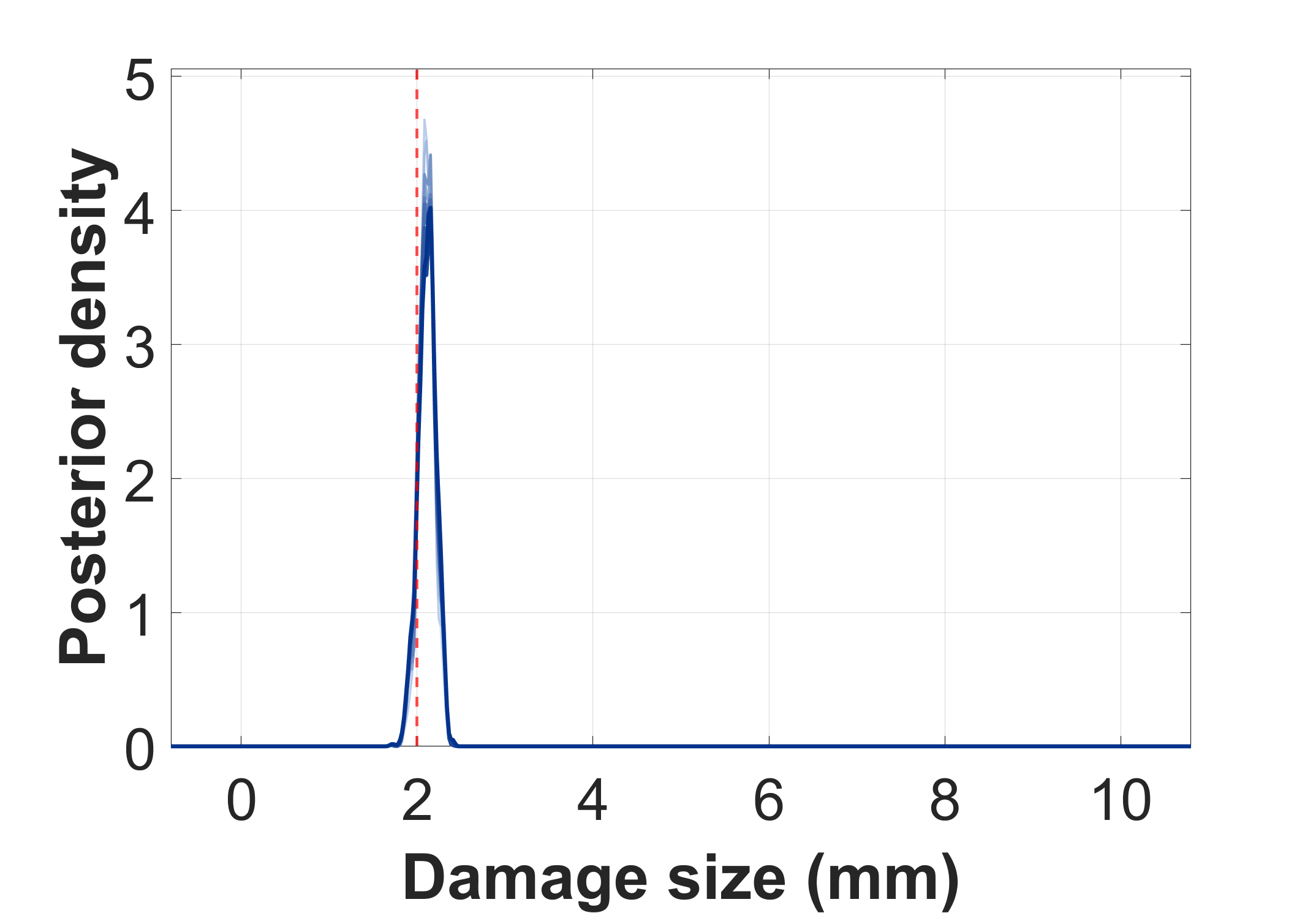}
    \caption{}
    \label{fig:bayes_state_gyrx_m1_2mm}
\end{subfigure}
\hfill
\begin{subfigure}[t]{0.32\textwidth}
    \centering
    \includegraphics[width=\textwidth]{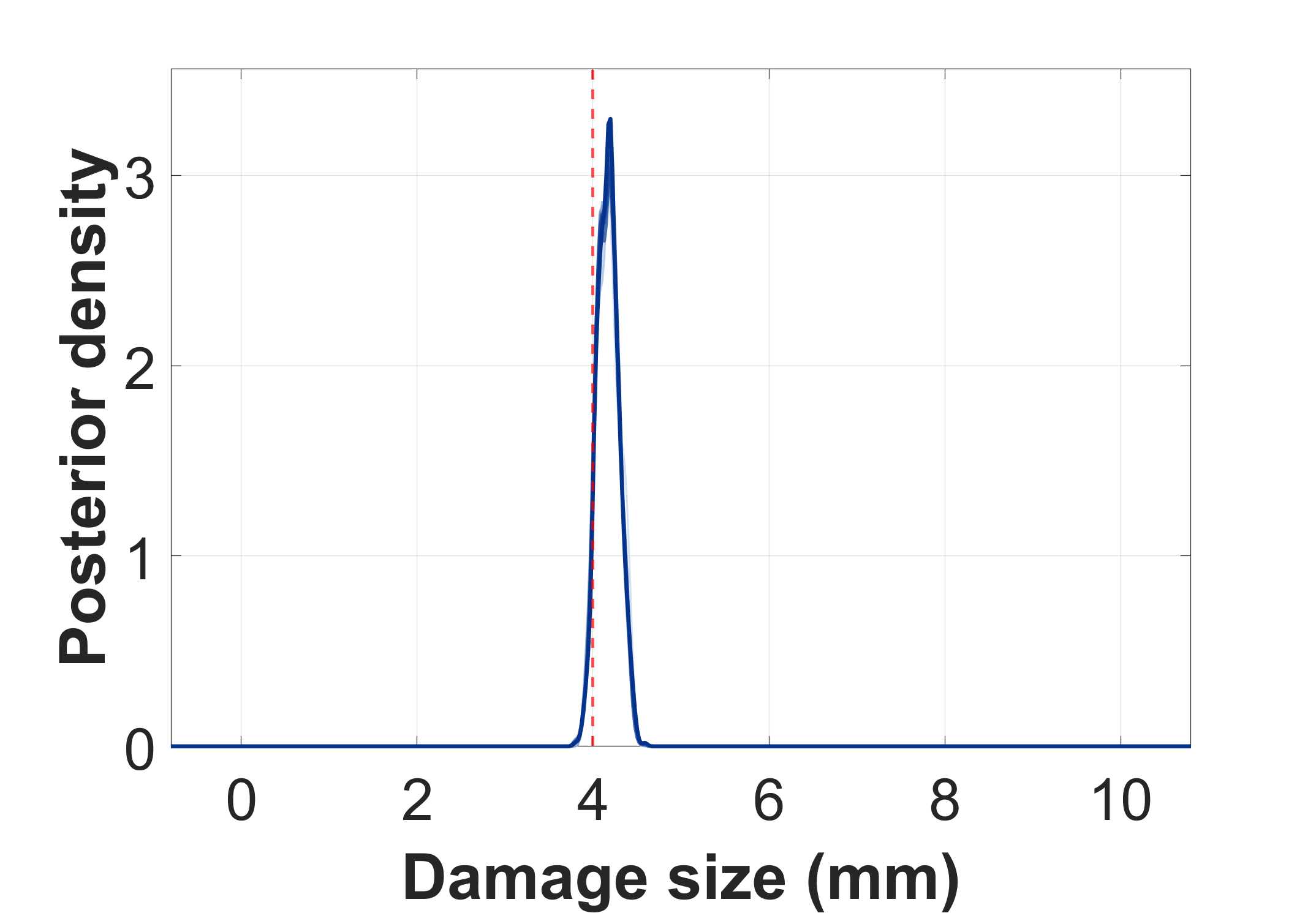}
    \caption{}
    \label{fig:bayes_state_accz_m1_4mm}
\end{subfigure}
\hfill
\begin{subfigure}[t]{0.32\textwidth}
    \centering
    \includegraphics[width=\textwidth]{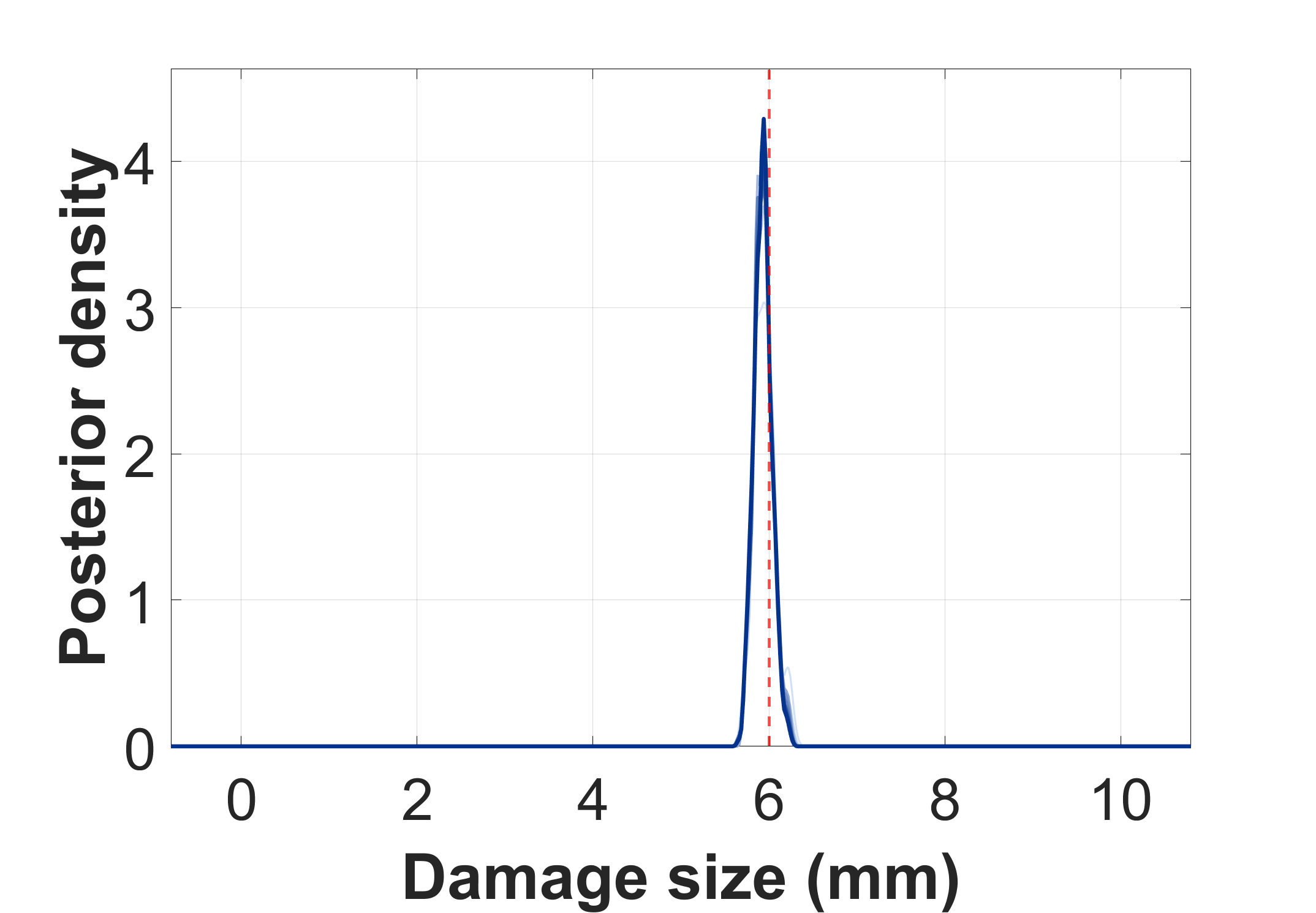}
    \caption{}
    \label{fig:bayes_state_accy_m1_6mm}
\end{subfigure}

\vspace{0.8em}

\begin{subfigure}[t]{0.32\textwidth}
    \centering
    \includegraphics[width=\textwidth]{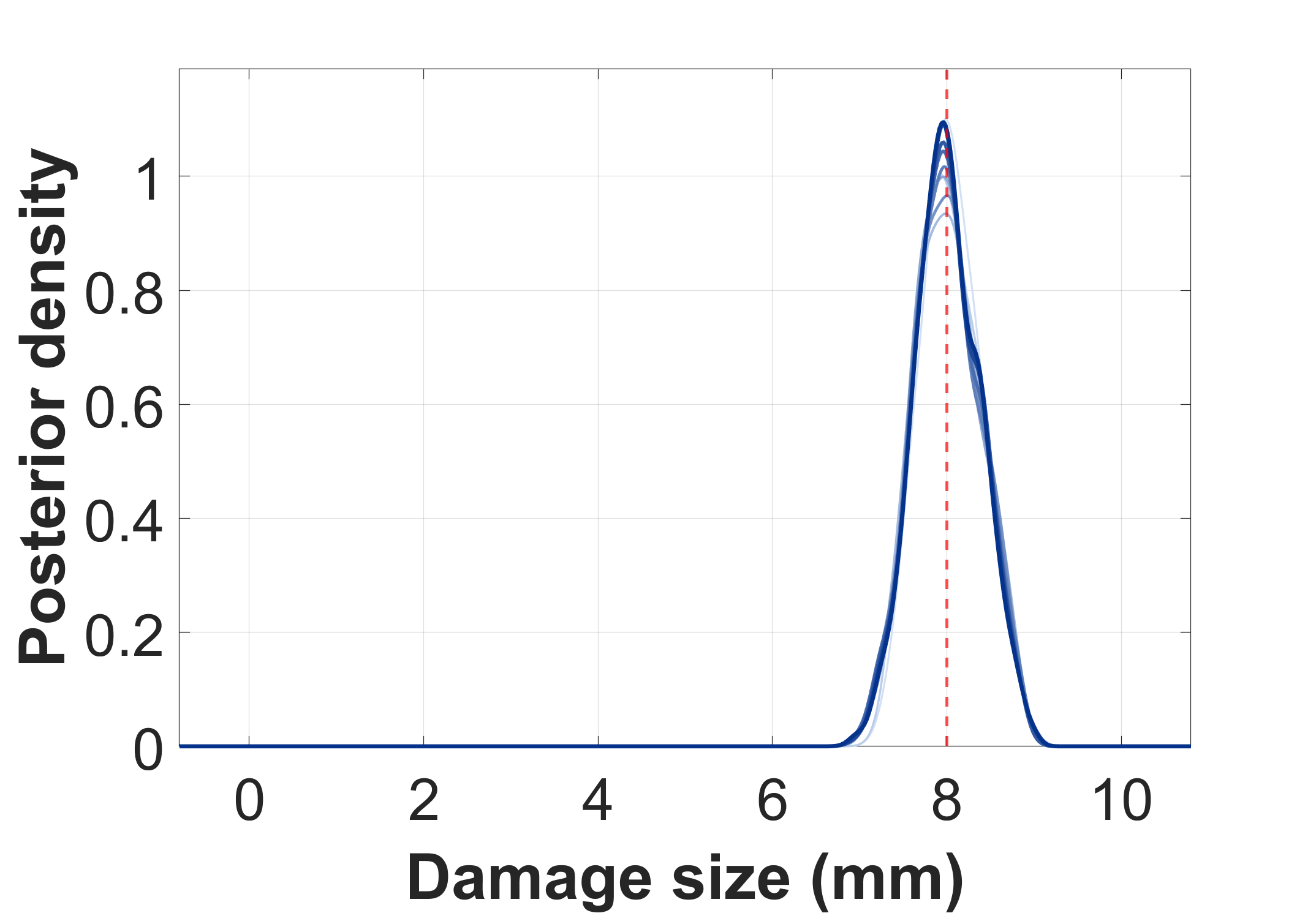}
    \caption{}
    \label{fig:bayes_state_accy_m3_8mm}
\end{subfigure}
\hfill
\begin{subfigure}[t]{0.32\textwidth}
    \centering
    \includegraphics[width=\textwidth]{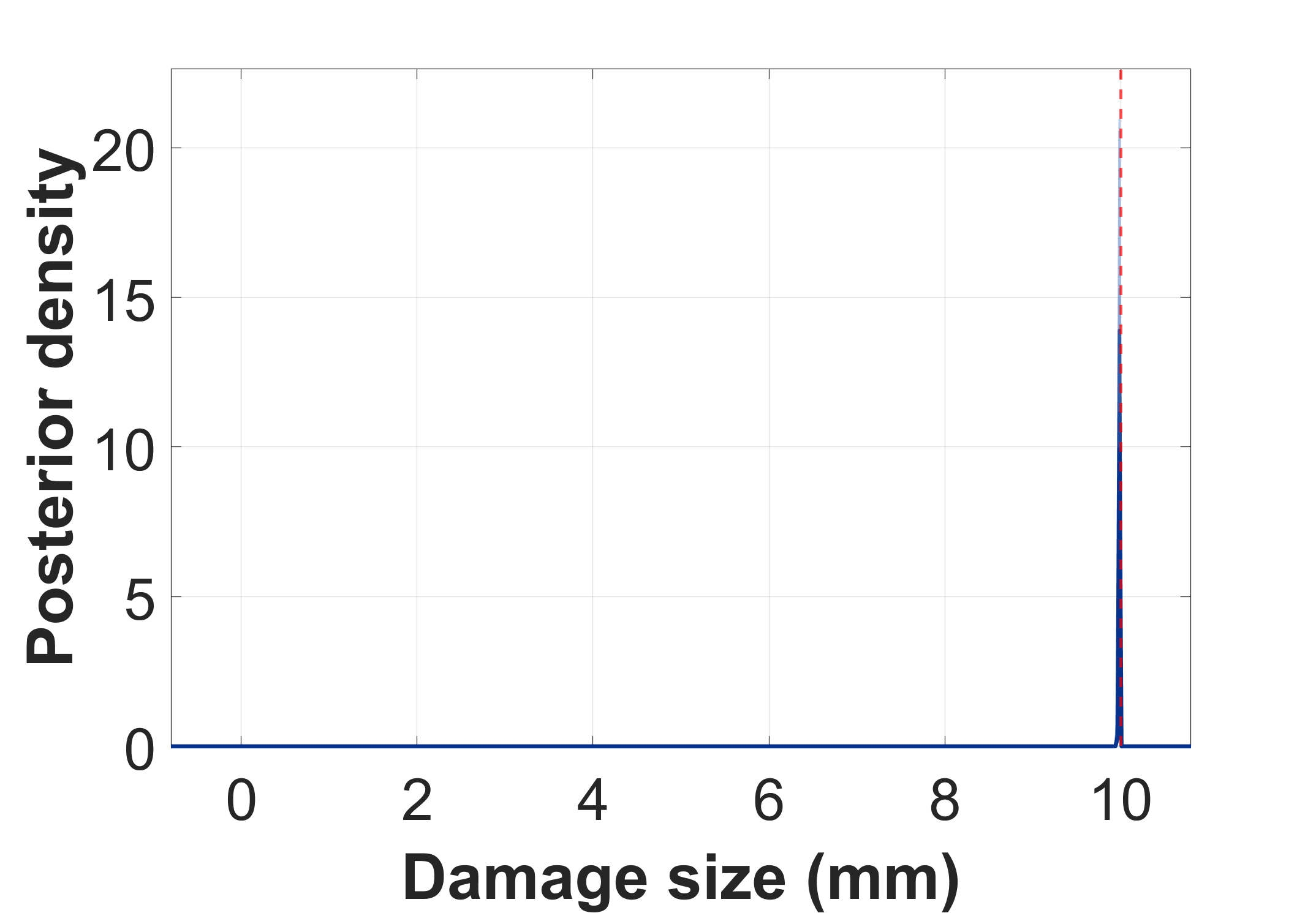}
    \caption{}
    \label{fig:bayes_state_accx_m1_10mm}
\end{subfigure}
\hfill
\begin{subfigure}[t]{0.32\textwidth}
    \centering
    \includegraphics[width=\textwidth]{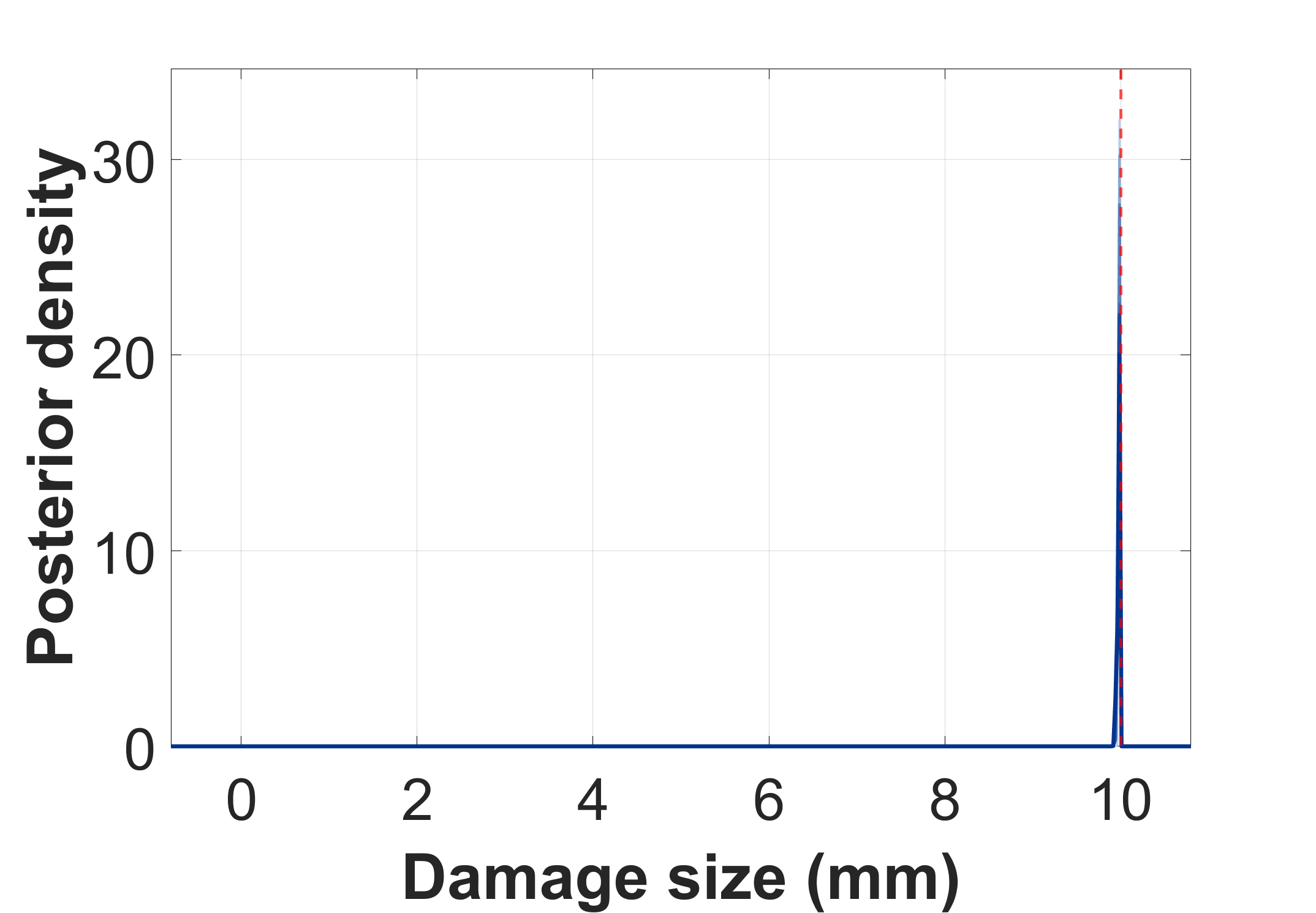}
    \caption{}
    \label{fig:bayes_state_gyrx_m6_10mm}
\end{subfigure}

\caption{Posterior density (KDE) examples of damage magnitude obtained using
the sequential state-informed prior. The examples are selected from cases where
the broad uniform prior produced biased estimates. The red dashed line indicates
the actual damage level, while progressively darker curves represent posterior
densities accumulated over more MCMC chains. Panels: (a) GyrX-M1, $k=2$ mm;
(b) AccZ-M1, $k=4$ mm; (c) AccY-M1, $k=6$ mm; (d) GyrX-M3, $k=8$ mm;
(e) AccX-M1, $k=10$ mm; (f) GyrX-M6, $k=10$ mm.}
\label{fig:bayes_state_prior_examples}
\end{figure}

Fig.~\ref{fig:bayes_state_prior_examples} further illustrates the effect of the sequential state-informed prior using six representative cases selected from different damage levels. These cases correspond to configurations where the broad uniform prior produced biased estimates, while the state-informed prior shifted the posterior mass closer to the actual damage level. As shown in the figure, the posterior densities become more localized within the state-informed prior range, and the remote modes observed under the broad prior are effectively suppressed.

The selected examples cover damage levels of 2 to 10 mm. For the 2 mm case
(GyrX-M1), the estimate is corrected from an overestimated value under the broad
prior to a value close to the actual damage level. For the 4 and 6 mm cases
(AccZ-M1 and AccY-M1), the state-informed prior reduces the posterior spread and
moves the posterior mass toward the actual damage level. For the 8 mm case
(GyrX-M3), the broad-prior estimate is strongly biased toward a lower damage
level, whereas the state-informed prior concentrates the posterior around 8 mm.
For the 10 mm cases (AccX-M1 and GyrX-M6), the broad prior leads to severe
underestimation, while the state-informed prior suppresses lower-damage modes
and recovers posterior estimates close to 10 mm.

These representative posterior distributions support the quantitative results
in Table~\ref{tab:bayesian2}. Overall, the state-informed prior improves
the robustness of Bayesian damage quantification by incorporating sequential
damage-state information. Nevertheless, the complete table also indicates that
some channel-dependent bias remains, particularly for the less sensitive
signal configurations.

Overall, Tables~\ref{tab:dot-error-matrix-nohealthy}, \ref{tab:bayesian1}, and
\ref{tab:bayesian2} indicate that the nonlinear inverse optimization approach
achieves accurate and repeatable estimates in the high-sensitivity cases, especially
for large damage levels, but exhibits larger variability for some small and
intermediate damage cases. The broad-prior Bayesian formulation captures the
multi-modal uncertainty of the inverse problem, although its posterior mean can
be biased when competing modes are present. The sequential state-informed prior
gives the most stable and physically consistent estimates by suppressing
unlikely modes and concentrating the posterior around the locally admissible
damage range. These results demonstrate that the Bayesian framework is most
effective when sequential prior state information is available, combining
damage-size estimation with explicit uncertainty quantification.

\begin{table*}[b!]
\centering
\footnotesize
\setlength{\tabcolsep}{4pt}
\caption{Bayesian damage size estimation results using the sequential
state-informed prior for each signal--motor combination. Values are reported as
mean $\pm$ standard deviation over 16 test data sets.}
\label{tab:bayesian2}
\begin{tabularx}{\textwidth}{l l *{5}{>{\centering\arraybackslash}X}}
\toprule
 & & \multicolumn{5}{c}{\textbf{Estimated damage size (mm)}} \\
\cmidrule(lr){3-7}
\textbf{Signal} & \textbf{Motor} & \textbf{2} & \textbf{4} & \textbf{6} & \textbf{8} & \textbf{10} \\
\midrule
      & M1 & 1.95 ± 0.86 & 4.32 ± 0.75  & 5.44 ± 0.58 & 8.13 ± 1.32 & 9.94 ± 0.17   \\
AccX & M3 & 1.83 ± 0.65  & 5.08 ± 0.51  & 4.82 ± 0.48 & 8.36 ± 0.48 & 9.83 ± 0.15 \\
  & M6 & 1.98 ± 0.85 &  4.15 ± 0.92  & 5.11 ± 0.57 & 7.12 ± 0.56 & 9.94 ± 0.09  \\
\midrule
  & M1 & 1.89 ± 0.35 & 3.76 ± 0.62  & 6.01 ± 0.67  & 7.87 ± 0.72 & 9.47 ± 0.30 \\
AccY & M3 & 2.02 ± 0.92 &  3.74 ± 0.82 & 5.58 ± 0.51  & 8.04 ± 0.48  & 9.86 ± 0.17\\
  & M6 & 2.81 ± 0.71 & 3.91 ± 0.92 & 5.78 ± 0.88 & 7.18 ± 0.65 & 9.82 ± 0.21 \\
\midrule
  & M1 & 1.53 ± 0.73 & 4.22 ± 0.62 & 5.77 ± 0.76 & 7.96 ± 1.38 & 9.92 ± 0.19\\
AccZ & M3  & 2.06 ± 1.08 & 3.36 ± 0.74 & 4.87 ± 0.61 & 7.63 ± 0.78 & 9.83 ± 0.17 \\
  & M6 & 1.83 ± 0.92 & 4.22 ± 1.06 & 5.43 ± 0.59 & 7.65 ± 0.51 & 9.94 ± 0.06 \\
\midrule
  & M1 & 2.13 ± 0.47 & 3.97 ± 0.94 & 6.08 ± 0.71 & 7.37 ± 0.65 & 9.93 ± 0.09 \\
GyrX & M3 & 1.82 ± 1.03  & 3.47 ± 1.27 & 4.26 ± 0.37 & 8.13 ± 1.64 & 9.74 ± 0.34 \\
  & M6 & 1.99 ± 0.66 & 4.73 ± 0.82  & 6.11 ± 1.12 & 7.58 ± 0.50& 9.93 ± 0.04 \\
\midrule
  & M1 & 2.50 ± 0.40 & 4.81 ± 0.70 & 6.38 ± 0.45 & 8.23 ± 0.62& 9.98 ± 0.07 \\
GyrY & M3 & 2.06 ± 0.54 & 4.18 ± 1.12 & 5.30 ± 0.53 & 8.37 ± 0.50  & 9.83 ± 0.20  \\
  & M6 & 1.98 ± 0.51 & 4.31 ± 0.67 & 6.03 ± 0.59 & 7.81 ± 0.31 & 9.95 ± 0.04 \\
\midrule
  & M1 & 1.88 ± 0.65 & 4.48 ± 0.99 & 5.35 ± 0.34 & 7.15 ± 0.83 & 9.94 ± 0.15 \\
GyrZ & M3 & 2.74 ± 1.12 & 2.42 ± 0.45 & 4.23 ± 0.58 & 7.02 ± 1.05 & 9.94 ± 0.02 \\
 & M6 & 2.36 ± 1.03 & 4.26 ± 1.07 & 5.81 ± 0.94 & 7.12 ± 0.45 & 9.94 ± 0.07 \\
\bottomrule
\end{tabularx}
\end{table*}

\section{Concluding remarks} \label{sec:conclusions}

This paper presented a unified, statistically grounded framework for multicopter propeller damage diagnosis based on FP-AR models, critically assessed via an extensive experimental flight test campaign. The framework integrates---within a single methodology---three inspection tasks that are often treated in isolation: (i) health monitoring and damage detection, (ii) damaged motor identification (damage localization), and (iii) damage magnitude quantification. The approach is deliberately interpretable and data-efficient: short windows of standard onboard IMU signals are summarized by parametric time-domain models whose parameters vary smoothly with the damage level through low-dimensional functional projections. A parameter pooling strategy then aggregates information across multiple flight segments to produce a mission-level model that captures the shared dynamics while attenuating segment idiosyncrasies. Finally, a Bayesian layer converts the residual landscape of the parametric fit into posterior distributions for the damage size, yielding credible intervals suitable for safety-critical decision making.

The assessment was carried out under realistic operating conditions: long missions ($\sim$250 s) following a 101-waypoint figure-eight trajectory, repeated on different days and under varying ambient conditions, and executed with the flight controller actively stabilizing the vehicle. The framework was assessed across all six IMU channels (AccX/Y/Z and GyrX/Y/Z), enabling a head-to-head comparison of the sensing modalities. The non-parametric analysis provided qualitative evidence of damage-induced dynamic changes, while the FP-AR models translated that evidence into quantitative decision statistics for detection and identification, and into continuous estimates of the damage magnitude.
The main lessons learned and conclusions drawn from this study can be summarized as follows.
\begin{enumerate}[label=(\roman*)]
  \item \textit{Interpretable models for effective damage diagnosis.} Compact AR structures, fitted on $\approx$4 s windows, enable consistent damage detection, damaged motor identification, and magnitude estimation with statistically controlled false alarm risk, supporting near real-time operation.
  \item \textit{Multi-signal generality with comparative performance.} A common FP-AR formulation applies across all six IMU channels, enabling a meaningful ranking of the sensing modalities; in the present data set, AccX and GyrY generally yield tighter estimates, whereas AccZ/GyrZ can be less informative in some regimes.
  \item \textit{Cross-flight consistency under realistic variability.} Models trained and evaluated over repeated missions on different days display stable decision statistics without retuning, indicating robustness to environmental drift, controller actions, and moderate operating-point changes.
  \item \textit{Operating complexity beyond hovering.} The methodology remains effective on dynamic waypoint-tracking trajectories (101-waypoint figure-eight paths over $\sim$250 s missions), rather than only on hover or straight-line flight, which is critical for practical autonomous operations.
  \item \textit{Cross-segment parameter pooling improves robustness.} Pooling the projection coefficients across multiple segments yields mission-level models with smoother residual landscapes and more stable minima. While the local and pooled models give comparable point estimates in benign cases, the pooled variant reduces the sensitivity to the segment choice and improves the healthy-state behavior.
 \item \textit{Bayesian quantification provides explicit uncertainty characterization and highlights the importance of prior design.} While the broad uniform prior can expose the multi-modal uncertainty in ambiguous cases, the sequential state-informed prior suppresses physically unlikely modes and produces more stable, physically consistent damage size estimates, especially at intermediate and large damage levels.
  \item \textit{Controller-compensated scenarios remain diagnosable.} Even when the autopilot partially masks the effects of damage, the residual-based statistics retain their discriminative power, allowing diagnosis without disabling the control loops.
  \item \textit{No additional sensing requirements and computationally lightweight.} The procedure requires only standard IMU measurements and linear-algebraic solvers, making it amenable to embedded implementation and easy integration with existing avionics.
\end{enumerate}

Future work will address simultaneous and interacting damage scenarios, adaptive and online model updating across missions and platforms, principled sensor fusion leveraging the most informative channels, and closed-loop integration in which the diagnostic uncertainty informs mission planning and fault-tolerant control. Overall, the postulated FP-AR and Bayesian framework offers an interpretable, data-efficient, and deployment-ready path toward onboard, uncertainty-aware damage diagnosis for autonomous aerial systems.

\bibliographystyle{elsarticle-num} 

\bibliography{references} 

\section{Appendix}
\subsection*{A. Basis functions}
The univariate polynomials used in this study are the shifted Chebyshev polynomials of the second kind (Type II Chebyshev polynomials), which belong to the broader family of Chebyshev orthogonal polynomials. These polynomials obey the following recurrence relation:
\begin{equation}
G_{n+1}(x) = (4x-2)\,G_n(x) - G_{n-1}(x),
\qquad x \in [0, 1] \subset \mathbb{R},
\end{equation}
initialized with \( G_0(x) = 1 \) and \( G_1(x) = 4x-2 \).

Hence, the first five basis polynomials employed in this study are:
\begin{align*}
G_0 &= 1 \\
G_1 &= -2 + 4x \\
G_2 &= 3 - 16x + 16x^2 \\
G_3 &= -4 + 40x - 96x^2 + 64x^3 \\
G_4 &= 5 - 80x + 336x^2 - 512x^3 + 256x^4
\end{align*}

In the present framework, the independent variable is the damage level $k$, mapped onto the natural domain of the polynomials via the normalization:
\begin{equation}
x = \tilde{k} = \frac{k}{k_{\text{max}}} \in [0,1] \subset \mathbb{R}.
\end{equation}

\subsection*{B. Cram\'er--Rao lower bound for the operating parameter vector estimates}

Consider the estimation of the operating parameter vector $\mathbf{k}\in\mathbb{R}^{d}$ from a response signal $y_u[t]$, $t=1,\ldots,N$, obtained from the system in an unknown state, based on the identified FP-AR model:
\begin{equation}
    y_u[t] + \sum_{i=1}^{n_a} a_i(\mathbf{k})\, y_u[t - i] = e_u[t, \mathbf{k}],
\end{equation}
or, equivalently, in the regression form of Eq.~\eqref{eq:sample}, $e_u[t,\mathbf{k}]=y_u[t]-\big(\boldsymbol{\varphi}_u[t]\otimes\boldsymbol{g}(\mathbf{k})\big)^{\!\top}\boldsymbol{\theta}$. The NLS estimate and the associated residual variance estimate are
\begin{equation}
    \widehat{\mathbf{k}} = \arg\min_{\mathbf{k}\in\mathbb{R}^{d}} \sum_{t=n_a+1}^{N} e_u^2[t, \mathbf{k}],
\qquad
    \widehat\sigma_{e,u}^2(\widehat{\mathbf{k}}) = \frac{1}{N-n_a} \sum_{t=n_a+1}^{N} e_u^2[t, \widehat{\mathbf{k}}].
\end{equation}

As $N \to \infty$, the estimator $\widehat{\mathbf{k}}$ is asymptotically Gaussian distributed, $\widehat{\mathbf{k}}\ \dot\sim\ \mathcal{N}(\mathbf{k}, \boldsymbol{\Sigma}_{\mathbf{k}})$.

Under the Gaussian innovations assumption, the log-likelihood function of the $N$ samples of the unknown signal is given by:
\begin{equation}
    \ln \mathcal{L}(\mathbf{k}, \sigma^2) = -\frac{N}{2} \ln(2\pi) - \frac{N}{2}\ln \sigma^2
     - \frac{1}{2\sigma^2} \sum_{t=1}^{N} e_u^2[t, \mathbf{k}].
\end{equation}

The Cram\'er--Rao lower bound for the covariance of any unbiased estimator of $\mathbf{k}$ is given by the inverse of the Fisher information matrix:
\begin{equation}
    \boldsymbol{\Sigma}_{\mathrm{CRLB}} = \left[ -\mathbb{E} \left[ \frac{\partial^2 \ln \mathcal{L}(\mathbf{k}, \sigma^2)}{\partial \mathbf{k}\, \partial \mathbf{k}^{\top}} \right] \right]^{-1}.
\end{equation}
Defining the residual sensitivity vector $\boldsymbol{\epsilon}[t,\mathbf{k}]\in\mathbb{R}^{d}$ with respect to the operating parameters,
\begin{equation}
    \boldsymbol{\epsilon}[t,\mathbf{k}] = \frac{\partial e_u[t, \mathbf{k}]}{\partial \mathbf{k}}
    = \frac{\partial}{\partial \mathbf{k}} \left( y_u[t] - \big(\boldsymbol{\varphi}_u[t] \otimes \boldsymbol{g}(\mathbf{k})\big)^{\!\top} \boldsymbol{\theta} \right)
    = -\left(\boldsymbol{\varphi}_u[t] \otimes \frac{\partial \boldsymbol{g}(\mathbf{k})}{\partial \mathbf{k}^{\top}}\right)^{\!\top} \boldsymbol{\theta},
\end{equation}
where $\partial \boldsymbol{g}(\mathbf{k})/\partial \mathbf{k}^{\top}\in\mathbb{R}^{r\times d}$ designates the Jacobian of the functional basis, the Fisher information matrix takes the form $\frac{1}{\sigma^2}\sum_{t}\boldsymbol{\epsilon}[t,\mathbf{k}]\,\boldsymbol{\epsilon}[t,\mathbf{k}]^{\top}$, and the Cram\'er--Rao lower bound for $\widehat{\mathbf{k}}$ can be written as:
\begin{equation}
    \boldsymbol{\Sigma}_{\mathrm{CRLB}}(\mathbf{k}) = \sigma^2 \left[ \sum_{t=1}^{N} \boldsymbol{\epsilon}[t,\mathbf{k}]\, \boldsymbol{\epsilon}[t,\mathbf{k}]^{\top} \right]^{-1},
\end{equation}
evaluated in practice at $\mathbf{k}=\widehat{\mathbf{k}}$, with $\sigma^2$ replaced by the residual variance estimate $\widehat\sigma_{e,u}^2(\widehat{\mathbf{k}})$. In the scalar case ($d=1$, $\mathbf{k}\equiv k$), the above reduces to the simplified form of Eq.~\eqref{eq:det-cov-scalar} employed in the detection stage.

\newpage

\subsection*{C. Additional Results}

\begin{figure}[th!]
    \centering 
    
    \begin{subfigure}{.3\textwidth} 
        \centering
        \includegraphics[width=\textwidth]{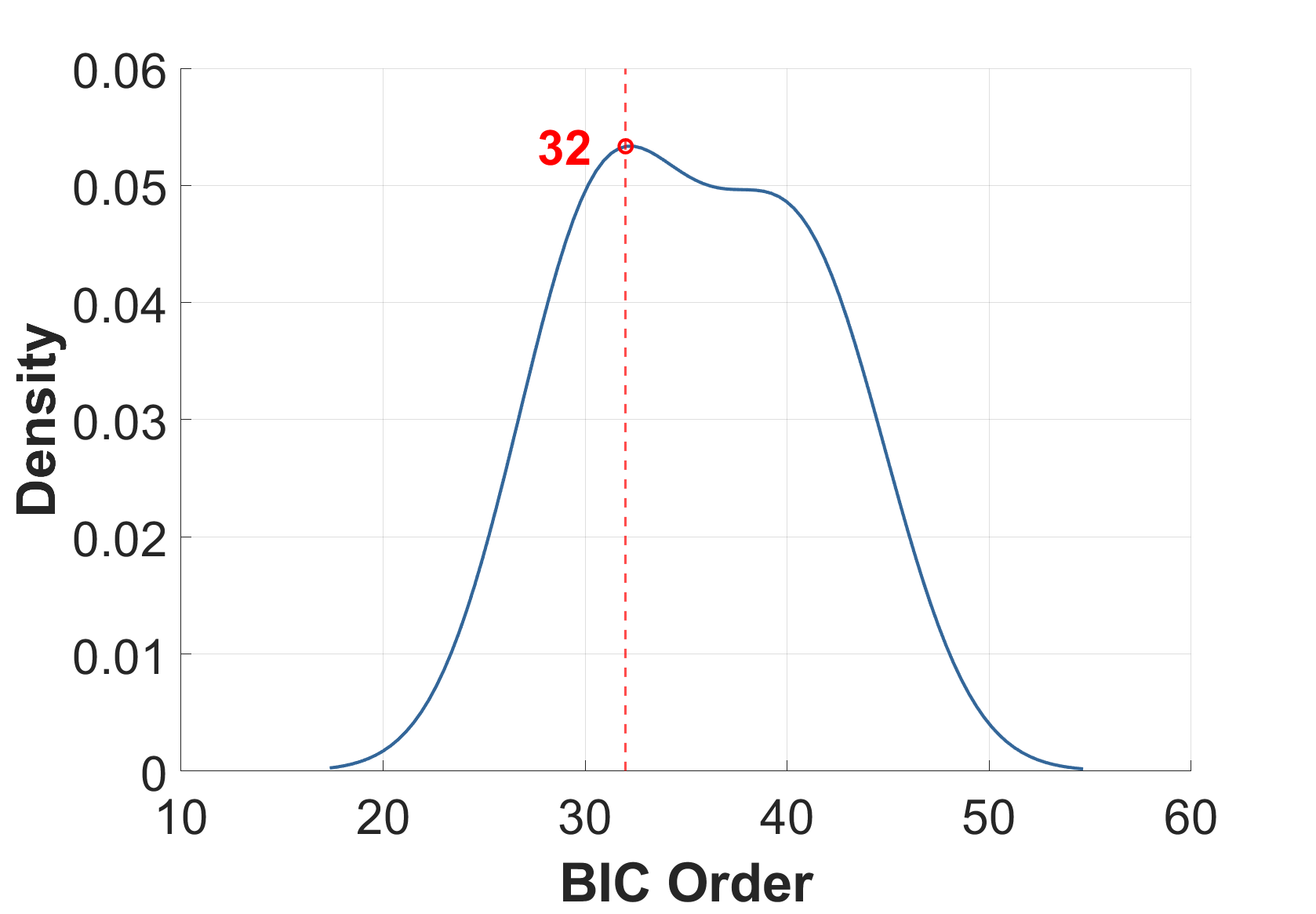}
        \caption{AccX}
    \end{subfigure}
    \hfill 
    \begin{subfigure}{.3\textwidth}
        \centering
        \includegraphics[width=\textwidth]{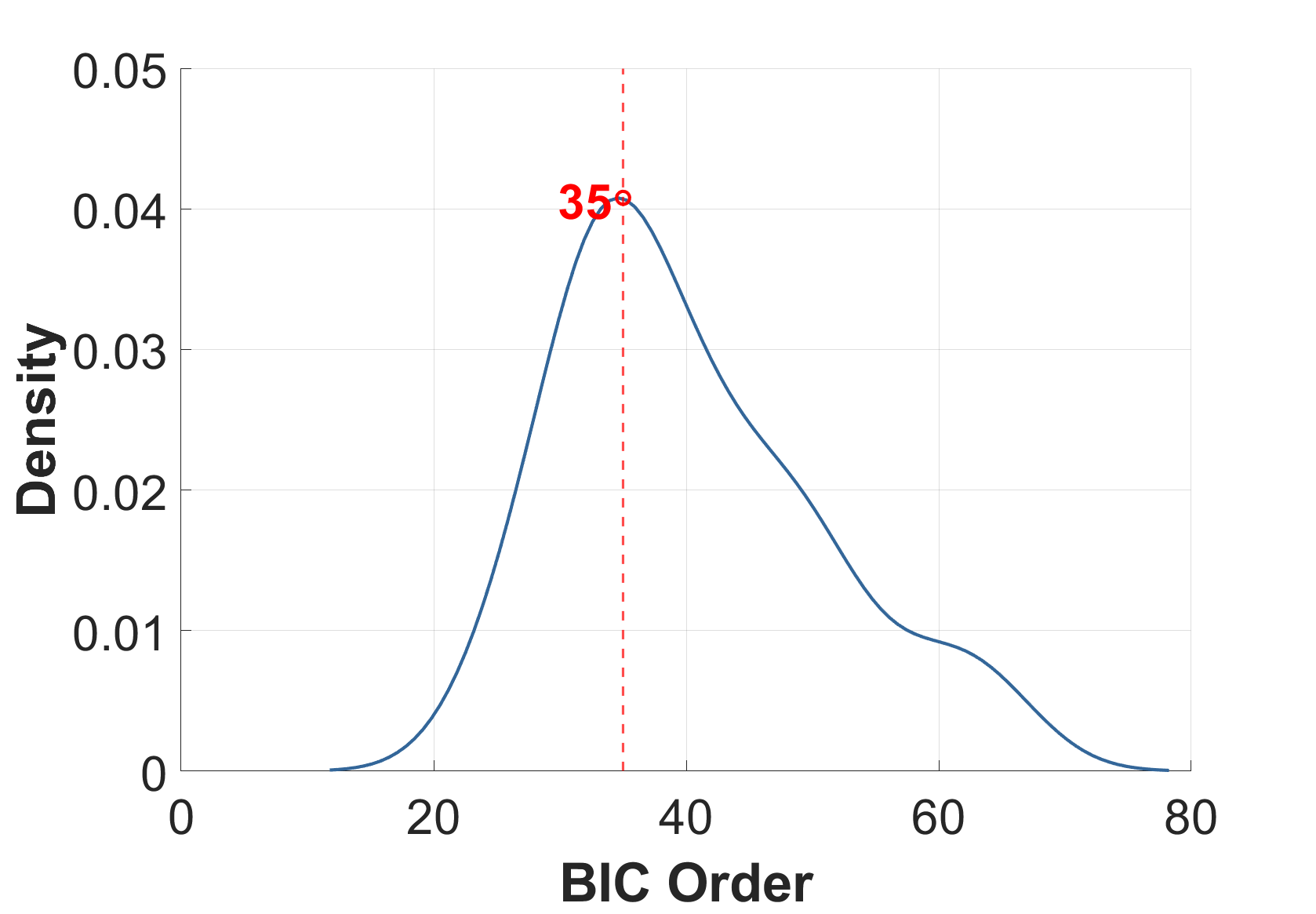}
        \caption{AccY}
    \end{subfigure}
    \hfill
    \begin{subfigure}{.3\textwidth}
        \centering
        \includegraphics[width=\textwidth]{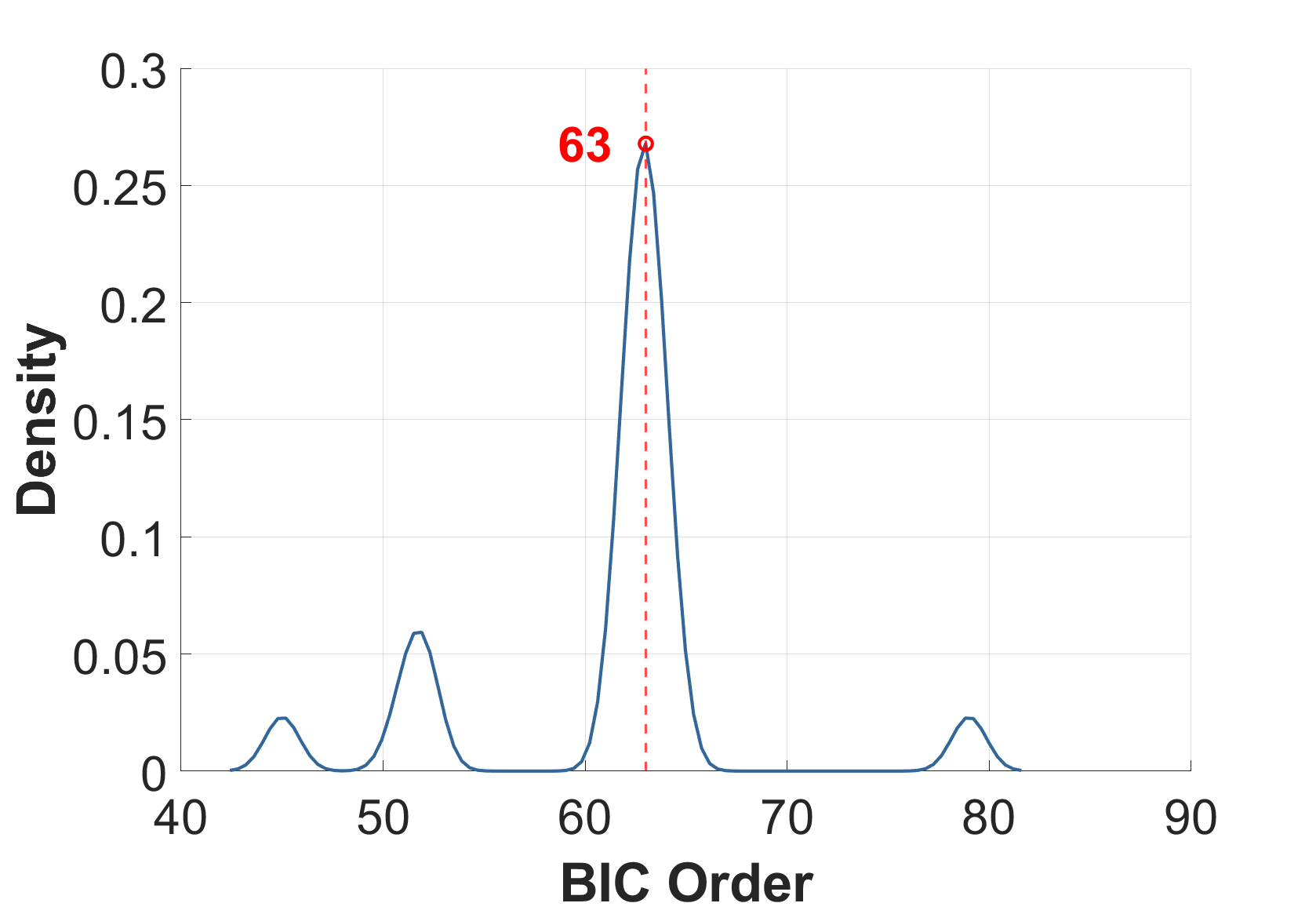}
        \caption{AccZ}
    \end{subfigure}

    \begin{subfigure}{.3\textwidth}
        \centering
        \includegraphics[width=\textwidth]{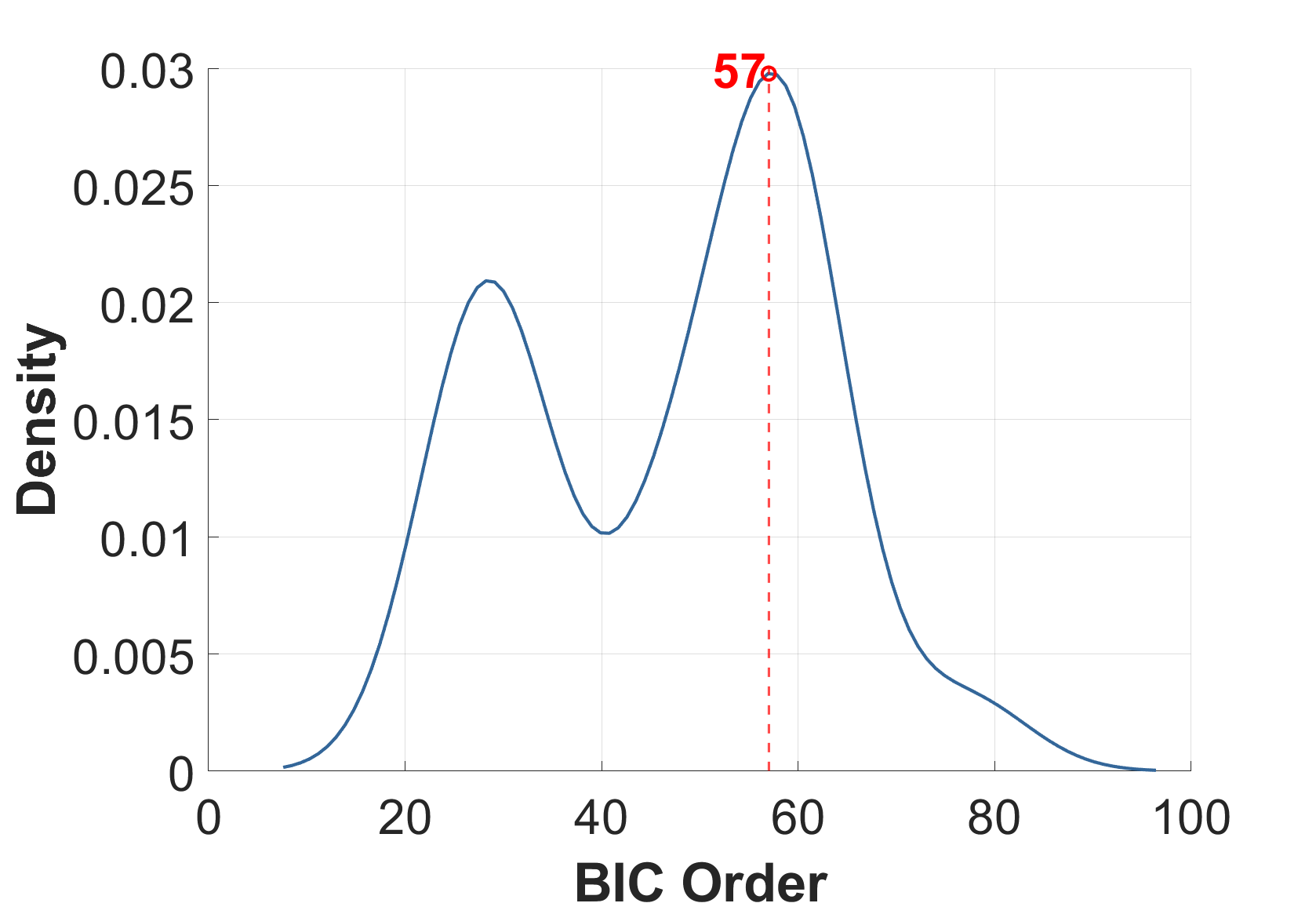}
        \caption{GyrX}
    \end{subfigure}
    \hfill
    \begin{subfigure}{.3\textwidth}
        \centering
        \includegraphics[width=\textwidth]{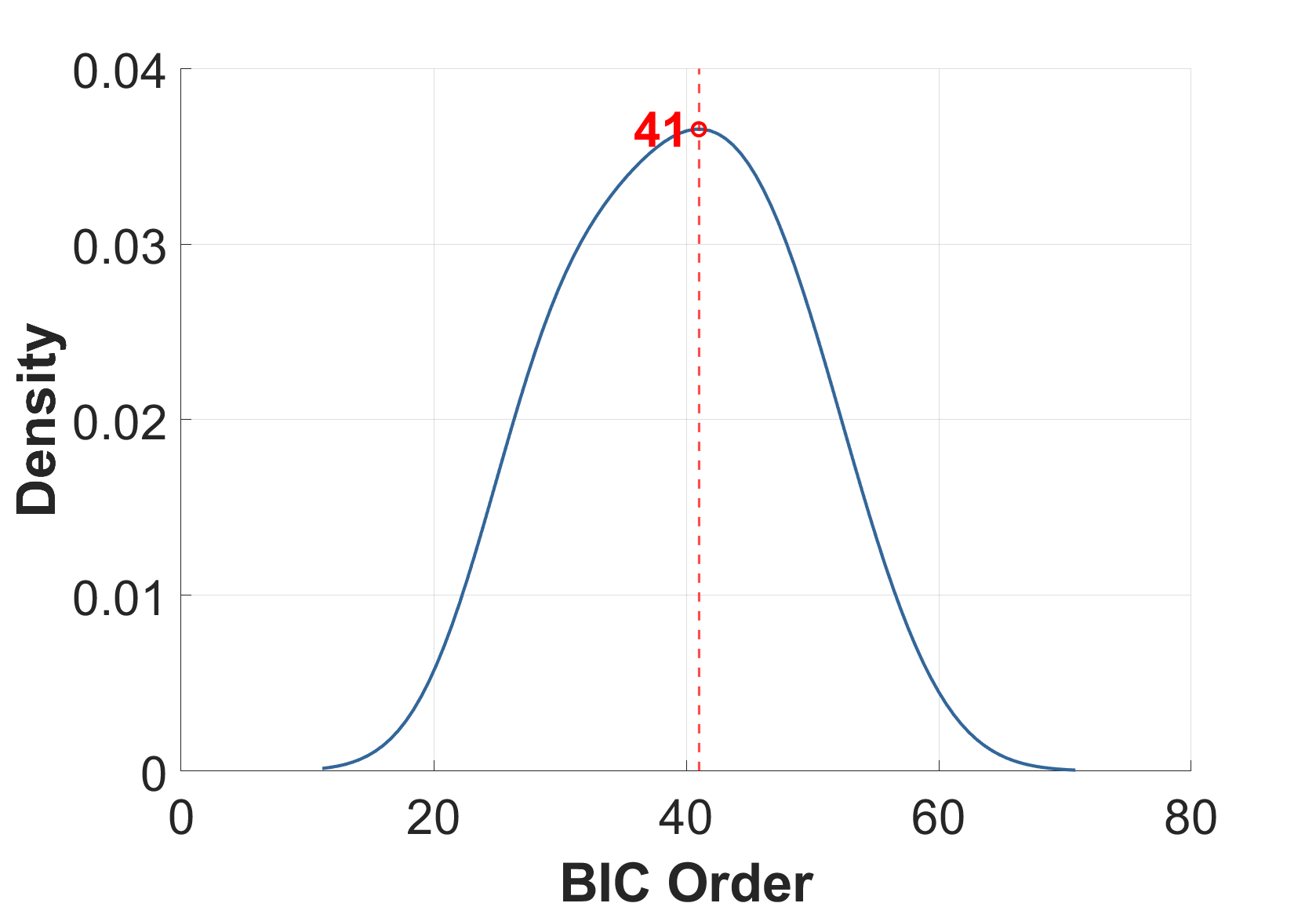}
        \caption{GyrY}
    \end{subfigure}
    \hfill
    \begin{subfigure}{.3\textwidth}
        \centering
        \includegraphics[width=\textwidth]{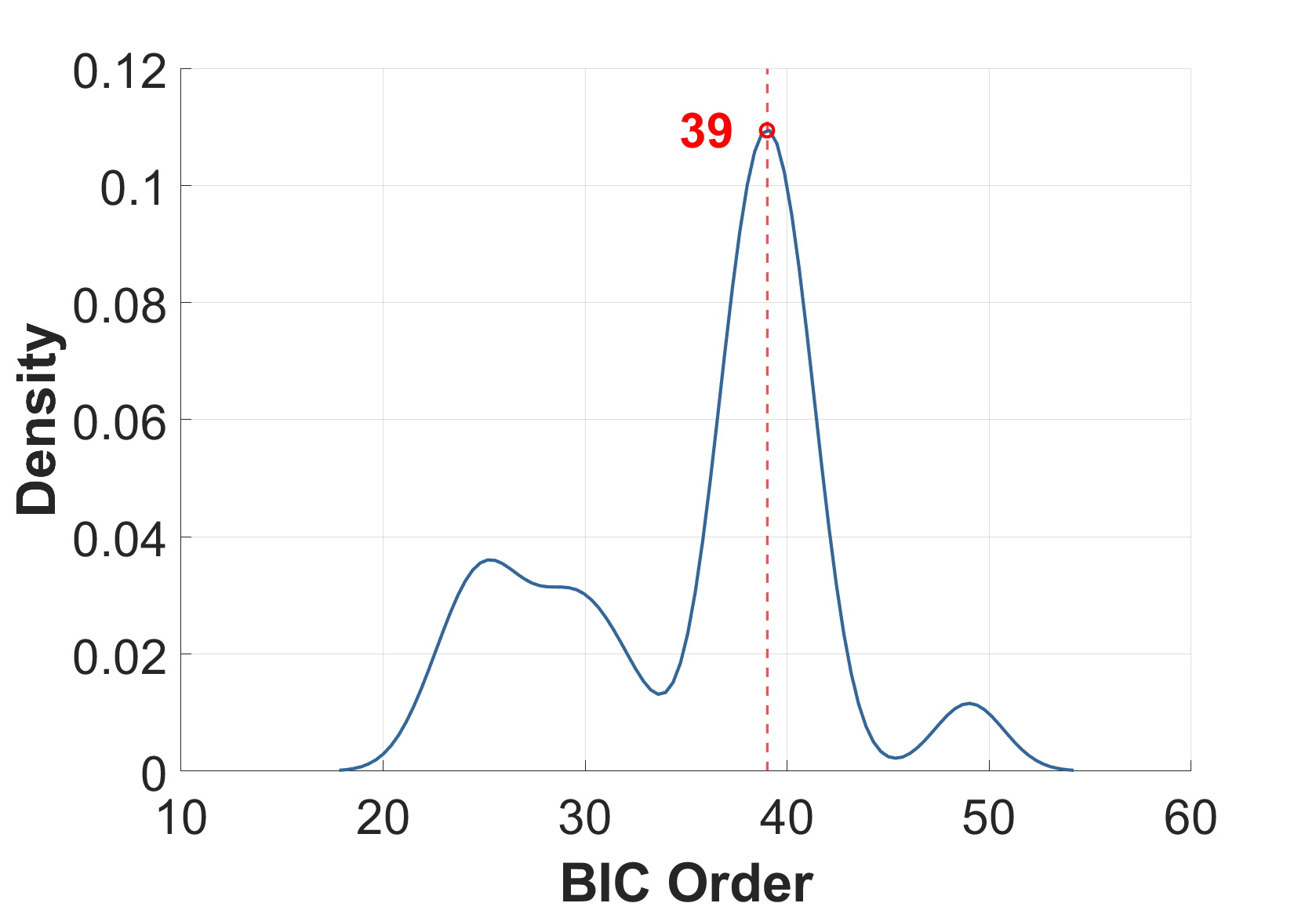}
        \caption{GyrZ}
    \end{subfigure}
    \caption{Gaussian KDEs of the BIC-optimal model orders for the six signals of Motor 3. The red vertical line indicates the peak of the density curve, representing the most probable model order selected for the pooled FP-AR model.}
    \label{fig:basis density M3}
\end{figure}

\begin{figure}[th!]
    \centering 
    
    \begin{subfigure}{.3\textwidth} 
        \centering
        \includegraphics[width=\textwidth]{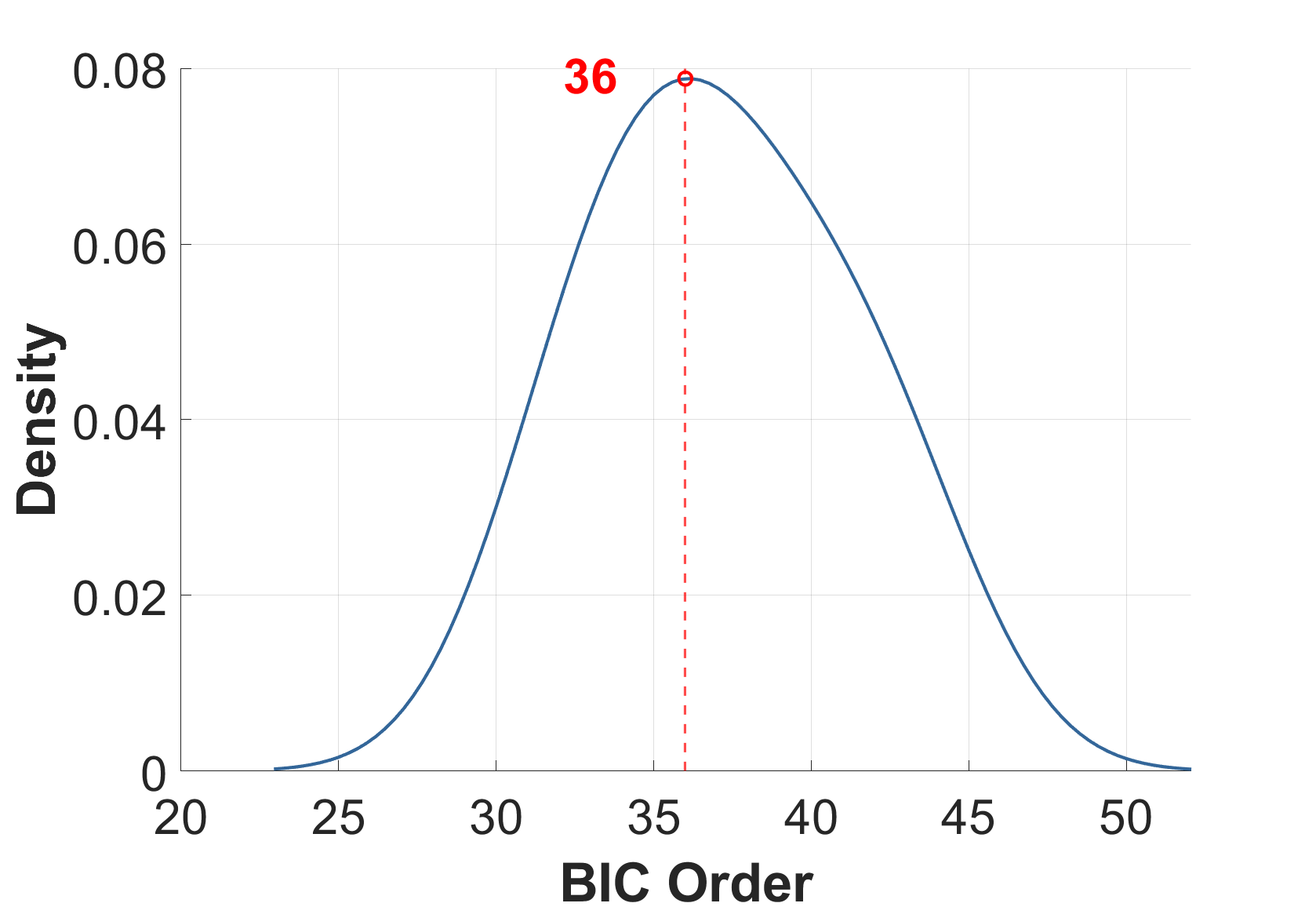}
        \caption{AccX}
    \end{subfigure}
    \hfill 
    \begin{subfigure}{.3\textwidth}
        \centering
        \includegraphics[width=\textwidth]{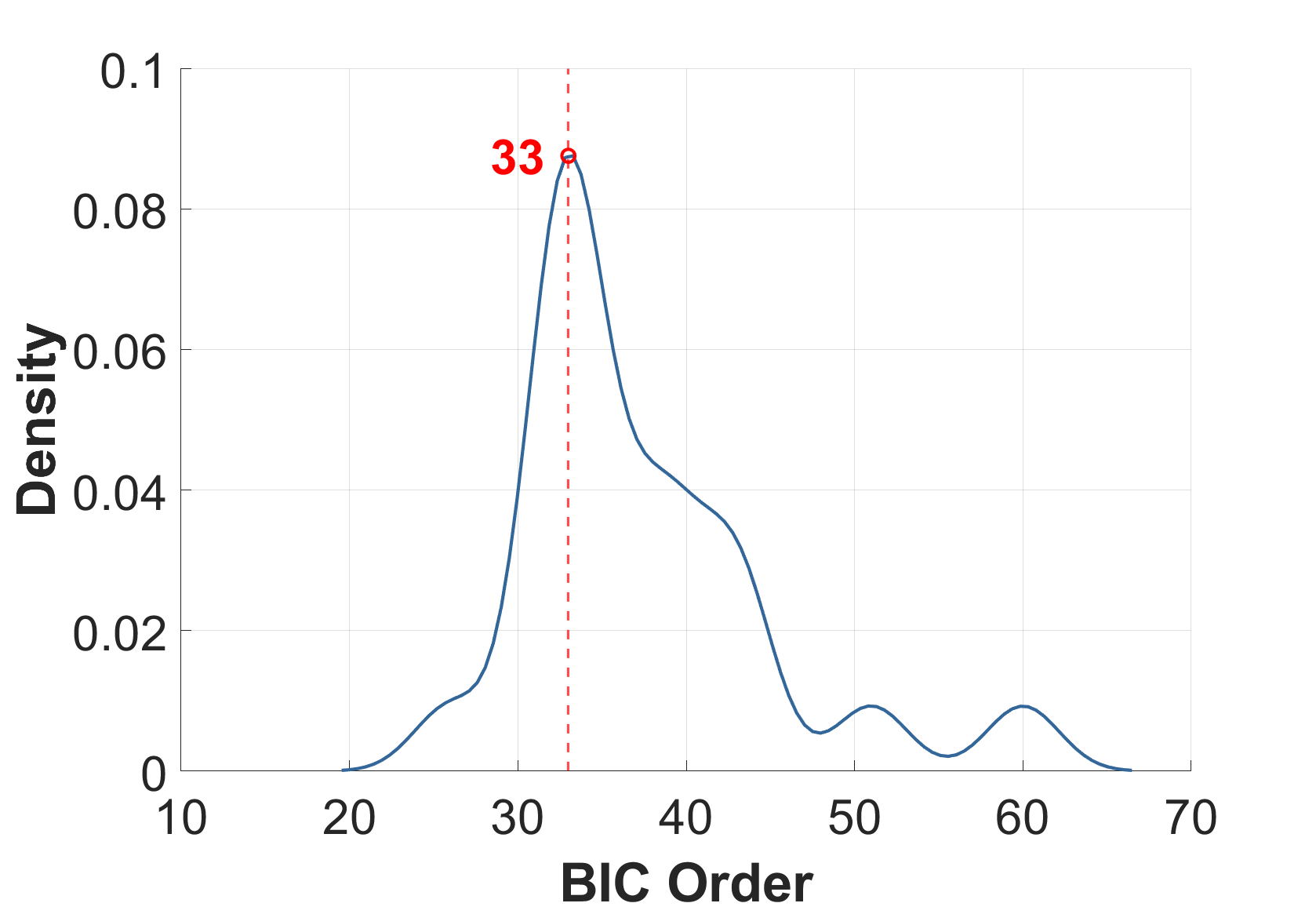}
        \caption{AccY}
    \end{subfigure}
    \hfill
    \begin{subfigure}{.3\textwidth}
        \centering
        \includegraphics[width=\textwidth]{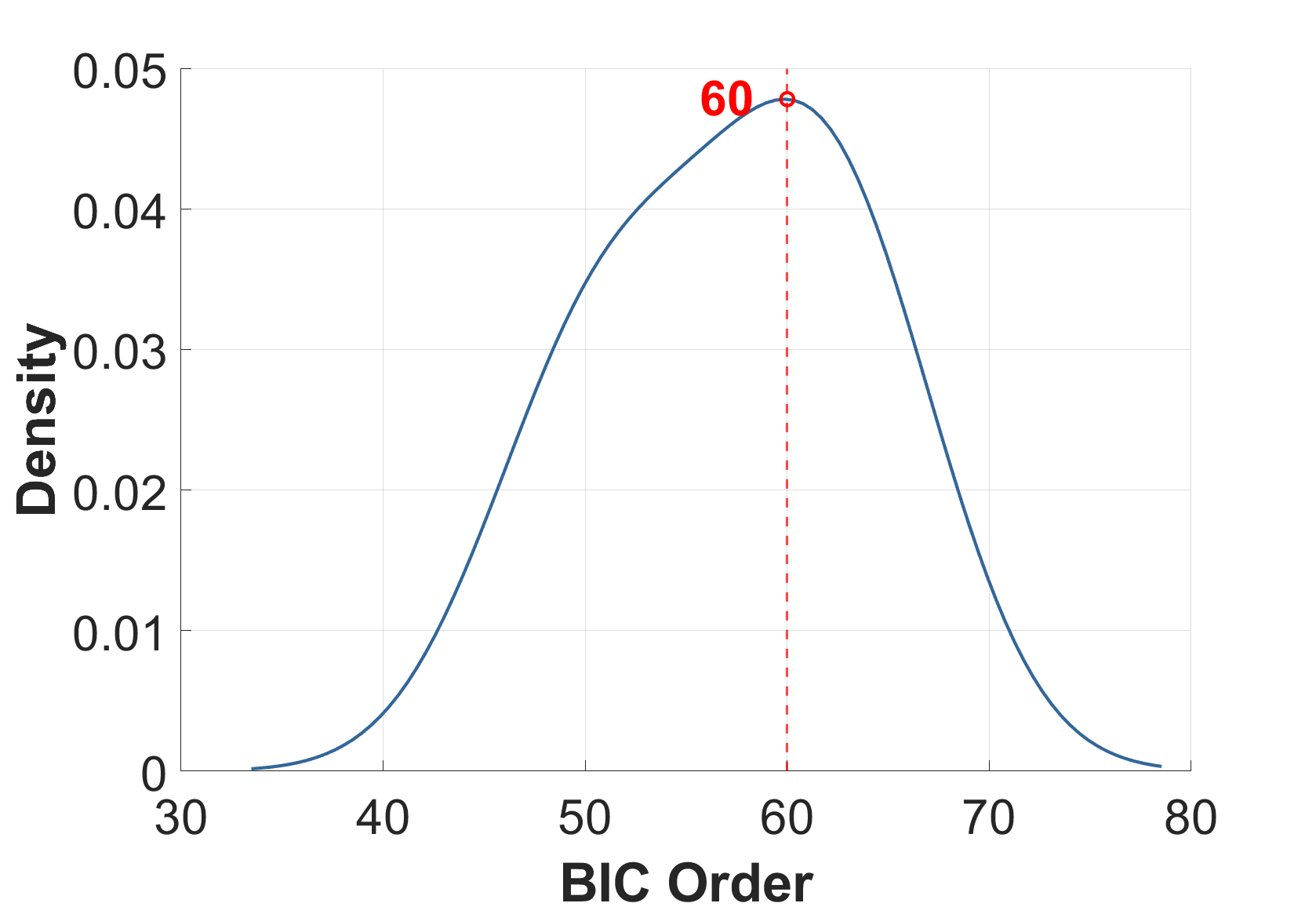}
        \caption{AccZ}
    \end{subfigure}

    \begin{subfigure}{.3\textwidth}
        \centering
        \includegraphics[width=\textwidth]{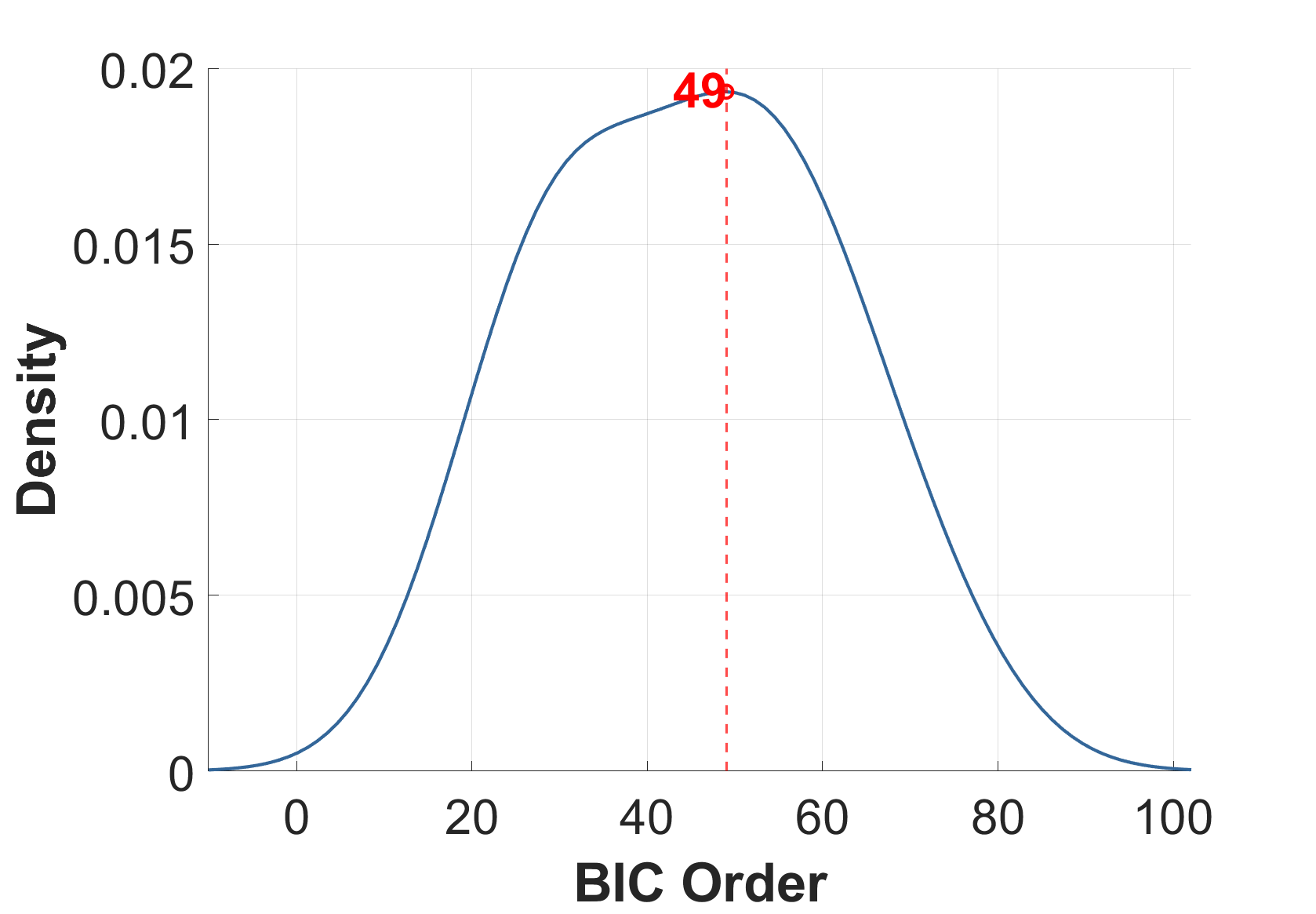}
        \caption{GyrX}
    \end{subfigure}
    \hfill
    \begin{subfigure}{.3\textwidth}
        \centering
        \includegraphics[width=\textwidth]{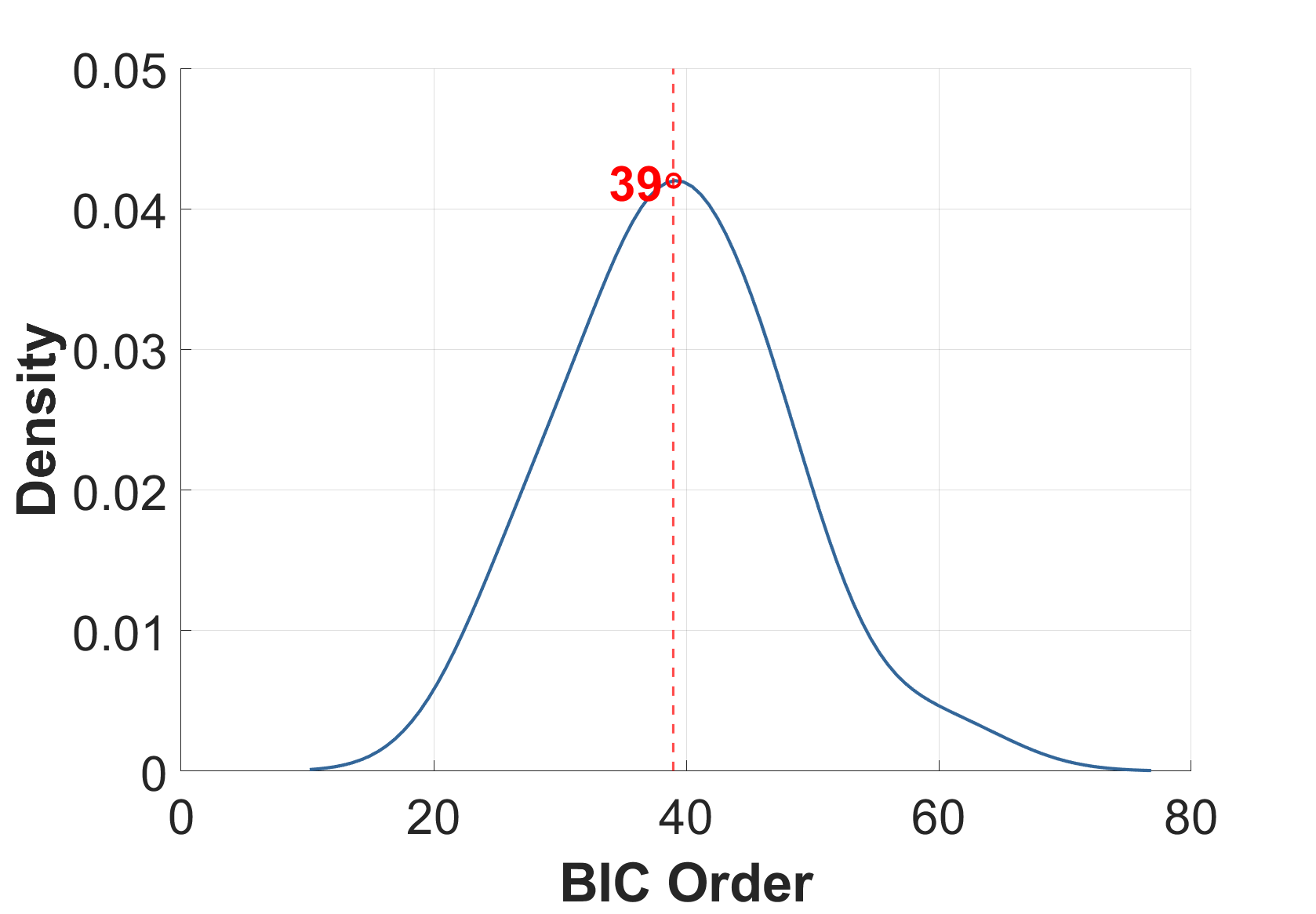}
        \caption{GyrY}
    \end{subfigure}
    \hfill
    \begin{subfigure}{.3\textwidth}
        \centering
        \includegraphics[width=\textwidth]{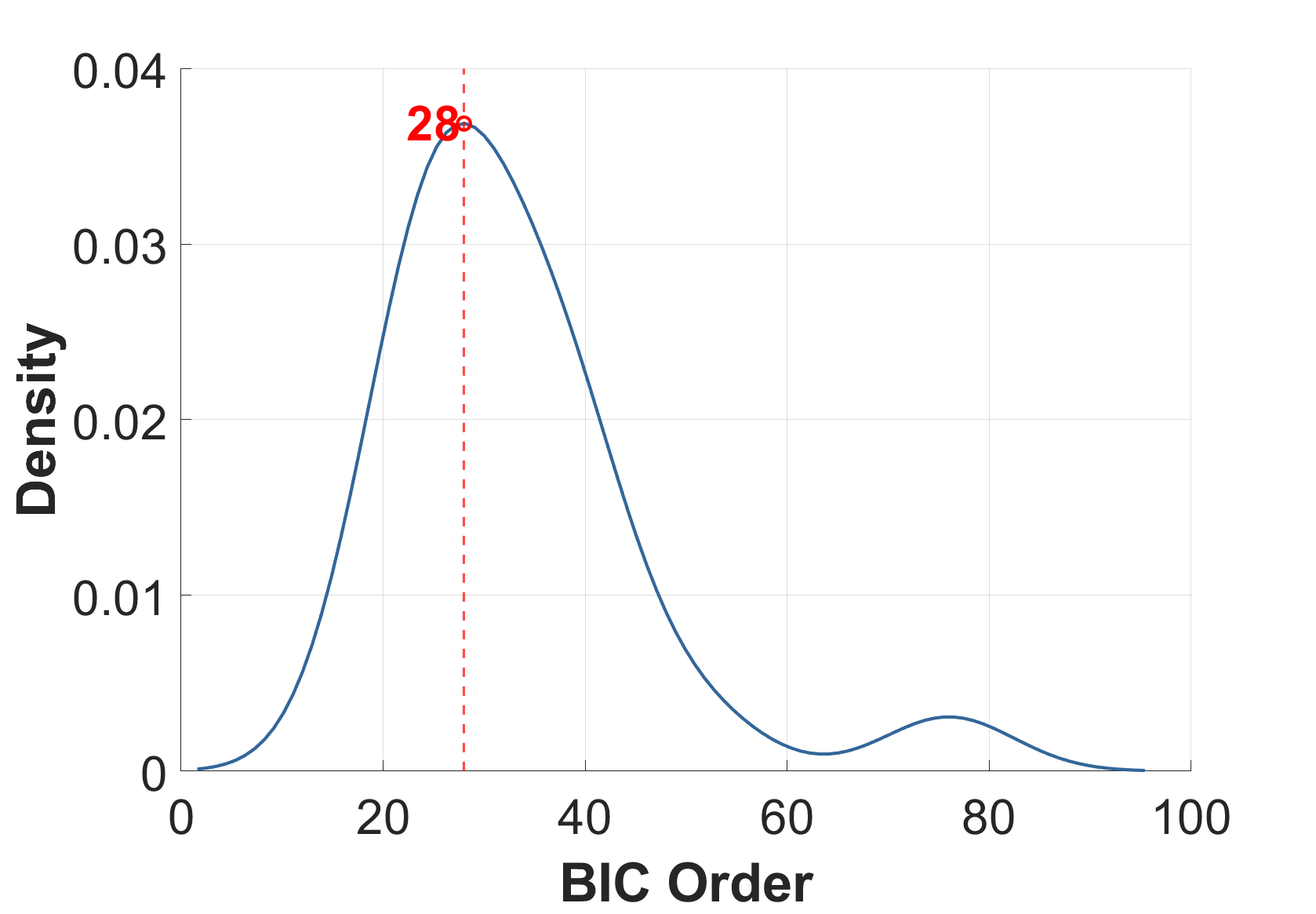}
        \caption{GyrZ}
    \end{subfigure}
    \caption{Gaussian KDEs of the BIC-optimal model orders for the six signals of Motor 6. The red vertical line indicates the peak of the density curve, representing the most probable model order selected for the pooled FP-AR model.}
    \label{fig:basis density M6}
\end{figure}

\begin{figure}[th!]
    \centering 
    
    \begin{subfigure}{.3\textwidth} 
        \centering
        \includegraphics[width=\textwidth]{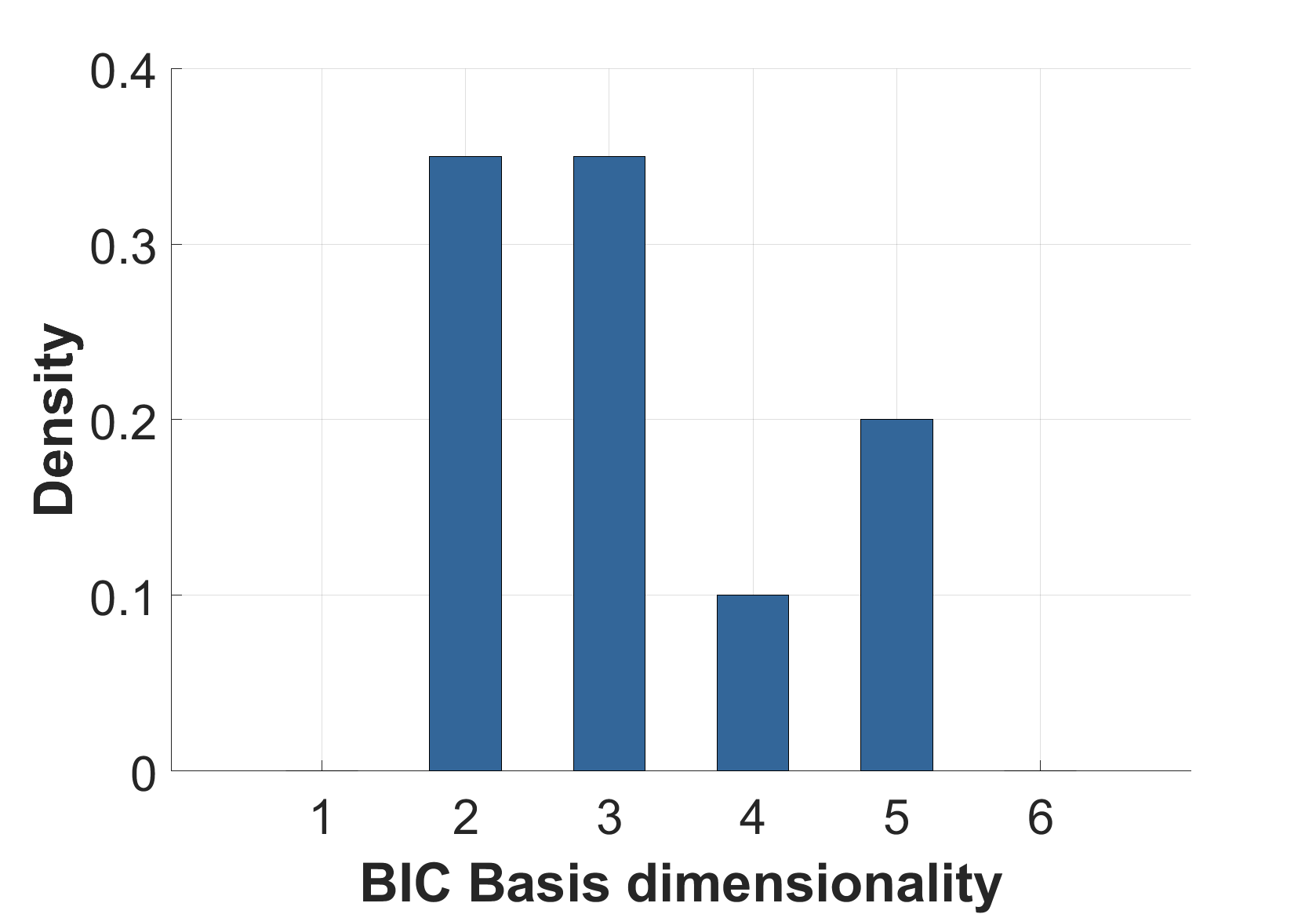}
        \caption{AccX}
    \end{subfigure}
    \hfill 
    \begin{subfigure}{.3\textwidth}
        \centering
        \includegraphics[width=\textwidth]{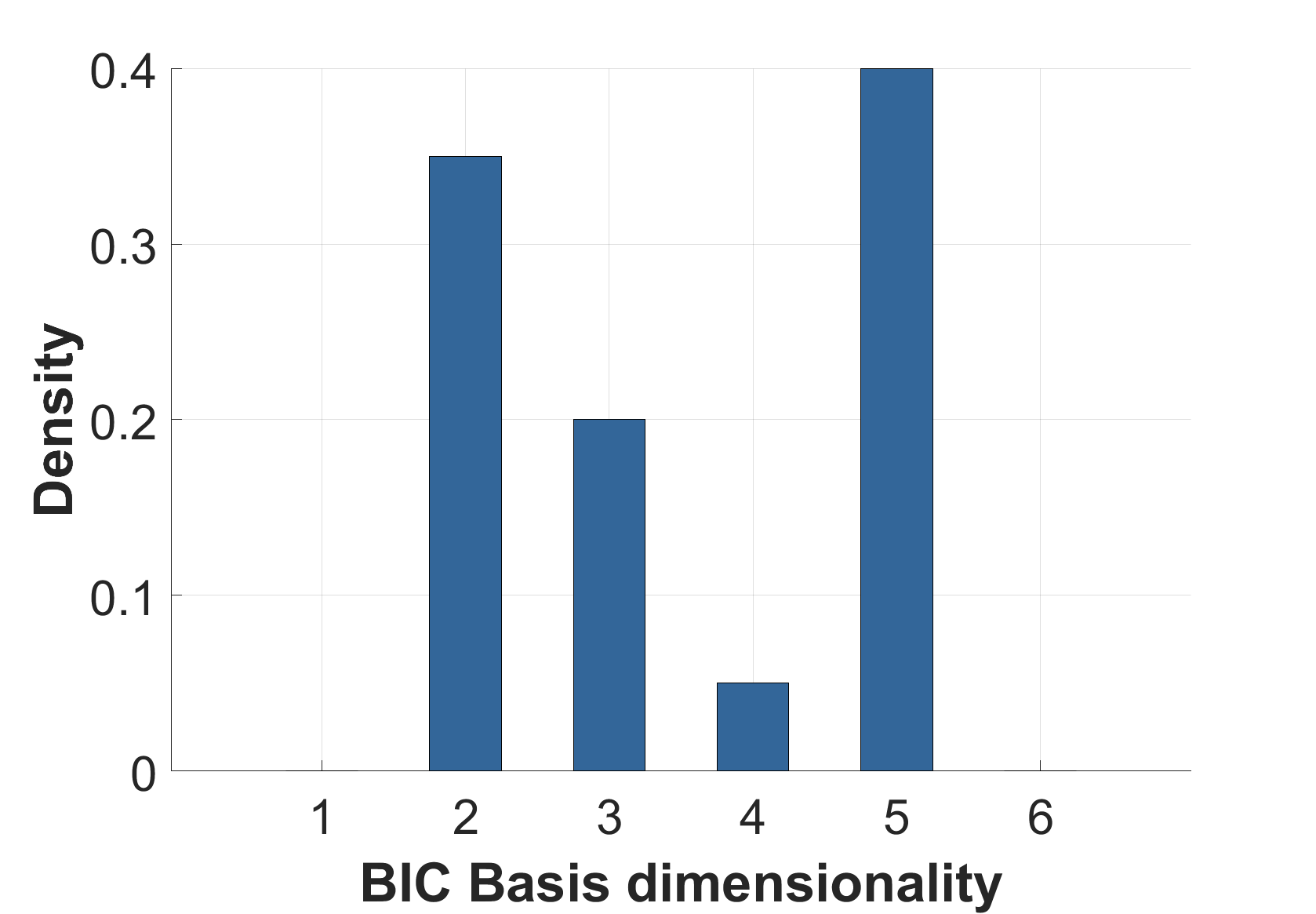}
        \caption{AccY}
    \end{subfigure}
    \hfill
    \begin{subfigure}{.3\textwidth}
        \centering
        \includegraphics[width=\textwidth]{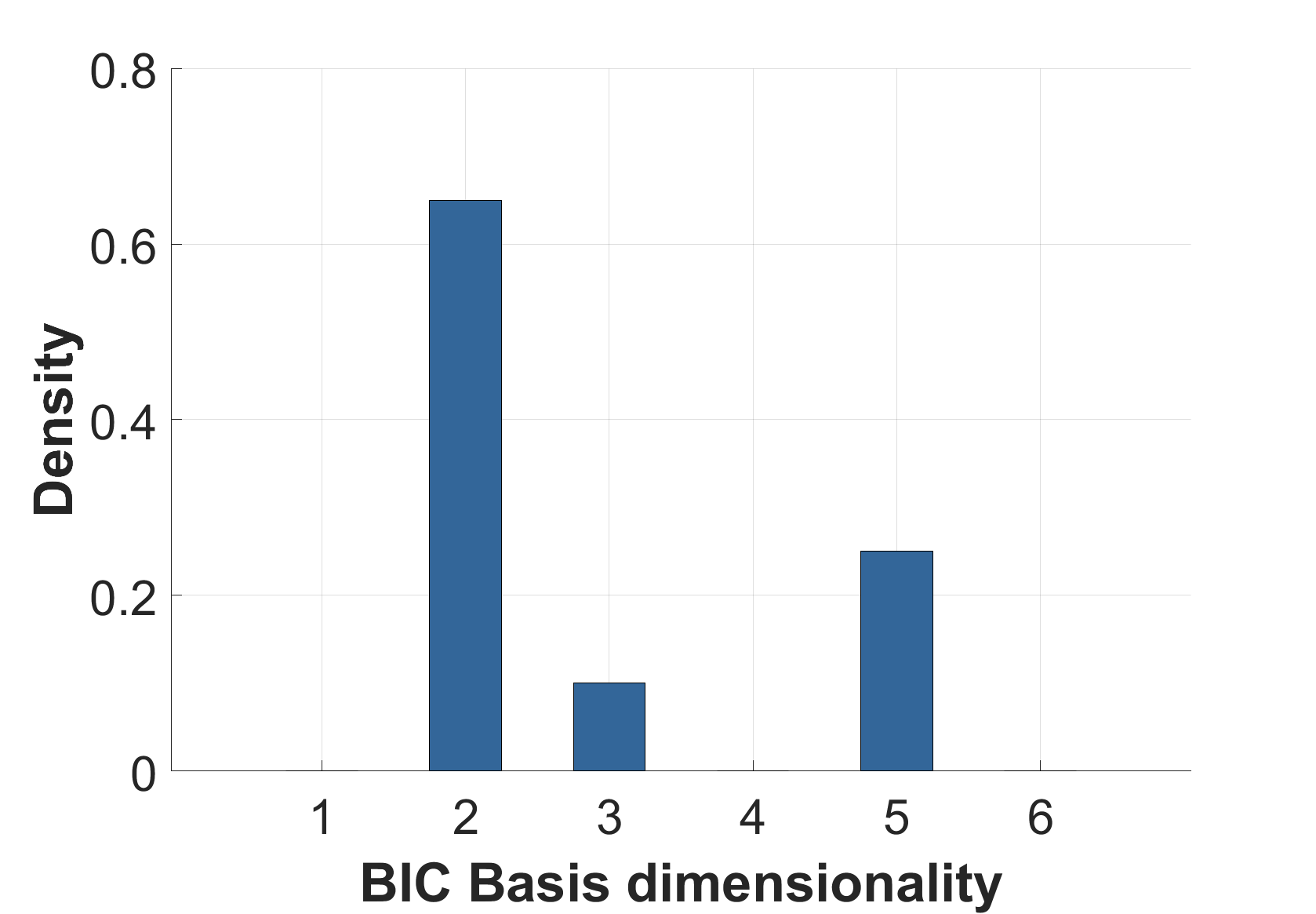}
        \caption{AccZ}
    \end{subfigure}

    \begin{subfigure}{.3\textwidth}
        \centering
        \includegraphics[width=\textwidth]{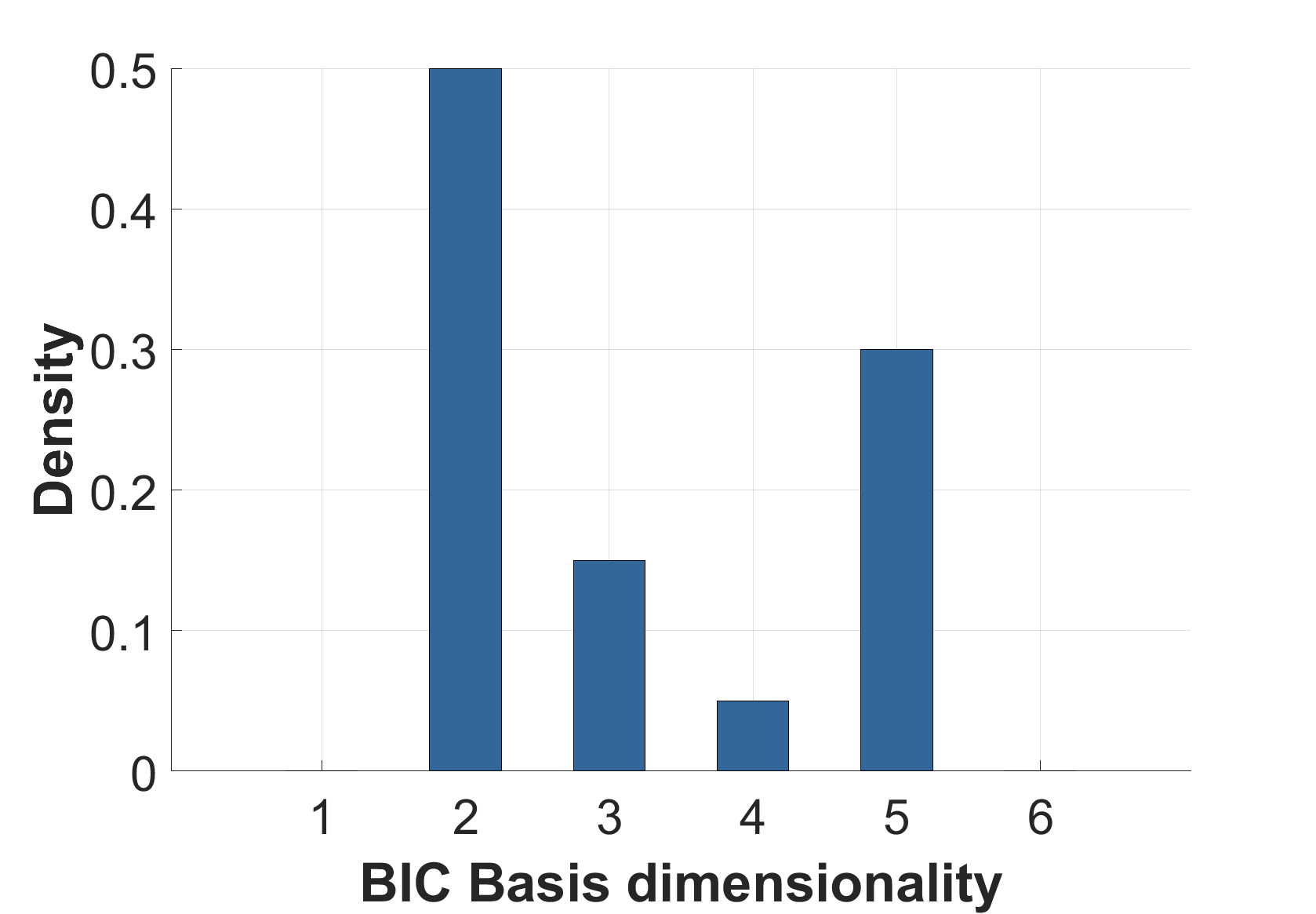}
        \caption{GyrX}
    \end{subfigure}
    \hfill
    \begin{subfigure}{.3\textwidth}
        \centering
        \includegraphics[width=\textwidth]{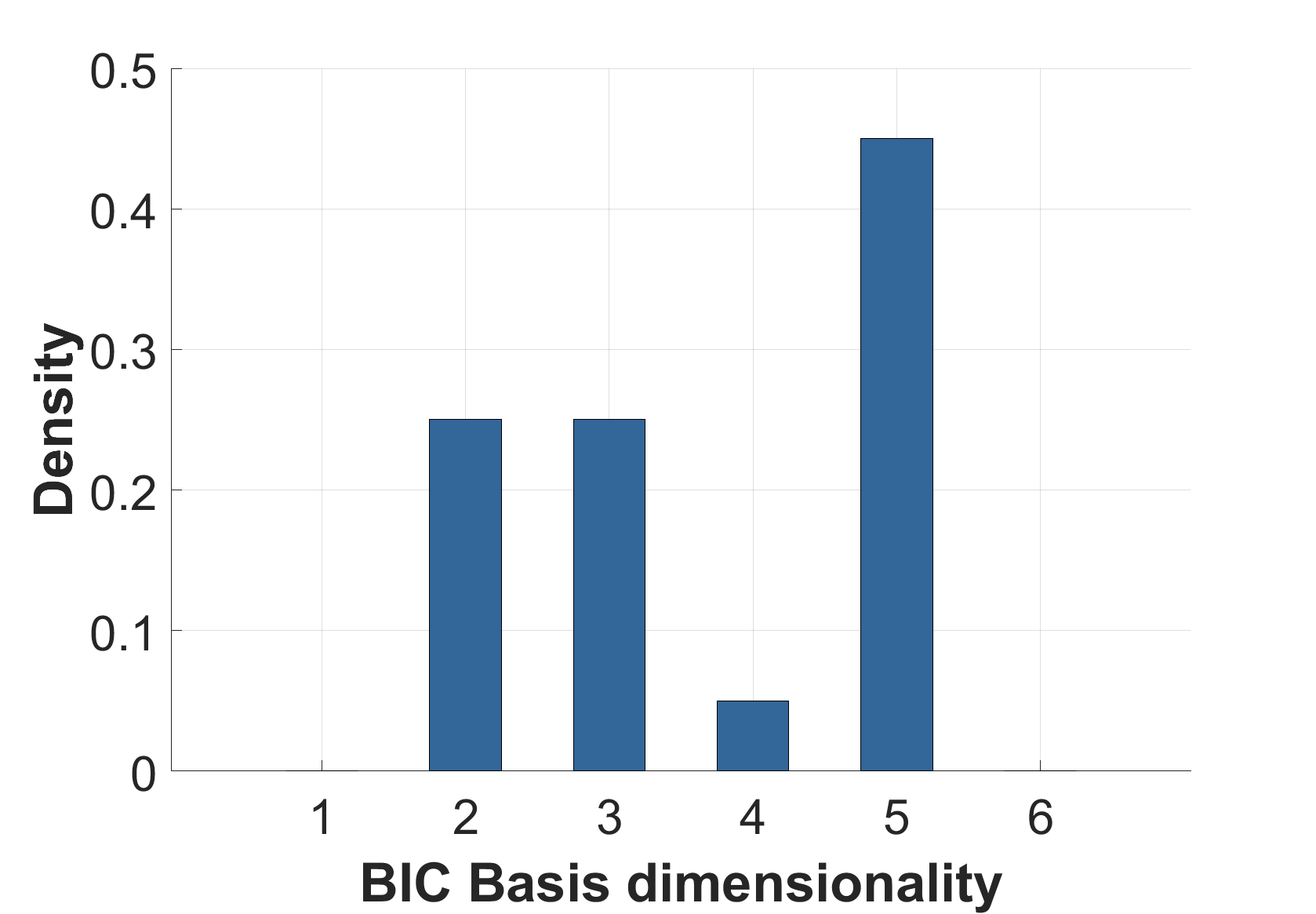}
        \caption{GyrY}
    \end{subfigure}
    \hfill
    \begin{subfigure}{.3\textwidth}
        \centering
        \includegraphics[width=\textwidth]{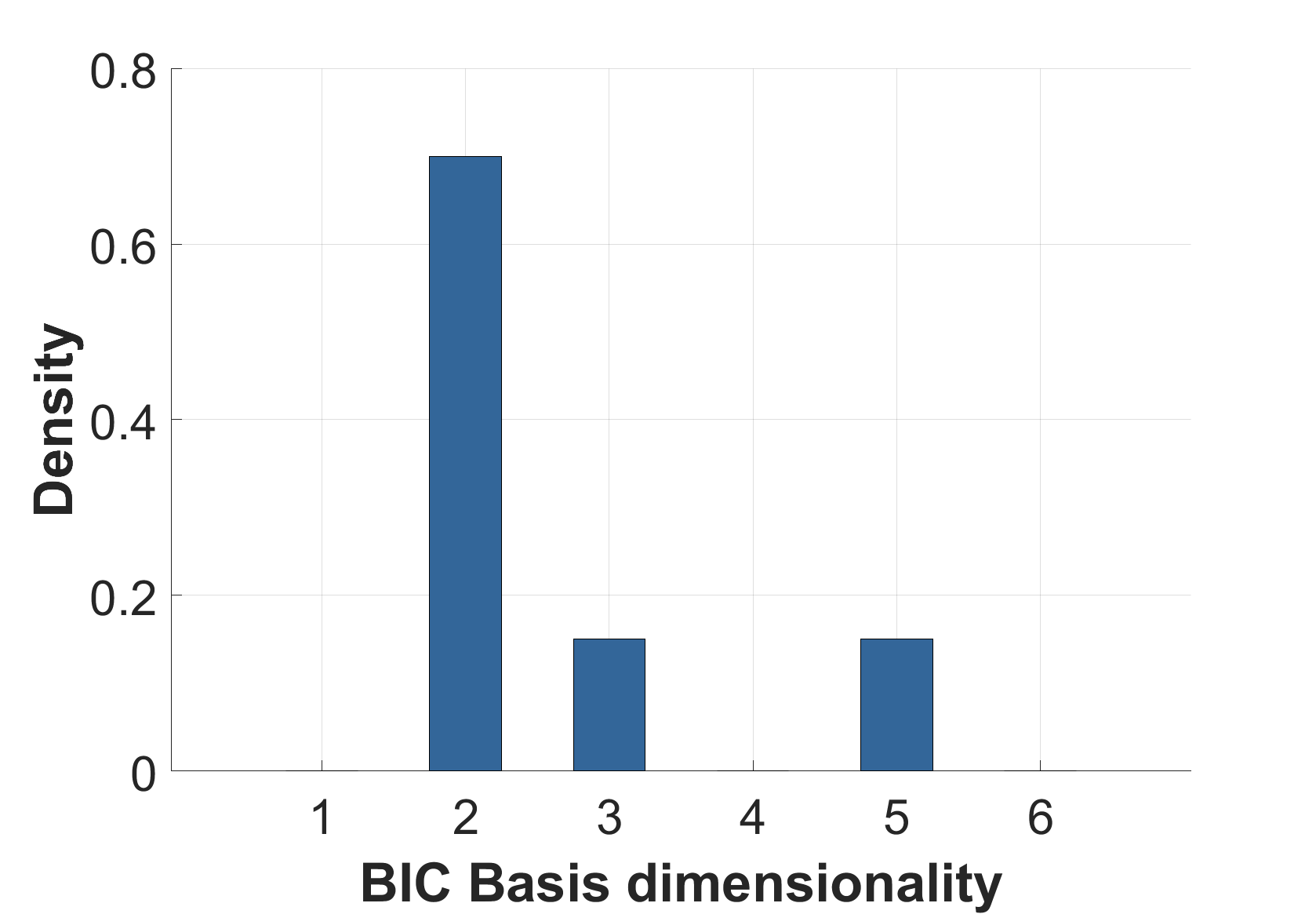}
        \caption{GyrZ}
    \end{subfigure}
    \caption{Histograms of the BIC-optimal basis dimensionalities for the six signals of Motor 3. The histograms show the frequency of the basis dimensionalities obtained from the training segments, with the most frequent value selected for the pooled FP-AR model.}
    \label{fig:order density M3}
\end{figure}

\begin{figure}[th!]
    \centering 
    
    \begin{subfigure}{.3\textwidth} 
        \centering
        \includegraphics[width=\textwidth]{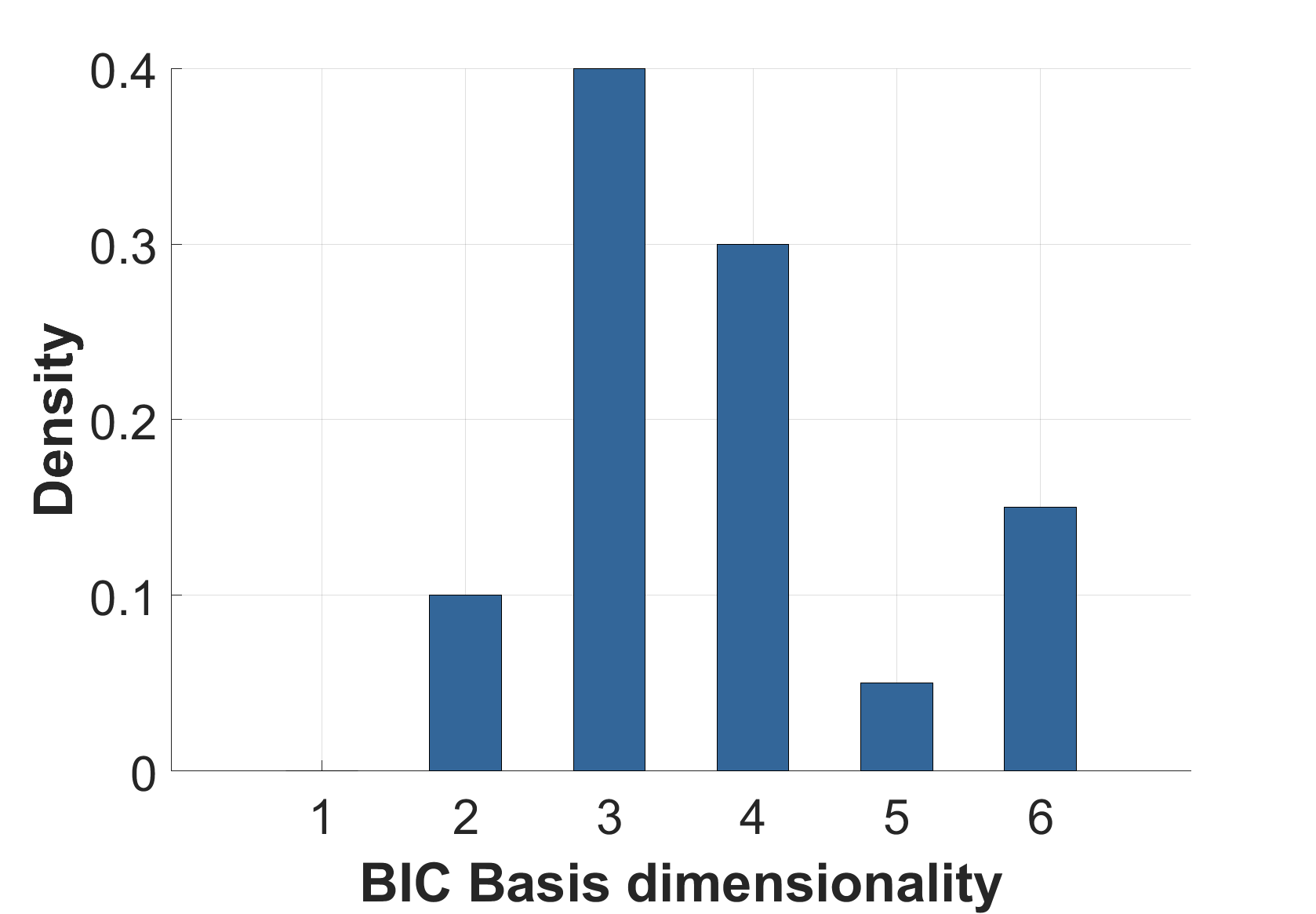}
        \caption{AccX}
    \end{subfigure}
    \hfill 
    \begin{subfigure}{.3\textwidth}
        \centering
        \includegraphics[width=\textwidth]{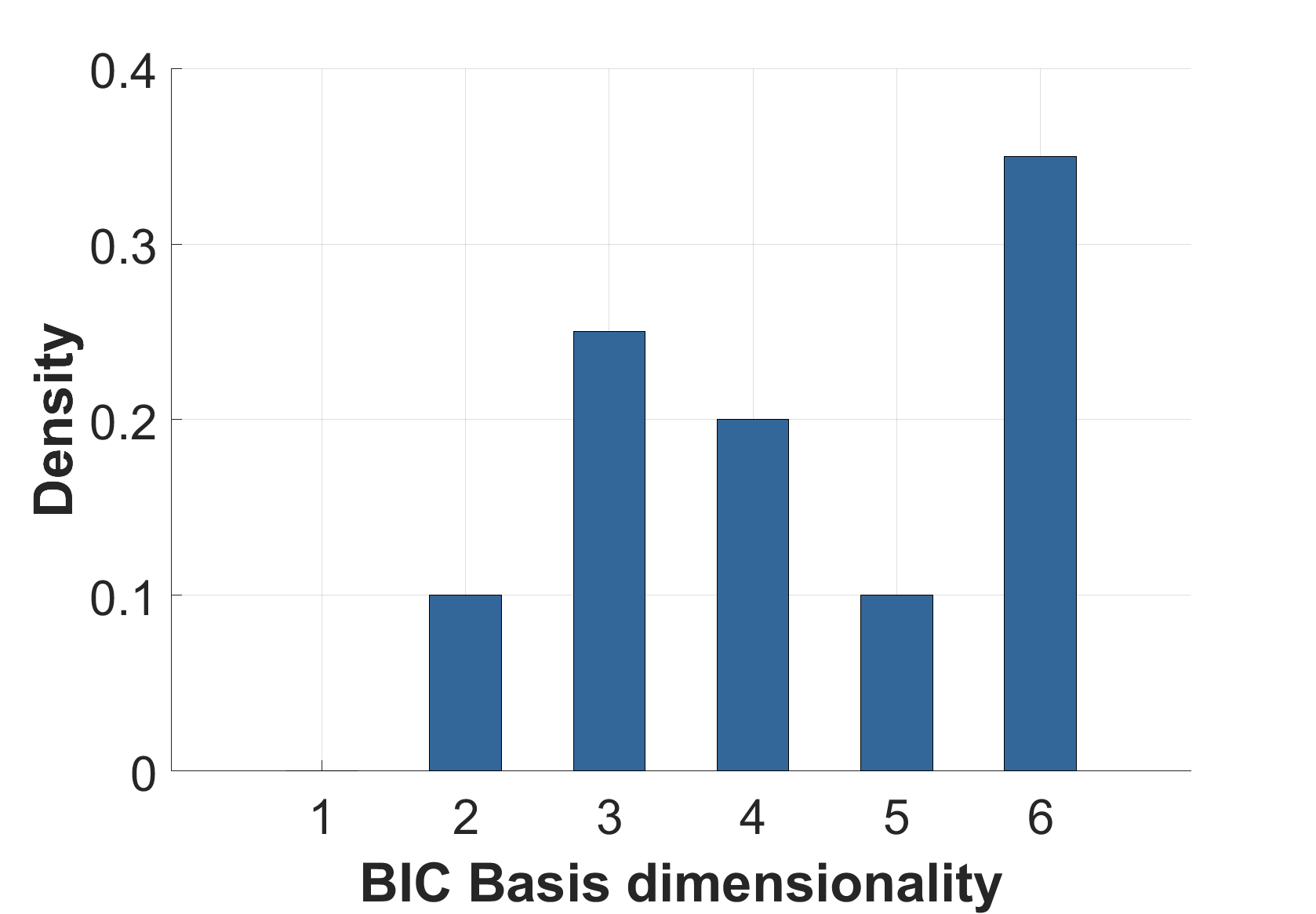}
        \caption{AccY}
    \end{subfigure}
    \hfill
    \begin{subfigure}{.3\textwidth}
        \centering
        \includegraphics[width=\textwidth]{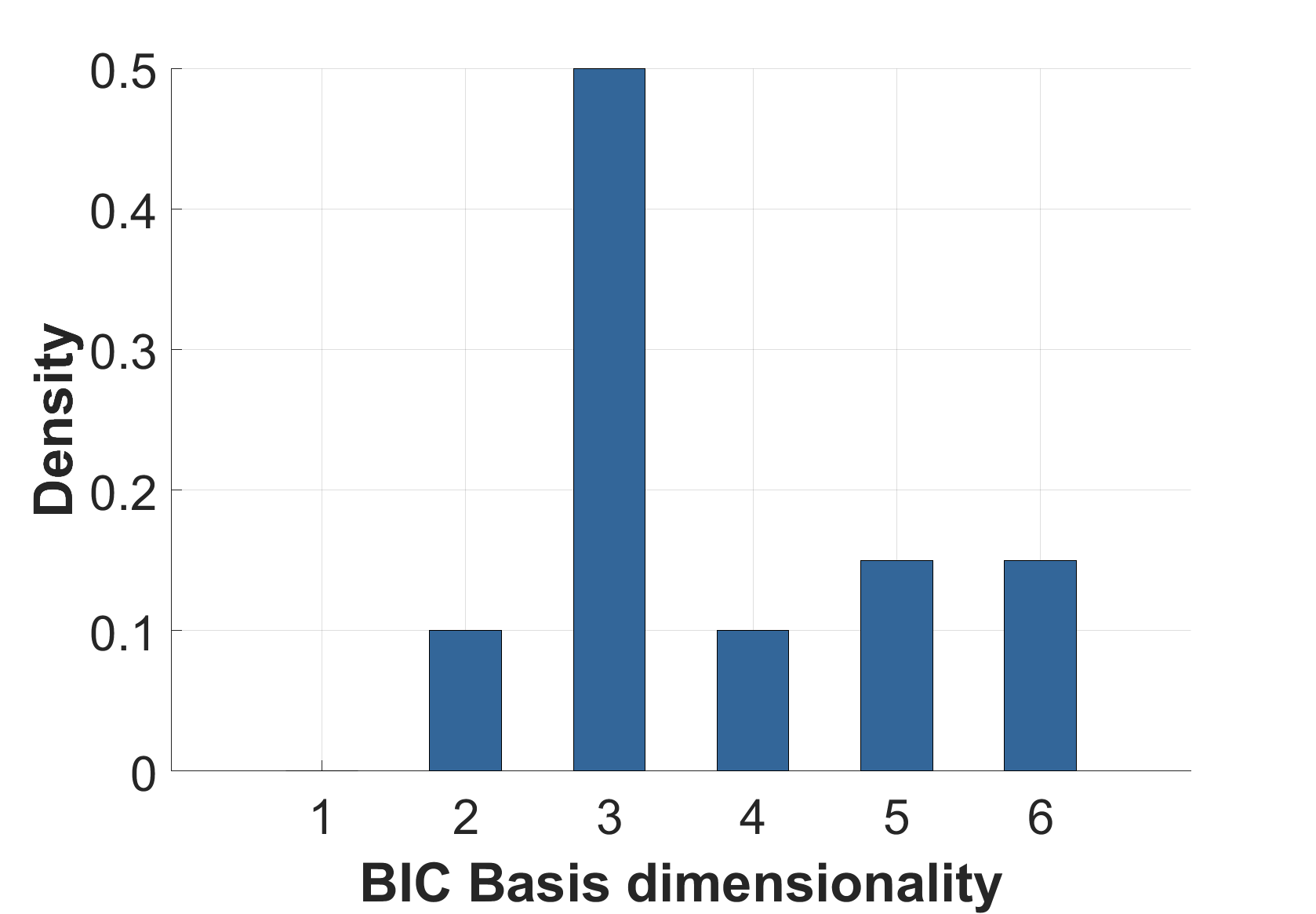}
        \caption{AccZ}
    \end{subfigure}

    \begin{subfigure}{.3\textwidth}
        \centering
        \includegraphics[width=\textwidth]{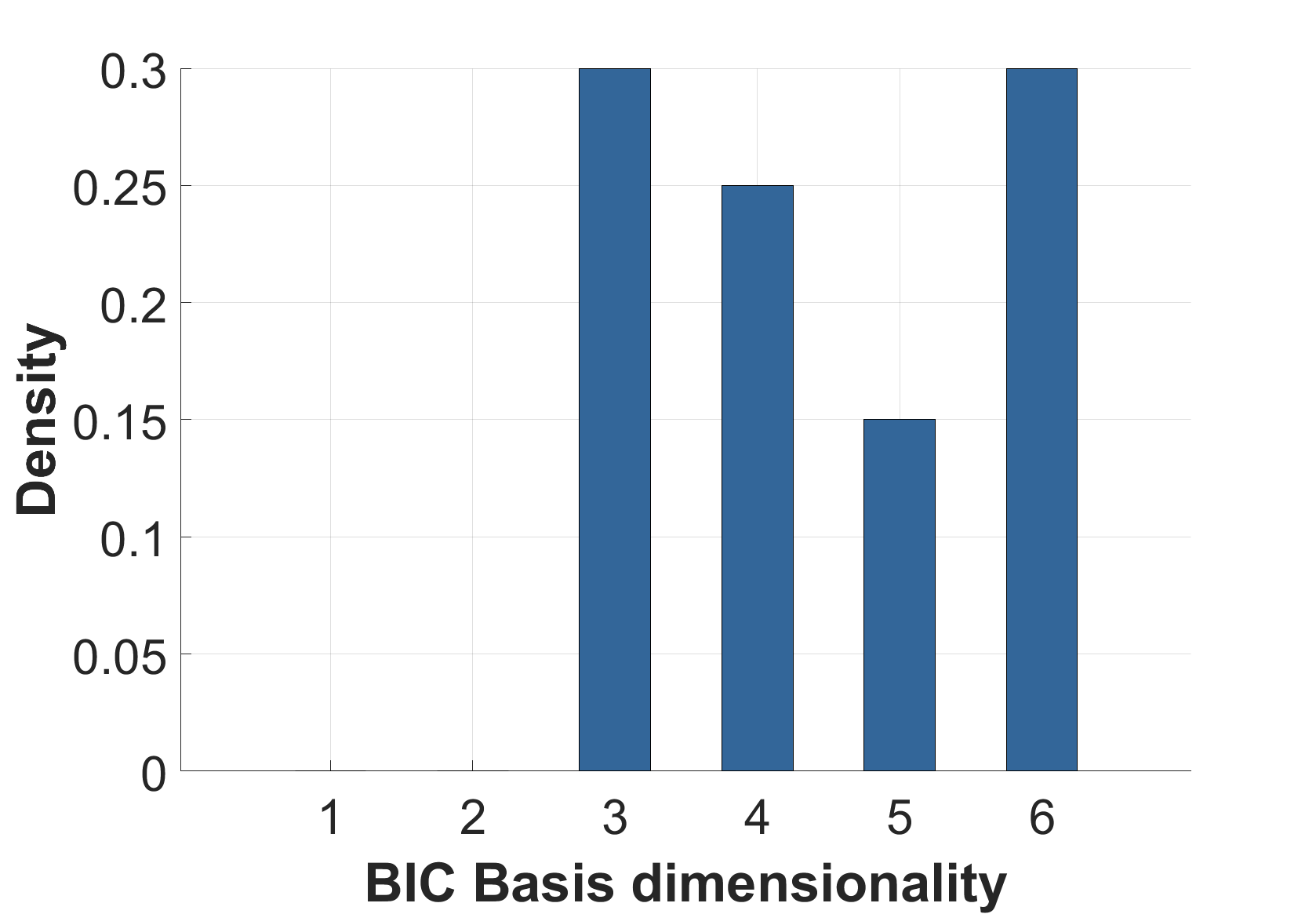}
        \caption{GyrX}
    \end{subfigure}
    \hfill
    \begin{subfigure}{.3\textwidth}
        \centering
        \includegraphics[width=\textwidth]{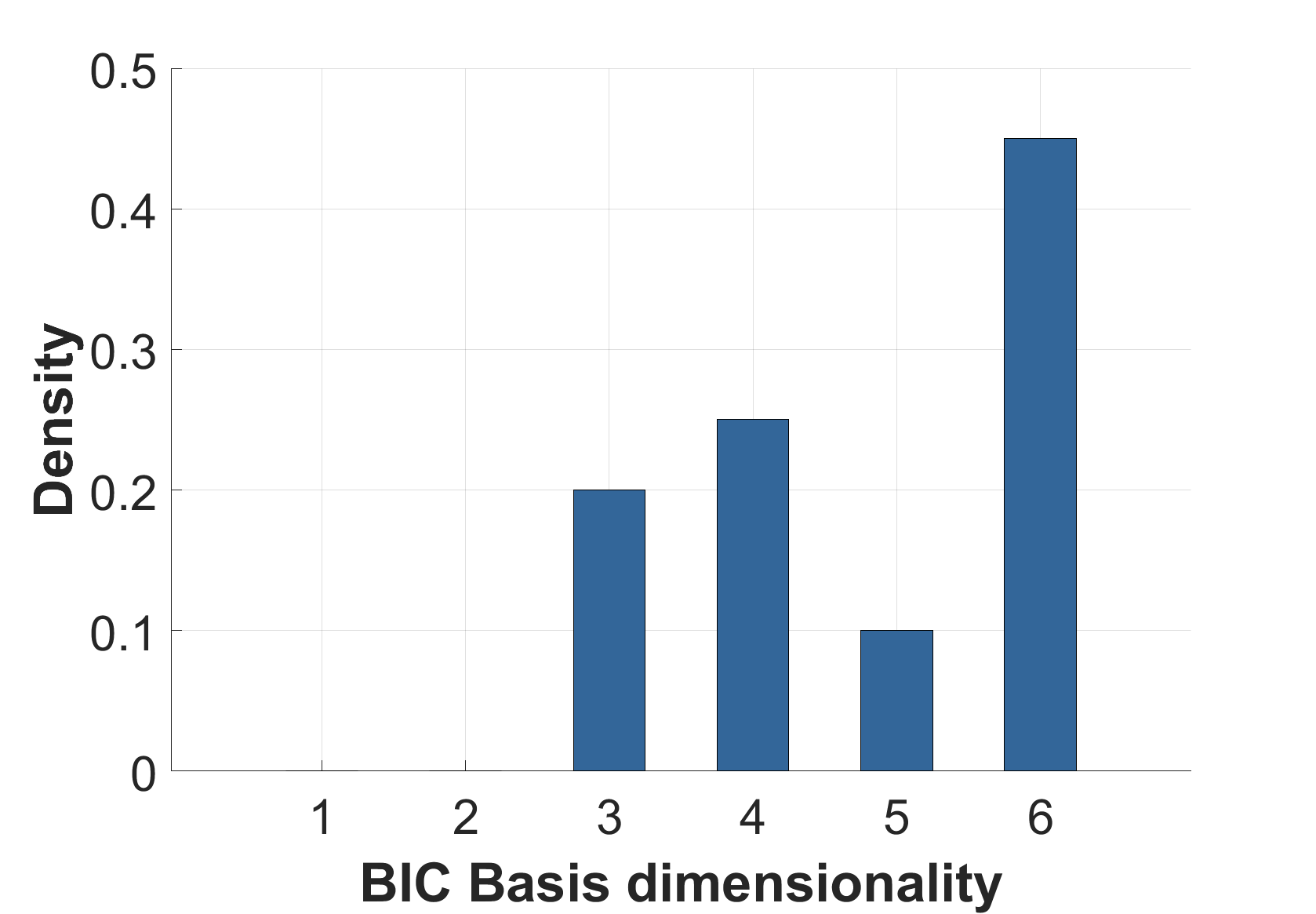}
        \caption{GyrY}
    \end{subfigure}
    \hfill
    \begin{subfigure}{.3\textwidth}
        \centering
        \includegraphics[width=\textwidth]{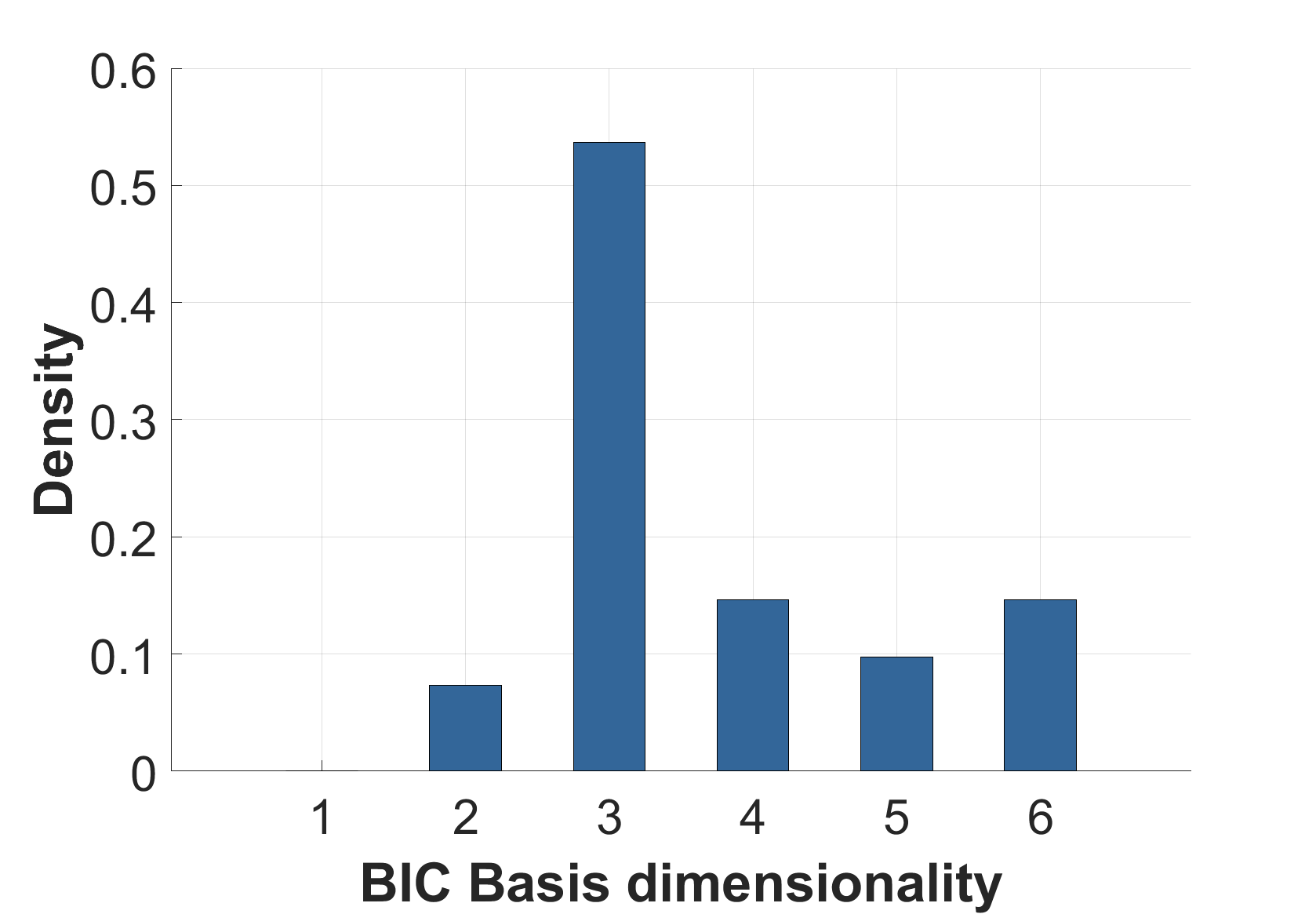}
        \caption{GyrZ}
    \end{subfigure}
    \caption{Histograms of the BIC-optimal basis dimensionalities for the six signals of Motor 6. The histograms show the frequency of the basis dimensionalities obtained from the training segments, with the most frequent value selected for the pooled FP-AR model.}
    \label{fig:order density M6}
\end{figure}

\begin{figure*}[t!]
	\centering
	\begin{subfigure}{.32\textwidth}
		\centering
		\captionsetup{width=\linewidth}
		\includegraphics[width=\linewidth]{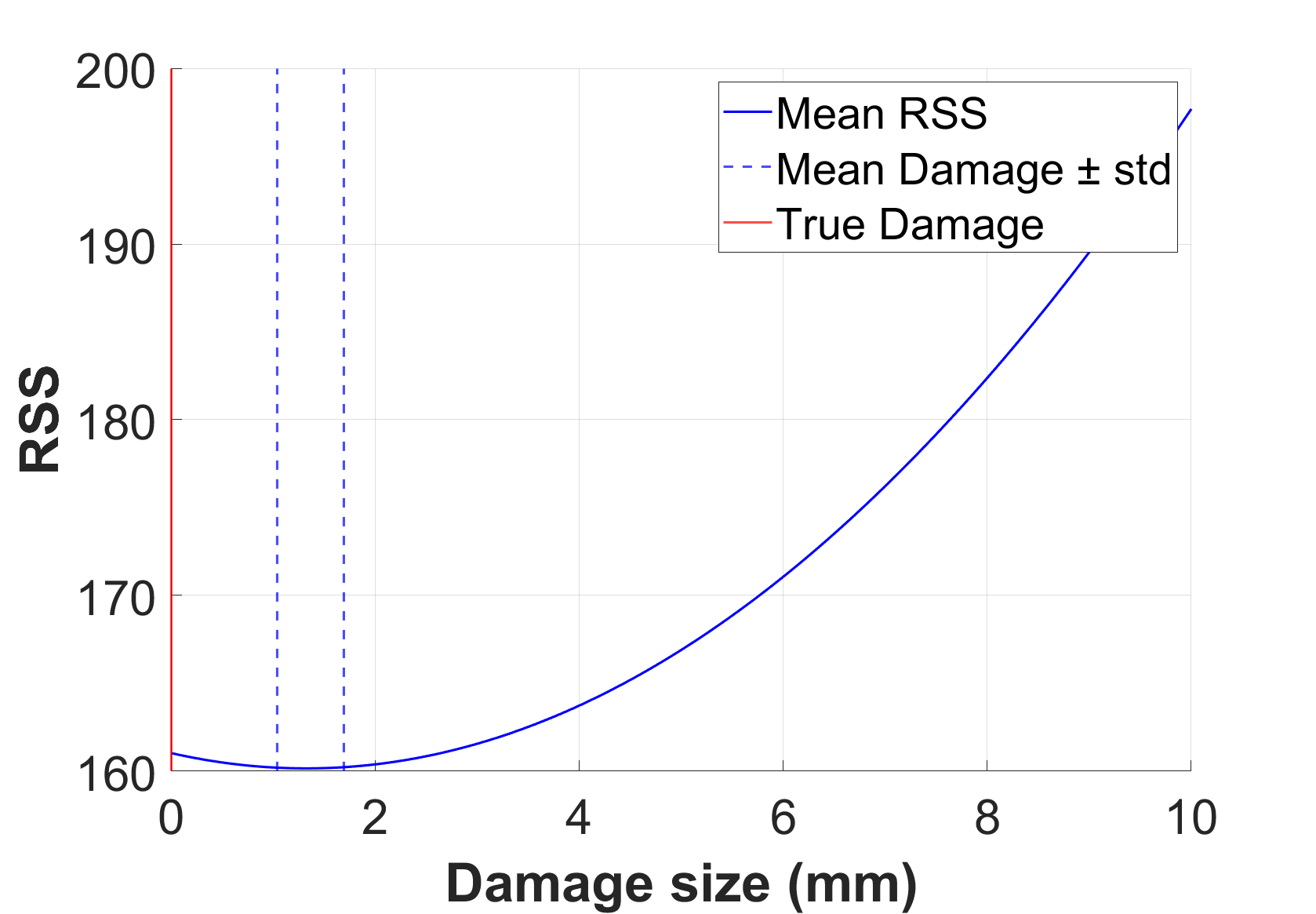}
		\caption{M3, 0 mm damage}
	\end{subfigure}
	\begin{subfigure}{.32\textwidth}
		\centering
		\captionsetup{width=\linewidth}
		\includegraphics[width=\linewidth]{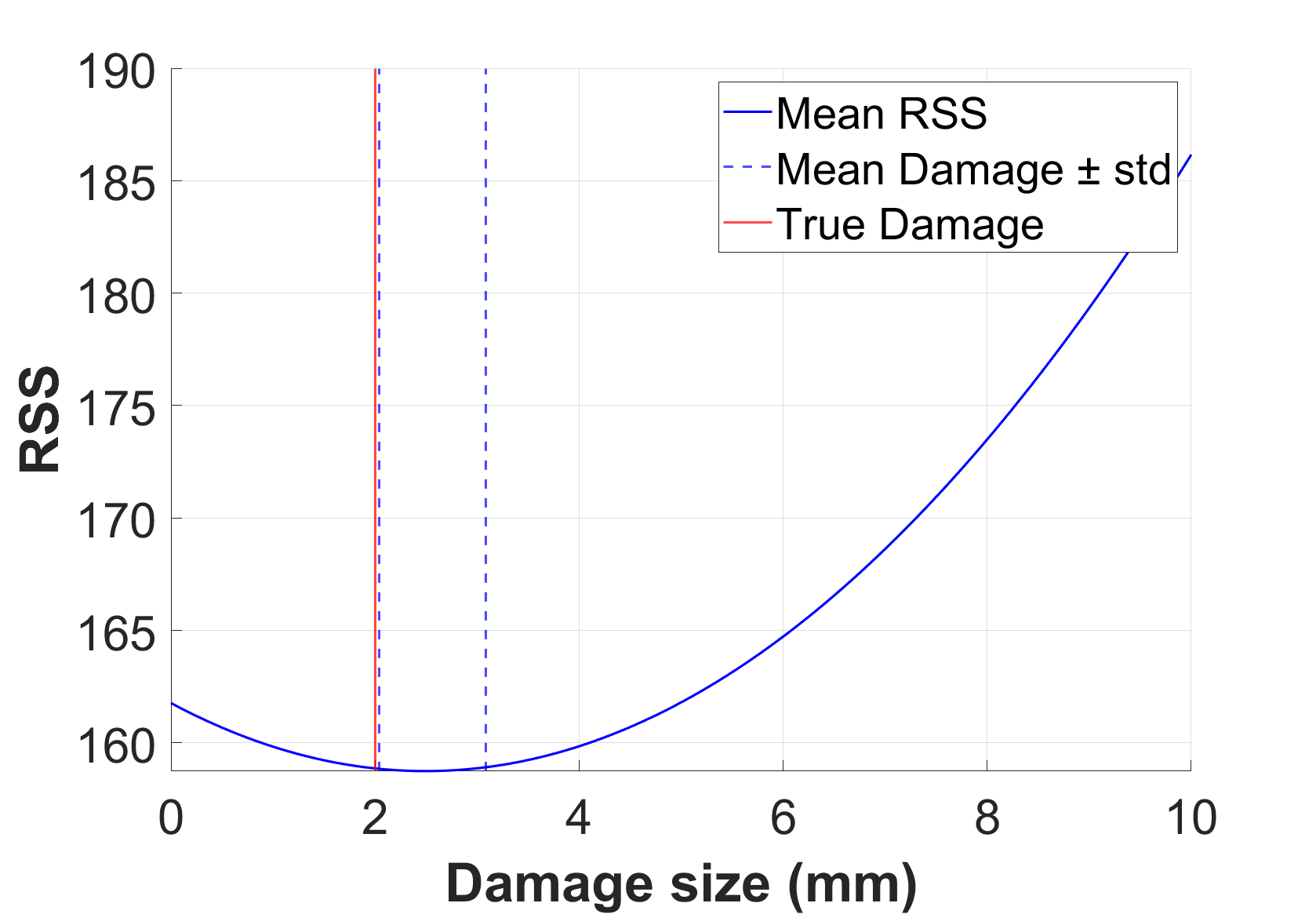}
		\caption{M3, 2 mm damage}
	\end{subfigure}
    \begin{subfigure}{.32\textwidth}
		\centering
		\captionsetup{width=\linewidth}
		\includegraphics[width=\linewidth]{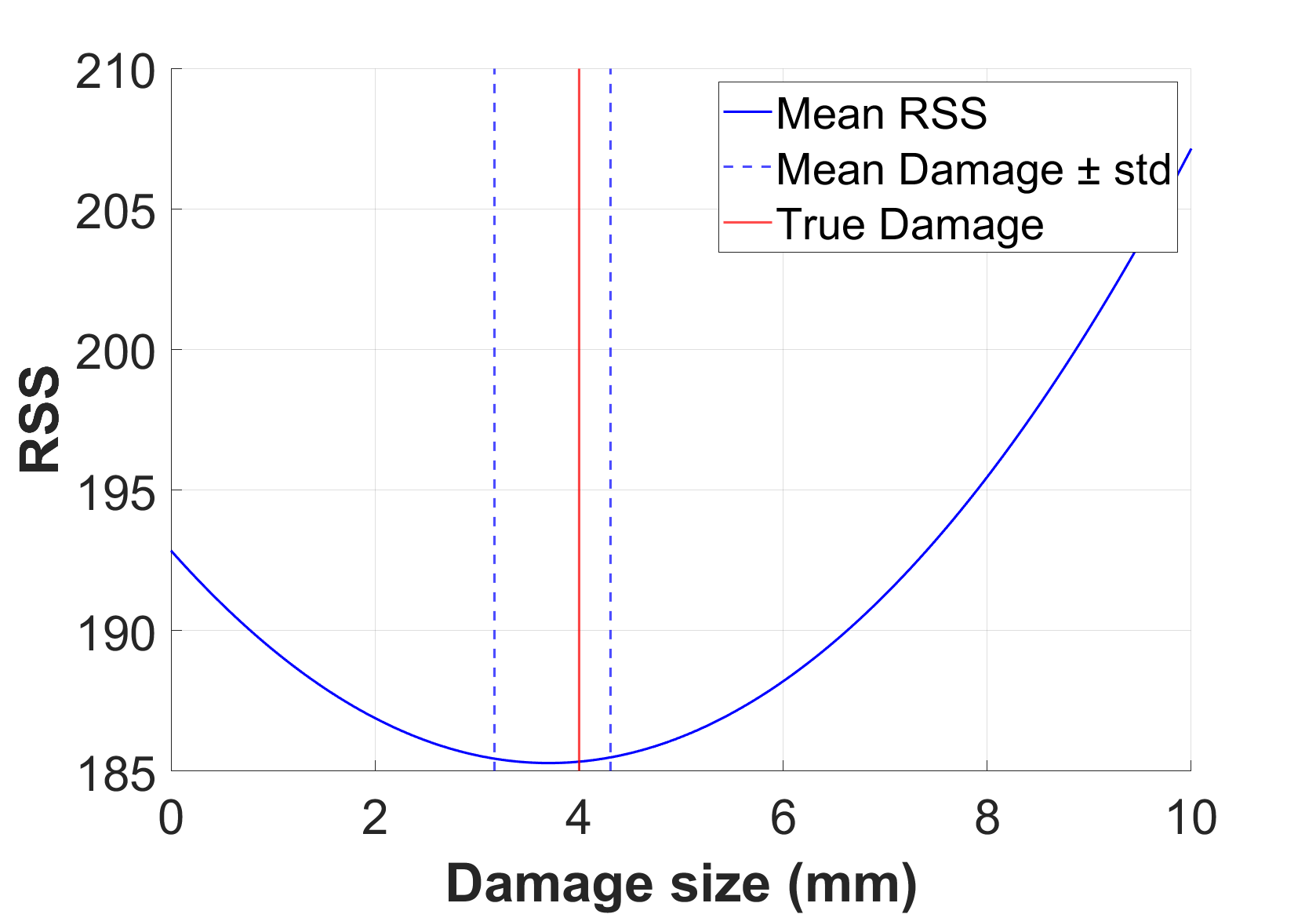}
		\caption{M3, 4 mm damage}
	\end{subfigure}
 
	\begin{subfigure}{.32\textwidth}
		\centering
		\captionsetup{width=\linewidth}
		\includegraphics[width=\linewidth]{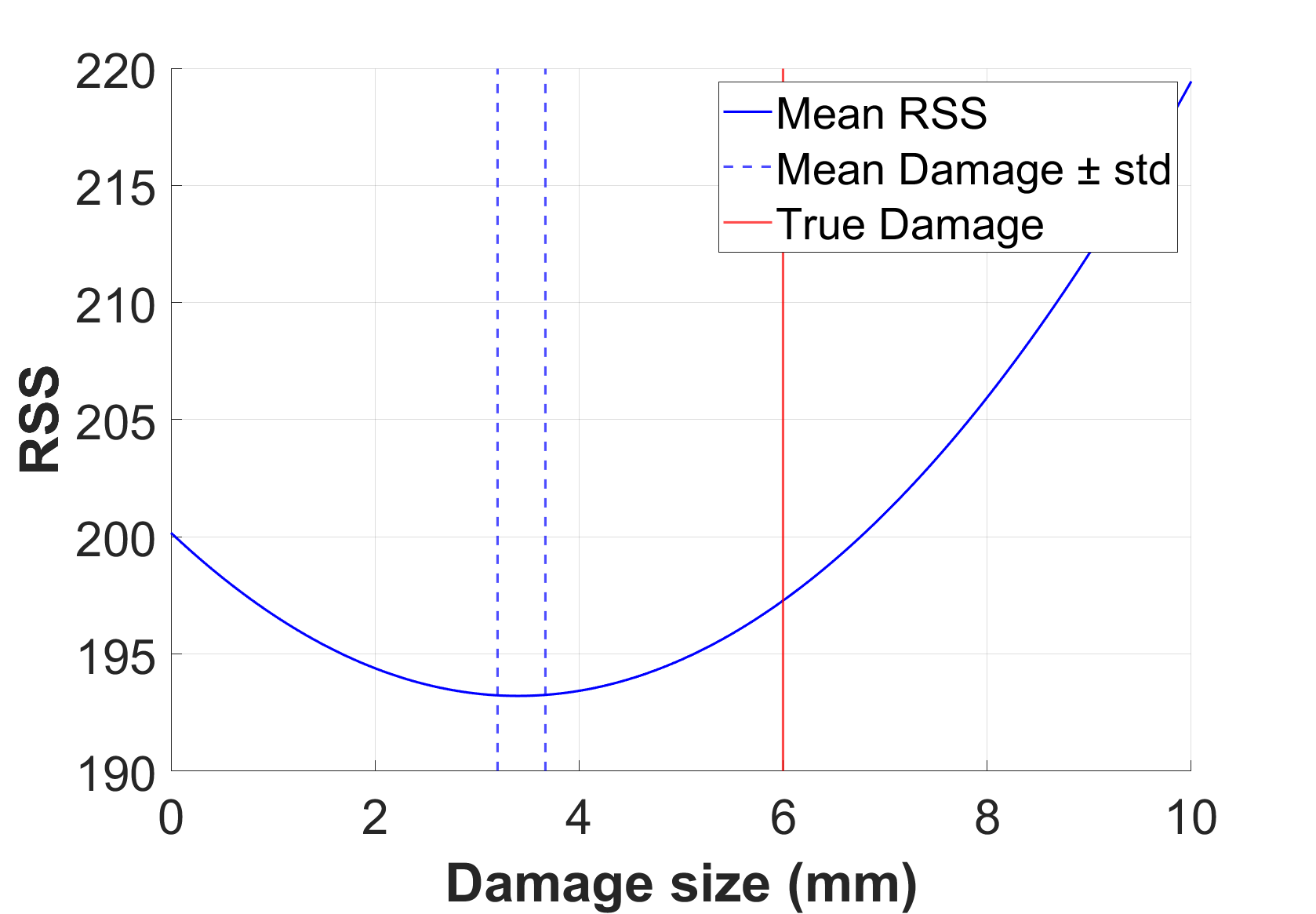}
		\caption{M3, 6 mm damage}
	\end{subfigure}
	\begin{subfigure}{.32\textwidth}
		\centering
		\captionsetup{width=\linewidth}
		\includegraphics[width=\linewidth]{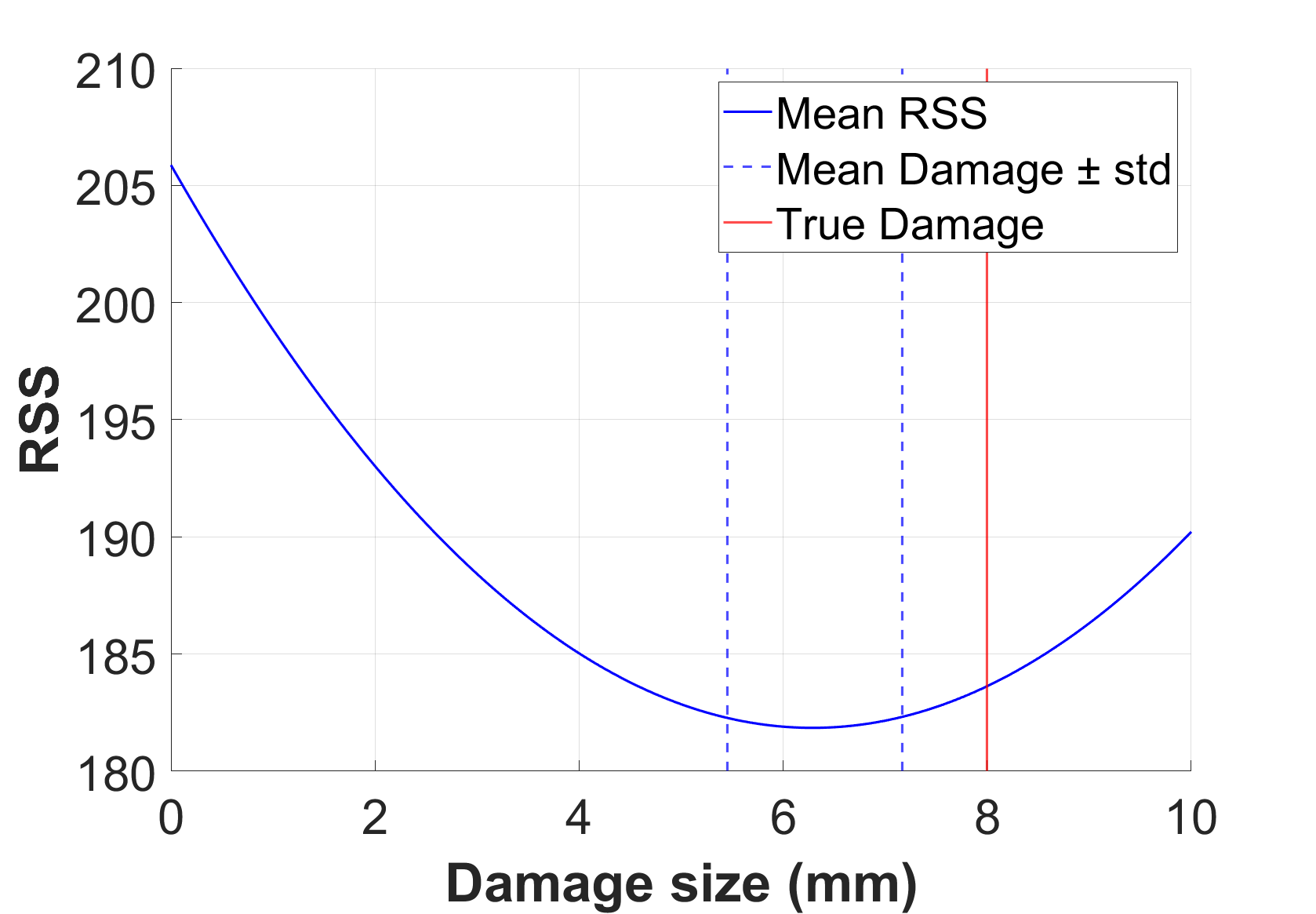}
		\caption{M3, 8 mm damage}
	\end{subfigure}
    \begin{subfigure}{.32\textwidth}
		\centering
		\captionsetup{width=\linewidth}
		\includegraphics[width=\linewidth]{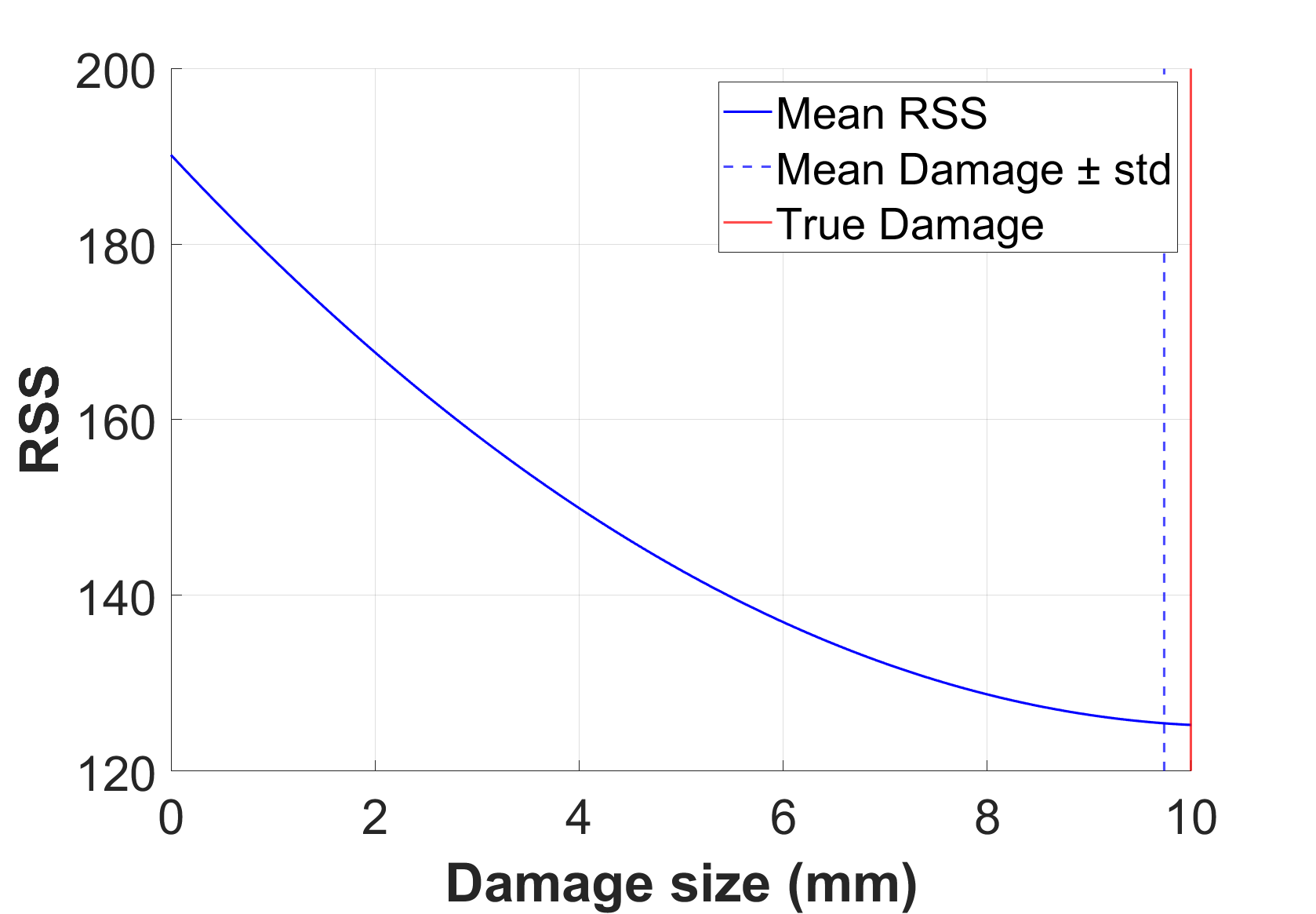}
		\caption{M3, 10 mm damage}
	\end{subfigure}
	\caption{Local (unpooled) FP-AR RSS-based damage size estimation using the AccX signal of Motor 3, across all damage levels (healthy to 10 mm).}
 \label{fig:Unpooled damage estimation M3}
\end{figure*}


\begin{figure*}[t!]
	\centering
	\begin{subfigure}{.32\textwidth}
		\centering
		\captionsetup{width=\linewidth}
		\includegraphics[width=\linewidth]{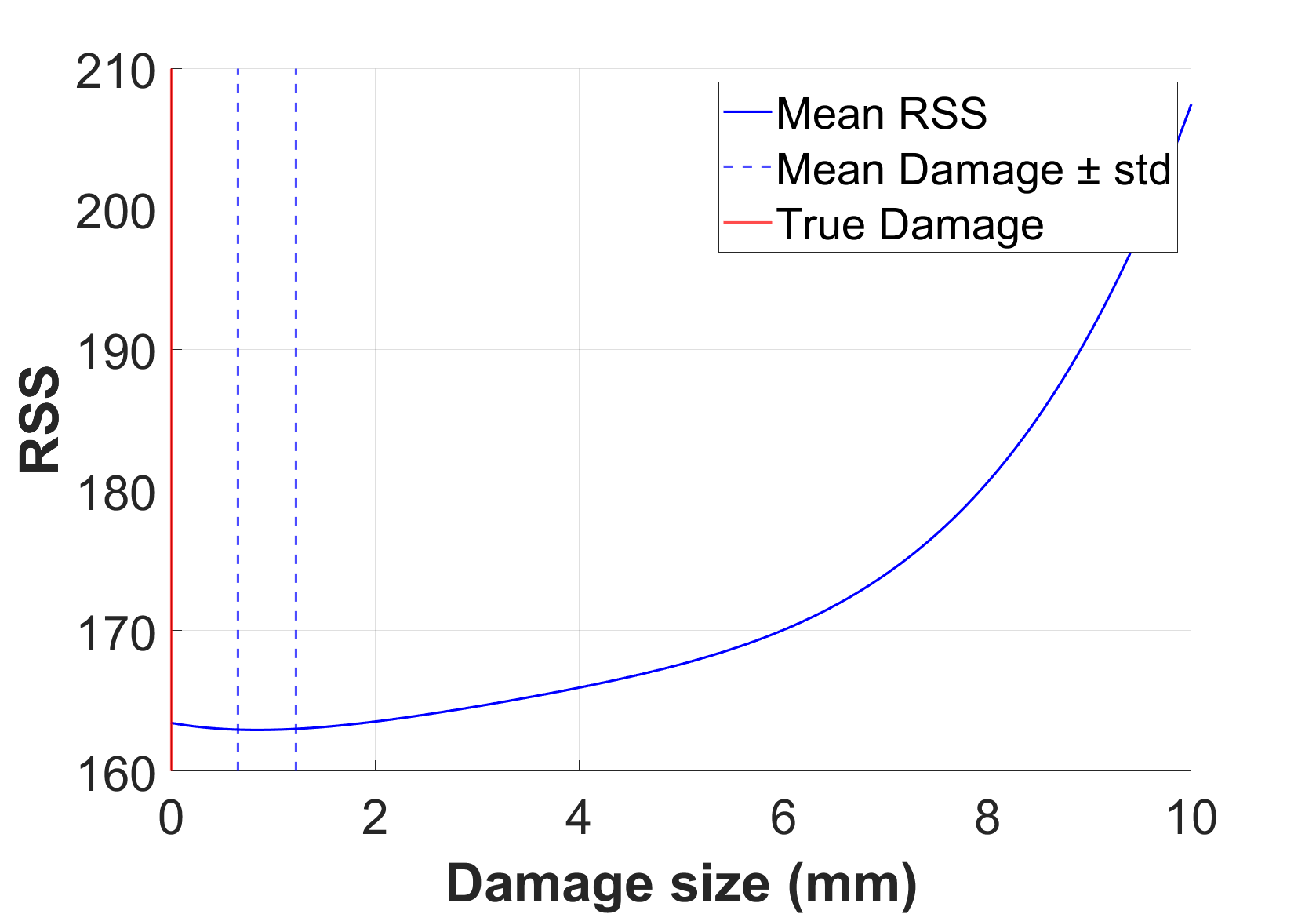}
		\caption{M3, 0 mm damage}
	\end{subfigure}
	\begin{subfigure}{.32\textwidth}
		\centering
		\captionsetup{width=\linewidth}
		\includegraphics[width=\linewidth]{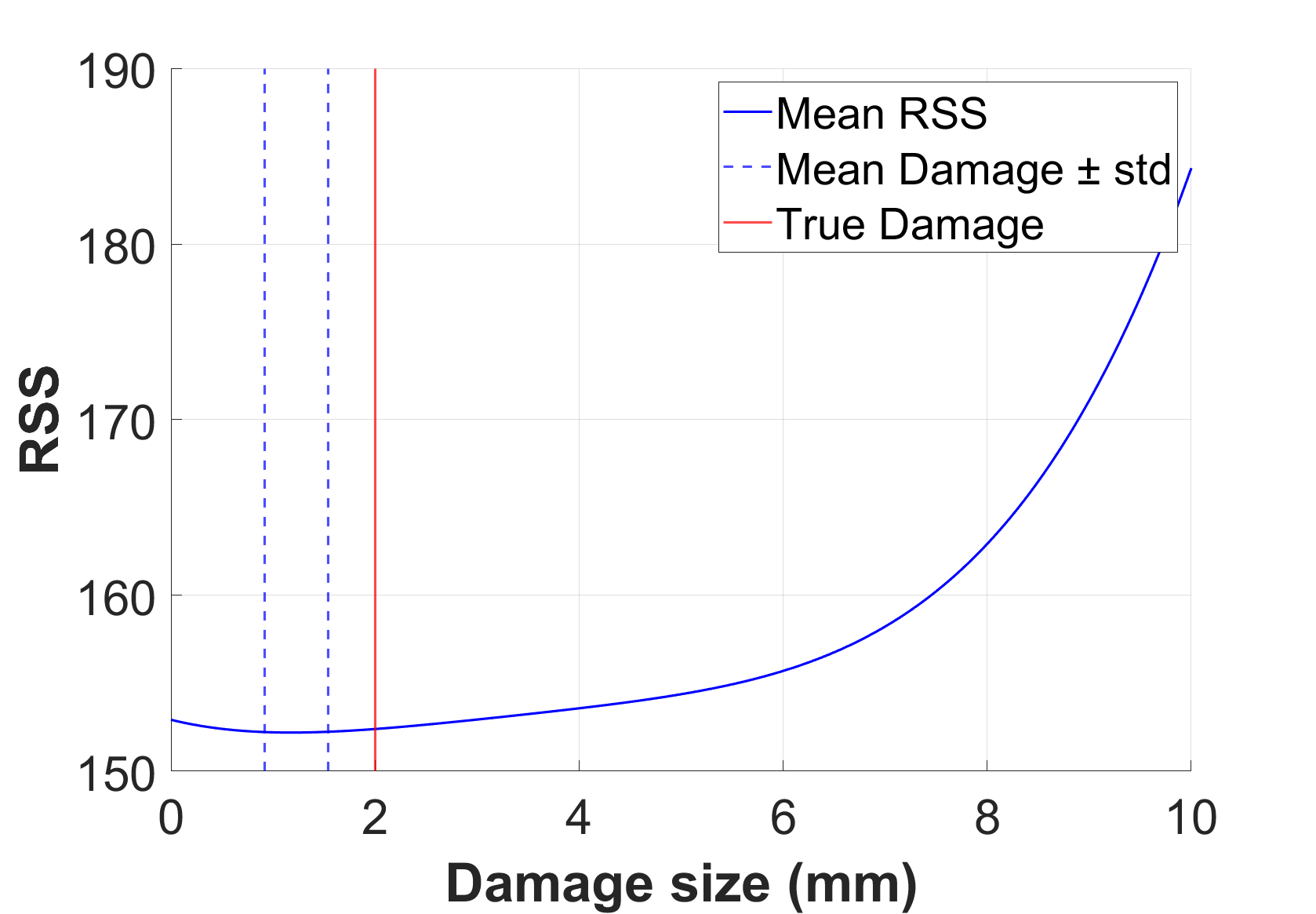}
		\caption{M3, 2 mm damage}
	\end{subfigure}
    \begin{subfigure}{.32\textwidth}
		\centering
		\captionsetup{width=\linewidth}
		\includegraphics[width=\linewidth]{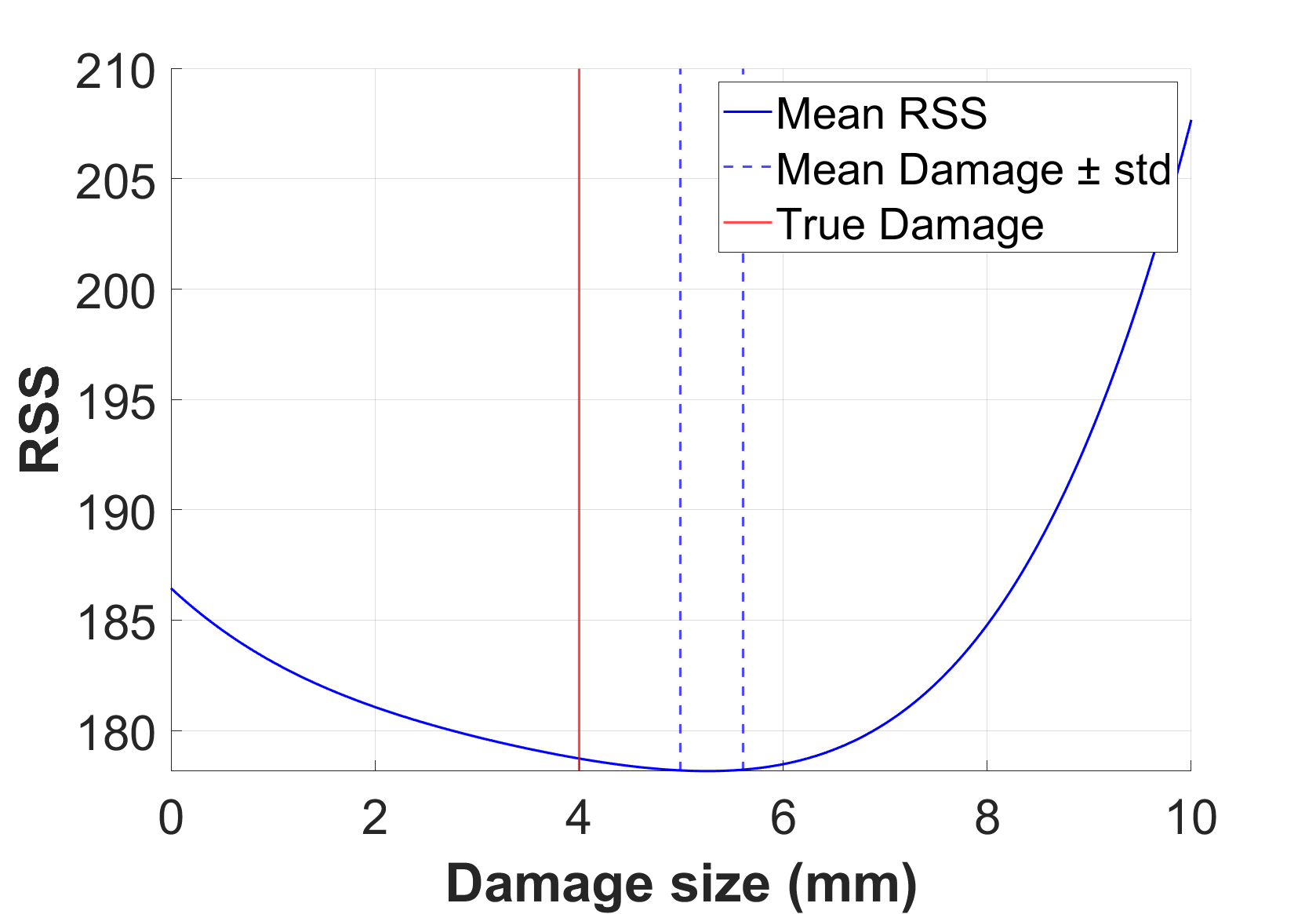}
		\caption{M3, 4 mm damage}
	\end{subfigure}
 
	\begin{subfigure}{.32\textwidth}
		\centering
		\captionsetup{width=\linewidth}
		\includegraphics[width=\linewidth]{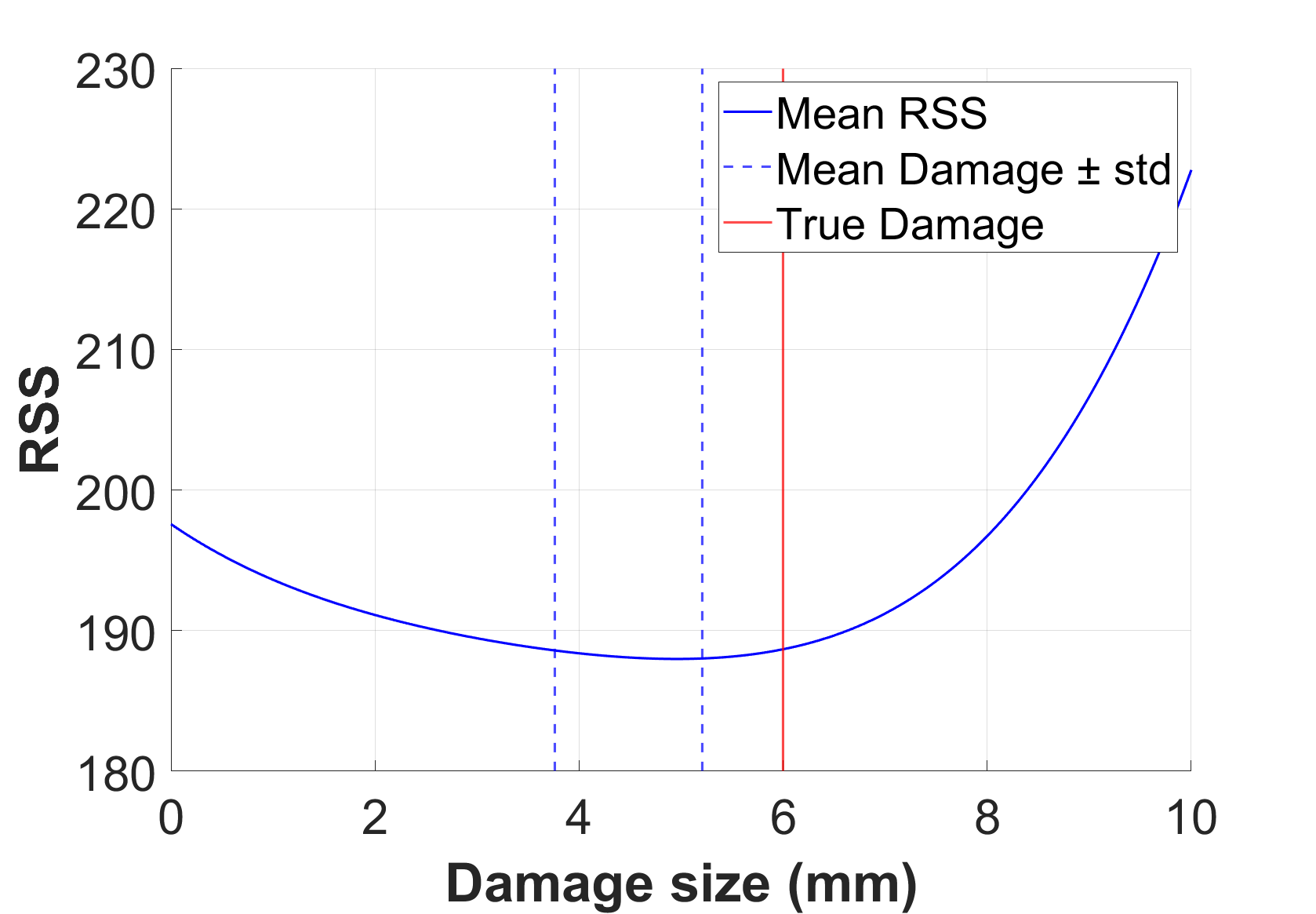}
		\caption{M3, 6 mm damage}
	\end{subfigure}
	\begin{subfigure}{.32\textwidth}
		\centering
		\captionsetup{width=\linewidth}
		\includegraphics[width=\linewidth]{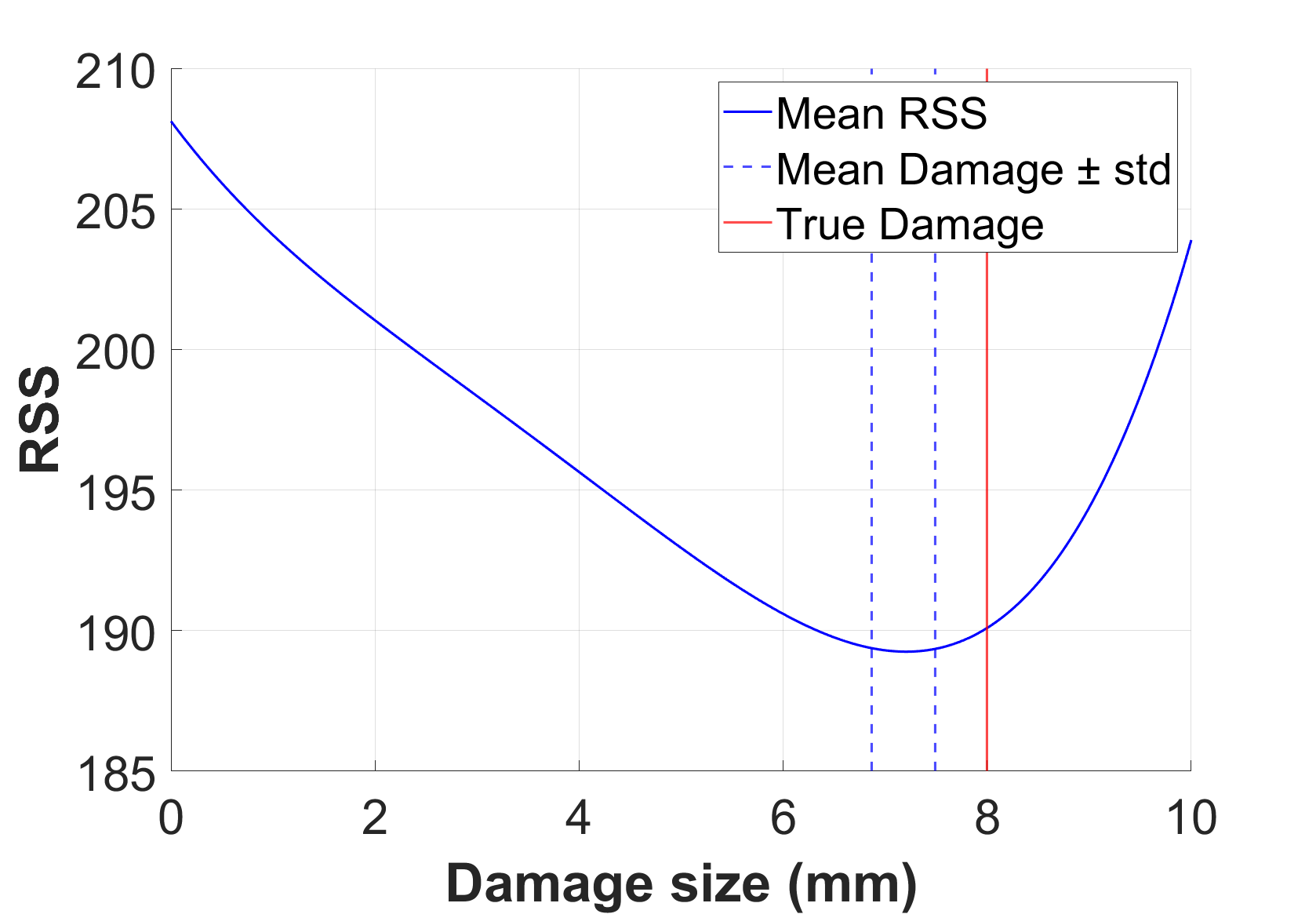}
		\caption{M3, 8 mm damage}
	\end{subfigure}
    \begin{subfigure}{.32\textwidth}
		\centering
		\captionsetup{width=\linewidth}
		\includegraphics[width=\linewidth]{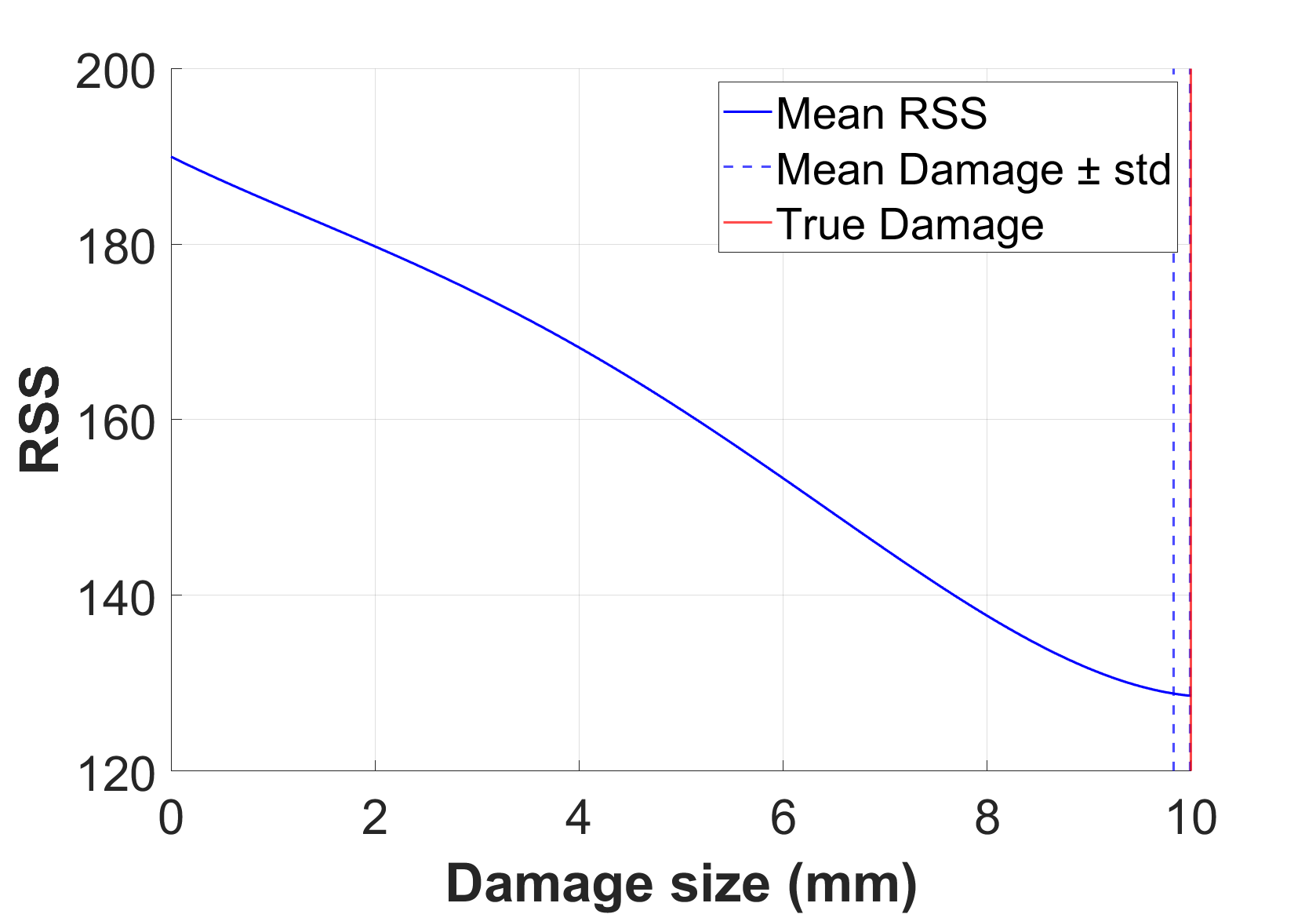}
		\caption{M3, 10 mm damage}
	\end{subfigure}
	\caption{Pooled FP-AR RSS-based damage size estimation using the AccX signal of Motor 3, across all damage levels (healthy to 10 mm).}
 \label{fig:Global damage estimation RSS M3}
\end{figure*}

\end{document}